\documentclass[11pt]{article}
\usepackage{verbatim,amsmath,amssymb}
\usepackage{epsfig,float,color}
\usepackage{epstopdf}
\usepackage{geometry}
\usepackage{setspace}
\usepackage{wrapfig}
\usepackage{hyperref}
\usepackage[utf8]{inputenc}
\usepackage{graphicx}
\usepackage{flushend}
\usepackage[linesnumbered,ruled]{algorithm2e}

\usepackage[font={footnotesize}]{caption}
\usepackage[font={footnotesize}]{subcaption}
\usepackage{footnote}
\usepackage{multicol}
\usepackage{multirow}
\usepackage{enumitem}   
\usepackage{soul}
\usepackage{framed}
\usepackage[titletoc,title]{appendix}
\usepackage{bm}
\usepackage{siunitx}
\usepackage{cite}
\usepackage{mathrsfs}
\usepackage[makeroom]{cancel} 
\usepackage[table]{xcolor}
\usepackage{hyperref}
\definecolor{darkblue}{rgb}{0,0,1}
\hypersetup{pdftex=true, colorlinks=true, breaklinks=true, linkcolor=magenta, menucolor=magenta, citecolor=magenta, urlcolor=magenta}

\def\BibTeX{{\rm B\kern-.05em{\sc i\kern-.025em b}\kern-.08em
		T\kern-.1667em\lower.7ex\hbox{E}\kern-.125emX}}
\usepackage{balance}
\usepackage{hyperref}
\definecolor{beaublue}{rgb}{0.94, 0.97, 1.0}
\definecolor{darkblue}{rgb}{0, 0, 1}
\definecolor{LightCyan}{rgb}{0.88,1,1}
\definecolor{mygreen}{RGB}{28,172,0}
\definecolor{mylilas}{RGB}{170,55,241}
\definecolor{grayblue}{RGB}{220,230,240}
\definecolor{topressbg}{rgb}{0.95,0.95,0.92}
\definecolor{topressgreen}{rgb}{0,0.5,0}
\definecolor{topressgray}{rgb}{0.5,0.5,0.5}
\definecolor{topresspurple}{rgb}{0.58,0,0.82}
\definecolor{topressblue}{rgb}{0,0,1}
\hypersetup{
	pdftex=true,
	colorlinks=true,
	breaklinks=true,
	linkcolor=darkblue,
	menucolor=darkblue,
	pagecolor=darkblue,
	citecolor=darkblue,
	urlcolor=darkblue
}
\usepackage{pgfplots}
\usetikzlibrary{positioning, arrows.meta}
\tikzset{%
	myarrow/.style = {-Stealth, shorten >=5pt}
}

\definecolor{darkblue}{rgb}{0,0,1}
\hypersetup{pdftex=true, colorlinks=true, breaklinks=true, linkcolor=darkblue, menucolor=darkblue, pagecolor=darkblue, citecolor=darkblue, urlcolor=darkblue}

\usepackage{array}
\usepackage{pgfplotstable}

\usepackage{hyperref}

\usetikzlibrary{calc, pgfplots.groupplots, backgrounds}
\usepgfplotslibrary{colorbrewer}
\usepackage{standalone}

\usepackage{makecell}
\usepackage{caption}
\usepackage{empheq}
\usepackage{pdflscape}
\usepackage{mwe}
\newcolumntype{C}[1]{>{\centering\arraybackslash}m{#1}}
\definecolor{white}{rgb}{1,1,1}
\definecolor{NavyBlue}{rgb}{0, 0.2, 0.4}
\definecolor{Teal}{rgb}{0, 0.5, 0.5}
\definecolor{DarkGreen}{rgb}{0, 0.3, 0}
\definecolor{ForestGreen}{rgb}{0, 0.4, 0.1}
\definecolor{LimeGreen}{rgb}{0.5, 1.0, 0.5}
\definecolor{DarkOrange}{rgb}{1.0, 0.55, 0}
\definecolor{Peach}{rgb}{1.0, 0.8, 0.6}
\definecolor{Maroon}{rgb}{0.5, 0, 0}
\definecolor{Burgundy}{rgb}{0.4, 0, 0.1}
\definecolor{HotPink}{rgb}{1.0, 0.4, 0.7}
\definecolor{DeepPink}{rgb}{0.8, 0.1, 0.4}
\definecolor{Indigo}{rgb}{0.29, 0, 0.51}
\definecolor{SlateGray}{rgb}{0.44, 0.5, 0.56}
\definecolor{LightGray}{rgb}{0.8, 0.8, 0.8}
\definecolor{SteelBlue}{rgb}{0.27, 0.51, 0.71}
\definecolor{Turquoise}{rgb}{0.25, 0.88, 0.82}
\definecolor{olivegreen}{rgb}{0.75, 0.75, 0}
\definecolor{greenyellow}{rgb}{0.75, 1, 0}
\definecolor{violet}{rgb}{0.4940, 0.1840, 0.5560}
\definecolor{yellowochre}{rgb}{0.9290, 0.6940, 0.1250}
\definecolor{redorange}{rgb}{0.8500, 0.3250, 0.0980}
\definecolor{brownred}{rgb}{0.6350, 0.0780, 0.1840}
\definecolor{skyblue}{rgb}{0.3010, 0.7450, 0.9330}
\definecolor{shafgreen}{rgb}{0.4660, 0.6740, 0.1880}
\definecolor{red}{rgb}{1,0,0}
\definecolor{blue}{rgb}{0,0,1}
\definecolor{green}{rgb}{0,1,0}
\definecolor{yellow}{rgb}{1,1,0}
\definecolor{magenta}{rgb}{1,0,1}
\definecolor{cyan}{rgb}{0,1,1}
\definecolor{black}{rgb}{0,0,0}
\definecolor{gray}{rgb}{0.6,0.6,0.6}
\usepackage{float}
\usepackage{tikz-dimline}
\usetikzlibrary{patterns}
\usetikzlibrary{arrows.meta}
\usepackage{placeins}

\graphicspath{{Figure1/}}
\definecolor{nika}{rgb}{0,0,1}

\definecolor{nikr}{rgb}{1,0,0}

\definecolor{nikg}{rgb}{0,1,0}

\begin{document}
	
	\begin{center}
		\Large{\bf{A generalized shape function approach for multimaterial topology optimization}}\\
		
	\end{center}
	
	\begin{center}
		Swagatam Islam Sarkar$^*$, Nikhil Singh$^\dagger$, Anupam Saxena$^\dagger$, Prabhat Kumar$^{*,}$$\footnote{pkumar@mae.iith.ac.in}$
		
		\vspace{4mm}
		\small{$*$\textit{Department Mechanical and Aerospace Engineering. India Institute of Technology Hyderbad, Telangana 502285, India}}\\
		\small{$\dagger$\textit{Department Mechanical Engineering. India Institute of Technology Kanpur, Uttar Pradesh 208016, India}}\\
			\vspace{1mm}
			Published\footnote{This pdf is the personal version of an article whose final publication is available at \href{https://onlinelibrary.wiley.com/doi/abs/10.1002/nme.70401}{https://onlinelibrary.wiley.com/journal/10970207}} 
		in \textit{International Journal for Numerical Methods in Engineering}, 
		\href{https://onlinelibrary.wiley.com/doi/abs/10.1002/nme.70401}{DOI:10.1002/nme.70401} \\
		Submitted on 28~January 2026, Revised on 13 July 2026, Accepted on 30 July 2026
	\end{center}
	
	\vspace{3mm}
	\rule{\linewidth}{.15mm}
	{\bf Abstract:}
{This paper presents a generalized shape function (gSF) approach for multi-material topology optimization that utilizes a compact design space to produce optimized configurations featuring a large number of materials. Building upon 1D (linear), 2D (bilinear), and 3D (trilinear)  shape functions, generalized nD (n-linear) shape functions are conceptualized to map the multi-material simplex domain. Natural coordinates of these shape functions are considered the design variables used to determine the material densities. These densities are mathematically proven to satisfy the essential barycentric properties, guaranteeing a physically valid material interpolation space. Furthermore, we demonstrate that applying density filtering directly to the natural coordinates is mathematically equivalent to filtering the densities themselves, and that the tailored projection scheme preserves these vital barycentric properties in the projected states. The versatility, efficacy, and success of the gSF approach are demonstrated across various 2D and 3D stiff-structure (SS) and compliant-mechanism (CM) design problems. Strain energy is minimized for SS, whereas a multicriteria objective is minimized for CMs with given volume constraints. Sensitivity analysis is performed using the adjoint variable method, and the optimization problem is solved using the method of moving asymptotes. Results for SSs and CMs in 2D and 3D, respectively, up to 24 and 15 different materials, are presented. Objective history plots indicate smooth convergence. The proposed approach removes practical restrictions on the number of candidate materials, offering excellent scalability for large-scale engineering applications. \\
	

	\vspace{-4mm}
	\rule{\linewidth}{.15mm}
	
  \section{Introduction}\label{sec1}
Topology optimization (TO), a systematic design technique, provides an innovative and efficient optimized material layout by smartly deciding where to place the material and where to make the holes for a given design problem by extremizing the formulated objective subject to physical and/or geometric constraints. The design domain is discretized using standard finite elements (FEs)~\cite{bendsoe1988generating,sigmund2013topology} or polygonal elements~\cite{sanders2018polymat,kumar2023honeytop90,banh2024comprehensive, singh2024three} to solve the associated boundary value problems using the finite element method to determine the related state variables. To perform the optimization step for a single material case, each element is generally assigned a design variable (material density variable) $\rho \ (0\le\rho\le1)$, which gets updated as TO progresses. However, a set of different materials is used in many applications, such as automobiles, aviation, medical, construction, etc., to achieve high-performance, cost-effective, low-weight designs. In addition, due to the unprecedented development of additive manufacturing (3D Printing), it has steadily become feasible to fabricate multilateral structures with complex and intricate geometry~\cite{bandyopadhyay2018additive}. Further, permitting multi-material conditions within the TO framework increases the design search space, leading to relatively more efficient optimized designs for numerous applications~\cite{sivapuram2021design,kumar2022topology}. In general, the number of design variables increases per element as the number of candidate materials increases. Additionally, design variables and constraints get coupled, which increases the nonlinear nature of the problem setting. Herein, we propose a generalized shape function (gSF) approach for multimaterial topology optimization (MMTO) in a density-based TO framework. The method provides crisp optimized 2D and 3D stiff structures (SS) and compliant mechanisms (CMs) with many materials while using only a few design variables, as depicted in the result section of the paper. 

Impressive and pioneering work by Bends{\o}e and Kikuchi~\cite{bendsoe1988generating} opened the field of TO, attracting many researchers. Eventually, many methods, e.g., density-based~\cite{sigmund2013topology}, level set-based~\cite{van2013level}, evolutionary structural optimization method~\cite{xie1993simple}, phase-field method~\cite{takezawa2010shape}, feature-based method~\cite{wein2020review}, binary structures method~\cite{sivapuram2018topology} and normalized filed product with embedded length scale~\cite{singh2025normalized}, have been proposed.  Among many TO methods that use a single material in the 0-1 setting, the density-based TO remains the most popular because of its ease of implementation~\cite{sigmund2013topology}. Thomsen \cite{thomsen1992topology} is the first to introduce MMTO, in which the method optimizes stiffness composed of one or two isotropic materials. Sigmund and Torquato~\cite{sigmund1997design} designate two variables to each element and calculate the local stiffness of the element in terms of these two variables to design a microstructure with extreme thermoelastic properties. The proposed method, which is also called the ``three-phase mixing scheme," opens the door to extending the Solid Isotropic Material with Penalization (SIMP) approach for multimaterial cases in a recursive manner~\cite{gao2011mass}. The extended SIMP method is widely employed with various approaches for different applications, for example, see Refs.~\cite {sivapuram2021design,kumar2022topology,gaynor2014multiple, roper2018multi,li2018multi,yun2017multi,banh2024comprehensive,banh2024frequency,nguyen2025buckling,nguyen2025robust}. However, to the best of the authors' knowledge, they are limited to showcasing optimized designs with 2/3/4 materials only. This extended SIMP approach typically requires $m$ design variables per element for $m$ candidate materials; thus, the computational cost increases with the number of candidate materials. Furthermore, the mass of each material is strongly coupled to the design variables, leading to numerical difficulties and increased nonlinear behavior~\cite{gao2011mass}. Gao and Zhang~\cite{gao2011mass} propose a uniform multiphase material interpolation strategy to decouple the mass constraints to circumvent the latter issue. Stegmann and Lund~\cite{stegmann2005discrete} mention that the extended SIMP formulation tends to get stuck in local minima for $m>=4$. Yang and Li~\cite{yang2018discrete} present an approach that solves a series of two-material subproblems with mass constraints. Li and Kim~\cite{li2018multi} introduce a minimum-weight MMTO method using the extended SIMP model to achieve lightweight, optimized multimaterial designs.   Yin and Ananthasuresh \cite{yin2001topology} propose the peak function-based approach to designing multi-material CMs using an unrestricted single variable. To achieve multiple peaks (multiple materials), gradual parameter adjustment is needed, which can become challenging as the number of candidate materials increases. They demonstrate results only up to 3-material CMs.

Tavakoli and Mohseni \cite{tavakoli2014alternating} demonstrate a novel alternating active-phase (AAP) algorithm to solve multiphase problems, wherein, for $n$-phase\footnote{Number of phases = number of candidate materials + 1} material cases, $n(n-1)/2$ binary phase problems are solved sequentially, leading to high computational cost. The AAP approach for multi-material is also employed by Doan and Lee~\cite{doan2017optimum} with buckling constraint. Zuo and Saitou \cite{zuo2017multi} introduce an ordered SIMP interpolation method that achieves results with up to 3 materials without additional variables to represent material selection; therefore, the computational cost is independent of the number of materials considered. However, the proposed scheme exhibits non-differentiability at material transitions, which pose challenges for gradient-based optimizers. The ordered SIMP approach has been adopted in Refs.~\cite {gu2022improved,ramnath2020multi} for their applications. Sanders et al.~\cite{sanders2018multi} present the MMTO approach on unstructured polygonal meshes that uses the method presented by Stegmann and Lund~\cite{stegmann2005discrete} for material interpolation with linearly separable volume constraints. Additionally, they provide MATLAB code for educational purposes in~\cite{sanders2018polymat}, where the approach is demonstrated by designing multi-material structures up to 15 materials. Jia et al.~\cite{jia2025multimaterial} propose a multimaterial and multiobjective topology optimization approach to create composite metallic structures with optimized elastoplastic characteristics.  A unified material interpolation is presented by Yi et al.~\cite{yi2023unified}. Various approaches using the level set method for multi-material designs are presented in~\cite{wang2004color}.

\begin{figure*}
	\centering
	\begin{subfigure}{0.45\textwidth}
		\includegraphics[width=0.8\textwidth]{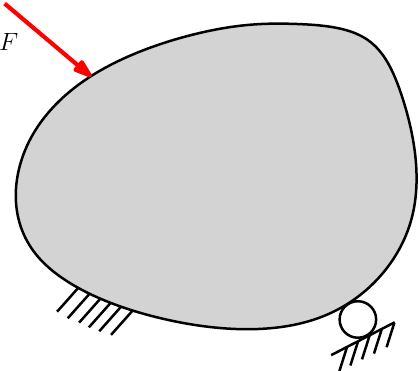}		
		\caption{}
		\label{Schematic1a}
	\end{subfigure}
	\begin{subfigure}{0.45\textwidth}
		\includegraphics[width=0.8\textwidth]{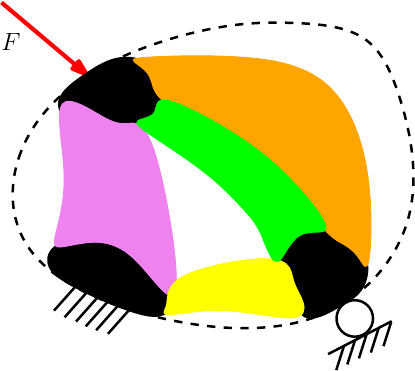}		
		\caption{}
		\label{Schematic1b}
	\end{subfigure}
	\caption{A schematic diagram of an arbitrary-shaped structure with a point load $F$. (a) Design domain, (b) representative optimized design with five materials. Different colors represent different materials.}
	\label{Schematic1}
\end{figure*}

Bruyneel~\cite{bruyneel2011sfp} presents a shape function with penalization (SFP) approach, where shape functions of FEM are used to interpolate mechanical properties to design composite structures. Bruyneel et al.~\cite{bruyneel2011extensions} extend the SFP approach to include more candidate materials. Working on the similar idea, Gao et al.~\cite{gao2012bi} propose a bi-value approach for designing composite structures, reducing the number of design variables. Cherri{\`e}re et al.~\cite{cherriere2022multi} use Wachpress shape functions to design a 3-phase electrical machine with the filtering technique. Singh and Saxena~\cite{singh2023projected} use the SFP approach for optimizing multi-material designs. A schematic diagram
for a multi-material problem and its representative solution are depicted in Fig.~\ref{Schematic1}.

Based on the insights gathered from the state-of-the-art discussed above, a novel MMTO approach should be able to deliver the following merits. 
\begin{itemize}
    \item [M$_1$]: The approach should be computationally inexpensive; with fewer design variables, the method should provide more candidate materials for the optimized designs readily.
    \item [M$_2$]: The approach should provide close to crisp solutions that, in turn, reduce post-processing steps for optimized results.
    \item [M$_3$]: The approach should facilitate separable mass/volume constraints for each material, that is, design variables and mass/volume constraints can be decoupled. This reduces the nonlinear nature of the MMTO problems. 
    \item [M$_4$]: The approach should be readily extendable to 3D problems and to various applications.
\end{itemize}
$M_i|_{i =1,\,2,\,3,\,4}$ indicate merits.

We propose a generalized shape function (gSF) approach that provides a viable way to include all the above-mentioned merits. The approach generalizes  1D (linear), 2D (bilinear), and 3D (trilinear) shape functions to nD (n-linear) shape functions for material modeling. The natural coordinates associated with the shape functions are used as design variables to determine the element material densities. As a result, the gSF approach enables modeling up to $2^n-1$ distinct materials ($2^n$ phases) by assigning $n$ design variables per element, thereby providing merit M$_1$. This capability is demonstrated through numerical examples up to 24 materials for both SSs and CMs. The formulation defines an $n$-dimensional multilinear manifold embedded in the classical material simplex, and we prove that the associated material densities satisfy the barycentric properties. To promote near-discrete (0-1) material distributions with minimum feature size, density filtering, and the conceptualized projection scheme are incorporated within the optimization framework (M$_2$). Since design variables (natural coordinates associated with the shape functions) lie between $-1\,\text{and}\, 1$, the filtered design variables are first proven to satisfy the barycentric properties, and then the conceptualized projection scheme is employed that retains the barycentric properties of the projected variables to achieve solutions close to 0-1. We demonstrate the usefulness of the density and the conceptualized projection filters Sec.~\ref{sec4}. An individual volume constraint is imposed for each candidate material without inter-material coupling; that is, the gSF approach avoids decoupling, thus providing merit M$_3$. The formulation is readily extended to three-dimensional multi-material SSs and CMs, demonstrating the robustness and generality of the proposed gSF approach (M$_4$). To summarize, the present manuscript provides the following new contributions:

\begin{itemize}
    \item \textbf{A generalized shape function (gSF) framework is introduced}: We generalize linear, bilinear, and trilinear shape functions to a fully consistent $n$-linear formulation to achieve robust and scalable material modeling. The essential barycentric properties of the formulated material density have been rigorously verified.
    \item \textbf{Number of design variables:} The proposed approach uses $n$ design variables per element to successfully represent and optimize designs with up to $2^n-1$ distinct materials.
    \item \textbf{Clear geometric mapping without loss of optimality:} We establish a direct mathematical relationship between the classical material simplex 	space $\mathcal{S}$ and gSF material space $\mathcal{H}\in [-1,\,1]^n$. While the gSF framework maps the design space to a lower-dimensional manifold embedded within the simplex, it is mathematically vertex-complete. This ensures that all optimal, pure material states are perfectly preserved and reachable, meaning the reduced search space filters out redundant intermediate mixtures without excluding superior physical solutions.
    \item  \textbf{Customized material modeling, filtering and projection for crisp 0-1 solutions:} The proposed approach, together with a tailored modification of the SIMP method to suit the gSF framework, a classical density filter (demonstrated to preserve the barycentric nature of the shape functions), and a developed projection filter derived from its conventional counterpart to ensure barycentric consistency, yields near 0-1 discrete solutions. The role and effectiveness of the density and projection filters in the gSF framework are demonstrated.
    \item \textbf{Proven versatility in 2D:} The robustness and versatility of the method are demonstrated through various 2D numerical examples, including both self-adjoint (e.g.,  compliance minimization) and non-self-adjoint (e.g., compliant mechanism) problems, involving up to 24 materials.
    \item \textbf{Smooth extension to 3D:}	The proposed gSF framework is readily extended to 3D multimaterial TO problems involving both stiff structures and compliant mechanisms; thus, demonstrating its generality and applicability.
\end{itemize}

 The remainder of this paper is structured as follows: Section~\ref{sec2} introduces the proposed generalized shape functions approach, a relationship between the classical material simplex and material space of the gSF approach, and evaluation of the filtered and projected material design variables and material density. Section~\ref{sec3} details the TO problem formulation for designing SS and CMs, and the sensitivity analyses. Section~\ref{sec4} presents a study related to the usefulness of the density and the conceptualized projection filters, various results of numerical experiments for 2D (up to 24 materials) and 3D (up to 15 materials) SSs and CMs, and discussions.  Lastly, concluding remarks are mentioned in Section~\ref{sec5}.

  \section{Generalized Shape Function approach}\label{sec2}
The SFP approach~\cite{bruyneel2011sfp,bruyneel2011extensions} improves the quality of the solution and structural performance by introducing penalization in the interpolation of the properties of the material. This method uses the properties of finite element shape functions to obtain the optimized material distribution within a design domain. Suppose that a material element\footnote{A material element indicates the element used to define the shape functions (material density functions) for material interpolation} contains $m$ vertices. In a standard FEM setting, each vertex is associated with a shape function; that is, the $m^\text{th}$ vertex is associated with shape function $N_m$, where $N_m$ indicates the $m^\text{th}$ shape function. These shape functions follow the partition of unity properties; the sum of the shape functions at any point within the element equals 1. Individually, the shape functions are non-negative. Together, this is also known as the barycentric nature of the shape functions. Another property says that each shape function has a value of 1 at its corresponding node and 0 at all other nodes; this is called the Kronecker delta property of the shape functions. In a multi-material TO formulation, the presence of one material in an FE implies the absence of all other materials in the same FE, which resembles the Kronecker delta property. Likewise, the sum of a function that indicates the presence and absence of materials in any FE equals 1; that is, it satisfies the barycentric property. Therefore, a shape function can be used to determine the material density of an FE. In a standard FEM, these shape functions are defined in terms of the natural coordinates. As these coordinates change, the shape functions change; thus, they can be used as design variables to update the material density (shape functions) of the elements.

Say, $\chi_i|_{i=1,\,2,\,3,\,\cdots}$ ($\chi_{i} \in [-1, 1]$) indicate natural coordinates that defines shape function $N_m$, we now call, $\rho_m$, to indicate material density. Using the properties of the natural coordinates and shape functions, we have
\begin{equation}\label{Eq: desingvaraible}
	\begin{aligned}
		-1 \leq \chi_i \leq 1,  \quad (i=1,2,\dots,n) ;\quad \, 
		0 \leq \rho_m \leq 1,  \quad (m=1,2,\dots,2^n) 
	\end{aligned}
\end{equation}
\begin{center}
\begin{minipage}{\linewidth}
\centering
\resizebox{\linewidth}{!}{
	\begin{tabular}[t]{|c|c|c|}
		\hline
		\textbf{SF} & \textbf{Shape function schematics (Material elements)} & \textbf{Material densities} \\
		\hline
		1D &
		\raisebox{-0.5\height}{
			\begin{tikzpicture}[scale=1]
				\fill [black] (-1.05,-0.05) rectangle (1.05,0.05);
				\filldraw[red] (-1,0) circle (2.5pt) node[anchor=north east]{\large 1};
				\filldraw[red] (1,0) circle (2.5pt) node[anchor=north west]{\large 2};
				\draw[blue, thick, -Stealth] (0,0) -- (2.25,0) node[anchor=south]{$\chi_1$};
			\end{tikzpicture}
		} & 
        \makecell[l]{
		$\displaystyle \rho_1=\frac{(1-\chi_1)}{2},\quad \rho_2=\frac{(1+\chi_1)}{2}$ \\
        i.e.,\,\,$\rho_m=\frac{(1+x_{1m}\chi_1)}{2}$, $\displaystyle m \in\{1,2\}$
        } \\
		\hline
		2D &
		\raisebox{-0.5\height}{
			\begin{tikzpicture}[scale=1]
				\fill [black] (-1.05,-1.05) rectangle (1.05,-0.95);
				\fill [black] (1.05,-0.95) rectangle (0.95,1.05);
				\fill [black] (0.95,1.05) rectangle (-1.05,0.95);
				\fill [black] (-1.05,0.95) rectangle (-0.95,-1.05);
				\filldraw[red] (-1,-1) circle (2.5pt) node[anchor=north east]{\large 1};
				\filldraw[red] (1,-1) circle (2.5pt) node[anchor=north west]{\large 2};
				\filldraw[red] (1,1) circle (2.5pt) node[anchor=south west]{\large 3};
				\filldraw[red] (-1,1) circle (2.5pt) node[anchor=south east]{\large 4};
				\draw[blue, thick, -Stealth] (0,0) -- (2.25,0) node[anchor=south]{$\chi_1$};
				\draw[blue, thick, -Stealth] (0,0) -- (0,2.25) node[anchor=west]{$\chi_2$};
			\end{tikzpicture}
		} & 
		\makecell[l]{
			$\displaystyle \rho_m=\frac{(1+x_{1m}\chi_1)(1+x_{2m}\chi_2)}{2^2},$ \\
			$\displaystyle m \in\{1,\dots,4\}$
		} \\
		\hline
		3D &
		\raisebox{-0.5\height}{
			\begin{tikzpicture}[scale=1]
				\fill [black] (-1.05,-1.05) rectangle (1.05,-0.95);
				\fill [black] (1.05,-0.95) rectangle (0.95,1.05);
				\fill [black] (0.95,1.05) rectangle (-1.05,0.95);
				\fill [black] (-1.05,0.95) rectangle (-0.95,-1.05);
				
				\draw[black, very thick] (-0.25,-0.5) -- (1.75,-0.5);
				\draw[black, very thick] (1.75,-0.5) -- (1.75,1.5);
				\draw[black, very thick] (1.75,1.5) -- (-0.25,1.5);
				\draw[black, very thick] (-0.25,1.5) -- (-0.25,-0.5);
				
				\draw[black, very thick] (-0.25,-0.50) -- (-1,-1);
				\draw[black, very thick] (1.75,-0.50) -- (1,-1);
				\draw[black, very thick] (1.75,1.5) -- (1,1);
				\draw[black, very thick] (-0.25,1.5) -- (-1,1);
				\filldraw[red] (-1,-1) circle (2.5pt) node[anchor=north east]{\large 1};
				\filldraw[red] (1,-1) circle (2.5pt) node[anchor=north west]{\large 2};
				\filldraw[red] (1,1) circle (2.5pt) node[anchor=south west]{\large 3};
				\filldraw[red] (-1,1) circle (2.5pt) node[anchor=south east]{\large 4};
				
				\filldraw[red] (-0.25,-0.5) circle (2.5pt) node[anchor=south east]{\large 5};
				\filldraw[red] (1.75,-0.5) circle (2.5pt) node[anchor=south east]{\large 6};
				\filldraw[red] (1.75,1.5) circle (2.5pt) node[anchor=south east]{\large 7};
				\filldraw[red] (-0.25,1.5) circle (2.5pt) node[anchor=south east]{\large 8};
				\draw[blue, thick, -Stealth] (0.375,0.25) -- (2.625,0.25) node[anchor=south]{$\chi_1$};
				\draw[blue, thick, -Stealth] (0.375,0.25) -- (0.375,2.5) node[anchor=west]{$\chi_2$};
				\draw[blue, thick, -Stealth] (0.375,0.25) -- (-1.375,-1) node[anchor=south]{$\chi_3$};
			\end{tikzpicture}
		} & 
		\makecell[l]{
			$\displaystyle 
			\begin{aligned}
				\rho_m &= \frac{(1+x_{1m}\chi_1)(1+x_{2m}\chi_2)(1+x_{3m}\chi_3)}{2^3}, 
			\end{aligned}$ \\
			$\displaystyle m \in\{1,\dots,8\}$
		} \\
		\hline
		4D &
		\raisebox{-0.5\height}{
			\begin{tikzpicture}[scale=1]
				\fill [black] (-1.05,-1.05) rectangle (1.05,-0.95);
				\fill [black] (1.05,-0.95) rectangle (0.95,1.05);
				\fill [black] (0.95,1.05) rectangle (-1.05,0.95);
				\fill [black] (-1.05,0.95) rectangle (-0.95,-1.05);
				
				\draw[black, very thick] (-0.25,-0.5) -- (1.75,-0.5);
				\draw[black, very thick] (1.75,-0.5) -- (1.75,1.5);
				\draw[black, very thick] (1.75,1.5) -- (-0.25,1.5);
				\draw[black, very thick] (-0.25,1.5) -- (-0.25,-0.5);
				
				\draw[black, very thick] (-0.25,-0.50) -- (-1,-1);
				\draw[black, very thick] (1.75,-0.50) -- (1,-1);
				\draw[black, very thick] (1.75,1.5) -- (1,1);
				\draw[black, very thick] (-0.25,1.5) -- (-1,1);
				
				\draw[black, very thick] (-1.25,-1.5) -- (2.75,-1.5);
				\draw[black, very thick] (2.75,-1.5) -- (2.75,2.5);
				\draw[black, very thick] (2.75,2.5) -- (-1.25,2.5);
				\draw[black, very thick] (-1.25,2.5) -- (-1.25,-1.5);
				
				\draw[black, very thick] (-1.25,-1.5) -- (-0.25,-0.5);
				\draw[black, very thick] (2.75,-1.5) -- (1.75,-0.5);
				\draw[black, very thick] (2.75,2.5) -- (1.75,1.5);
				\draw[black, very thick] (-1.25,2.5) -- (-0.25,1.5);
				
				\fill [black] (-2.05,-2.05) rectangle (2.05,-1.95);
				\fill [black] (2.05,-1.95) rectangle (1.95,2.05);
				\fill [black] (1.95,2.05) rectangle (-2.05,1.95);
				\fill [black] (-2.05,1.95) rectangle (-1.95,-2.05);
				
				\draw[black, very thick] (-2,-2) -- (-1,-1);
				\draw[black, very thick] (2,-2) -- (1,-1);
				\draw[black, very thick] (2,2) -- (1,1);
				\draw[black, very thick] (-2,2) -- (-1,1);
				
				\draw[black, very thick] (-2,-2) -- (-1.25,-1.5);
				\draw[black, very thick] (2,-2) -- (2.75,-1.5);
				\draw[black, very thick] (2,2) -- (2.75,2.5);
				\draw[black, very thick] (-2,2) -- (-1.25,2.5);
				
				\filldraw[red] (-1,-1) circle (2.5pt);
				\filldraw[red] (1,-1) circle (2.5pt);
				\filldraw[red] (1,1) circle (2.5pt);
				\filldraw[red] (-1,1) circle (2.5pt);
				
				\filldraw[red] (-0.25,-0.5) circle (2.5pt);
				\filldraw[red] (1.75,-0.5) circle (2.5pt);
				\filldraw[red] (1.75,1.5) circle (2.5pt);
				\filldraw[red] (-0.25,1.5) circle (2.5pt);
				
				\filldraw[red] (-2,-2) circle (2.5pt);
				\filldraw[red] (2,-2) circle (2.5pt);
				\filldraw[red] (2,2) circle (2.5pt);
				\filldraw[red] (-2,2) circle (2.5pt);
				
				\filldraw[red] (-1.25,-1.5) circle (2.5pt);
				\filldraw[red] (2.75,-1.5) circle (2.5pt);
				\filldraw[red] (2.75,2.5) circle (2.5pt);
				\filldraw[red] (-1.25,2.5) circle (2.5pt);
				
				\draw[blue, thick, -Stealth] (0.375,0.25) -- (3.625,0.25) node[anchor=south]{$\chi_1$};
				\draw[blue, thick, -Stealth] (0.375,0.25) -- (0.375,3.5) node[anchor=west]{$\chi_2$};
				\draw[blue, thick, -Stealth] (0.375,0.25) -- (-2.75,-1.75) node[anchor=south]{$\chi_3$};
				
				\draw[blue, thick, -Stealth] (-1.2,-1) -- (-2.2,-2) node[anchor=north]{$\chi_4$};
				\draw[blue, thick, -Stealth] (1.2,-1) -- (2.2,-2) node[anchor=west]{$\chi_4$};
				\draw[blue, thick, -Stealth] (0.8,1) -- (1.8,2) node[anchor=south]{$\chi_4$};
				\draw[blue, thick, -Stealth] (-1.2,1) -- (-2.2,2) node[anchor=east]{$\chi_4$};
				
				\draw[blue, thick, -Stealth] (-0.05,-0.5) -- (-1.05,-1.5) node[anchor=west]{$\chi_4$};
				\draw[blue, thick, -Stealth] (1.95,-0.5) -- (2.95,-1.5) node[anchor=south west]{$\chi_4$};
				\draw[blue, thick, -Stealth] (1.95,1.5) -- (2.95,2.5) node[anchor=west]{$\chi_4$};
				\draw[blue, thick, -Stealth] (-0.45,1.5) -- (-1.45,2.5) node[anchor=east]{$\chi_4$};
				
			\end{tikzpicture}
		} & 
		\makecell[l]{
			$\displaystyle 
			\begin{aligned}
				\rho_m &= \frac{(1+x_{1m}\chi_1)(1+x_{2m}\chi_2)(1+x_{3m}\chi_3)(1+x_{4m}\chi_4)}{2^4},
			\end{aligned}$ \\
			$ m \in\{1,\dots,16\}$
		} \\
		\hline
		\dots &
		\dots & \dots\\
		\hline
		nD &
		\raisebox{-0.5\height}{

            \begin{tikzpicture}[scale=0.7]
				\fill [black] (-1.05,-1.05) rectangle (1.05,-0.95);
				\fill [black] (1.05,-0.95) rectangle (0.95,1.05);
				\fill [black] (0.95,1.05) rectangle (-1.05,0.95);
				\fill [black] (-1.05,0.95) rectangle (-0.95,-1.05);

				\filldraw[black] (-1.7,1) circle (1.5pt);
				\filldraw[black] (-2.4,1) circle (1.5pt);
				\filldraw[black] (-3.1,1) circle (1.5pt);

                \filldraw[black] (-1,-1.6125) circle (1.5pt);
				\filldraw[black] (-1,-2.225) circle (1.5pt);
				\filldraw[black] (-1,-2.8375) circle (1.5pt);

                \filldraw[black] (1.7,-1) circle (1.5pt);
				\filldraw[black] (2.4,-1) circle (1.5pt);
				\filldraw[black] (3.1,-1) circle (1.5pt);
                
                \filldraw[black] (1,1.6125) circle (1.5pt);
				\filldraw[black] (1,2.225) circle (1.5pt);
				\filldraw[black] (1,2.8375) circle (1.5pt);
				
				\draw[black, very thick] (-0.25,-0.5) -- (1.75,-0.5);
				\draw[black, very thick] (1.75,-0.5) -- (1.75,1.5);
				\draw[black, very thick] (1.75,1.5) -- (-0.25,1.5);
				\draw[black, very thick] (-0.25,1.5) -- (-0.25,-0.5);

				\filldraw[black] (-0.95,-0.55) circle (1.5pt);
				\filldraw[black] (-1.65,-0.6) circle (1.5pt);
				\filldraw[black] (-2.35,-0.65) circle (1.5pt);

                \filldraw[black] (1.775,-1.1125) circle (1.5pt);
				\filldraw[black] (1.8,-1.725) circle (1.5pt);
				\filldraw[black] (1.825,-2.3375) circle (1.5pt);

                \filldraw[black] (2.45,1.525) circle (1.5pt);
				\filldraw[black] (3.15,1.55) circle (1.5pt);
				\filldraw[black] (3.85,1.575) circle (1.5pt);
                
                \filldraw[black] (-0.275,2.1) circle (1.5pt);
				\filldraw[black] (-0.3,2.7) circle (1.5pt);
				\filldraw[black] (-0.325,3.3) circle (1.5pt);
				
				\draw[black, very thick] (-0.25,-0.50) -- (-1,-1);
				\draw[black, very thick] (1.75,-0.50) -- (1,-1);
				\draw[black, very thick] (1.75,1.5) -- (1,1);
				\draw[black, very thick] (-0.25,1.5) -- (-1,1);
				
				\draw[black, very thick] (-1.25,-1.5) -- (2.75,-1.5);
				\draw[black, very thick] (2.75,-1.5) -- (2.75,2.5);
				\draw[black, very thick] (2.75,2.5) -- (-1.25,2.5);
				\draw[black, very thick] (-1.25,2.5) -- (-1.25,-1.5);
				
				\draw[black, very thick] (-1.25,-1.5) -- (-0.25,-0.5);
				\draw[black, very thick] (2.75,-1.5) -- (1.75,-0.5);
				\draw[black, very thick] (2.75,2.5) -- (1.75,1.5);
				\draw[black, very thick] (-1.25,2.5) -- (-0.25,1.5);
				
				\fill [black] (-2.05,-2.05) rectangle (2.05,-1.95);
				\fill [black] (2.05,-1.95) rectangle (1.95,2.05);
				\fill [black] (1.95,2.05) rectangle (-2.05,1.95);
				\fill [black] (-2.05,1.95) rectangle (-1.95,-2.05);
				
				\draw[black, very thick] (-2,-2) -- (-1,-1);
				\draw[black, very thick] (2,-2) -- (1,-1);
				\draw[black, very thick] (2,2) -- (1,1);
				\draw[black, very thick] (-2,2) -- (-1,1);
				
				\draw[black, very thick] (-2,-2) -- (-1.25,-1.5);
				\draw[black, very thick] (2,-2) -- (2.75,-1.5);
				\draw[black, very thick] (2,2) -- (2.75,2.5);
				\draw[black, very thick] (-2,2) -- (-1.25,2.5);
				
				\draw[black, very thick] (-2.75,-3.15) -- (5.25,-3.15);
				\draw[black, very thick] (5.25,-3.15) -- (5.25,4.85);
				\draw[black, very thick] (5.25,4.85) -- (-2.75,4.85);
				\draw[black, very thick] (-2.75,4.85) -- (-2.75,-3.15);
                
				\fill [black] (-4.05,-4.05) rectangle (4.05,-3.95);
				\fill [black] (4.05,-3.95) rectangle (3.95,4.05);
				\fill [black] (3.95,4.05) rectangle (-4.05,3.95);
				\fill [black] (-4.05,3.95) rectangle (-3.95,-4.05);
				
				\filldraw[black] (-1.5,-1.75) circle (1.5pt);
				\filldraw[black] (3,-1.75) circle (1.5pt);
				\filldraw[black] (3,2.75) circle (1.5pt);
				\filldraw[black] (-1.5,2.75) circle (1.5pt);
				
				\filldraw[black] (-1.75,-2) circle (1.5pt);
				\filldraw[black] (3.25,-2) circle (1.5pt);
				\filldraw[black] (3.25,3) circle (1.5pt);
				\filldraw[black] (-1.75,3) circle (1.5pt);
				
				\filldraw[black] (-2,-2.25) circle (1.5pt);
				\filldraw[black] (3.5,-2.25) circle (1.5pt);
				\filldraw[black] (3.5,3.25) circle (1.5pt);
				\filldraw[black] (-2,3.25) circle (1.5pt);

                \filldraw[black] (-1.3,-2.775) circle (1.5pt);
				\filldraw[black] (-0.25,-2.95) circle (1.5pt);
                \draw[black, very thick, dashed] (-0.25,-2.95) -- (1.85,-2.95);
				\filldraw[black] (1.85,-2.95) circle (1.5pt);
				\filldraw[black] (2.9,-2.775) circle (1.5pt);
                \filldraw[black] (4.2,-1.55) circle (1.5pt);
				\filldraw[black] (4.55,-0.5) circle (1.5pt);
                \draw[black, very thick, dashed] (4.55,-0.5) -- (4.55,1.6);
				\filldraw[black] (4.55,1.6) circle (1.5pt);
				\filldraw[black] (4.2,2.65) circle (1.5pt);
                \filldraw[black] (2.8,3.775) circle (1.5pt);
				\filldraw[black] (1.75,3.9) circle (1.5pt);
                \draw[black, very thick, dashed] (1.75,3.9) -- (-0.35,3.9);
				\filldraw[black] (-0.35,3.9) circle (1.5pt);
				\filldraw[black] (-1.4,3.775) circle (1.5pt);
                \filldraw[black] (-2.7,2.55) circle (1.5pt);
				\filldraw[black] (-3.05,1.4) circle (1.5pt);
                \draw[black, very thick, dashed] (-3.05,1.4) -- (-3.05,-0.7);
				\filldraw[black] (-3.05,-0.7) circle (1.5pt);
				\filldraw[black] (-2.7,-1.75) circle (1.5pt);

                \draw[black, very thick, dashed] (-2,-2.25) -- (-2.75,-3.15);
                \draw[black, very thick, dashed] (-1.3,-2.775) -- (-1.75,-3.9);
                \draw[black, very thick, dashed] (-0.25,-2.95) -- (-0.25,-4.15);
                \draw[black, very thick, dashed] (1.85,-2.95) -- (2.75,-4.15);
                \draw[black, very thick, dashed] (2.9,-2.775) -- (4.25,-3.9);
                
                \draw[black, very thick, dashed] (3.5,-2.25) -- (5.25,-3.15);
                \draw[black, very thick, dashed] (4.2,-1.55) -- (6.25,-2.15);
                \draw[black, very thick, dashed] (4.55,-0.5) -- (6.75,-0.65);
                \draw[black, very thick, dashed] (4.55,1.6) -- (6.75,2.35);
                \draw[black, very thick, dashed] (4.2,2.65) -- (6.25,3.85);

                \draw[black, very thick, dashed] (3.5,3.25) -- (5.25,4.85);
                \draw[black, very thick, dashed] (2.8,3.775) -- (4.25,5.6);
                \draw[black, very thick, dashed] (1.75,3.9) -- (2.75,5.85);
                \draw[black, very thick, dashed] (-0.35,3.9) -- (-0.25,5.85);
                \draw[black, very thick, dashed] (-1.4,3.775) -- (-1.75,5.6);
                
                \draw[black, very thick, dashed] (-2,3.25) -- (-2.75,4.85);
                \draw[black, very thick, dashed] (-2.7,2.55) -- (-3.75,3.85);
                \draw[black, very thick, dashed] (-3.05,1.4) -- (-4.25,2.35);
                \draw[black, very thick, dashed] (-3.05,-0.7) -- (-4.25,-0.65);
                \draw[black, very thick, dashed] (-2.7,-1.75) -- (-3.75,-2.15);
				
				\filldraw[black] (-2.25,-2.25) circle (1.5pt);
				\filldraw[black] (2.25,-2.25) circle (1.5pt);
				\filldraw[black] (2.25,2.25) circle (1.5pt);
				\filldraw[black] (-2.25,2.25) circle (1.5pt);
				
				\filldraw[black] (-2.5,-2.5) circle (1.5pt);
				\filldraw[black] (2.5,-2.5) circle (1.5pt);
				\filldraw[black] (2.5,2.5) circle (1.5pt);
				\filldraw[black] (-2.5,2.5) circle (1.5pt);
				
				\filldraw[black] (-2.75,-2.75) circle (1.5pt);
				\filldraw[black] (2.75,-2.75) circle (1.5pt);
				\filldraw[black] (2.75,2.75) circle (1.5pt);
				\filldraw[black] (-2.75,2.75) circle (1.5pt);

                \filldraw[black] (-2.05,-3.275) circle (1.5pt);
				\filldraw[black] (-1,-3.45) circle (1.5pt);
                \draw[black, very thick, dashed] (-1,-3.45) -- (1.1,-3.45);
				\filldraw[black] (1.1,-3.45) circle (1.5pt);
				\filldraw[black] (2.15,-3.275) circle (1.5pt);
                \filldraw[black] (3.45,-2.05) circle (1.5pt);
				\filldraw[black] (3.8,-1) circle (1.5pt);
                \draw[black, very thick, dashed] (3.8,-1) -- (3.8,1.1);
				\filldraw[black] (3.8,1.1) circle (1.5pt);
				\filldraw[black] (3.45,2.15) circle (1.5pt);
                \filldraw[black] (2.05,3.275) circle (1.5pt);
				\filldraw[black] (1,3.45) circle (1.5pt);
                \draw[black, very thick, dashed] (1,3.45) -- (-1.1,3.45);
				\filldraw[black] (-1.1,3.45) circle (1.5pt);
				\filldraw[black] (-2.15,3.275) circle (1.5pt);
                \filldraw[black] (-3.45,2.05) circle (1.5pt);
				\filldraw[black] (-3.8,1) circle (1.5pt);
                \draw[black, very thick, dashed] (-3.8,1) -- (-3.8,-1.1);
				\filldraw[black] (-3.8,-1.1) circle (1.5pt);
				\filldraw[black] (-3.45,-2.15) circle (1.5pt);

                \draw[black, very thick, dashed] (-2.75,-2.75) -- (-4,-4);
                \draw[black, very thick, dashed] (-2.05,-3.275) -- (-3,-4.75);
                \draw[black, very thick, dashed] (-1,-3.45) -- (-1.5,-5);
                \draw[black, very thick, dashed] (1.1,-3.45) -- (1.5,-5);
                \draw[black, very thick, dashed] (2.15,-3.275) -- (3,-4.75);
                
                \draw[black, very thick, dashed] (2.75,-2.75) -- (4,-4);
                \draw[black, very thick, dashed] (3.45,-2.05) -- (5,-3);
                \draw[black, very thick, dashed] (3.8,-1) -- (5.5,-1.5);
                \draw[black, very thick, dashed] (3.8,1.1) -- (5.5,1.5);
                \draw[black, very thick, dashed] (3.45,2.15) -- (5,3);

                \draw[black, very thick, dashed] (2.75,2.75) -- (4,4);
                \draw[black, very thick, dashed] (2.05,3.275) -- (3,4.75);
                \draw[black, very thick, dashed] (1,3.45) -- (1.5,5);
                \draw[black, very thick, dashed] (-1.1,3.45) -- (-1.5,5);
                \draw[black, very thick, dashed] (-2.15,3.275) -- (-3,4.75);
                
                \draw[black, very thick, dashed] (-2.75,2.75) -- (-4,4);
                \draw[black, very thick, dashed] (-3.45,2.05) -- (-5,3);
                \draw[black, very thick, dashed] (-3.8,1) -- (-5.5,1.5);
                \draw[black, very thick, dashed] (-3.8,-1.1) -- (-5.5,-1.5);
                \draw[black, very thick, dashed] (-3.45,-2.15) -- (-5,-3);

                \draw[black, very thick] (-4,-4) -- (-5,-3);
                \draw[black, very thick] (-5,-3) -- (-5.5,-1.5);
                \draw[black, very thick, dashed] (-5.5,-1.5) -- (-5.5,1.5);
                \draw[black, very thick] (-5.5,1.5) -- (-5,3);
                \draw[black, very thick] (-5,3) -- (-4,4);

                \draw[black, very thick] (-4,-4) -- (-3,-4.75);
                \draw[black, very thick] (-3,-4.75) -- (-1.5,-5);
                \draw[black, very thick, dashed] (-1.5,-5) -- (1.5,-5);
                \draw[black, very thick] (1.5,-5) -- (3,-4.75);
                \draw[black, very thick] (3,-4.75) -- (4,-4);

                \draw[black, very thick] (4,-4) -- (5,-3);
                \draw[black, very thick] (5,-3) -- (5.5,-1.5);
                \draw[black, very thick, dashed] (5.5,-1.5) -- (5.5,1.5);
                \draw[black, very thick] (5.5,1.5) -- (5,3);
                \draw[black, very thick] (5,3) -- (4,4);

                \draw[black, very thick] (4,4) -- (3,4.75);
                \draw[black, very thick] (3,4.75) -- (1.5,5);
                \draw[black, very thick, dashed] (1.5,5) -- (-1.5,5);
                \draw[black, very thick] (-1.5,5) -- (-3,4.75);
                \draw[black, very thick] (-3,4.75) -- (-4,4);

                \draw[black, very thick] (-2.75,-3.15) -- (-3.75,-2.15);
                \draw[black, very thick] (-3.75,-2.15) -- (-4.25,-0.65);
                \draw[black, very thick, dashed] (-4.25,-0.65) -- (-4.25,2.35);
                \draw[black, very thick] (-4.25,2.35) -- (-3.75,3.85);
                \draw[black, very thick] (-3.75,3.85) -- (-2.75,4.85);

                \draw[black, very thick] (-2.75,4.85) -- (-1.75,5.6);
                \draw[black, very thick] (-1.75,5.6) -- (-0.25,5.85);
                \draw[black, very thick, dashed] (-0.25,5.85) -- (2.75,5.85);
                \draw[black, very thick] (2.75,5.85) -- (4.25,5.6);
                \draw[black, very thick] (4.25,5.6) -- (5.25,4.85);

                \draw[black, very thick] (5.25,4.85) -- (6.25,3.85);
                \draw[black, very thick] (6.25,3.85) -- (6.75,2.35);
                \draw[black, very thick, dashed] (6.75,2.35) -- (6.75,-0.65);
                \draw[black, very thick] (6.75,-0.65) -- (6.25,-2.15);
                \draw[black, very thick] (6.25,-2.15) -- (5.25,-3.15);

                \draw[black, very thick] (5.25,-3.15) -- (4.25,-3.9);
                \draw[black, very thick] (4.25,-3.9) -- (2.75,-4.15);
                \draw[black, very thick, dashed] (2.75,-4.15) -- (-0.25,-4.15);
                \draw[black, very thick] (-0.25,-4.15) -- (-1.75,-3.9);
                \draw[black, very thick] (-1.75,-3.9) -- (-2.75,-3.15);
				
				\draw[black, very thick] (-4,-4) -- (-2.75,-3.15);
				\draw[black, very thick] (4,-4) -- (5.25,-3.15);
				\draw[black, very thick] (4,4) -- (5.25,4.85);
				\draw[black, very thick] (-4,4) -- (-2.75,4.85);

                \draw[black, very thick] (-3,-4.75) -- (-1.75,-3.9);
                \draw[black, very thick] (-1.5,-5) -- (-0.25,-4.15);
                \draw[black, very thick] (1.5,-5) -- (2.75,-4.15);
                \draw[black, very thick] (3,-4.75) -- (4.25,-3.9);
                \draw[black, very thick] (5,-3) -- (6.25,-2.15);
                \draw[black, very thick] (5.5,-1.5) -- (6.75,-0.65);
                \draw[black, very thick] (5.5,1.5) -- (6.75,2.35);
                \draw[black, very thick] (5,3) -- (6.25,3.85);
                \draw[black, very thick] (3,4.75) -- (4.25,5.6);
                \draw[black, very thick] (1.5,5) -- (2.75,5.85);
                \draw[black, very thick] (-1.5,5) -- (-0.25,5.85);
                \draw[black, very thick] (-3,4.75) -- (-1.75,5.6);
                \draw[black, very thick] (-5,3) -- (-3.75,3.85);
                \draw[black, very thick] (-5.5,1.5) -- (-4.25,2.35);
                \draw[black, very thick] (-5.5,-1.5) -- (-4.25,-0.65);
                \draw[black, very thick] (-5,-3) -- (-3.75,-2.15);
                
				\filldraw[red] (-1,-1) circle (2.5pt);
				\filldraw[red] (1,-1) circle (2.5pt);
				\filldraw[red] (1,1) circle (2.5pt);
				\filldraw[red] (-1,1) circle (2.5pt);
				
				\filldraw[red] (-0.25,-0.5) circle (2.5pt);
				\filldraw[red] (1.75,-0.5) circle (2.5pt);
				\filldraw[red] (1.75,1.5) circle (2.5pt);
				\filldraw[red] (-0.25,1.5) circle (2.5pt);
				
				\filldraw[red] (-2,-2) circle (2.5pt);
				\filldraw[red] (2,-2) circle (2.5pt);
				\filldraw[red] (2,2) circle (2.5pt);
				\filldraw[red] (-2,2) circle (2.5pt);
				
				\filldraw[red] (-1.25,-1.5) circle (2.5pt);
				\filldraw[red] (2.75,-1.5) circle (2.5pt);
				\filldraw[red] (2.75,2.5) circle (2.5pt);
				\filldraw[red] (-1.25,2.5) circle (2.5pt);
				
				\filldraw[red] (-4,-4) circle (2.5pt);
				\filldraw[red] (4,-4) circle (2.5pt);
				\filldraw[red] (4,4) circle (2.5pt);
				\filldraw[red] (-4,4) circle (2.5pt);

                \filldraw[red] (-2.75,-3.15) circle (2.5pt);
				\filldraw[red] (5.25,-3.15) circle (2.5pt);
				\filldraw[red] (5.25,4.85) circle (2.5pt);
				\filldraw[red] (-2.75,4.85) circle (2.5pt);
				
                \filldraw[red] (-5,-3) circle (2.5pt);
                \filldraw[red] (-5.5,-1.5) circle (2.5pt);
                \filldraw[red] (-5,3) circle (2.5pt);
                \filldraw[red] (-5.5,1.5) circle (2.5pt);
                
                \filldraw[red] (-3,-4.75) circle (2.5pt);
                \filldraw[red] (-1.5,-5) circle (2.5pt);
                \filldraw[red] (1.5,-5) circle (2.5pt);
                \filldraw[red] (3,-4.75) circle (2.5pt);

                \filldraw[red] (5,-3) circle (2.5pt);
                \filldraw[red] (5.5,-1.5) circle (2.5pt);
                \filldraw[red] (5,3) circle (2.5pt);
                \filldraw[red] (5.5,1.5) circle (2.5pt);

                \filldraw[red] (-3,4.75) circle (2.5pt);
                \filldraw[red] (-1.5,5) circle (2.5pt);
                \filldraw[red] (1.5,5) circle (2.5pt);
                \filldraw[red] (3,4.75) circle (2.5pt);

				\filldraw[red] (-1.75,-3.9) circle (2.5pt);
                \filldraw[red] (-0.25,-4.15) circle (2.5pt);
                \filldraw[red] (2.75,-4.15) circle (2.5pt);
                \filldraw[red] (4.25,-3.9) circle (2.5pt);

                \filldraw[red] (6.25,-2.15) circle (2.5pt);
                \filldraw[red] (6.75,-0.65) circle (2.5pt);
                \filldraw[red] (6.75,2.35) circle (2.5pt);
                \filldraw[red] (6.25,3.85) circle (2.5pt);
                
                \filldraw[red] (-1.75,5.6) circle (2.5pt);
                \filldraw[red] (-0.25,5.85) circle (2.5pt);
                \filldraw[red] (2.75,5.85) circle (2.5pt);
                \filldraw[red] (4.25,5.6) circle (2.5pt);
                
                \filldraw[red] (-3.75,-2.15) circle (2.5pt);
                \filldraw[red] (-4.25,-0.65) circle (2.5pt);
                \filldraw[red] (-4.25,2.35) circle (2.5pt);
                \filldraw[red] (-3.75,3.85) circle (2.5pt);
                
				\draw[blue, thick, -Stealth] (0.375,0.25) -- (7.625,0.25) node[anchor=south]{$\chi_1$};
				\draw[blue, thick, -Stealth] (0.375,0.25) -- (0.375,6.5) node[anchor=west]{$\chi_2$};
				\draw[blue, thick, -Stealth] (0.375,0.25) -- (-5.5,-3.5) node[anchor=east]{$\chi_3$};
				
				\draw[blue, thick, -Stealth] (-1.2,-1) -- (-2.2,-2) node[anchor=south]{$\chi_4$};
				\draw[blue, thick, -Stealth] (1.2,-1) -- (2.2,-2) node[anchor=west]{$\chi_4$};
				\draw[blue, thick, -Stealth] (0.8,1) -- (1.8,2) node[anchor=south]{$\chi_4$};
				\draw[blue, thick, -Stealth] (-1.2,1) -- (-2.2,2) node[anchor=north]{$\chi_4$};
				
				\draw[blue, thick, -Stealth] (-0.05,-0.5) -- (-1.05,-1.5) node[anchor=north]{$\chi_4$};
				\draw[blue, thick, -Stealth] (1.95,-0.5) -- (2.95,-1.5) node[anchor=south]{$\chi_4$};
				\draw[blue, thick, -Stealth] (1.95,1.5) -- (2.95,2.5) node[anchor=north]{$\chi_4$};
				\draw[blue, thick, -Stealth] (-0.45,1.5) -- (-1.45,2.5) node[anchor=west]{$\chi_4$};
				
				\draw[blue, thick, -Stealth] (-2.95,-2.75) -- (-4.25,-4.05) node[anchor=east]{$\chi_n$};
                \draw[blue, thick, -Stealth] (-2.25,-3.275) -- (-3.2,-4.75) node[anchor=east]{$\chi_n$};
                \draw[blue, thick, -Stealth] (-1.2,-3.45) -- (-1.7,-5) node[anchor=north]{$\chi_n$};
                \draw[blue, thick, -Stealth] (1.3,-3.45) -- (1.7,-5) node[anchor=north]{$\chi_n$};
                \draw[blue, thick, -Stealth] (2.35,-3.275) -- (3.2,-4.75) node[anchor=north]{$\chi_n$};
                \draw[blue, thick, -Stealth] (2.55,-2.75) -- (3.8,-4) node[anchor=north]{$\chi_n$};
                \draw[blue, thick, -Stealth] (3.45,-2.25) -- (5,-3.2) node[anchor=east]{$\chi_n$};
                \draw[blue, thick, -Stealth] (3.8,-1.2) -- (5.5,-1.7) node[anchor=west]{$\chi_n$};
                \draw[blue, thick, -Stealth] (3.8,1.3) -- (5.5,1.7) node[anchor=west]{$\chi_n$};
                \draw[blue, thick, -Stealth] (3.45,1.95) -- (5,2.9) node[anchor=west]{$\chi_n$};
                \draw[blue, thick, -Stealth] (2.55,2.85) -- (3.8,4.1) node[anchor=east]{$\chi_n$};
                \draw[blue, thick, -Stealth] (2.25,3.275) -- (3.2,4.75) node[anchor=west]{$\chi_n$};
                \draw[blue, thick, -Stealth] (1.2,3.45) -- (1.7,5) node[anchor=west]{$\chi_n$};
                \draw[blue, thick, -Stealth] (-0.9,3.45) -- (-1.3,5) node[anchor=west]{$\chi_n$};
                \draw[blue, thick, -Stealth] (-2.35,3.275) -- (-3.2,4.75) node[anchor=south]{$\chi_n$};
                \draw[blue, thick, -Stealth] (-2.55,2.75) -- (-3.7,4) node[anchor=west]{$\chi_n$};
                \draw[blue, thick, -Stealth] (-3.45,2.25) -- (-5,3.2) node[anchor=east]{$\chi_n$};
                \draw[blue, thick, -Stealth] (-3.8,1.2) -- (-5.5,1.7) node[anchor=east]{$\chi_n$};
                \draw[blue, thick, -Stealth] (-3.8,-1.3) -- (-5.5,-1.7) node[anchor=east]{$\chi_n$};
                \draw[blue, thick, -Stealth] (-3.45,-1.85) -- (-5,-2.8) node[anchor=south]{$\chi_n$};

                \draw[blue, thick, -Stealth] (-1.8,-2.25) -- (-2.55,-3.15) node[anchor=west]{$\chi_n$};
                \draw[blue, thick, -Stealth] (-1.1,-2.775) -- (-1.55,-3.9) node[anchor=south]{$\chi_n$};
                \draw[blue, thick, -Stealth] (-0.05,-2.95) -- (-0.05,-4.15) node[anchor=north]{$\chi_n$};
                \draw[blue, thick, -Stealth] (1.65,-2.95) -- (2.55,-4.15) node[anchor=north]{$\chi_n$};
                \draw[blue, thick, -Stealth] (3.1,-2.775) -- (4.45,-3.9) node[anchor=north]{$\chi_n$};
                \draw[blue, thick, -Stealth] (3.5,-2.05) -- (5.25,-2.95) node[anchor=south]{$\chi_n$};
                \draw[blue, thick, -Stealth] (4.2,-1.75) -- (6.25,-2.35) node[anchor=west]{$\chi_n$};
                \draw[blue, thick, -Stealth] (4.55,-0.7) -- (6.75,-0.85) node[anchor=west]{$\chi_n$};
                \draw[blue, thick, -Stealth] (4.55,1.8) -- (6.75,2.55) node[anchor=west]{$\chi_n$};
                \draw[blue, thick, -Stealth] (4.2,2.85) -- (6.25,4.05) node[anchor=west]{$\chi_n$};
                \draw[blue, thick, -Stealth] (3.5,3.45) -- (5.25,5.05) node[anchor=west]{$\chi_n$};
                \draw[blue, thick, -Stealth] (3,3.775) -- (4.45,5.6) node[anchor=west]{$\chi_n$};
                \draw[blue, thick, -Stealth] (1.95,3.9) -- (2.95,5.85) node[anchor=west]{$\chi_n$};
                \draw[blue, thick, -Stealth] (-0.55,3.9) -- (-0.45,5.85) node[anchor=south]{$\chi_n$};
                \draw[blue, thick, -Stealth] (-1.6,3.775) -- (-1.95,5.6) node[anchor=south]{$\chi_n$};
                \draw[blue, thick, -Stealth] (-1.8,3.25) -- (-2.55,4.85) node[anchor=south]{$\chi_n$};
                \draw[blue, thick, -Stealth] (-2.7,2.35) -- (-3.75,3.65) node[anchor=north]{$\chi_n$};
                \draw[blue, thick, -Stealth] (-3.05,1.2) -- (-4.25,2.15) node[anchor=east]{$\chi_n$};
                \draw[blue, thick, -Stealth] (-3.05,-0.5) -- (-4.25,-0.45) node[anchor=east]{$\chi_n$};
                \draw[blue, thick, -Stealth] (-2.7,-1.55) -- (-3.75,-1.95) node[anchor=east]{$\chi_n$};
			\end{tikzpicture}
		} & 
		\makecell[l]{
			$\displaystyle \rho_m=\frac{(1+x_{1m}\chi_1)(1+x_{2m}\chi_2)(1+x_{3m}\chi_3)\dots(1+x_{nm}\chi_n)}{2^n},$ \\
			$\displaystyle m \in\{1,\dots,2^n\}$
		} \\
		\hline
	\end{tabular}
    }
    \captionof{table}{Shape function schematics for material interpolation}
	\label{SF_schematics}
    \end{minipage}
\end{center}
\noindent where $\chi_{i}$ is the $i^{th}$ design variable, $n$ is the total number of design variables, and $\rho_m$ is the density of material index $m$. The presented approach provides solutions with up to $2^n$ materials, including void, employing only $n$ design variables per element as discussed below. Additionally, the presence of one material implies the absence of another, that is (Kronecker delta property),
\begin{equation}
    \rho_m = 1,\quad \rho_l = 0; \quad \forall m \neq l, \quad m,l \in\{1,\dots,2^n\}
\end{equation}
Moreover, the sum of the presence and absence of material at any point equals 1, that is,  we have (barycentric property)
\begin{equation}\label{eq:Barycentric_properties}
	\begin{split}
		\sum_{m=1}^{2^n} \rho_m = 1,
	\end{split}
\end{equation}
where $2^n$ is the number of candidate materials (including the void).
One writes the material density functions using a 1D-material element with $\chi_1$ design variable as
\begin{equation}\label{eq:1DMD}
	 \rho_m = \frac{1+x_{1m}\chi_1}{2},\quad \, m = 1,\,2
\end{equation}
where $\rho_m$ is the material density function for the candidate material $m$ within an element with $x_{11} =-1$ and $x_{12} =1$.
Likewise, for a 2D-material element with $\chi_1$ and $\chi_2$ design variables, the material density functions $\rho_m$ (bilinear shape functions) are 
\begin{equation}\label{eq:2DMD}
\rho_m=\frac{(1+x_{1m}\chi_1)(1+x_{2m}\chi_2)}{2\times2},\quad \, m= 1,\,2,\,3,\,4
\end{equation}
where $(x_{11},x_{21})=(-1,-1), (x_{12},x_{22})=(1,-1)$, $(x_{13},x_{23})=(1,1), (x_{14},x_{24})=(-1,1)$. $(x_{1m},x_{2m})$ are the coordinates of the vertices or nodes of a square (for $m=4$) with side length of $2$ units whose centroid is located at the origin (Table~\ref{SF_schematics}). 
This material density function can be used to obtain structures with a 4-material phase. 
For a 3D-material element with $\chi_1$, $\chi_2$, and $\chi_3$ design variables, the material density functions $\rho_m$ (trilinear shape functions) are
\begin{equation}\label{eq:3DMD}
	\rho_m=\frac{(1+x_{1m}\chi_1)(1+x_{2m}\chi_2)(1+x_{3m}\chi_3)}{2\times 2\times 2},\quad \, m = 1,\,2,\,3,\,\cdots,2^3
\end{equation}
where $(x_{1m}, x_{2m}, x_{3m})$ denote the coordinates of the vertices of a cuboid with side length of 2 units (see Table~\ref{SF_schematics}), corresponding to the candidate materials or nodes, with the origin located at its centroid. One can achieve optimized structures up to an 8-material phase. Generalizing these shape functions further for higher dimensions, we write the material density $\rho_m$ for a 4D-material element with  $\chi_1$, $\chi_2$, $\chi_3$, and $\chi_4$ design variables as
\begin{equation}\label{eq:4DMD}
	\begin{aligned}
	\rho_m =\frac{(1+x_{1m}\chi_1)(1+x_{2m}\chi_2)(1+x_{3m}\chi_3)(1+x_{4m}\chi_4)}{2\times 2\times 2\times 2},\quad \, m = 1,\,2,\,3,\cdots,2^4
	\end{aligned}
\end{equation}
where $(x_{1m},x_{2m},x_{3m},x_{4m})$ represented the coordinate of the vertices of a 4D-element with $2$ units side length (Table \ref{SF_schematics}), corresponding to the candidate materials or nodes, with the origin located at its centroid. As we advance,  the generalized definition for the material density $\rho_m$ for an nD-material element can be written as

 \begin{equation}\label{eq:nDMD}
 \begin{aligned}
\rho_m = \frac{(1 + x_{1m}\chi_1)\cdots(1 + x_{nm}\chi_n)}{2^n}
= \prod_{i=1}^{n} \frac{(1 + x_{im}\chi_i)}{2}
= \frac{1}{2^n} \prod_{i=1}^{n} \left(1 + x_{im}\chi_i\right),\\
\quad i = 1,2,\dots,n; \quad m = 1,2,\dots,2^n.
 \end{aligned}
\end{equation}
where  $(x_{1m},x_{2m},x_{3m},\cdots, x_{nm})$ denote the vertices of an nD-element (nD-hypercube) with $2$ units side length, corresponding to the candidate materials, with centroid at $(0,\,0)$, that is, $x_{im} \in \{-1, 1\}$. Utilizing $n$-design variables gives results for $2^n$ material systems ($2^n$ number of candidate materials, including voids). Table~\ref{SF_schematics} depicts schematic diagrams of material density functions (shape functions) for 1D, 2D, 3D, 4D, and nD-material elements. The node numbers are shown in red and associated with red dots, and the axes' names with corresponding directional arrows are shown in blue. 

Based on the patterns of 1D, 2D, and 3D shape functions, a generalized shape function for nD-material elements, $n$-dimensional hypercube is proposed in Eq.~\ref{eq:nDMD}. We call it the ``generalized shape functions (gSF)" approach.

\subsection{Mathematical verification of barycentric properties of $\rho_m$}
	We demonstrate that the proposed $n$-linear shape functions, the material densities $\rho_m$, satisfy barycentric properties. This requires fulfilling two conditions: $0 \le \rho_m \le 1$ (non-negativity) and $\displaystyle\sum_{m=1}^{2^n} \rho_m = 1$ (partition of unity). Meeting these conditions ensures that the proposed formulation yields a physically valid multi-material interpolation space.

\subsubsection{Proof of bounded non-negativity}
	As discussed above, the design variables $\chi_{i}$ are strictly constrained within a bounded hypercube domain such that $\chi_{i} \in [-1, 1]$, and the vertex coordinates are defined as $x_{im} \in \{-1, 1\}$. Consequently, the linear component for any dimension $i$ is strictly non-negative and bounded as:
\begin{equation}
		0 \le \frac{1 + x_{im}\chi_{i}}{2} \le 1
	\end{equation}
	Because each $\rho_{m}$ is defined in Eq.~\ref{eq:nDMD} as a product of these individual bounded components, the product of strictly non-negative fractions must itself be non-negative and bounded by 1; thus, the condition $0 \le \rho_{m} \le 1$ is naturally satisfied.

\subsubsection{Proof of partition of unity property}
	Next, we prove that the sum of all available material densities at any given spatial location identically equals 1 (Eq.~\ref{eq:Barycentric_properties}). We use mathematical induction on the number of design variables $n$ to prove this property.

In view of Eq.~\ref{eq:nDMD},  Eq.~\ref{eq:Barycentric_properties} can be written as:
	\begin{equation}\label{Eq:MathInduct}
		\sum_{m=1}^{2^n} \rho_{m} = 	\sum_{m=1}^{2^n} \left( \frac{1}{2^n} \prod_{i=1}^{n} (1 + x_{im}\chi_{i}) \right).
	\end{equation}
	For the base case where $n=1$ (a single design variable), the framework provides $2^1 = 2$ material phases with vertex coordinates $x_{11} = 1$ and $x_{12} = -1$. For this case, Eq.~\ref{Eq:MathInduct} yields:
\begin{equation}
		\begin{split}
			\sum_{m=1}^{2^1} \rho_{m} =\frac{1}{2^1}\sum_{m=1}^{2^1}\prod_{i=1}^{1}(1+x_{im}\chi_i)
			&=\frac{1}{2}(1+x_{11}\chi_1)+\frac{1}{2}(1+x_{12}\chi_1) \\
			&=\frac{1}{2}(1+\chi_1)+\frac{1}{2}(1-\chi_1) = 1.
		\end{split}
	\end{equation}
	Next, we assume that the partition of unity holds true for $n=k$ design variables, meaning:
	\begin{equation}\label{Eq:MathInductk}
		\sum_{m=1}^{2^k} \rho_{m} = \sum_{m=1}^{2^k} \left( \frac{1}{2^k} \prod_{i=1}^{k} (1 + x_{im}\chi_{i}) \right) = 1.
	\end{equation}
	Now, considering a system with $n=k+1$ design variables, Eq.~\ref{Eq:MathInduct} transforms to:
	\begin{equation}\label{Eq:MathInductk+1}
		\sum_{m=1}^{2^{k+1}} {\rho}_{m} = 	\sum_{m=1}^{2^{k+1}} \left( \frac{1}{2^{k+1}} \prod_{i=1}^{k+1} (1 + x_{im}\chi_{i}) \right).
	\end{equation}
	We expand the LHS of Eq.~\ref{Eq:MathInductk+1} by splitting the single summation into two consecutive parts:
	\begin{equation}\label{Eq:MathInductk+1LHS}
		\sum_{m=1}^{2^{k+1}} {\rho}_{m} = \sum_{m=1}^{2^{k}} {\rho}_{m} + \sum_{m=2^{k}+1}^{2^{k+1}} {\rho}_{m}.
	\end{equation}
	To evaluate each part, we isolate the $(k+1)$-th component from the product rule of ${\rho}_{m}$ for $n=k+1$ variables:
	\begin{equation}\label{Eq:MathInductk+1LHSSingle}
		{\rho}_{m} = \prod_{i=1}^{k+1} \frac{1 + x_{im}{\chi}_{i}}{2} =\left( \prod_{i=1}^{k} \frac{1 + x_{im}{\chi}_{i}}{2} \right) \cdot \left( \frac{1 + x_{k+1,m}{\chi}_{k+1}}{2} \right).
	\end{equation}
	Substituting Eq.~\ref{Eq:MathInductk+1LHSSingle} into Eq.~\ref{Eq:MathInductk+1LHS} yields:
	\begin{equation}\label{Eq:MathInductk+1LHSRHS}
		\sum_{m=1}^{2^{k+1}} {\rho}_{m} = 	\sum_{m=1}^{2^{k}} \left( \prod_{i=1}^{k} \frac{1 + x_{im}{\chi}_{i}}{2} \cdot \frac{1 + x_{k+1,m}{\chi}_{k+1}}{2} \right) + 	\sum_{m=2^{k}+1}^{2^{k+1}} \left( \prod_{i=1}^{k} \frac{1 + x_{im}{\chi}_{i}}{2} \cdot \frac{1 + x_{k+1,m}{\chi}_{k+1}}{2} \right).
	\end{equation}
	Due to the binary symmetry of the hypercube, the index limits of the second summation block can be shifted from $[2^k + 1 \to 2^{k+1}]$ back to $[1 \to 2^k]$ without altering the values of the first $k$ product terms. Additionally, the coordinate signs for the new dimension are fixed at $x_{k+1,m} = 1$ for the first block and $x_{k+1,m} = -1$ for the second block. Applying these substitutions simplifies the expression to:
	\begin{equation}\label{Eq:MathInductk+1LHSRHSmod}
		\begin{split}
			\sum_{m=1}^{2^{k+1}} {\rho}_{m} = &	\sum_{m=1}^{2^{k}} \left( \prod_{i=1}^{k} \frac{1 + x_{im}{\chi}_{i}}{2} \cdot \frac{1 + (1){\chi}_{k+1}}{2} \right) + 		\sum_{m=1}^{2^{k}} \left( \prod_{i=1}^{k} \frac{1 + x_{im}{\chi}_{i}}{2} \cdot \frac{1 + (-1){\chi}_{k+1}}{2} \right)\\
			=& \frac{1 + \chi_{k+1}}{2} \underbrace{\sum_{m=1}^{2^k} \left( \frac{1}{2^k} \prod_{i=1}^{k} (1 + x_{im}\chi_{i}) \right)}_{=1 \text{ (by induction assumption, Eq.~\ref{Eq:MathInductk})}} + \frac{1 - \chi_{k+1}}{2} \underbrace{\sum_{m=1}^{2^k} \left( \frac{1}{2^k} \prod_{i=1}^{k} (1 + x_{im}\chi_{i}) \right)}_{=1 \text{ (by induction assumption, Eq.~\ref{Eq:MathInductk})}}\\
			=&\left( \frac{1 + \chi_{k+1}}{2} \right) (1) + \left( \frac{1 - \chi_{k+1}}{2} \right) (1) = 1			
		\end{split}
	\end{equation}
	By mathematical induction, the partition of unity property is shown to be satisfied for any arbitrary number of design variables $n$. Therefore, the proposed material density fields are mathematically guaranteed to retain consistent barycentric properties across the entire optimization domain. As mentioned above, the gSF approach provide $2^n$ material phases while using only $n-$design variables, a relationship between the proposed gSF approach material space and classical material simplex is discussed next.

\subsection{Relationship between the classical material simplex and gSF approach material space}
The classical multimaterial TO method (extended-SIMP/DMO) typically operates within an $(2^n-1)$-dimensional linear simplex $\mathcal{S}$, where
	\begin{equation}
		\mathcal{S} = \left\{ \boldsymbol{\rho} \in \mathbb{R}^{2^n} \middle|\; \sum_{m=1}^{2^n} \rho_m = 1, \rho_m \geq 0 \right\}.
	\end{equation}
	The proposed gSF approach maps $n-$dimensional hypercube of design variables $\mathcal{H} = [-1,\, 1]^n$ into $\mathcal{S}$ by $n-$linear shape functions. That is, the gSF defines a multilinear manifold
\begin{equation}
\mathcal{M}
=
\left\{
\boldsymbol{\rho}\in\mathbb{R}^{2^n}
\;\middle|\;
\rho_m
=
\frac{1}{2^n}
\prod_{i=1}^{n}
\left(1+x_{im}\chi_i\right),
\;
\boldsymbol{\chi}\in\mathcal{H},
\;
x_{im}\in\{-1,1\},
\;
m=1,\ldots,2^n
\right\}.
\label{eq:manifold}
\end{equation}
embedded within $\mathcal{S}$. While this manifold is a lower-dimensional subset (for $n>1$), we show in Appendix~\ref{AppA} that the mapping is surjective onto the set of pure material states, i.e., the vertices of the simplex. In an MMTO, the main goal is to obtain a discrete, well-defined distribution of (close to) pure material states. Elements with intermediate material states indicate either material mixing or artificial states, both of which are undesirable. In a density-based TO, the design variables corresponding to the optimized designs must reside at the extreme points (vertices) of the material simplex, say, denoted by $\text{ext}(\mathcal{S})$. The proposed gSF $n$-linear mapping functions $\Phi: \mathcal{H} \to \mathcal{S}$ are formulated such that the set of vertices of the hypercube $\mathcal{H} \in [-1, 1]^n$ maps bijectively onto the set of pure material vertices of the simplex $\mathcal{S}$, mathematically ensuring that $\text{ext}(\mathcal{S}) \subset \mathcal{M}$. Because the resulting embedded manifold $\mathcal{M}$ is vertex-complete for original, filtered and projected design variables\footnote{The latter two also satisfy the barycentric properties as shown in Sec.~\ref{Sec:EFilMd} and Sec.~\ref{Sec:EprojMd}}, the optimizer is not restricted from reaching any pure material phase or any discrete combination thereof. Consequently, all the superior results, in which the design variables must approach the vertices, are within the perfect reach of the proposed gSF. Therefore, the proposed approach does not preclude rather steers towards the achievement of superior solutions.

We use a density filter to implicitly control the minimum member size, subdue checkerboard patterns (Sec.~\ref{subsec8}), and avoid point connections (Sec.~\ref{subsec8}) in the optimized designs. A modified projection filter is implemented to achieve optimized, crisp solutions. Next, these implementations are discussed in the gSF framework while maintaining the desired barycentric properties of the material density functions.

\subsection{Evaluation of filtered material density}\label{Sec:EFilMd}
 Using the density filtering, the filtered material density can be determined  in a continuum setting as 
\begin{equation}\label{eq:FilteredMD}
     \Tilde{\rho}_m (\bm{x})= \frac{\int_{\mathbb{N}(\bm{x})} w(\bm{x},\,\bm{y}) \rho_m(\bm{y})  dy} {\int_{\mathbb{N}(\bm{x})}  w(\bm{x},\,\bm{y})  dy}
\end{equation}
where $\mathbb{N}(\bm{x})$ indicates the chosen neighboring region for $\bm{x}$. The filter domain is typically circular for 2D problems and spherical for 3D problems. $w(\bm{x},\,\bm{y})$ ($> 0$) is a weight function~\cite{bruns2001topology}. As $0\le \rho_m \le 1$, in view of Eq.~\ref{eq:FilteredMD}, $0\le \Tilde{\rho}_m \le 1$. Now, to access the barycentric property of the filtered material density,  one writes
\begin{equation}\label{eq:filteredDensity}
   \begin{split}
       \sum_{m=1}^{2^n} \Tilde{\rho}_m =  \sum_{m=1}^{2^n} \left(\frac{\int_{\mathbb{N}(\bm{x})} \rho_m  w(\bm{x},\,\bm{y})  dy} {\int_{\mathbb{N}(\bm{x})}  w(\bm{x},\,\bm{y})  dy}\right) =   \frac{\int_{\mathbb{N}(\bm{x}) } \left(\sum_{m=1}^{2^n} \rho_m \right) w(\bm{x},\,\bm{y})  dy} {\int_{\mathbb{N}(\bm{x})} w(\bm{x},\,\bm{y}) dy}= \frac{\int_{\mathbb{N}(\bm{x})}  w(\bm{x},\,\bm{y}) dy} {\int_{\mathbb{N}(\bm{x})} w(\bm{x},\,\bm{y}) dy} =1
   \end{split} 
\end{equation}

 This implies that filtered material density functions satisfy the barycentric property. Furthermore, Eq.~\ref{eq:filteredDensity} is true for any generic material density and weight functions~\cite{singh2023projected}. As $\chi_i|_{i =1,\,2,\,\cdots,n}$ (Eq.~\ref{Eq: desingvaraible}) are the design variables, one can determine the corresponding filtered $\Tilde{\chi}_i$; thus
\begin{equation}\label{eq:filteredgeneralization}
\begin{split}
    \tilde{\rho}_m=\frac{(1+x_{1m}\tilde{\chi}_1) (1+x_{2m}\tilde{\chi}_2)\dots(1+x_{nm}\tilde{\chi}_n)}{2^n},\quad \, m = 1,\,2,\,3,\,\cdots,\, 2^n
\end{split}
\end{equation}
Consider a 2D case for simplicity to demonstrate the determination of $\tilde{\chi}_1$ and $\tilde{\chi}_2$, hence the generalization. Using Eq.~\ref{eq:filteredgeneralization}, one writes for 2D case as
\begin{equation}\label{eq:filteredgeneralization2D}
    \tilde{\rho}_m=\frac{(1+x_{1m}\tilde{\chi}_1) (1+x_{2m}\tilde{\chi}_2)}{2^2},\quad \, m = 1,\,2,\,3,\,4
\end{equation}
We get the following in the context of Eq.~\ref{eq:filteredgeneralization2D}.
\begin{equation}\label{eq:seperateFilter2D}
    \tilde{\chi}_1  = 2(\tilde{\rho}_2 + \tilde{\rho}_3) -1,\quad
     \tilde{\chi}_2  = 2(\tilde{\rho}_3 + \tilde{\rho}_4) -1
\end{equation}
Now, using Eq.~\ref{eq:2DMD}, Eq.~\ref{eq:FilteredMD}, and Eq.~\ref{eq:seperateFilter2D}, we get
\begin{equation}
	\label{eq:FilteredDV2D}
		\tilde{\chi}_1(\bm{x})
        =
		\frac{\displaystyle \int_{\mathbb{N}(\bm{x})}
			w(\bm{x},\bm{y})\,\chi_1(\bm{y})\, d\bm{y}}
		{\displaystyle \int_{\mathbb{N}(\bm{x})}
			w(\bm{x},\bm{y})\, d\bm{y}},
        \quad \quad 
		\tilde{\chi}_2(\bm{x})
        =
		\frac{\displaystyle \int_{\mathbb{N}(\bm{x})}
			w(\bm{x},\bm{y})\,\chi_2(\bm{y})\, d\bm{y}}
		{\displaystyle \int_{\mathbb{N}(\bm{x})}
			w(\bm{x},\bm{y})\, d\bm{y}}.
\end{equation}
Likewise, for $nD$ cases, one determines the filtered design variables as
\begin{equation}\label{eq:nDFilteredDVs}
     \Tilde{\chi}_n (\bm{x})= \frac{\int_{\mathbb{N}(\bm{x})} w(\bm{x},\,\bm{y}) \chi_n(\bm{y})  dy} {\int_{\mathbb{N}(\bm{x})}  w(\bm{x},\,\bm{y})  dy}
\end{equation}
 In view of the above equations, $-1 \le\chi_n (\bm{x})\le 1$ implies $-1 \le\Tilde{\chi}_n (\bm{x})\le 1$. In a discrete setting, to associate the design variables with their respective elements, we introduce two indices for clarity: the first indicates the position of the design variable, and the second the element number. Now, the filtered design variables are defined as follows 
\begin{equation} \label{Filtering}
	\tilde{\chi}_{ij} = \frac{\sum_{k \in N_e} \hat{H}_{jk} \chi_{ik}}{\sum_{k \in N_e} \hat{H}_{jk}},
\end{equation}
where $\tilde{\chi}_{ij}$ denotes the $i$\textsuperscript{th} filtered design variable for element $j$, and $\chi_{ik}$ represents the $i$\textsuperscript{th} design variable for element $k$. $N_e$ is the set of elements $k$ for which the center-to-center distance $\text{Dis}(j,k)$ to element $j$ is less than or equal to the filter radius $r_{\text{min}}$. The weight factor $\hat{H}_{jk}$ is determined as:
\begin{equation} \label{Weight_factor}
	\hat{H}_{jk} = \max({0,r_\text{min}-\text{Dis}(j,k)}).
\end{equation}
In the matrix form, Eq.~\ref{Filtering} transpires to 
\begin{equation} \label{Filtering_matrix}
	\tilde{\bm{\chi}} = \mathbf{H} \bm{\chi},
\end{equation}
where $\tilde{\bm{\chi}}$ and $\bm{\chi}$ are the filtered and actual design variable vectors, respectively. $\mathbf{H}$ denotes the constant filter matrix, which is determined once before the optimization process for the discretized design domain.

Although the filtered implementation circumvents geometrical singularities in TO, the optimized solutions contain gray elements~\cite{bruns2001topology}. Therefore, we apply a projection filter to obtain solutions close to 0-1, as discussed next.
\subsection{Projected material density evaluation} \label{Sec:EprojMd}
The projection filter is typically employed to achieve solutions close to 0-1, i.e., optimized results with distinct boundaries. Since $-1 \le\Tilde{\chi}_{ij} (\bm{x})\le 1$, we modify the projection filter as
\begin{equation} \label{Projection}
	\bar{\chi}_{ij}=2\frac{\tanh \beta_p\eta_p+\tanh \beta_p((\tilde{\chi}_{ij}+1)/2-\eta_p)}{\tanh \beta_p\eta_p+\tanh \beta_p(1-\eta_p)}-1,\quad (i=1,2,...,n)
\end{equation}
where $\beta_p$ and $\eta_p$ control the design's discreteness and smoothness. $\bar{\chi}_{ij}$ denotes the $i$\textsuperscript{th} projected filtered design variable for element $j$, whereas $\tilde{\chi}_{ij}$ is its corresponding filtered design variable. Eq.~\ref{Projection} provides $-1 \le\bar{\chi}_{ij} (\bm{x})\le 1$, therefore the projected design variable, $0\le\bar{\rho}_{mj}\le 1$, is determined:
\begin{equation} \label{Projected_filtered_density}
\bar{\rho}_{mj}=\prod_{i=1}^{n}\frac{(1+x_{im}\bar{\chi}_{ij})}{2},
\end{equation}
for material $m \in\{1,...,2^n\}$ within an element $j$ for $n$ numbers of design variables. Note that in context of Eq.~\ref{Projected_filtered_density}, $0\le\bar{\rho}_{mj}\le 1$ and $-1 \le\bar{\chi}_{ij} (\bm{x})\le 1$ (Eq.~\ref{Projection}), the projected design variables $\bar{\rho}_{mj}$ are also barycentric. 
With the projected design variables  $-1\le\bar{\chi}_n\le 1$, the determined $\bar{\rho}_m$ in Eq.~\ref{Projected_filtered_density} is analogous to ${\rho}_m$ (Eq.~\ref{eq:nDMD}); thus,  $\bar{\rho}_m$ satisfy the essential barycentric properties. Therefore, because the original, filtered, and projected design variables are strictly confined to the $[-1, 1]^n$ hypercube, the resulting material densities automatically satisfy the required barycentric properties across all mathematical states of the gSF approach. As the projection parameter is progressively increased using a continuation strategy during the optimization process, the variables $\bar{\chi}_1, \bar{\chi}_2, \dots, \bar{\chi}_n$ are steered toward the vertices of the hypercube, providing solutions close to 0-1.

To check how close the solution is to `0-1' (binary), we use the measure of non-discreteness $(M\textsubscript{nd})$. Mathematically
\begin{equation}\label{Eq:Mnd_gSF}
    {M_\text{nd}} =\frac{\displaystyle \sum_{j=1}^{N_e} 4\bar{\rho}_{ij}^m\left(1 - \bar{\rho}_{ij}^m\right)}{N_e}\times 100\%,
\end{equation}
where $\bar{\rho}_{ij}^m=\sum_{i=1}^{m} \bar{\rho}_{ij}$, $m$ is the number of materials used, $N_e$ is the number of discretized elements, and $\bar{\rho}_{ij}$ is the physical  density of $i$\textsuperscript{th} material of the $j$\textsuperscript{th} element. The closer its value is to 0, the closer the solution is to a binary.

  \section{Problem Formulation}\label{sec3}
This section presents the material interpolation technique used, a general problem statement for the problems presented, and the sensitivity analyses for the objectives and constraints. 

\subsection{Material Interpolation}\label{subsec1} 
The most commonly used material interpolation function in TO is the SIMP approach. Herein, we use the same; however, it is modified to cater to the proposed gSF approach. $E_j$ of element $j$ is determined as :
\begin{equation} \label{Material_interpolation}
	E_j = E_\text{min}+\sum_{m=1}^{2^n} (\bar{\rho}_{mj})^p(E_m-E_\text{min}).
\end{equation}
 $E_m$ represents the elastic stiffness of material $m$. As mentioned above, the total number of candidate materials is $2^n$ (including voids), where $n$ is the number of design variables. $E_\text{min}$ refers to a very small stiffness assigned to the void regions/elements to avoid singularity in the stiffness matrix. The penalization factor, $p$, is set to 3. $\bar{\rho}_{mj}$ denotes the projected filtered density corresponding to candidate material $m$ of element $j$, determined using the aforementioned filtering and projection techniques.

\subsection{Problem Statement}\label{subsec4}
We solve two different TO problems related to stiff structure and compliant mechanism with many materials to demonstrate the versatility and effectiveness of the presented gSF approach. For SS, MBB beam and cantilever beams are solved, whereas an inverter and a gripper are taken for CMs. Both 2D and 3D examples are solved. One writes the generic optimization formulation with volume constraints as:
\begin{equation} \label{Optimization_eqn.} 
	\begin{rcases}
		\begin{split}
			& \min:\quad f_0 \\
			&\text{subjected to:}\\
			&\bm{\lambda}:\,\,\mathbf{K} \mathbf{U} - \mathbf{F} = \mathbf{0}\\
			&\bm{\mu}:\,\,g_m=\frac{\sum_{j=1}^{N_e}v_{j}\bar{\rho}_{mj}}{{V}_{m}}-1 \le 0,\,m = 1,\,2,\,3,\,\cdots,\,2^n-1 \\
			&\quad\,\,\,\, -1 \leq \chi_{1j},...,\chi_{nj} \leq 1 \quad(j=1,\,2,...,\,N_e)
		\end{split}
	\end{rcases},
\end{equation}
where $f_0$ is the objective function. $\mathbf{U}$, $\mathbf{F}$, and  $\mathbf{K}$ are the global displacement vector, force vector, and stiffness matrix, respectively. $g_m$ denotes the inequality constraint associated with the volume of material $m$. ${V}_{m}$  and  $\sum_{j=1}^{N_e}v_{j}\bar{\rho}_{mj}$ are the permitted and current volumes for the material $m$ assigned for the designs, respectively; where $v_j$ represents the volume of element $j$. For the $m$ number of materials, there are $m$ volume constraints. $\bm{\lambda}$ (a vector) and $\mu$ (a vector) are the Lagrange multipliers corresponding to the state equation and the volume constraints, respectively. $\chi_{1j},...,\chi_{nj}$ are the design variables of the $j$\textsuperscript{th} element. $N_e$ is the total number of FEs used to discretize the domain.

\subsection{Sensitivity analysis}\label{subsec5}
 Herein, we use the method of moving asymptotes (MMA, cf.~\cite{svanberg1987method}), a gradient-based optimizer, for updating the design variables. Therefore, the gradients of the objective and the constraints are required, which are determined as follows.

The sensitivity of the objective function in an element is evaluated as :
\begin{equation} \label{Eq:Sensitivity} 
	\frac{\partial f_0}{\partial \chi_{ik}}=\sum_{j=1}^{N_e}\left(\frac{\partial f_0}{\partial E_{j}}\right)\left(\frac{\partial E_{j}}{\partial \chi_{ik}}\right),\quad (i=1,2,...,n;\,\,k,j=1,2,...,N_e)
\end{equation}
where, $E_{j}$ is the interpolated Young's modulus of $j$\textsuperscript{th} element. $\chi_{ik}$ represents the $i$\textsuperscript{th} design variable of $k$\textsuperscript{th} element. The term $\left(\frac{\partial E_{j}}{\partial \chi_{ik}}\right)$ can be calculated using Eq.~\ref{Material_interpolation} as follows :
\begin{equation} \label{Sensitivity1} 
	\frac{\partial E_{j}}{\partial \chi_{ik}}=\sum_{m=1}^{2^n}p(\bar{\rho}_{mj})^{p-1}(E_m-E_\text{min})\left(\frac{\partial \bar{\rho}_{mj}}{\partial \chi_{ik}}\right)
\end{equation}
Now, we employ the chain-rule to find $\left(\frac{\partial \bar{\rho}_{mj}}{\partial \chi_{ik}}\right)$ as:
\begin{equation} \label{Sensitivity2} 
	\frac{\partial \bar{\rho}_{mj}}{\partial \chi_{ik}}=\sum_{l=1}^{N_e}\left(\frac{\partial \bar{\rho}_{mj}}{\partial \bar{\chi}_{il}}\right)\left(\frac{\partial \bar{\chi}_{il}}{\partial \chi_{ik}}\right)=\left(\frac{\partial \bar{\rho}_{mj}}{\partial \bar{\chi}_{ij}}\right)\left(\frac{\partial \bar{\chi}_{ij}}{\partial \chi_{ik}}\right)
\end{equation}
Using Eq.~\ref{Projected_filtered_density}, we can write the term $\left(\frac{\partial \bar{\rho}_{mj}}{\partial \bar{\chi}_{ij}}\right)$ as :
\begin{equation} \label{Sensitivity3} 
	\frac{\partial \bar{\rho}_{mj}}{\partial \bar{\chi}_{ij}}=\frac{x_{im}}{2}\prod_{r=1,r\ne i}^{n}\frac{(1+x_{rm}\bar{\chi}_{rj})}{2}
\end{equation}
Note that while summation (Eq.~\ref{Sensitivity1}) contains $2^n$ terms, the partial derivative of $\frac{\partial \bar{\rho}_{mj}}{\partial \bar{\chi}_{ij}}$ (Eq.~\ref{Sensitivity3}) for each material is a product of the other $(n-1)$ coordinates. This ensures that the gradient does not vanish or explode as $n$ increases, since the natural coordinates are bounded by [-1, 1]. Additionally, during the optimization process the normalization of the objective function; thus, sensitivity are performed using $n_f = \min\left({\frac{10}{\text{obj}},\,\num{1e2}}\right)|_{\mathtt{loop} =1}$, wherein $\mathtt{loop}$ indicates the MMA iteration number (see Sec.~\ref{sec4}).
Also, using the chain-rule $\left(\frac{\partial \bar{\chi}_{ij}}{\partial \chi_{ik}}\right)$ is  calculated as :
\begin{equation} \label{Sensitivity4} 
	\frac{\partial \bar{\chi}_{ij}}{\partial \chi_{ik}}=\sum_{l=1}^{N_e}\left(\frac{\partial \bar{\chi}_{ij}}{\partial \tilde{\chi}_{il}}\right)\left(\frac{\partial \tilde{\chi}_{il}}{\partial \chi_{ik}}\right)=\left(\frac{\partial \bar{\chi}_{ij}}{\partial \tilde{\chi}_{ij}}\right)\left(\frac{\partial \tilde{\chi}_{ij}}{\partial \chi_{ik}}\right)
\end{equation}
Now, $\left(\frac{\partial \bar{\chi}_{ij}}{\partial \tilde{\chi}_{ij}}\right)$ is derived from Eq.~\ref{Projection} and written as :
\begin{equation} \label{Projection_elemental} 
	\frac{\partial \bar{\chi}_{ij}}{\partial \tilde{\chi}_{ij}}=\beta_p\frac{1-\tanh^2 \beta_p((\tilde{\chi}_{ij}+1)/2-\eta_p)}{\tanh \beta_p\eta_p+\tanh \beta_p(1-\eta_p)}
\end{equation}
From Eq.~\ref{Filtering_matrix}, we evaluate $\left(\frac{\partial \tilde{\chi}_{ij}}{\partial \chi_{ik}}\right)$ as
\begin{equation} \label{Filter_elemental} 
	\frac{\partial \tilde{\chi}_{ij}}{\partial \chi_{ik}}=H_{jk}
\end{equation}
where $H_{jk}$ is the element at $ j$\textsuperscript {th} row and $ k$\textsuperscript {th} column of constant filter matrix $\mathbf{H}$, which is evaluated only once prior to the optimization process. We use the adjoint-variable method to find $\frac{\partial f_0}{\partial E_{j}}$ for SS and CMs as follows, as per their objective functions used.

\paragraph{\textbf{Stiff Structures}:}
The objective function for the stiff structures is $f_0=\frac{1}{2}\mathbf{U}^T\mathbf{KU}$ (strain energy). The augmented performance function using the objective function and its equality constraint is written as~\cite{kumar2023honeytop90}:
\begin{equation} \label{APF_SS} 
	\mathscr{L}=\frac{1}{2}\mathbf{U}^T\mathbf{KU} + \bm{\lambda}^T (\mathbf{KU-F})
\end{equation}
The derivative of this function with respect to $E_j$ yields

\begin{align} \label{Sensitivity_Lagrangian} 
    \begin{split}
        \frac{\partial \mathscr{L}}{\partial E_{j}} = \left(\mathbf{U}^T + \bm{\lambda}^T \right)\mathbf{K}\left(\frac{\partial \mathbf{U}}{\partial E_{j}}\right) + \left(\frac{1}{2}\mathbf{U}^T + \bm{\lambda}^T \right)\left(\frac{\partial \mathbf{K}}{\partial E_{j}}\right)\mathbf{U}, \, \quad \quad \left[\text{as} \, \, \frac{\partial \mathbf F}{\partial E_{j}} = 0\right]
    \end{split}
\end{align}
Using the principles of the adjoint-variable method, we get $\bm{\lambda}^T = -\mathbf{U}^T$. Therefore, the value of $\frac{\partial f_0}{\partial E_{j}}$ is finally be written as :
\begin{equation} \label{Eq:Sensitivity_MBB_Beam} 
	\frac{\partial f_0}{\partial E_{j}} = \frac{\partial \mathscr{L}}{\partial E_{j}} = -\frac{1}{2}\mathbf{U}^T\left(\frac{\partial \mathbf{K}}{\partial E_{j}}\right)\mathbf{U}
\end{equation}

\paragraph{\textbf{Compliant Mechanisms}:}
The objective function for the compliant mechanisms is taken as $f_0=-\alpha \frac{\mathbf{L}^T\mathbf{U}}{\mathbf{F}^T\mathbf{U}}$ with a user-defined scaling factor $\alpha$. Like the SS case, we write the augmented performance function  as :
\begin{equation} \label{APF_CM} 
	\mathscr{L}=-\alpha \frac{\mathbf{L}^T\mathbf{U}}{\mathbf{F}^T\mathbf{U}} + \bm{\lambda}_1^T (\mathbf{KU-F}) 
\end{equation}
The derivative of this function with respect to $E_j$ yields

\begin{align} \label{Eq:Sensitivity_Lagrangian_CM} 
    \begin{split}
        \frac{\partial \mathscr{L}}{\partial E_{j}} = \left(-\frac{\alpha}{\mathbf{F}^T\mathbf{U}} \mathbf{L}^T\mathbf{K}^{-1} - \frac{f_0}{\mathbf{F}^T\mathbf{U}}\mathbf{U}^T + \bm{\lambda}_1^T \right)\mathbf{K}\frac{\partial \mathbf{U}}{\partial E_{j}} + \bm{\lambda}_1^T \frac{\partial \mathbf{K}}{\partial E_{j}} \mathbf{U} 
    \end{split}
\end{align}
Using the principles of adjoint-variable method, we get $\bm{\lambda}_1^T = \left(\frac{\alpha}{\mathbf{F}^T\mathbf{U}} \mathbf{L}^T \mathbf{K}^{-1} + \frac{f_0}{\mathbf{F}^T\mathbf{U}}\mathbf{U}^T \right)$. Substituting $\bm{\lambda}_1$ in Eq.~\ref{Eq:Sensitivity_Lagrangian_CM}, we get 
\begin{equation} \label{Sensitivity_CM} 
	\frac{\partial f_0}{\partial E_{j}} = \frac{\partial \mathscr{L}}{\partial E_{j}} = \left(\alpha \mathbf{L}^T\mathbf{K}^{-1}\left(\frac{\partial \mathbf{K}}{\partial E_{j}}\right)\mathbf{U}+f_0 \mathbf{U}^T\left(\frac{\partial \mathbf{K}}{\partial E_{j}}\right)\mathbf{U}\right)\frac{1}{\mathbf{F}^T\mathbf{U}}
\end{equation}
Therefore, $\frac{\partial f_0}{\partial \chi_{ik}}$ is evaluated using the above expressions. Next, the gradients of the volume constraints (i.e. constraint sensitivities) are calculated from Eq.~\ref{Optimization_eqn.} as :
\begin{equation} \label{Constraint_sensitivity} 
	\frac{\partial g_m}{\partial \chi_{ik}}=\frac{\sum_{j=1}^{N_e}v_j\frac{\partial \bar{\rho}_{mj}}{\partial \chi_{ik}}}{{V}_{m}}.
\end{equation}

\begin{figure}
\centering
\begin{subfigure}{0.4\textwidth}
\begin{tikzpicture}[scale=0.5]
			\makeatletter
			\pgfdeclarepatternformonly[\LineSpace,\LineThickness]{custom hatch}
			{\pgfqpoint{0pt}{0pt}}{\pgfqpoint{\LineSpace}{\LineSpace}}
			{\pgfqpoint{\LineSpace}{\LineSpace}}
			{
				\pgfsetlinewidth{\LineThickness}
				\pgfpathmoveto{\pgfqpoint{0pt}{0pt}}
				\pgfpathlineto{\pgfqpoint{\LineSpace}{\LineSpace}}
				\pgfusepath{stroke}
			}
			\makeatother
			
			\tikzset{
				LineSpace/.store in=\LineSpace,
				LineSpace=2pt,
				LineThickness/.store in=\LineThickness,
				LineThickness=0.4pt
			}
			
			\fill[pattern=custom hatch] (8.7,-0.6) rectangle (9.3,-0.4);
			\fill[pattern=custom hatch] (-0.3,-0.6) rectangle (0.3,-0.4);
			\draw[color=black, very thick](9.0,-0.2) circle (0.2);
			\fill [gray!50] (0,0) rectangle (9,3);
			\draw[black, very thick] (0,0) -- (9,0);
			\draw[black, very thick] (9,0) -- (9,3);
			\draw[black, very thick] (9,3) -- (0,3);
			\draw[black, very thick] (0,3) -- (0,0);
			\draw[black, very thick] (8.7,-0.4) -- (9.3,-0.4);
			\filldraw[black] (0,0) circle (0.05);
			\draw[thick] (-0.3,-0.4) -- (0,0) -- (0.3,-0.4) -- cycle;
			\dimline[color=blue,extension start length=5pt, extension end length=5pt,label style={fill=none, yshift=6pt}]{(4.5,3.6)}{(9,3.6)}{$L_x/2$};
			\dimline[color=blue,extension start length=2pt, extension end length=2pt,label style={fill=none, yshift=-6pt}]{(0,-0.8)}{(9,-0.8)}{$L_x$};
			\draw[red, very thick, -Stealth] (4.5,4.4) -- (4.5,3);
			\dimline[color=blue,extension start length=2pt, extension end length=2pt,label style={fill=none, yshift=6pt}]{(9.4,3)}{(9.4,0)}{$L_y$};
			\node at (4.1,3.8) {$F$};	
		\end{tikzpicture}
        \caption{MBB beam}
		\label{MBB_Beam_2D}
\end{subfigure}
\begin{subfigure}{0.4\textwidth}
\centering
\begin{tikzpicture}[scale=0.75]			
			\tikzset{
				LineSpace/.store in=\LineSpace,
				LineSpace=2pt,
				LineThickness/.store in=\LineThickness,
				LineThickness=0.4pt
			}
			
			\fill[pattern=custom hatch] (-0.2,0) rectangle (0,3);
			\fill [gray!50] (0,0) rectangle (6,3);
			\draw[black, very thick] (0,0) -- (6,0);
			\draw[black, very thick] (6,0) -- (6,3);
			\draw[black, very thick] (6,3) -- (0,3);
			\draw[black, very thick] (0,3) -- (0,0);
			\dimline[color=blue,extension start length=2pt, extension end length=2pt,label style={fill=none, yshift=-6pt}]{(0,-0.4)}{(6,-0.4)}{$L_x$};
			\draw[red, very thick, -Stealth] (6,0) -- (6,-1.2);
			\dimline[color=blue,extension start length=2pt, extension end length=2pt,label style={fill=none, yshift=6pt}]{(6.3,3)}{(6.3,0)}{$L_y$};
			\node at (6.5,-1) {$F$};	
		\end{tikzpicture}
        \caption{Cantilever}
		\label{Cantilever_2D}
\end{subfigure}
\caption{2D design domain of (a) an MBB beam, and (b) a cantilever}
\label{Structure_design_domain}
\end{figure}
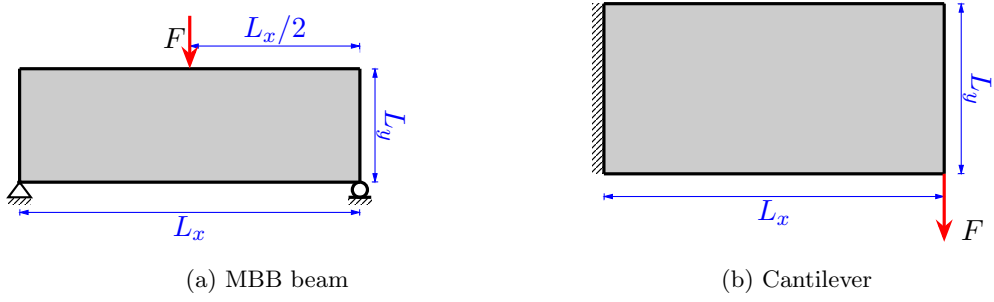

\section{Numerical examples and discussion}\label{sec4}
In this section, we present two categories of problems: (a) SS, (b) CMs.  We chose the MBB beam problem to solve 2D and 3D structures, and the cantilever problem to solve a 2D structure for the SS demonstration. On the other hand, 2D and 3D inverter problems and 2D gripper problems are optimized for CMs. The objectives and corresponding sensitivity analyses for these problems are mentioned above.  We consider the Young's modulus (E) to be a normalized value of $1$ unit for the material with the highest Young's modulus for corresponding SFs (i.e., 1D, 2D, 3D, 4D, 5D). The maximum number of MMA iterations is set to 400. The filtering and projection parameters for the problems are mentioned in the specific problem definition section. We use bilinear quadrilateral FEs to parameterize the design domains. The parameters (normalized Young's moduli, volume fractions, and color codes) mentioned in Table~\ref{3Mtable}, Table~\ref{7Mtable}, Table~\ref{15Mtable}, and Table~\ref{24Mtable} are used in this paper for 2D, 3D, 4D, and 5D SFs, respectively, unless otherwise stated. We normalize the objective, and thus sensitivity using a multiplication factor in every iteration, $n_f =\left.\min \left(\frac{10}{\mathrm{obj}},1 \times 10^2\right)\right|_{\mathrm{loop}=1}$, wherein $\mathtt{loop}$ indicates MMA iteration number.

\subsection{Stiff structure problems}\label{subsec6}
A well-known optimization problem in TO is the Messerschmitt-Bolkow-Blohm (MBB) beam. This problem falls into the category of a simply supported beam with a point load (for 2D, cf. Fig.~\ref{MBB_Beam_2D}) or a distributed load (for 3D, cf. Fig.~\ref{MBB_Beam_3D}). Another example is the optimization of a cantilever beam. Fig.~\ref{MBB_Beam_2D} illustrates the 2D MBB beam subjected to a downward point load applied at the midpoint of its upper edge. Fig.~\ref{Cantilever_2D} represents a 2D cantilever beam subjected to a downward point load applied at the free end of its bottom-right corner. These optimization problems aim to determine the optimal material distribution while restricting the total usage of materials and minimizing the strain energy of the system. 

The dimensions in \(x\) and \(y\)  directions are indicated via  \(L_x\) and \(L_y\), respectively. For the 2D MBB beam, \(L_x:L_y = 6:1\), whereas \(L_x:L_y = 2:1\) is taken for the 2D cantilever beam. An external force (\(F\)) of magnitude $10^{-3}$ unit is applied. 

\subsubsection{MBB Beam}\label{subsec8}
The 2D MBB beam problem is solved here. The domain is parameterized using \(N_{ex}\)$=180$ FEs and \(N_{ey}\)$=30$ FEs, where \(N_{ex}\) and \(N_{ey}\) indicate the number of elements in $x$ and $y$ directions, respectively. The right symmetric half section is optimized, and for visual representation, we show the full optimized design of the MBB beam from 4 material onward. The filter radius, $r_{\text{min}}$ is set to 3.6.  $\eta_p=0.5$ is taken. $\beta_p$ starts from 1 and is doubled every 75 iterations until it reaches 32.\\
\textbf{Qualifying the use of density and introduced projection filters in gSF approach:} 

Here to qualify the use of density and the introduced projection filters within the proposed gSF approach, we solve the MBB beam with $m=1, \,2\,,\text{and}, 3$ materials with three cases, CASE I: No density filtering, CASE II: With density filtering, and CASE III: With density and projection filters. The results are presented in Table~\ref{Tab:ImpactofFilterProjection}. For CASE I (Table~\ref{Tab:ImpactofFilterProjection}, Column 3), the multimaterial-optimized designs contain checkerboard patterns and alternate void and material elements. Geometrical singularities typically appear due to poor numerical modeling of the stiffness of checkerboards~\cite{jog1996stability}. To circumvent this issue, we use the density filter~\cite{}, and the results are depicted in column 4 of Table~\ref{Tab:ImpactofFilterProjection}. Though checkerboard patterns do not exist now, the optimized designs contain a large number of gray elements, as indicated by the measure of non-discreteness M$_{\text{nd}}$ (Eq.~\ref{Eq:Mnd_gSF}). This hinders the determination of the clear boundaries (between the different materials). Next, to achieve solutions close to 0-1, we introduce a projection filter for the gSF approach. Results for CASE III are provided in column 5 of Table~\ref{Tab:ImpactofFilterProjection}. The optimized designs are crisp, and their M$_{\text{nd}}$ values are comparatively very low. Additionally, they have improved final objective values. Therefore, the gSF approach is equipped with both density and projection filtering schemes to achieve well-performing, optimized, and crisp designs. Next, we present numerical design examples using distinct candidate materials while incorporating both density and projection filters.


\begin{table}[]
    \caption{Assessment of the role and effectiveness of density and projection filters in the gSF approach.}\label{Tab:ImpactofFilterProjection}
    \centering
    \begin{tabular}{|C{0.7cm}|C{0.7cm}|C{2cm}|C{3.5cm}C{3.5cm}C{3.5cm}|}
        \hline
        \textbf{SF} & $m$ & Results & CASE I & CASE II & CASE III \\ \hline
        \noalign{\hrule height 1.5pt}
        \multirow{3}{*}{1D} & \multirow{3}{*}{1} & \makecell{Optimized\\Design} & \vspace{0.2cm} \includegraphics[scale=0.30]{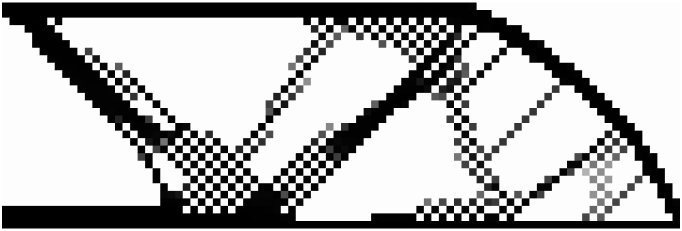} & \vspace{0.2cm} \includegraphics[scale=0.30]{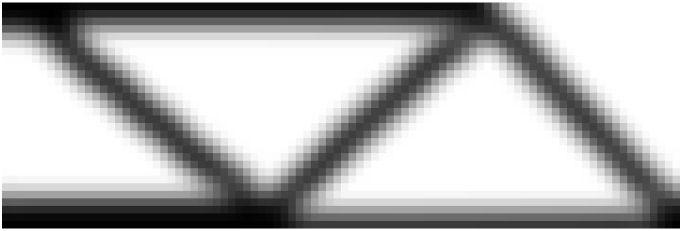} & \vspace{0.2cm} \includegraphics[scale=0.30]{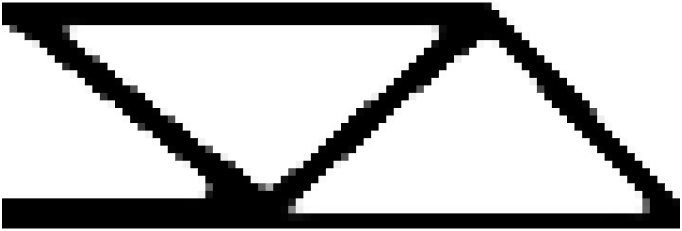}\\ \cline{3-6}
        & & $f_0$ & $0.0172$ & $0.0263$ & $0.0149$ \\ \cline{3-6}
        & & $M\textsubscript{nd}$ & $6.36\%$ & $39.38\%$ & $1.57\%$ \\ \hline
        \noalign{\hrule height 1.5pt}
        \multirow{6}{*}{2D} & \multirow{3}{*}{2} & \makecell{Optimized\\Design} & \vspace{0.2cm} \includegraphics[scale=0.30]{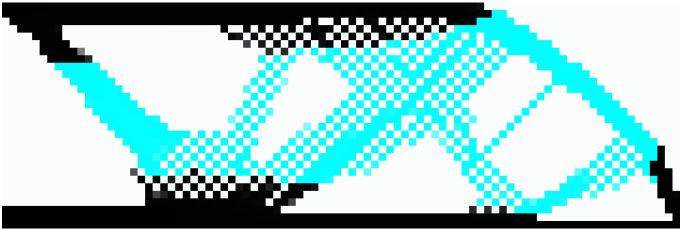} & \vspace{0.2cm} \includegraphics[scale=0.30]{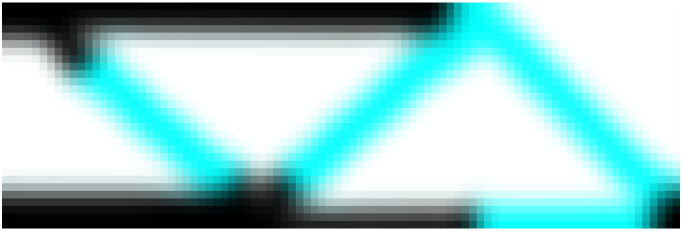} & \vspace{0.2cm} \includegraphics[scale=0.30]{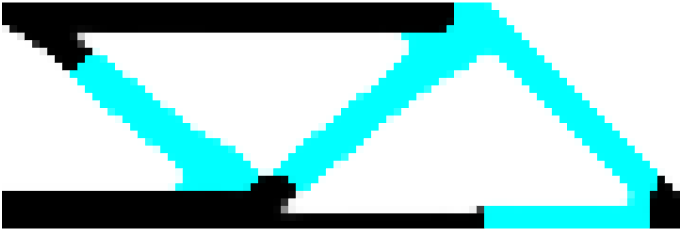} 
        \\ \cline{3-6}
        & & $f_0$ & $0.0145$ & $0.0208$ & $0.0130$ \\ \cline{3-6}
        & & $M\textsubscript{nd}$ & $4.15\%$ & $23.20\%$ & $0.78\%$ \\ \cline{2-6}
        & \multirow{3}{*}{3} & \makecell{Optimized\\Design} & \vspace{0.2cm} \includegraphics[scale=0.30]{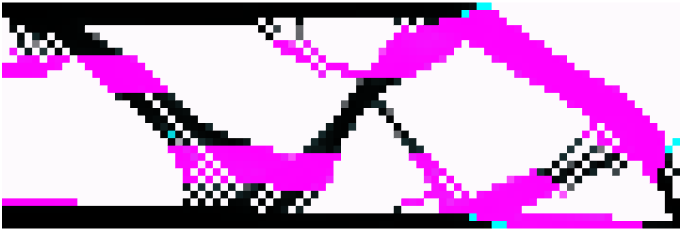} & \vspace{0.2cm} \includegraphics[scale=0.30]{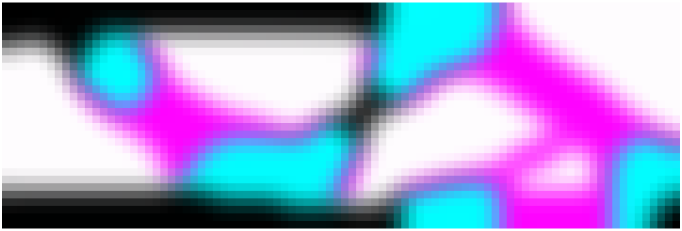} & \vspace{0.2cm} \includegraphics[scale=0.30]{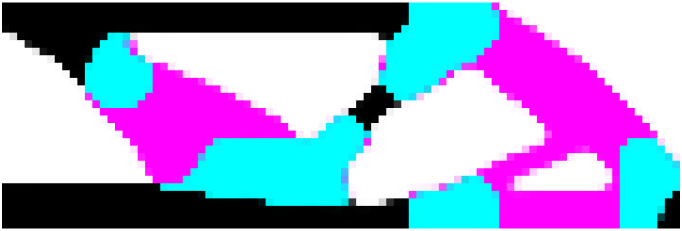} 
        \\ \cline{3-6}
        & & $f_0$ & $0.0175$ & $0.0180$ & $0.0112$ \\ \cline{3-6}
        & & $M\textsubscript{nd}$ & $3.51\%$ & $23.21\%$ & $1.37\%$ \\ \hline
    \end{tabular}
\end{table}

\begin{table}[]
    \caption{Material data and color scheme for 2D SFs}
	\label{3Mtable}
	\centering
\begin{tabular}{|c|c|c|}
		\hline
		\textbf{Colour codes} & \textbf{E (normalized)} & \textbf{Volume fraction} \\ \hline
		\cellcolor{magenta} 	& 1/3  & 0.20  \\ \hline
		\cellcolor{cyan}  		& 2/3  & 0.20  \\ \hline
		\cellcolor{black}  		& 1    & 0.20  \\ \hline
\end{tabular}
\end{table}

We take 1D, 2D, 3D, 4D, and 5D material elements to demonstrate optimized results for up to 24 materials. The last column of the Table~\ref{Tab:ImpactofFilterProjection} depicts the optimized structures obtained using 1D and 2D SFs. The volume fractions remain active at the end of the optimization. The optimized results resemble those obtained using the SIMP formulation for the one-material case using a conventional density-based method~\cite{sigmund2013topology}. For 1 material result, we use 1D SF with $v_f = 0.3$. 2-material and 3-material results obtained using 2D SFs. $v_f = 0.2$ is fixed for each candidate material, i.e., the total volume fraction sought for the optimized problem with 2 materials is 0.4 and with 3 materials is 0.6. The normalized Young's modulus E, volume fraction, and color codes are indicated in Table~\ref{3Mtable}. The portions with roller support and point of application are occupied by material with the highest Young's modulus; however, that may not always be the case; if the stiffness contrast between materials is not significant, then the optimizer may choose any of them to determine the optimum point. This may occur when we are solving problems with a high number of candidate materials. Additionally, the multi-material problem setting is highly non-convex; the optimizer may get stuck at a local minimum for the current material distribution.

Further, optimized results with 4, 5, 6, and 7 materials are obtained using 3D material elements and are shown in Fig.~\ref{MBB2DSF3D}. $v_f = 0.08$ is taken for each candidate material. Table~\ref{7Mtable} shows Young's normalized moduli of the used materials, color codes, and volume fractions. The optimized designs are depicted in  Fig.~\ref{MBB2DSF3D4Mvf08}, Fig.~\ref{MBB2DSF3D5Mvf08}, Fig.~\ref{MBB2DSF3D6Mvf08}, and Fig.~\ref{MBB2DSF3D7Mvf08} for  4, 5, 6, and 7 candidate materials, respectively. The portions with roller support and load application are occupied by the stiffest material (Fig.~\ref{MBB2DSF3D4Mvf08}). In contrast, those are occupied by the material with the second-highest stiffness in Fig.~\ref{MBB2DSF3D5Mvf08}, Fig.~\ref{MBB2DSF3D6Mvf08}, and Fig.~\ref{MBB2DSF3D7Mvf08}. We also depict the evolution of the design for the 4-material case in Fig.~\ref{Convergence_history} at  50, 75, 100, 200, 300, and 400 iterations. In the $50$\textsuperscript{th} iteration, the resulting design is blurry; however, as the optimization progresses, the blurriness is reduced by the projection filter and eventually yields a crisp solution near 400 MMA iterations.
\begin{table}
\caption{Material data and color scheme for 3D SFs. Vf indicates volume fraction}
\centering
\begin{subtable}[t]{0.45\textwidth}
\centering
\begin{tabular}{|c|c|c|}
\hline
\textbf{Colour codes} & \textbf{E (normalized)} & \textbf{Vf} \\ \hline
\cellcolor{red}   &  1/7 &  0.08 \\ \hline
\cellcolor{blue}  &  2/7 &  0.08 \\ \hline
\cellcolor{green} &  3/7 &  0.08 \\ \hline
\end{tabular}
\end{subtable}
\hfill
\begin{subtable}[t]{0.45\textwidth}
\centering
\begin{tabular}{|c|c|c|}
\hline
\textbf{Colour codes} & \textbf{E (normalized)} & \textbf{Vf} \\ \hline
\cellcolor{yellow}  &  4/7 &  0.08 \\ \hline
\cellcolor{magenta} &  5/7 &  0.08 \\ \hline
\cellcolor{cyan}    &  6/7 &  0.08 \\ \hline
\cellcolor{black}   &  1   &  0.08 \\ \hline
\end{tabular}
\end{subtable}
\label{7Mtable}
\end{table}

\begin{figure}[]
	\centering
	\begin{subfigure}[t]{0.45\textwidth}
		\centering
		\includegraphics[width=\textwidth]{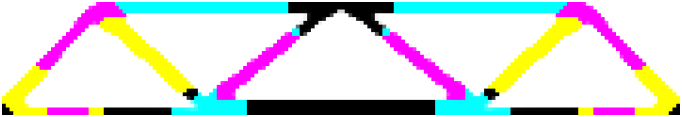}
		\caption{4 materials, $v_f^i= 0.08|_{i=1,\,2,\,3,\,4}$, $f_0$$= 0.0172$}
		\label{MBB2DSF3D4Mvf08}
	\end{subfigure}
	\hfill
	\begin{subfigure}[t]{0.45\textwidth}
		\centering
		\includegraphics[width=\textwidth]{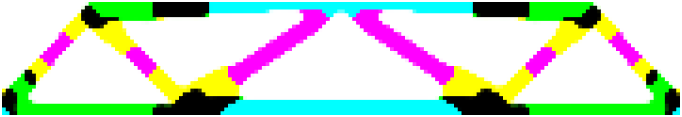}
		\caption{5 materials, $v_f^i= 0.08|_{i=1,\,\cdots,\,5}$, $f_0$$= 0.0167$}
		\label{MBB2DSF3D5Mvf08}
	\end{subfigure}
	\hfill
	\begin{subfigure}[t]{0.45\textwidth}
		\centering
		\includegraphics[width=\textwidth]{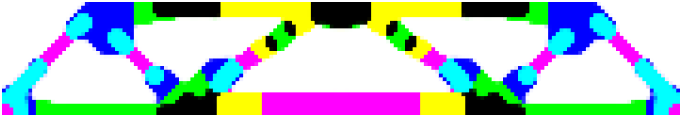}
		\caption{6 materials, $v_f^i= 0.08|_{i=1,\,\cdots,\,6}$, $f_0$$= 0.0162$}
		\label{MBB2DSF3D6Mvf08}
	\end{subfigure}
	\hfill
	\begin{subfigure}[t]{0.45\textwidth}
		\centering
		\includegraphics[width=\textwidth]{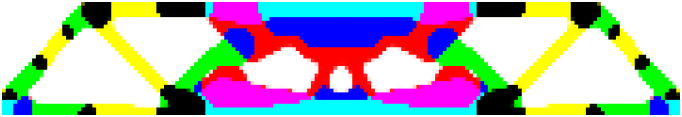}
		\caption{7 materials, $v_f^i= 0.08|_{i=1,\,\cdots,\,7}$, $f_0$$= 0.0145$}
		\label{MBB2DSF3D7Mvf08}
	\end{subfigure}
	\caption{Optimized results using 3D SFs}
	\label{MBB2DSF3D}
\end{figure}

\begin{table}[]
\caption{Material data and color scheme for 4D SFs}
\centering
\begin{subtable}[t]{0.45\textwidth}
\centering
\begin{tabular}{|c|c|c|}
\hline
\textbf{Colour codes} & \textbf{E (normalized)} & \textbf{Vf} \\ \hline
\cellcolor{olivegreen}  & 1/15   & 0.04  \\ \hline
\cellcolor{greenyellow} & 2/15   & 0.04  \\ \hline
\cellcolor{violet}  	& 3/15   & 0.04  \\ \hline
\cellcolor{yellowochre} & 4/15   & 0.04  \\ \hline
\cellcolor{redorange}  	& 5/15   & 0.04  \\ \hline
\cellcolor{brownred}  	& 6/15   & 0.04  \\ \hline
\cellcolor{skyblue}  	& 7/15   & 0.04  \\ \hline
\end{tabular}
\end{subtable}
\hfill
\begin{subtable}[t]{0.45\textwidth}
\centering
\begin{tabular}{|c|c|c|}
\hline
\textbf{Colour codes} & \textbf{E (normalized)} & \textbf{Vf} \\ \hline
\cellcolor{shafgreen}  	& 8/15   & 0.04  \\ \hline
\cellcolor{red} 		& 9/15   & 0.04  \\ \hline
\cellcolor{blue}  		& 10/15  & 0.04  \\ \hline
\cellcolor{green}  		& 11/15  & 0.04  \\ \hline
\cellcolor{yellow}  	& 12/15  & 0.04  \\ \hline
\cellcolor{magenta} 	& 13/15  & 0.04  \\ \hline
\cellcolor{cyan}  		& 14/15  & 0.04  \\ \hline
\cellcolor{black}  		& 1   	 & 0.04  \\ \hline
\end{tabular}
\end{subtable}
\label{15Mtable}
\end{table}

\begin{figure}
	\centering
	\begin{subfigure}[t]{0.45\textwidth}
		\centering
		\includegraphics[width=\textwidth]{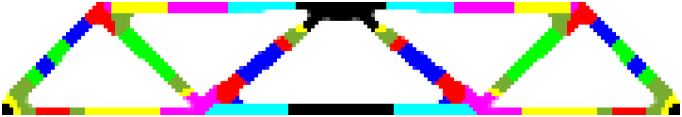} 
		\caption{8 Materials, $v_f^i= 0.04|_{i=1,\,\cdots,\,8}$, $f_0$$= 0.0183$}
		\label{MBB2DSF4D8Mvf04}
	\end{subfigure}
	\hfill
	\begin{subfigure}[h]{0.45\textwidth}
		\centering
		\includegraphics[width=\textwidth]{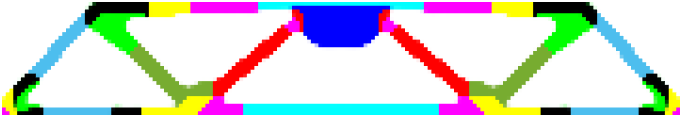} 
		\caption{9 Materials, $v_f^i= 0.04|_{i=1,\,\cdots,\,9}$, $f_0$$= 0.0184$}
		\label{MBB2DSF4D9Mvf04}
	\end{subfigure}
	\hfill
	\begin{subfigure}[t]{0.45\textwidth}
		\centering
		\includegraphics[width=\textwidth]{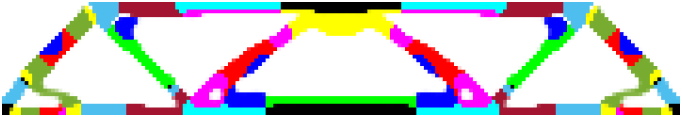}
		\caption{10 Materials, $v_f^i= 0.04|_{i=1,\,\cdots,\,10}$, $f_0$$= 0.0169$}
		\label{MBB2DSF4D10Mvf04}
	\end{subfigure}
	\hfill
	\begin{subfigure}[t]{0.45\textwidth}
		\centering
		\includegraphics[width=\textwidth]{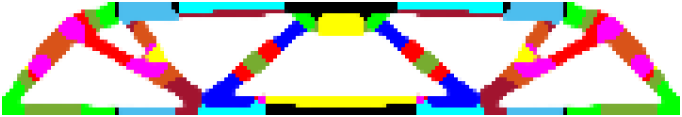}
		\caption{11 Materials, $v_f^i= 0.04|_{i=1,\,\cdots,\,11}$, $f_0$$= 0.0161$}
		\label{MBB2DSF4D11Mvf04}
	\end{subfigure}
	\begin{subfigure}[t]{0.45\textwidth}
		\centering
		\includegraphics[width=\textwidth]{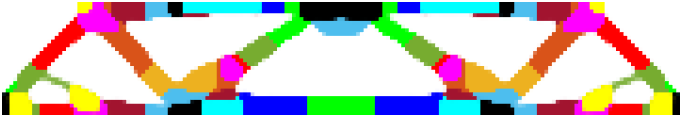}
		\caption{12 Materials,  $v_f^i= 0.04|_{i=1,\,\cdots,\,12}$, $f_0$$= 0.0164$}
		\label{MBB2DSF4D12Mvf04}
	\end{subfigure}
	\hfill
	\begin{subfigure}[t]{0.45\textwidth}
		\centering
		\includegraphics[width=\textwidth]{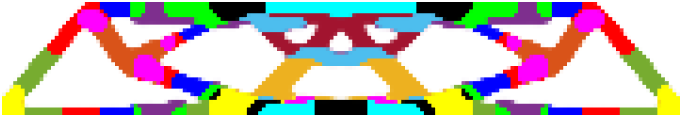}
		\caption{13 Materials, $v_f^i= 0.04|_{i=1,\,\cdots,\,13}$, $f_0$$= 0.0155$}
		\label{MBB2DSF4D13Mvf04}
	\end{subfigure}
	\hfill
	\begin{subfigure}[t]{0.45\textwidth}
		\centering
		\includegraphics[width=\textwidth]{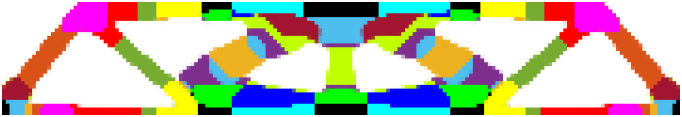}
		\caption{14 Materials, $v_f^i= 0.04|_{i=1,\,\cdots,\,14}$, $f_0$$= 0.0144$}
		\label{MBB2DSF4D14Mvf04}
	\end{subfigure}
	\hfill
	\begin{subfigure}[t]{0.45\textwidth}
		\centering
		\includegraphics[width=\textwidth]{MBB2D15Mbeta75}
		\caption{15 Materials, $v_f^i= 0.04|_{i=1,\,\cdots,\,15}$, $f_0$$= 0.0152$}
		\label{MBB2DSF4D15Mvf04}
	\end{subfigure}
	\caption{Optimized results using 4D SFs}
	\label{MBB2DSF4D}
\end{figure}

\begin{figure}
	\centering
	\includegraphics[scale=1]{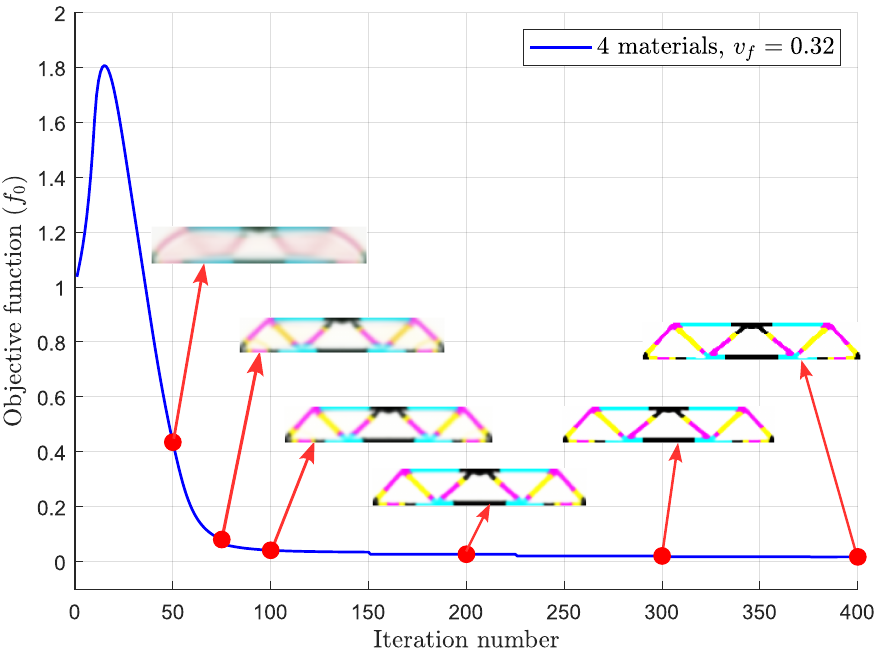}
	\caption{Convergence history for the 4 materials case with material evolution diagram at different instances of the optimization}
	\label{Convergence_history}
\end{figure}

\begin{table}[]
\caption{Material data and color scheme for 5D SFs}
\label{24Mtable}
\centering
\begin{subtable}[t]{0.45\textwidth}
\centering
\begin{tabular}{|c|c|c|}
\hline
\textbf{Colour codes} & \textbf{E (normalized)} & \textbf{Vf} \\ \hline
\cellcolor{Maroon}  	& 8/31   & 0.025  \\ \hline
\cellcolor{Burgundy}  	& 9/31   & 0.025  \\ \hline
\cellcolor{HotPink} 	& 10/31  & 0.025  \\ \hline
\cellcolor{DeepPink}  	& 11/31  & 0.025  \\ \hline
\cellcolor{Indigo}  	& 12/31  & 0.025  \\ \hline
\cellcolor{SlateGray}  	& 13/31  & 0.025  \\ \hline
\cellcolor{LightGray} 	& 14/31  & 0.025  \\ \hline
\cellcolor{SteelBlue}  	& 15/31  & 0.025  \\ \hline
\cellcolor{Turquoise}  	& 16/31  & 0.025  \\ \hline
\cellcolor{olivegreen}  & 17/31  & 0.025  \\ \hline
\cellcolor{greenyellow} & 18/31  & 0.025  \\ \hline
\cellcolor{violet}  	& 19/31  & 0.025  \\ \hline
\end{tabular}
\end{subtable}
\hfill
\begin{subtable}[t]{0.45\textwidth}
\centering
\begin{tabular}{|c|c|c|}
\hline
\textbf{Colour codes} & \textbf{E (normalized)} & \textbf{Vf} \\ \hline
\cellcolor{yellowochre} & 20/31  & 0.025  \\ \hline
\cellcolor{redorange}  	& 21/31  & 0.025  \\ \hline
\cellcolor{brownred}  	& 22/31  & 0.025  \\ \hline
\cellcolor{skyblue}  	& 23/31  & 0.025  \\ \hline
\cellcolor{shafgreen}  	& 24/31  & 0.025  \\ \hline
\cellcolor{red} 		& 25/31  & 0.025  \\ \hline
\cellcolor{blue}  		& 26/31  & 0.025  \\ \hline
\cellcolor{green}  		& 27/31  & 0.025  \\ \hline
\cellcolor{yellow}  	& 28/31  & 0.025  \\ \hline
\cellcolor{magenta} 	& 29/31  & 0.025  \\ \hline
\cellcolor{cyan}  		& 30/31  & 0.025  \\ \hline
\cellcolor{black}  		& 1   	 & 0.025  \\ \hline
\end{tabular}
\end{subtable}
\end{table}



\begin{figure}[]
	\centering
	\begin{subfigure}[t]{0.47\textwidth}
		\centering
		\includegraphics[width=\textwidth]{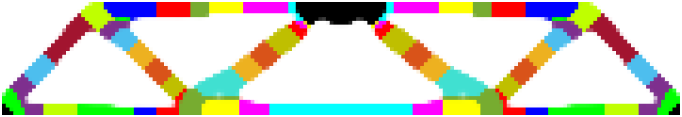}
		\caption{16 materials, $v_f^i= 0.025|_{i=1,\,\cdots,\,16}$, $f_0$$=  0.0092$}
		\label{MBB2DSF5D16Mvf025}
	\end{subfigure}
	\hfill
	\begin{subfigure}[t]{0.47\textwidth}
		\centering
		\includegraphics[width=\textwidth]{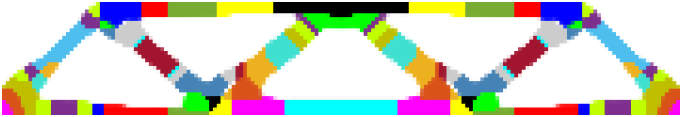}
		\caption{18 materials, $v_f^i= 0.025|_{i=1,\,\cdots,\,18}$, $f_0$$= 0.0090$}
		\label{MBB2DSF5D18Mvf025}
	\end{subfigure}
	\hfill
	\begin{subfigure}[t]{0.47\textwidth}
		\centering
		\includegraphics[width=\textwidth]{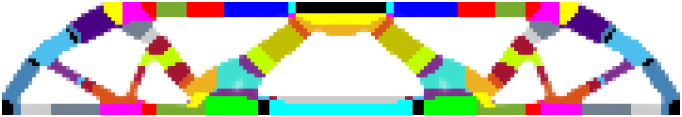}
		\caption{20 materials, $v_f^i= 0.025|_{i=1,\,\cdots,\,20}$, $f_0$$= 0.0091$}
		\label{MBB2DSF5D20Mvf025}
	\end{subfigure}
	\hfill
	\begin{subfigure}[t]{0.47\textwidth}
		\centering
		\includegraphics[width=\textwidth]{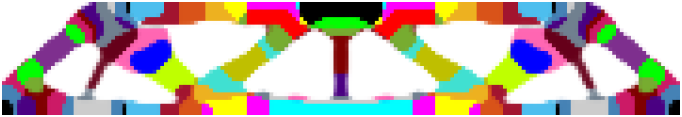}
		\caption{24 materials, $v_f^i= 0.025|_{i=1,\,\cdots,\,24}$, $f_0$$= 0.0093$}
		\label{MBB2DSF5D24Mvf025}
	\end{subfigure}
	\caption{Optimized results using 5D SFs}
	\label{MBB2DSF5D}
\end{figure}



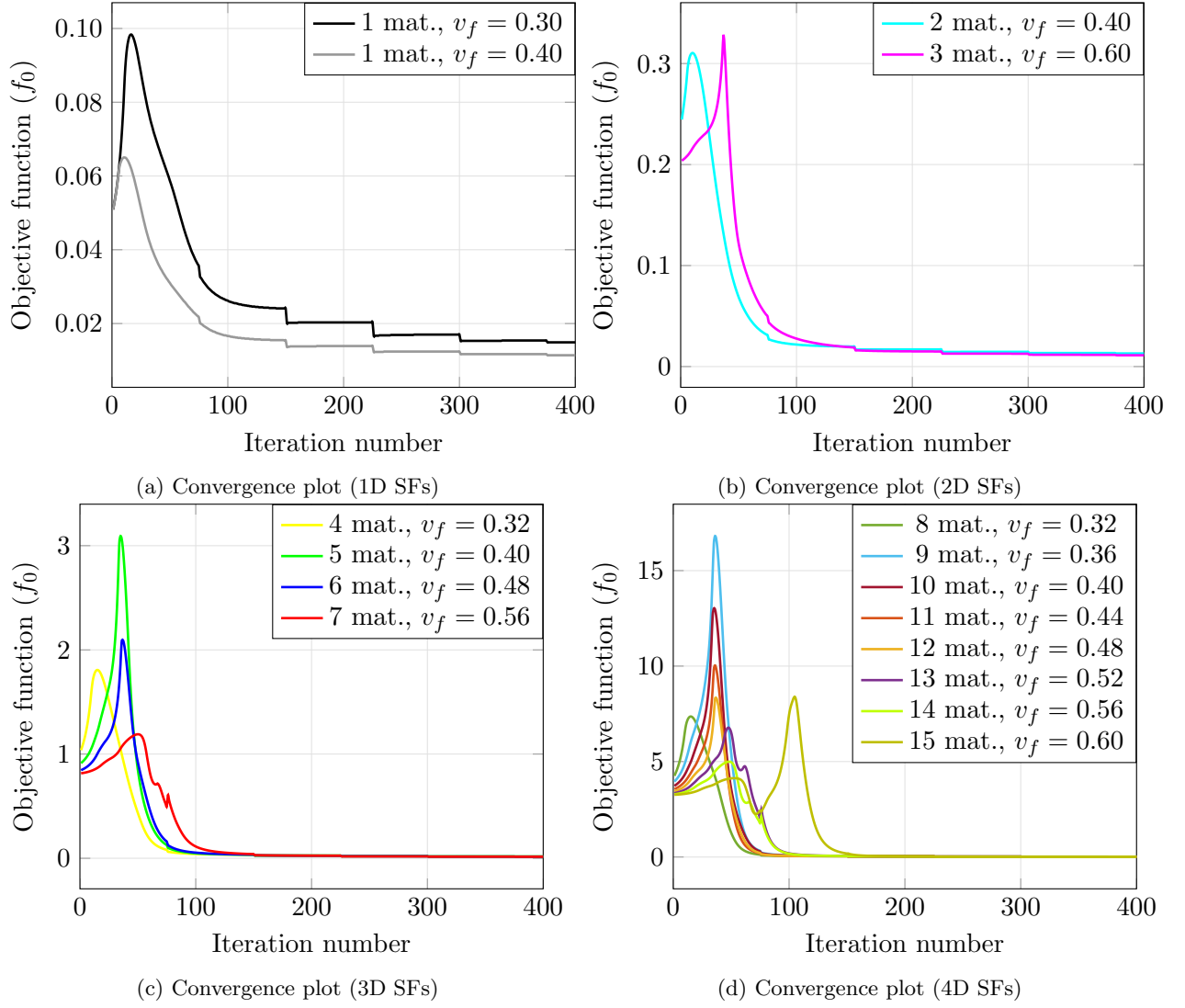
\begin{figure}
	\centering
	\begin{subfigure}[h!]{0.48\textwidth}
		\centering
		\begin{tikzpicture}[scale=1]
			\pgfplotsset{compat=1.9}
			\begin{axis}[
				width = 1\textwidth,
				xlabel=  Iteration number,
				ylabel=  Objective function $(f_0)$,
				xmin=0,xmax=400,
				ytick={0.02, 0.04, 0.06, 0.08, 0.10},
				yticklabels={
					0.02,
					0.04,
					0.06,
					0.08,
					0.10
				},
				grid=both,
				major grid style={line width=0.2pt, draw=gray!30},
				legend style={at={(1.0,1.0)}, anchor=north east}]	
				\pgfplotstableread{MBB2D1Mvf30beta75graph.txt}\mydata;
				\addplot[smooth,{black}, line width=1pt, mark = none]
				table {\mydata};
				\addlegendentry{1 mat., $v_f=0.30$}				
				\pgfplotstableread{MBB2D1Mvf40beta75graph.txt}\mydata;
				\addplot[smooth,{gray}, line width=1pt, mark = none]
				table {\mydata};
				\addlegendentry{1 mat., $v_f=0.40$}			
			\end{axis}
		\end{tikzpicture}
		\caption{Convergence plot (1D SFs)}
		\label{MBB2D_SF1D_graphs}
	\end{subfigure}
	\begin{subfigure}[h!]{0.48\textwidth}
		\centering
		\begin{tikzpicture}[scale=1]
			\pgfplotsset{compat=1.9}
			\begin{axis}[
				width = 1\textwidth,
				xlabel=  Iteration number,
				ylabel=  Objective function $(f_0)$,
				xmin=0,xmax=400,
				grid=both,
				major grid style={line width=0.2pt, draw=gray!30},
				legend style={at={(1.0,1.0)}, anchor=north east}]			
				\pgfplotstableread{MBB2D2Mbeta75graph.txt}\mydata;
				\addplot[smooth,{cyan}, line width=1pt, mark = none]
				table {\mydata};
				\addlegendentry{2 mat., $v_f=0.40$}			
				\pgfplotstableread{MBB2D3Mbeta75graph.txt}\mydata;
				\addplot[smooth,{magenta}, line width=1pt, mark = none]
				table {\mydata};
				\addlegendentry{3 mat., $v_f=0.60$}			
			\end{axis}
		\end{tikzpicture}
		\caption{Convergence plot (2D SFs)}
		\label{MBB2D_SF2D_graphs}
	\end{subfigure}
	\begin{subfigure}[h!]{0.48\textwidth}
		\centering
		\begin{tikzpicture}[scale=1]
			\pgfplotsset{compat=1.9}
			\begin{axis}[
				width = 1\textwidth,
				xlabel=  Iteration number,
				ylabel=  Objective function $(f_0)$,
				xmin=0,xmax=400,
				grid=both,
				major grid style={line width=0.2pt, draw=gray!30},
				legend style={at={(1.0,1.0)}, anchor=north east}]			
				\pgfplotstableread{MBB2D4Mbeta75graph.txt}\mydata;
				\addplot[smooth,{yellow}, line width=1pt, mark = none]
				table {\mydata};
				\addlegendentry{4 mat., $v_f=0.32$}
				
				\pgfplotstableread{MBB2D5Mbeta75graph.txt}\mydata;
				\addplot[smooth,{green}, line width=1pt, mark = none]
				table {\mydata};
				\addlegendentry{5 mat., $v_f=0.40$}
				
				\pgfplotstableread{MBB2D6Mbeta75graph.txt}\mydata;
				\addplot[smooth,{blue}, line width=1pt, mark = none]
				table {\mydata};
				\addlegendentry{6 mat., $v_f=0.48$}
				
				\pgfplotstableread{MBB2D7Mbeta75graph.txt}\mydata;
				\addplot[smooth,{red}, line width=1pt, mark = none]
				table {\mydata};
				\addlegendentry{7 mat., $v_f=0.56$}		
			\end{axis}
		\end{tikzpicture}
		\caption{Convergence plot (3D SFs)}
		\label{MBB2D_SF3D_graphs}
	\end{subfigure}
	\begin{subfigure}[h!]{0.48\textwidth}
		\centering
		\begin{tikzpicture}[scale=1]
			\pgfplotsset{compat=1.9}
			\begin{axis}[
				width = 1\textwidth,
				xlabel=  Iteration number,
				ylabel=  Objective function $(f_0)$,
				xmin=0,xmax=400,
				grid=both,
				major grid style={line width=0.2pt, draw=gray!30},
				legend style={at={(1.0,1.0)}, anchor=north east}]			
				\pgfplotstableread{MBB2D8Mbeta75graph.txt}\mydata;
				\addplot[smooth,{shafgreen}, line width=1pt, mark = none]
				table {\mydata};
				\addlegendentry{8 mat., $v_f=0.32$}				
				\pgfplotstableread{MBB2D9Mbeta75graph.txt}\mydata;
				\addplot[smooth,{skyblue}, line width=1pt, mark = none]
				table {\mydata};
				\addlegendentry{9 mat., $v_f=0.36$}				
				\pgfplotstableread{MBB2D10Mbeta75graph.txt}\mydata;
				\addplot[smooth,{brownred}, line width=1pt, mark = none]
				table {\mydata};
				\addlegendentry{10 mat., $v_f=0.40$}				
				\pgfplotstableread{MBB2D11Mbeta75graph.txt}\mydata;
				\addplot[smooth,{redorange}, line width=1pt, mark = none]
				table {\mydata};
				\addlegendentry{11 mat., $v_f=0.44$}				
				\pgfplotstableread{MBB2D12Mbeta75graph.txt}\mydata;
				\addplot[smooth,{yellowochre}, line width=1pt, mark = none]
				table {\mydata};
				\addlegendentry{12 mat., $v_f=0.48$}				
				\pgfplotstableread{MBB2D13Mbeta75graph.txt}\mydata;
				\addplot[smooth,{violet}, line width=1pt, mark = none]
				table {\mydata};
				\addlegendentry{13 mat., $v_f=0.52$}				
				\pgfplotstableread{MBB2D14Mbeta75graph.txt}\mydata;
				\addplot[smooth,{greenyellow}, line width=1pt, mark = none]
				table {\mydata};
				\addlegendentry{14 mat., $v_f=0.56$}				
				\pgfplotstableread{MBB2D15Mbeta75graph.txt}\mydata;
				\addplot[smooth,{olivegreen}, line width=1pt, mark = none]
				table {\mydata};
				\addlegendentry{15 mat., $v_f=0.60$}		
			\end{axis}
		\end{tikzpicture}
		\caption{Convergence plot (4D SFs)}
		\label{MBB2D_SF4D_graphs}
	\end{subfigure}
	\caption{Convergence plots for corresponding SFs}
	\label{MBB2D_convergency}
\end{figure}



Furthermore, we consider a 4D material element to obtain optimized structures for up to 15 materials, demonstrating the gSF's capability to handle many materials. Table~\ref{15Mtable} denotes the normalized Young's modulus E, volume fractions $v_f$, and the color codes for the candidate material for the 4D SFs case taken here. Figure \ref{MBB2DSF4D} depicts the optimized results obtained using 4D SFs. Figs.~\ref{MBB2DSF4D8Mvf04}, \ref{MBB2DSF4D9Mvf04}, \ref{MBB2DSF4D10Mvf04}, \ref{MBB2DSF4D11Mvf04}, \ref{MBB2DSF4D12Mvf04}, \ref{MBB2DSF4D13Mvf04}, \ref{MBB2DSF4D14Mvf04}, and \ref{MBB2DSF4D15Mvf04} show the optimized design with 8, 9, 10, 11, 12, 13, 14, and 15 materials, respectively. One notices that the optimizer selects different candidate materials for different locations while determining the optimized designs. The material occupies the portions with roller and hinge support and load application with the highest Young's modulus (Figs.~\ref{MBB2DSF4D8Mvf04}, \ref{MBB2DSF4D10Mvf04}, \ref{MBB2DSF4D12Mvf04} and  \ref{MBB2DSF4D14Mvf04}) However, in Figs.~\ref{MBB2DSF4D9Mvf04}, \ref{MBB2DSF4D11Mvf04}, \ref{MBB2DSF4D13Mvf04} and  \ref{MBB2DSF4D15Mvf04} they are occupied by material with the second or third highest stiffness. 5D SFs are used to get the optimized results with 16, 18, 20, and 24 candidate materials as shown in Fig.~\ref{MBB2DSF5D16Mvf025}, Fig.~\ref{MBB2DSF5D18Mvf025}, Fig.~\ref{MBB2DSF5D20Mvf025}, and Fig.~\ref{MBB2DSF5D24Mvf025}, respectively. $E$, $v_f$, and color schemers are provided in Table~\ref{24Mtable} for the 5D SFs case.  The material with the highest Young's modulus occupies the regions with supports and applied load, while the support regions in Fig.~\ref{MBB2DSF5D18Mvf025} are occupied by the third-highest stiffness.

The objective convergence plots for all the optimized designs presented above are shown in Fig.~\ref{MBB2D_convergency}, where Fig.~\ref{MBB2D_SF1D_graphs}, Fig.~\ref{MBB2D_SF2D_graphs}, Fig.~\ref{MBB2D_SF3D_graphs}, and Fig.~\ref{MBB2D_SF4D_graphs} depict the same for the 1D, 2D, 3D, and 4D SFs cases, respectively. The jumps in the plots indicate the iterations at which $\beta_p$ gets updated. These plots show a converging nature and suggest that using the same number of shape function variables to optimize for different numbers of materials leads to convergence after nearly the same number of iterations. Each case's volume constraints are predominantly active at the end of the optimization.  Next, the 2D cantilever beam problem is solved with up to 15 candidate materials.
\begin{figure}[b]
	\centering
	\begin{subfigure}[t]{0.23\textwidth}
		\centering
		\includegraphics[scale=0.25]{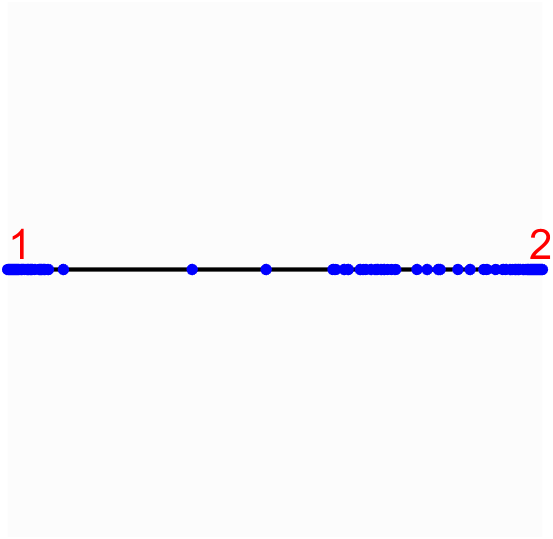}
		\caption{$\mathrm{M_{nd}}=1.57\%$}
		\label{ gf}
	\end{subfigure}
	\label{fgf}
	\hspace{0.1cm}
	\begin{subfigure}[t]{0.23\textwidth}
		\centering
		\includegraphics[scale=0.25]{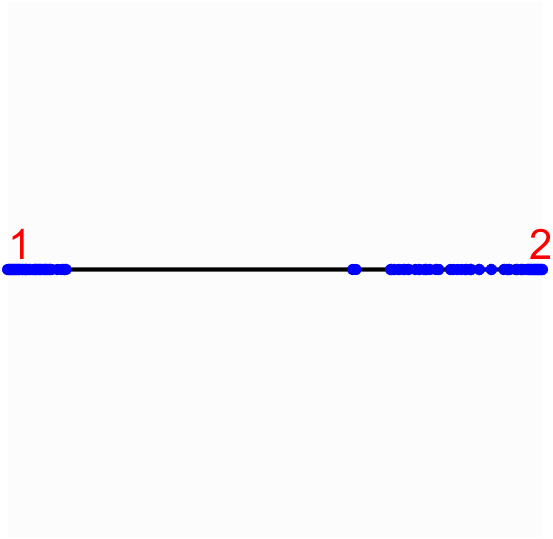}
		\caption{$\mathrm{M_{nd}}=1.63\%$}
	\end{subfigure}
	\label{fgf}
	\hspace{0.1cm}
	\begin{subfigure}[t]{0.23\textwidth}
		\centering
		\includegraphics[scale=0.28]{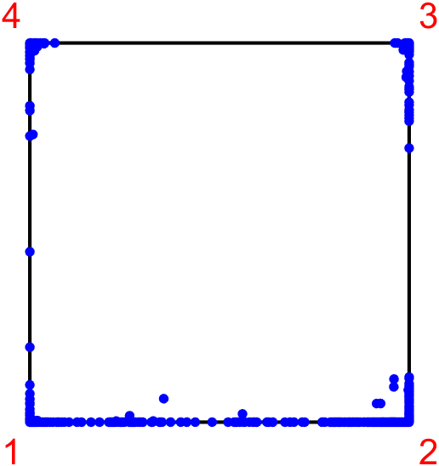}
		\caption{$\mathrm{M_{nd}}=1.39\%$}
	\end{subfigure}
	\label{fgf}
	\hspace{0.1cm}
	\begin{subfigure}[t]{0.23\textwidth}
		\centering
		\includegraphics[scale=0.28]{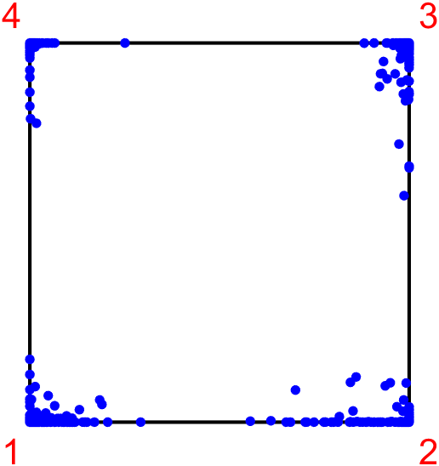}
		\caption{$\mathrm{M_{nd}}=2.03\%$}
	\end{subfigure}
	\caption{Material mapping for 1D ((a) and (b)) and 2D ((c) and (d)) SFs; the projected design variable plots for (a) 1 material case with $v_f=0.30$, (b) 1 material case with $v_f=0.40$, (c) 2 material case, and (d) 3 material case.}
	\label{Material_mapping_1D2D}
\end{figure}

\begin{figure}[]
	\centering
	\begin{subfigure}[t]{0.23\textwidth}
		\centering
		\includegraphics[scale=0.30]{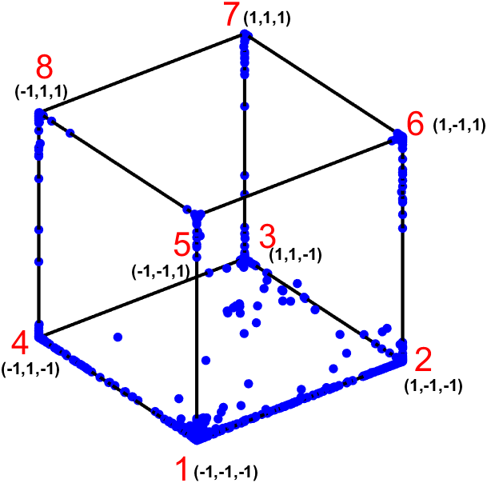}
		\caption{$\mathrm{M_{nd}}=1.60\%$}
	\end{subfigure}
	\label{fgf}
	\hspace{0.1cm}
	\begin{subfigure}[t]{0.23\textwidth}
		\centering
		\includegraphics[scale=0.30]{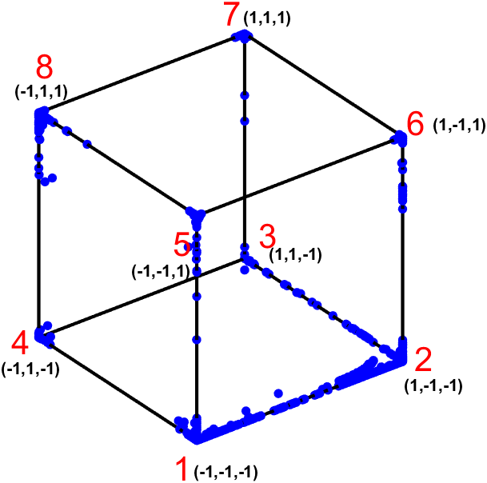}
		\caption{$\mathrm{M_{nd}}=1.60\%$}
	\end{subfigure}
	\label{fgf}
	\hspace{0.1cm}
	\begin{subfigure}[t]{0.23\textwidth}
		\centering
		\includegraphics[scale=0.30]{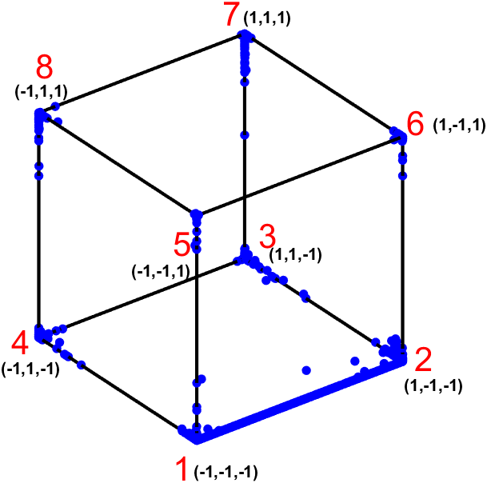}
		\caption{$\mathrm{M_{nd}}=1.55\%$}
	\end{subfigure}
	\hspace{0.1cm}
	\begin{subfigure}[t]{0.23\textwidth}
		\centering
		\includegraphics[scale=0.30]{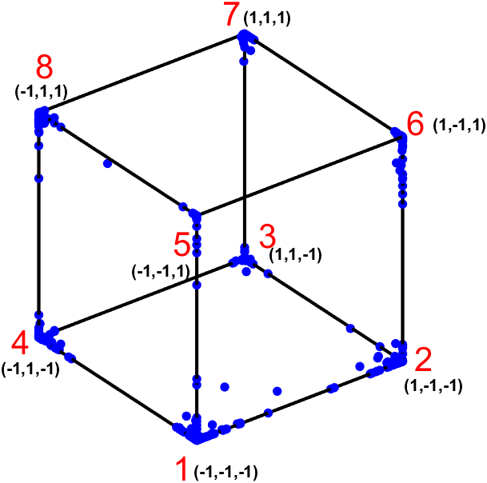}
		\caption{$\mathrm{M_{nd}}=1.69\%$}
	\end{subfigure}
	\caption{Material mapping for 3D SF; (a), (b), (c), and (d) are the projected design variable plots for the 4, 5, 6, and 7 material cases, respectively.}
	\label{Material_mapping_3D}
\end{figure}

\begin{figure}[]
	\centering
	\begin{subfigure}[t]{0.23\textwidth}
		\centering
		\includegraphics[scale=0.30]{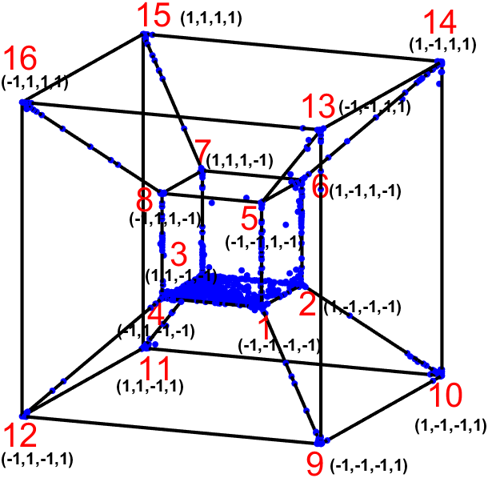}
		\caption{$\mathrm{M_{nd}}=1.53\%$}
	\end{subfigure}
	\label{fgf}
	\hspace{0.1cm}
	\begin{subfigure}[t]{0.23\textwidth}
		\centering
		\includegraphics[scale=0.30]{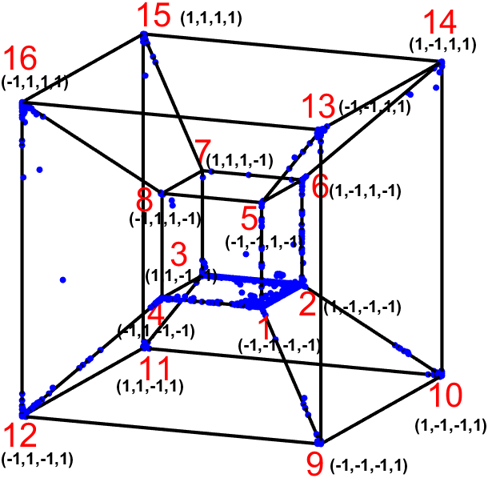}
		\caption{$\mathrm{M_{nd}}=1.38\%$}
	\end{subfigure}
	\label{fgf}
	\hspace{0.1cm}
	\begin{subfigure}[t]{0.23\textwidth}
		\centering
		\includegraphics[scale=0.30]{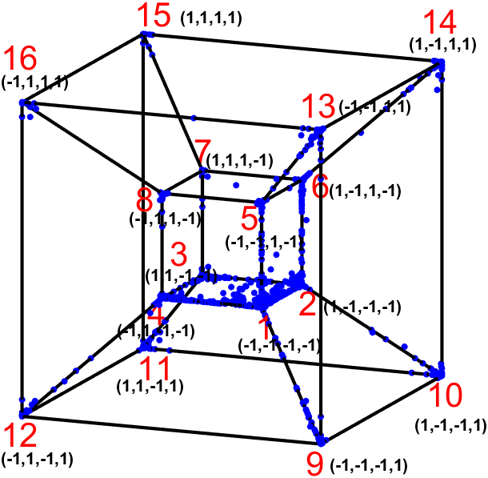}
		\caption{$\mathrm{M_{nd}}=2.48\%$}
	\end{subfigure}
	\label{fgf}
	\hspace{0.1cm}
	\begin{subfigure}[t]{0.23\textwidth}
		\centering
		\includegraphics[scale=0.30]{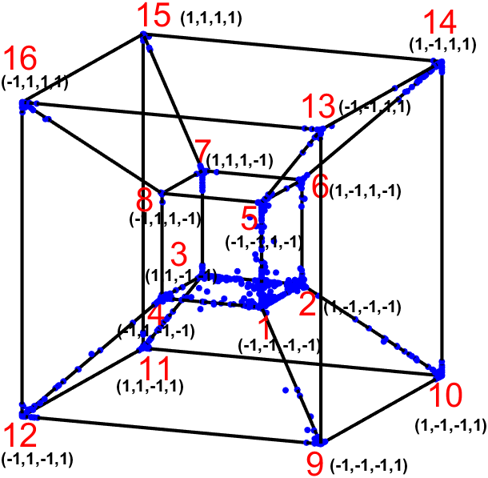}
		\caption{$\mathrm{M_{nd}}=2.46\%$}
	\end{subfigure}
	\label{fgf}
	\hspace{0.1cm}
	\begin{subfigure}[t]{0.23\textwidth}
		\centering
		\includegraphics[scale=0.30]{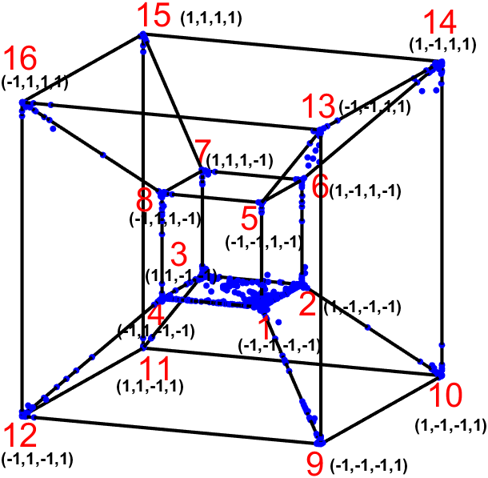}
		\caption{$\mathrm{M_{nd}}=2.08\%$}
	\end{subfigure}
	\label{fgf}
	\hspace{0.1cm}
	\begin{subfigure}[t]{0.23\textwidth}
		\centering
		\includegraphics[scale=0.30]{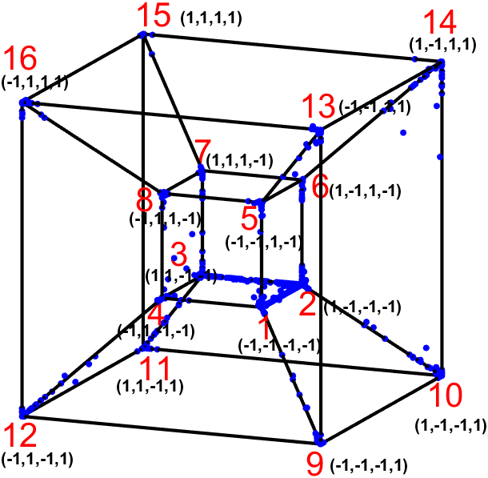}
		\caption{$\mathrm{M_{nd}}=2.76\%$}
	\end{subfigure}
	\label{fgf}
	\hspace{0.1cm}
	\begin{subfigure}[t]{0.23\textwidth}
		\centering
		\includegraphics[scale=0.30]{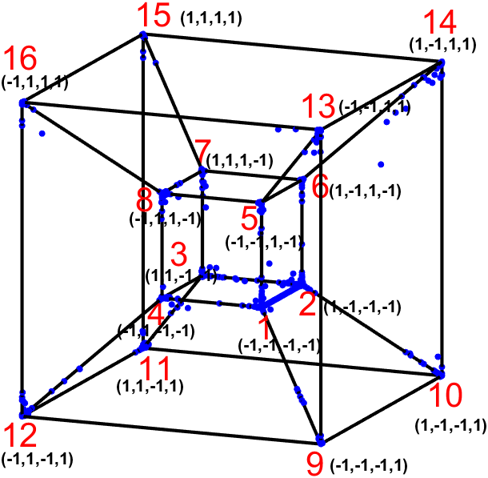}
		\caption{$\mathrm{M_{nd}}=2.29\%$}
	\end{subfigure}
	\label{fgf}
	\hspace{0.1cm}
	\begin{subfigure}[t]{0.23\textwidth}
		\centering
		\includegraphics[scale=0.30]{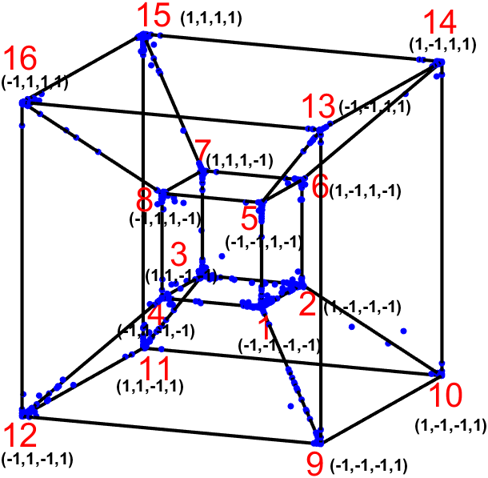}
		\caption{$\mathrm{M_{nd}}=3.05\%$}
	\end{subfigure}
	\caption{Material mapping for 4D SF; (a), (b), (c), (d), (e), (f), (g), and (h) are the projected design variable plots for the 8, 9, 10, 11, 12, 13, 14, and 15 material cases, respectively.}
	\label{Material_mapping_4D}
\end{figure}

The final positions of the physical design variables on their respective material spaces are illustrated in Fig.~\ref{Material_mapping_1D2D}, Fig.~\ref{Material_mapping_3D} and Fig.~\ref{Material_mapping_4D} for the MBB beam problem (Fig.~\ref{MBB_Beam_2D}) with 1D \& 2D, 3D, and 4D material interpolation elements, respectively. It is observed that these variables cluster near the vertices of the material space while remaining within the admissible domain. This behavior indicates that the solutions are close to discrete (0-1), as further supported by the discreteness measures $M\textsubscript{nd}$ reported in the figure captions. This also shows that the material subspace is being effectively utilized.

\begin{figure} 
	\centering
	\begin{subfigure}[H]{0.2\textwidth}
		\centering
		\includegraphics[width=\textwidth]{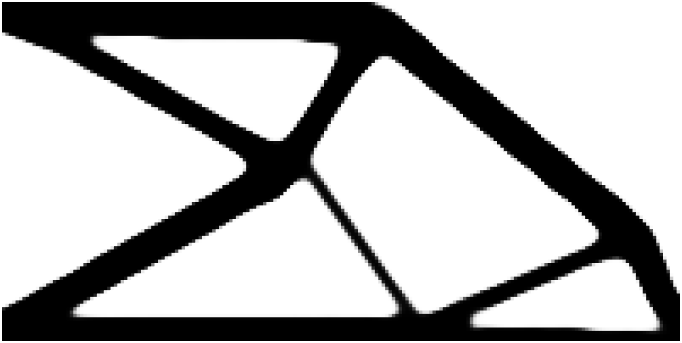}
		\captionsetup{justification=centering, labelformat=empty}
		\caption{\parbox{\linewidth}{\centering (a) 1 material,\\ $v_f= 0.30$,\\ $f_0$$= 0.0055$}}
		\label{C2DSF1D1Mvf3}
	\end{subfigure}
	\hfill
	\begin{subfigure}[H]{0.2\textwidth}
		\centering
		\includegraphics[width=\textwidth]{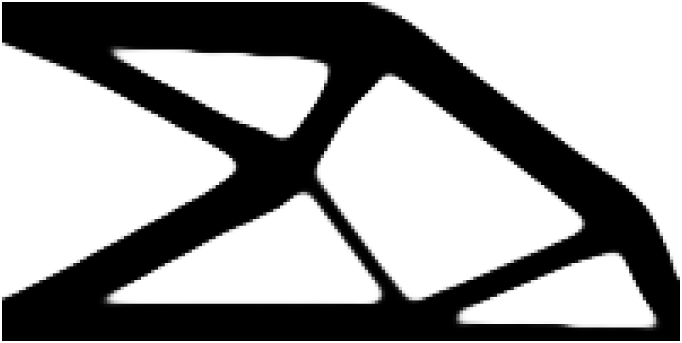}
		\captionsetup{justification=centering, labelformat=empty}
		\caption{\parbox{\linewidth}{\centering (b) 1 material,\\ $v_f= 0.40$,\\ $f_0$$= 0.0042$}}
		\label{C2DSF1D1Mvf4}
	\end{subfigure}
	\hfill
	\begin{subfigure}[H]{0.2\textwidth}
		\centering
		\includegraphics[width=\textwidth]{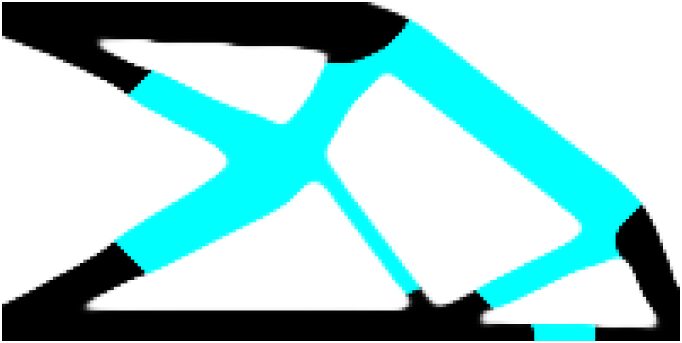}
		\captionsetup{justification=centering, labelformat=empty}
		\caption{\parbox{\linewidth}{\centering (c) 2 materials,\\ $v_f= 0.40$,\\ $f_0$$= 0.0049$}}
		\label{C2DSF2D2Mvf2}
	\end{subfigure}
	\hfill
	\begin{subfigure}[H]{0.2\textwidth}
		\centering
		\includegraphics[width=\textwidth]{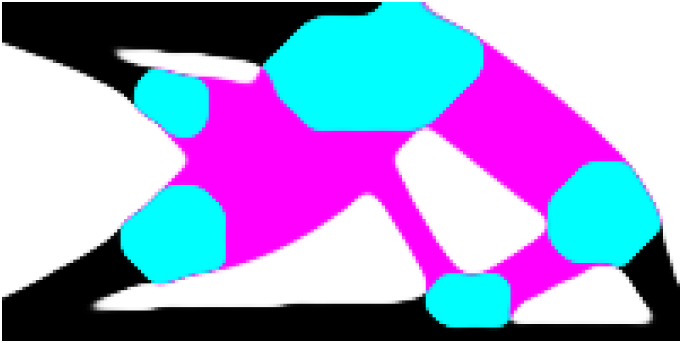}
		\captionsetup{justification=centering, labelformat=empty}
		\caption{\parbox{\linewidth}{\centering (d) 3 materials,\\$v_f= 0.60$,\\$f_0$$= 0.0042$}}
		\label{C2DSF2D3Mvf2}
	\end{subfigure}
	\caption{Optimized results using 1D and 2D SFs}
	\label{C2DSF1D2D}
\end{figure}

\begin{figure}
	\centering
	\begin{subfigure}[t]{0.2\textwidth}
		\centering
		\includegraphics[width=\textwidth]{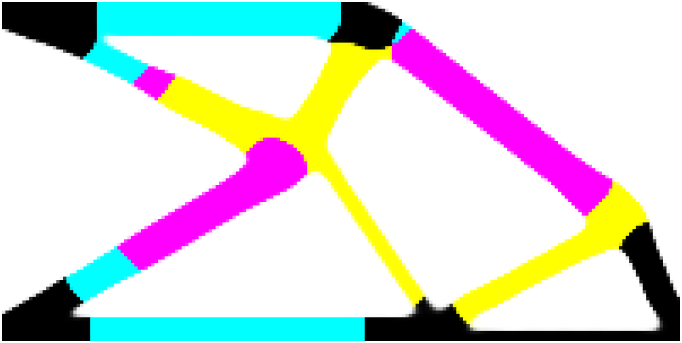}
		\captionsetup{justification=centering, labelformat=empty}
		\caption{\parbox{\linewidth}{\centering (a) 4 materials,\\ $v_f= 0.32$,\\ $f_0$$= 0.0063$}}
		\label{C2DSF3D4Mvf08}
	\end{subfigure}
	\hfill
	\begin{subfigure}[t]{0.2\textwidth}
		\centering
		\includegraphics[width=\textwidth]{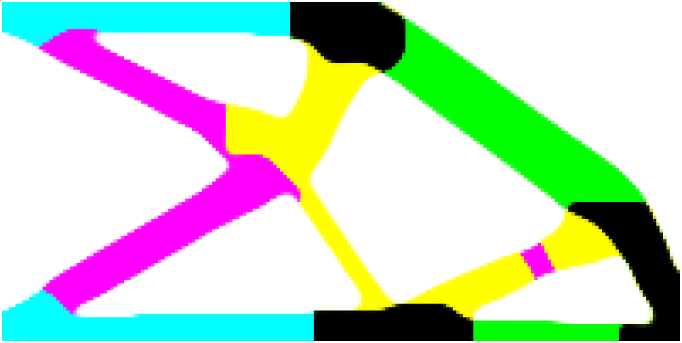}
		\captionsetup{justification=centering, labelformat=empty}
		\caption{\parbox{\linewidth}{\centering (b) 5 materials,\\ $v_f= 0.40$,\\ $f_0$$= 0.0060$}}
		\label{C2DSF3D5Mvf08}
	\end{subfigure}
	\hfill
	\begin{subfigure}[t]{0.2\textwidth}
		\centering
		\includegraphics[width=\textwidth]{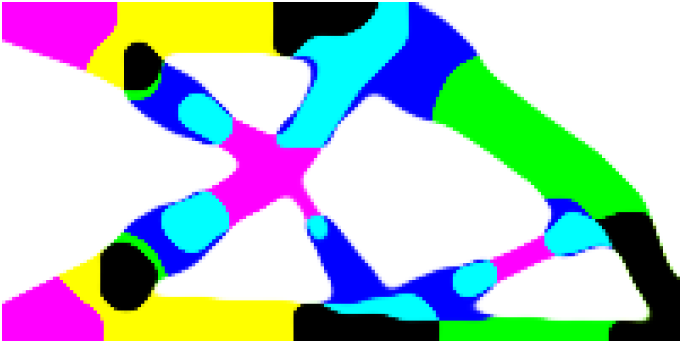}
		\captionsetup{justification=centering, labelformat=empty}
		\caption{\parbox{\linewidth}{\centering (c) 6 materials,\\ $v_f= 0.48$,\\ $f_0$$= 0.0059$}}
		\label{C2DSF3D6Mvf08}
	\end{subfigure}
	\hfill
	\begin{subfigure}[t]{0.2\textwidth}
		\centering
		\includegraphics[width=\textwidth]{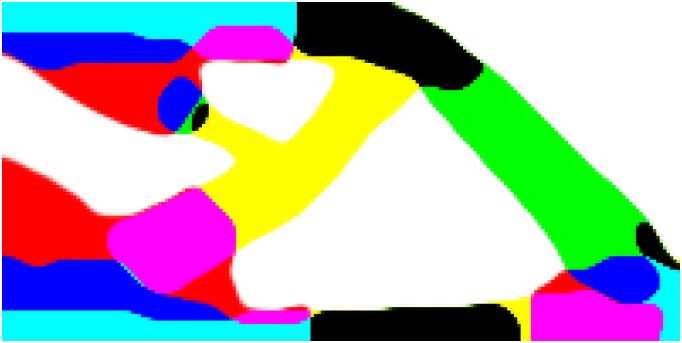}
		\captionsetup{justification=centering, labelformat=empty}
		\caption{\parbox{\linewidth}{\centering (d) 7 materials,\\ $v_f= 0.56$,\\ $f_0$$= 0.0056$}}
		\label{C2DSF3D7Mvf08}
	\end{subfigure}
	\caption{Optimized results using 3D SFs}
	\label{C2DSF3D}
\end{figure}
\subsubsection{Cantilever beam}\label{subsec9}
A 2D cantilever beam with a point load is solved and discussed here. The beam is discretized along the \(x\) and \(y\) directions by \(N_\text{ex}\)$=200$ FEs and \(N_\text{ey}\)$=100$ FEs, respectively. The filter radius $r_{\text{min}}$ is taken as 8. $\eta_p$ is set to 0.5, and $\beta_p$ is taken as 1 and doubled after every 50 iterations until it reaches 128. We use 1D, 2D, 3D, and 4D material elements to solve the problem and achieve solutions for up to 15 materials. The parameters listed in Table~\ref{3Mtable}, Table~\ref{7Mtable}, and Table~\ref{15Mtable} are used.

\begin{figure}
	\centering
	\begin{subfigure}[t]{0.2\textwidth}
		\centering
		\includegraphics[width=\textwidth]{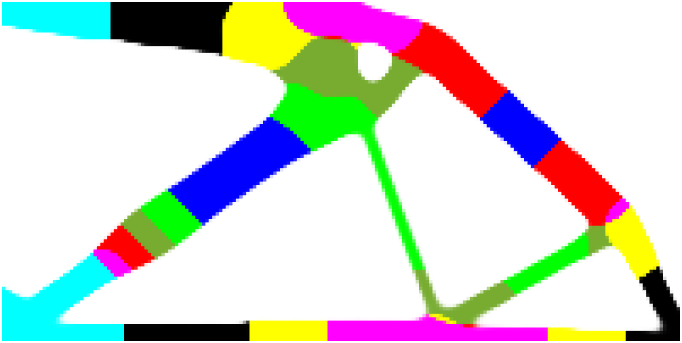}
		\captionsetup{justification=centering, labelformat=empty}
		\caption{\parbox{\linewidth}{\centering (a) 8 materials,\\ $v_f= 0.32$,\\ $f_0$$= 0.0183$}}
		\label{C2DSF4D8Mvf04}
	\end{subfigure}
	\hfill
	\begin{subfigure}[t]{0.2\textwidth}
		\centering
		\includegraphics[width=\textwidth]{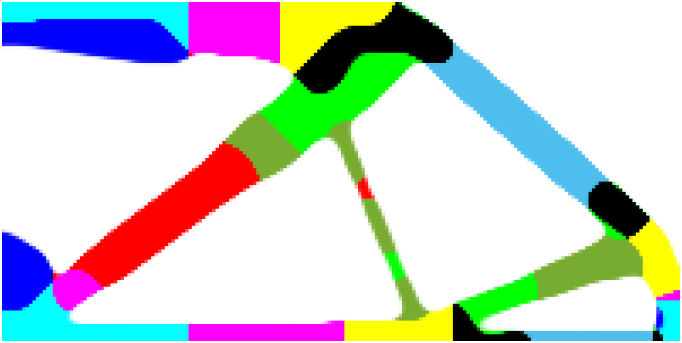}
		\captionsetup{justification=centering, labelformat=empty}
		\caption{\parbox{\linewidth}{\centering (b) 9 materials,\\ $v_f= 0.36$,\\ $f_0$$= 0.0184$}}
		\label{C2DSF4D9Mvf04}
	\end{subfigure}
	\hfill
	\begin{subfigure}[t]{0.2\textwidth}
		\centering
		\includegraphics[width=\textwidth]{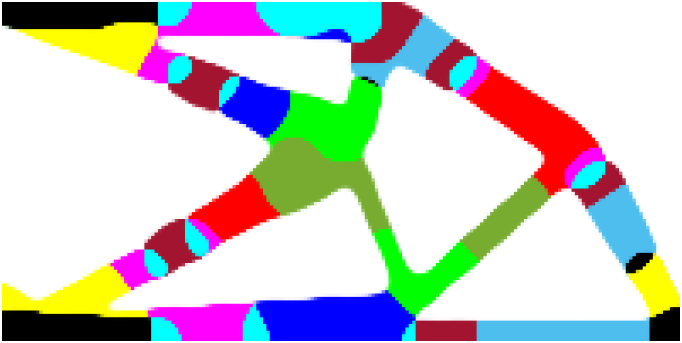}
		\captionsetup{justification=centering, labelformat=empty}
		\caption{\parbox{\linewidth}{\centering (c) 10 materials,\\ $v_f= 0.40$,\\ $f_0$$= 0.0169$}}
		\label{C2DSF4D10Mvf04}
	\end{subfigure}
	\hfill
	\begin{subfigure}[t]{0.2\textwidth}
		\centering
		\includegraphics[width=\textwidth]{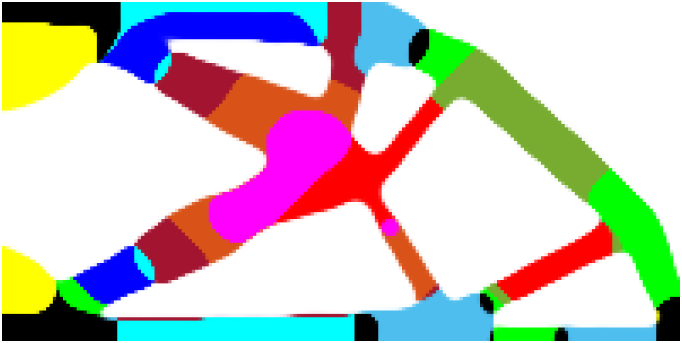}
		\captionsetup{justification=centering, labelformat=empty}
		\caption{\parbox{\linewidth}{\centering (d) 11 materials,\\ $v_f= 0.44$,\\ $f_0$$= 0.0060$}}
		\label{C2DSF4D11Mvf04}
	\end{subfigure}
	\begin{subfigure}[t]{0.2\textwidth}
		\centering
		\includegraphics[width=\textwidth]{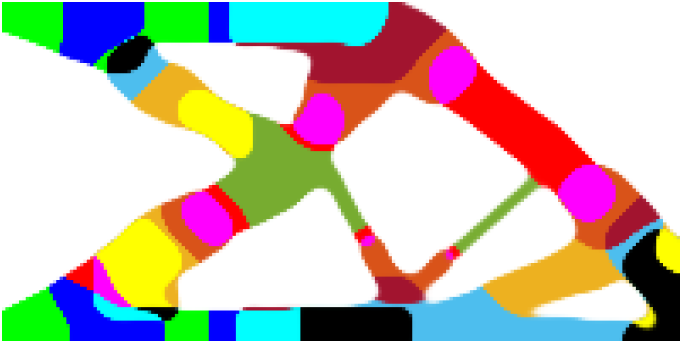}
		\captionsetup{justification=centering, labelformat=empty}
		\caption{\parbox{\linewidth}{\centering (e) 12 materials,\\ $v_f= 0.48$,\\ $f_0$$= 0.0059$}}
		\label{C2DSF4D12Mvf04}
	\end{subfigure}
	\hfill
	\begin{subfigure}[t]{0.2\textwidth}
		\centering
		\includegraphics[width=\textwidth]{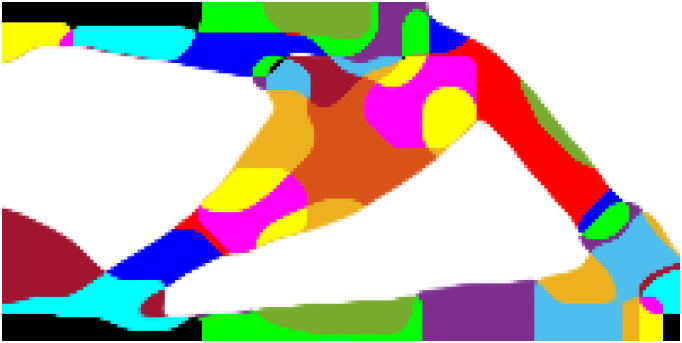}
		\captionsetup{justification=centering, labelformat=empty}
		\caption{\parbox{\linewidth}{\centering (f) 13 materials,\\ $v_f= 0.52$,\\ $f_0$$= 0.0059$}}
		\label{C2DSF4D13Mvf04}
	\end{subfigure}
	\hfill
	\begin{subfigure}[t]{0.2\textwidth}
		\centering
		\includegraphics[width=\textwidth]{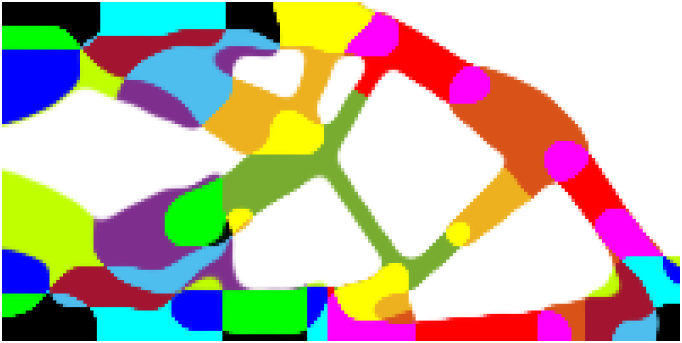}
		\captionsetup{justification=centering, labelformat=empty}
		\caption{\parbox{\linewidth}{\centering (g) 14 materials,\\ $v_f= 0.56$,\\ $f_0$$= 0.0056$}}
		\label{C2DSF4D14Mvf04}
	\end{subfigure}
	\hfill
	\begin{subfigure}[t]{0.2\textwidth}
		\centering
		\includegraphics[width=\textwidth]{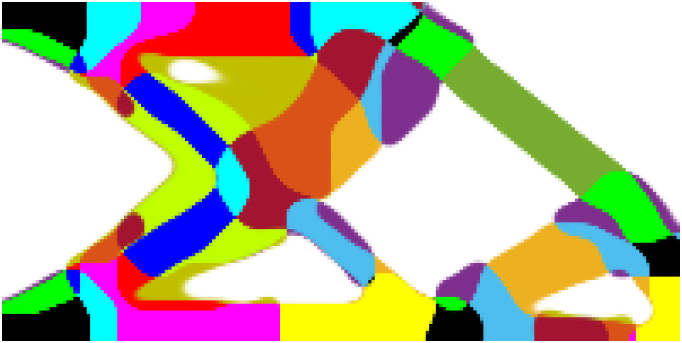}
		\captionsetup{justification=centering, labelformat=empty}
		\caption{\parbox{\linewidth}{\centering (h) 15 materials,\\ $v_f= 0.60$,\\ $f_0$$= 0.0054$}}
		\label{C2DSF4D15Mvf04}
	\end{subfigure}
	\caption{Optimized results using 4D SFs}
	\label{C2DSF4D}
\end{figure}

\begin{figure}
	\centering
	\begin{subfigure}[h!]{0.48\textwidth} 
		\centering
		\begin{tikzpicture}[scale=1]
			\pgfplotsset{compat=1.9}
			\begin{axis}[
				width = 1\textwidth,
				xlabel=  Iteration number,
				ylabel=  Objective function $(f_0)$,
				xmin=0,xmax=400,
                scaled y ticks=false, 
                y tick label style={/pgf/number format/fixed, /pgf/number format/precision=3},
				grid=both,
				major grid style={line width=0.2pt, draw=gray!30},
				legend style={at={(1.0,1.0)}, anchor=north east}]	
				\pgfplotstableread{C2D1Mvf30beta50graph.txt}\mydata;
				\addplot[smooth,{black}, line width=1pt, mark = none]
				table {\mydata};
				\addlegendentry{1 mat., $v_f=0.30$}				
				\pgfplotstableread{C2D1Mvf40beta50graph.txt}\mydata;
				\addplot[smooth,{gray}, line width=1pt, mark = none]
				table {\mydata};
				\addlegendentry{1 mat., $v_f=0.40$}			
			\end{axis}
		\end{tikzpicture}
		\caption{Convergence plot (1D SFs)}
		\label{C2D_SF1D_graphs}
	\end{subfigure}
	\begin{subfigure}[h!]{0.48\textwidth} 
		\centering
		\begin{tikzpicture}[scale=1]
			\pgfplotsset{compat=1.9}
			\begin{axis}[
				width = 1\textwidth,
				xlabel=  Iteration number,
				ylabel=  Objective function $(f_0)$,
				xmin=0,xmax=400,
                scaled y ticks=false, 
                y tick label style={/pgf/number format/fixed, /pgf/number format/precision=3},
				grid=both,
				major grid style={line width=0.2pt, draw=gray!30},
				legend style={at={(1.0,1.0)}, anchor=north east}]
				\pgfplotstableread{C2D2Mbeta50graph.txt}\mydata;
				\addplot[smooth,{cyan}, line width=1pt, mark = none]
				table {\mydata};
				\addlegendentry{2 mat., $v_f=0.40$}			
				\pgfplotstableread{C2D3Mbeta50graph.txt}\mydata;
				\addplot[smooth,{magenta}, line width=1pt, mark = none]
				table {\mydata};
				\addlegendentry{3 mat., $v_f=0.60$}			
			\end{axis}
		\end{tikzpicture}
		\caption{Convergence plot (2D SFs)}
		\label{C2D_SF2D_graphs}
	\end{subfigure}
	\begin{subfigure}[h!]{0.48\textwidth} 
		\centering
		\begin{tikzpicture}[scale=1]
			\pgfplotsset{compat=1.9}
			\begin{axis}[
				width = 1\textwidth,
				xlabel=  Iteration number,
				ylabel=  Objective function $(f_0)$,
				xmin=0,xmax=400,
				grid=both,
				major grid style={line width=0.2pt, draw=gray!30},
				legend style={at={(1.0,1.0)}, anchor=north east}]	
				\pgfplotstableread{C2D4Mbeta50graph.txt}\mydata;
				\addplot[smooth,{yellow}, line width=1pt, mark = none]
				table {\mydata};
				\addlegendentry{4 mat., $v_f=0.32$}				
				\pgfplotstableread{C2D5Mbeta50graph.txt}\mydata;
				\addplot[smooth,{green}, line width=1pt, mark = none]
				table {\mydata};
				\addlegendentry{5 mat., $v_f=0.40$}				
				\pgfplotstableread{C2D6Mbeta50graph.txt}\mydata;
				\addplot[smooth,{blue}, line width=1pt, mark = none]
				table {\mydata};
				\addlegendentry{6 mat., $v_f=0.48$}				
				\pgfplotstableread{C2D7Mbeta50graph.txt}\mydata;
				\addplot[smooth,{red}, line width=1pt, mark = none]
				table {\mydata};
				\addlegendentry{7 mat., $v_f=0.56$}		
			\end{axis}
		\end{tikzpicture}
		\caption{Convergence plot (3D SFs)}
		\label{C2D_SF3D_graphs}
	\end{subfigure}
	\begin{subfigure}[h!]{0.48\textwidth} 
		\centering
		\begin{tikzpicture}[scale=1]
			\pgfplotsset{compat=1.9}
			\begin{axis}[
				width = 1\textwidth,
				xlabel=  Iteration number,
				ylabel=  Objective function $(f_0)$,
				xmin=0,xmax=400,
				grid=both,
				major grid style={line width=0.2pt, draw=gray!30},
				legend style={at={(1.0,1.0)}, anchor=north east}]	
				\pgfplotstableread{C2D8Mbeta50graph.txt}\mydata;
				\addplot[smooth,{shafgreen}, line width=1pt, mark = none]
				table {\mydata};
				\addlegendentry{8 mat., $v_f=0.32$}				
				\pgfplotstableread{C2D9Mbeta50graph.txt}\mydata;
				\addplot[smooth,{skyblue}, line width=1pt, mark = none]
				table {\mydata};
				\addlegendentry{9 mat., $v_f=0.36$}				
				\pgfplotstableread{C2D10Mbeta50graph.txt}\mydata;
				\addplot[smooth,{brownred}, line width=1pt, mark = none]
				table {\mydata};
				\addlegendentry{10 mat., $v_f=0.40$}				
				\pgfplotstableread{C2D11Mbeta50graph.txt}\mydata;
				\addplot[smooth,{redorange}, line width=1pt, mark = none]
				table {\mydata};
				\addlegendentry{11 mat., $v_f=0.44$}				
				\pgfplotstableread{C2D12Mbeta50graph.txt}\mydata;
				\addplot[smooth,{yellowochre}, line width=1pt, mark = none]
				table {\mydata};
				\addlegendentry{12 mat., $v_f=0.48$}				
				\pgfplotstableread{C2D13Mbeta50graph.txt}\mydata;
				\addplot[smooth,{violet}, line width=1pt, mark = none]
				table {\mydata};
				\addlegendentry{13 mat., $v_f=0.52$}				
				\pgfplotstableread{C2D14Mbeta50graph.txt}\mydata;
				\addplot[smooth,{greenyellow}, line width=1pt, mark = none]
				table {\mydata};
				\addlegendentry{14 mat., $v_f=0.56$}				
				\pgfplotstableread{C2D15Mbeta50graph.txt}\mydata;
				\addplot[smooth,{olivegreen}, line width=1pt, mark = none]
				table {\mydata};
				\addlegendentry{15 mat., $v_f=0.60$}		
			\end{axis}
		\end{tikzpicture}
		\caption{Convergence plot (4D SFs)}
		\label{C2D_SF4D_graphs}
	\end{subfigure}
	\caption{Convergence plots (2D Cantilever)}
	\label{C2D_all_graphs}
\end{figure}
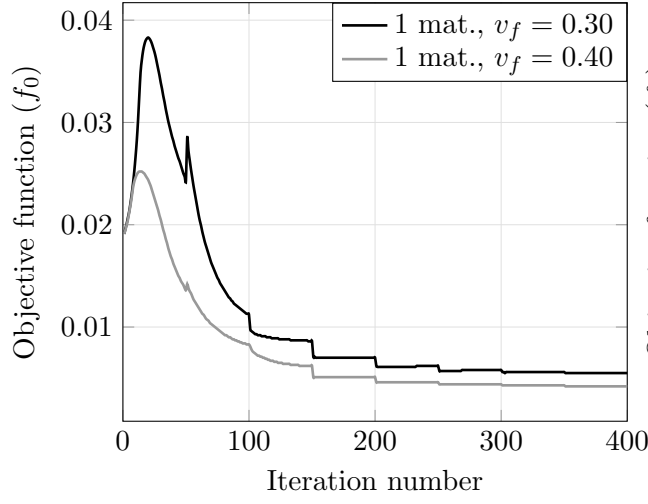
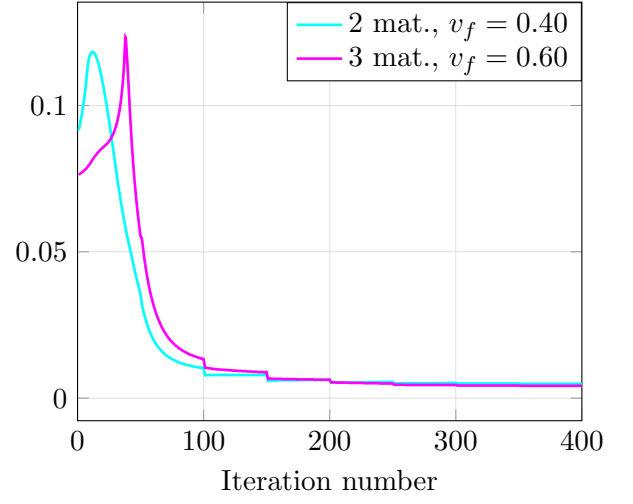
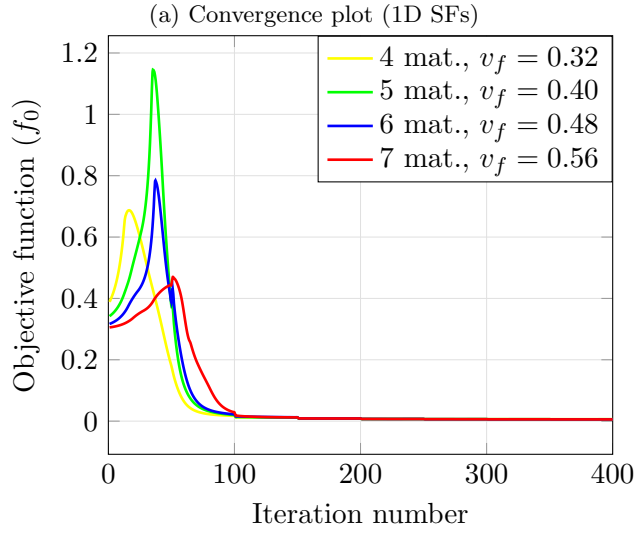
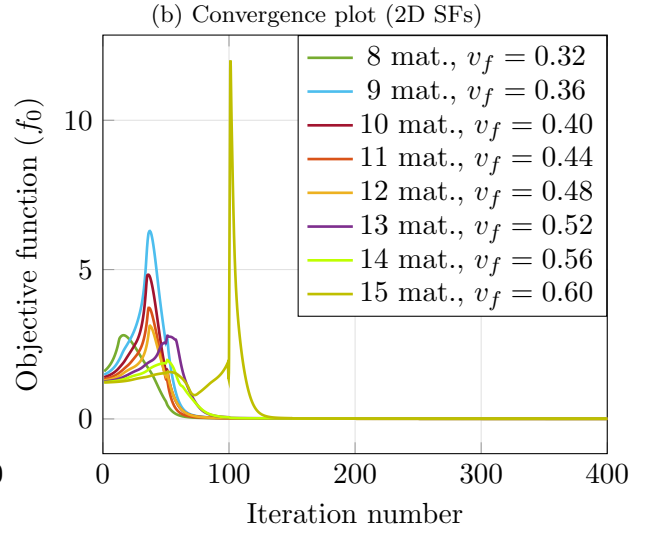

Figure~\ref{C2DSF1D1Mvf3} and Fig.~\ref{C2DSF1D1Mvf4} show the optimized results with 1D material elements with volume fractions of 0.3 and 0.4, respectively. These
optimized results resemble those obtained using the SIMP method with the one-material in conjunction with
the density-based method~\cite{sigmund2013topology}. Fig.~\ref{C2DSF2D2Mvf2} and Fig.~\ref{C2DSF2D3Mvf2} depict the results with 2 materials and 3 materials, respectively, with $v_f$ of 0.2 for each material.  

The optimized results with 4, 5, 6, and 7 materials using 3D SFs (3D material element) are shown in 
Figure \ref{C2DSF3D4Mvf08}, \ref{C2DSF3D5Mvf08}, \ref{C2DSF3D6Mvf08}, and \ref{C2DSF3D7Mvf08}. Volume fraction for each candidate material is fixed to 0.08. For achieving solutions up to 4D SFs (4D material element) are employed and optimized results are depicted in Figs.~\ref{C2DSF4D8Mvf04}, \ref{C2DSF4D9Mvf04}, \ref{C2DSF4D10Mvf04}, \ref{C2DSF4D11Mvf04}, \ref{C2DSF4D12Mvf04}, \ref{C2DSF4D13Mvf04}, \ref{C2DSF4D14Mvf04}, and \ref{C2DSF4D15Mvf04} with 8, 9, 10, 11, 12, 13, 14, and 15 materials, respectively. $v_f=0.04$ is set for each material. A similar observation, as reported in the MBB beam regarding material placement, is evident in each case. Objective function convergence plots are shown in Fig.~\ref{C2D_all_graphs}. Jumps observed in the plots are due to $\beta_p$ updation. Overall, these plots have a converging nature. Volume constraints are found to be active at the end of optimization, by and large. 

In compliance minimization (stiff-structure) problems, the optimizer naturally allocates materials with the highest or nearly highest Young's modulus to regions of highest strain energy density. The optimizer does this because the sensitivity of the objective function with respect to the design variables is directly related to the local strain energy. Therefore, the stiffest material is allocated where it can most effectively maximize stiffness — typically near the supports, where reaction forces are high, and at the points where loads are applied. Next, to demonstrate the robustness of the presented method, CMs are designed with many materials.

\subsection{Compliant mechanism problems}\label{subsec7}
A CM is a mechanism that transfers motion or energy from the deflection of its flexible members. Popular CM optimization problems in TO are the inverter and the gripper problems, which we consider here.

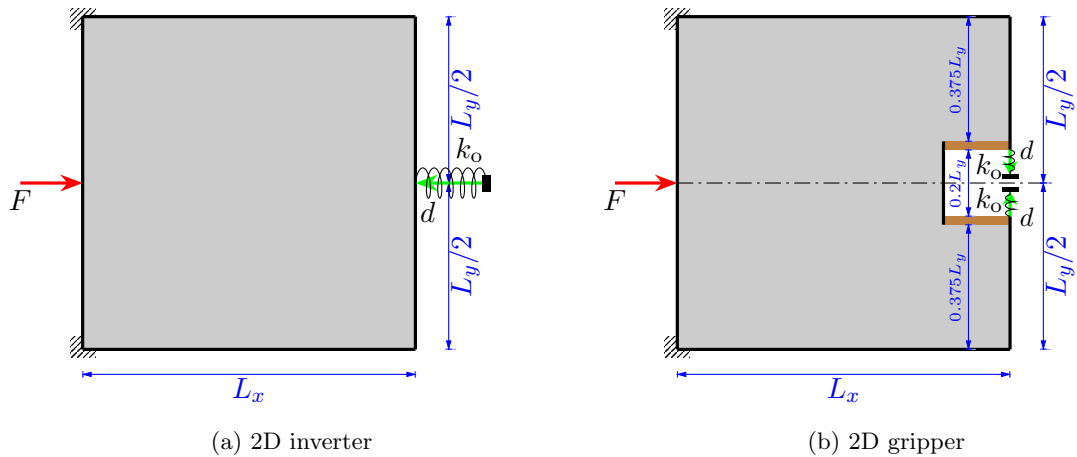
\begin{figure}[b]
\centering
	\begin{subfigure}{0.45\textwidth}
		\begin{tikzpicture}[scale=0.55]
			
			\tikzset{
				LineSpace/.store in=\LineSpace,
				LineSpace=2pt,
				LineThickness/.store in=\LineThickness,
				LineThickness=0.4pt
			}
			
			\fill[pattern=custom hatch] (-0.3,-0.2) rectangle (0.3,0);
			\fill[pattern=custom hatch] (-0.3,-0.2) rectangle (0,0.3);
			\fill[pattern=custom hatch] (-0.3,7.7) rectangle (0,8.2);
			\fill[pattern=custom hatch] (0,8) rectangle (0.3,8.2);
			\fill [gray!50] (0,0) rectangle (8,8);
			\draw[black, very thick] (0,0) -- (8,0);
			\draw[black, very thick] (8,0) -- (8,8);
			\draw[black, very thick] (8,8) -- (0,8);
			\draw[black, very thick] (0,8) -- (0,0);
			\dimline[color=blue,extension start length=2pt, extension end length=2pt,label style={fill=none, yshift=-6pt}]{(0,-0.6)}{(8,-0.6)}{$L_x$};
			\draw[red, very thick, -Stealth] (-1.5,4) -- (0,4);
			\dimline[color=blue,extension start length=5pt, extension end length=5pt,label style={fill=none, yshift=-6pt}]{(8.8,0)}{(8.8,4)}{$L_y/2$};
			\dimline[color=blue,extension start length=5pt, extension end length=5pt,label style={fill=none, yshift=-6pt}]{(8.8,4)}{(8.8,8)}{$L_y/2$};
			\draw[green, very thick, -Stealth] (9.8,4) -- (8,4);
            \draw[decorate, decoration={coil, aspect=0.3, segment length=1.5mm, amplitude=2mm}] (8,4) -- (9.8,4);
            \fill [black] (9.6,3.8) rectangle (9.8,4.2);
            \node at (9.3,4.8) {$k_{\text{o}}$};
			\node at (-1.5,3.6) {$F$};
			\node at (8.3,3.3) {$d$};
		\end{tikzpicture}
		\caption{2D inverter}
		\label{Inverter_2D}
	\end{subfigure}
	\begin{subfigure}{0.45\textwidth}
		\begin{tikzpicture}[scale=0.55]			
			\tikzset{
				LineSpace/.store in=\LineSpace,
				LineSpace=2pt,
				LineThickness/.store in=\LineThickness,
				LineThickness=0.4pt
			}
			
			\fill[pattern=custom hatch] (-0.3,-0.2) rectangle (0.3,0);
			\fill[pattern=custom hatch] (-0.3,-0.2) rectangle (0,0.3);
			\fill[pattern=custom hatch] (-0.3,7.7) rectangle (0,8.2);
			\fill[pattern=custom hatch] (0,8) rectangle (0.3,8.2);
			\fill [gray!50] (0,0) rectangle (8,8);
			\fill [white] (6.4,3) rectangle (8,5);
			\fill [brown] (6.4,3) rectangle (8,3.2);
			\fill [brown] (6.4,5) rectangle (8,4.8);
			
			\draw[black, very thick] (0,0) -- (8,0);
			\draw[black, very thick] (8,0) -- (8,3.2);
			\draw[black, very thick] (8,4.8) -- (8,8);
			\draw[black, very thick] (8,8) -- (0,8);
			\draw[black, very thick] (0,8) -- (0,0);
			\draw[black, very thick] (6.4,3) -- (6.4,5);
			
			\draw[dash pattern=on 6pt off 2pt on 1pt off 2pt] (0,4) -- (9,4);
			\dimline[color=blue,extension start length=2pt, extension end length=2pt,label style={fill=none, yshift=-6pt}]{(0,-0.6)}{(8,-0.6)}{$L_x$};
			\dimline[color=blue,extension start length=5pt, extension end length=5pt,label style={fill=none, yshift=-6pt}]{(8.8,0)}{(8.8,4)}{$L_y/2$};
			\dimline[color=blue,extension start length=5pt, extension end length=5pt,label style={fill=none, yshift=-6pt}]{(8.8,4)}{(8.8,8)}{$L_y/2$};
			\dimline[color=blue,extension start length=5pt, extension end length=5pt,label style={fill=none, yshift=4pt}]{(7,0)}{(7,3)}{\tiny$0.375L_y$};
			\dimline[color=blue,extension start length=5pt, extension end length=5pt,label style={fill=none, yshift=4pt}]{(7,5)}{(7,8)}{\tiny$0.375L_y$};
			\dimline[color=blue,extension start length=5pt, extension end length=5pt,label style={fill=none, yshift=4pt}]{(7,3.2)}{(7,4.8)}{\tiny$0.2L_y$};
			
			\draw[red, very thick, -Stealth] (-1.5,4) -- (0,4);
			\draw[green, very thick, -{Stealth[length=1.8mm, width=2mm]}] (8,3.2) -- (8,3.8);
            \draw[decorate, decoration={coil, aspect=0.3, segment length=1mm, amplitude=0.6mm}] (8,3.2) -- (8,3.8);
            \fill [black] (8.2,3.8) rectangle (7.8,3.9);
			\draw[green, very thick, -{Stealth[length=1.8mm, width=2mm]}] (8,4.8) -- (8,4.2);
            \draw[decorate, decoration={coil, aspect=0.3, segment length=1mm, amplitude=0.6mm}] (8,4.8) -- (8,4.2);
            \fill [black] (8.2,4.2) rectangle (7.8,4.1);
            \node at (8.4,3.2) {\small$d$};
            \node at (8.4,4.8) {\small$d$};
			\node at (-1.5,3.6) {$F$};
			\node at (7.5,3.6) {$k_{\text{o}}$};
			\node at (7.5,4.4) {$k_{\text{o}}$};
		\end{tikzpicture}
		\caption{2D gripper}
		\label{Gripper_2D}
	\end{subfigure}
	\caption{2D design domain of (a) an inverter mechanism, and (b) a gripper mechanism}
	\label{Compliant_mechanism_2D}
\end{figure}

Figure~\ref{Inverter_2D}  and Fig.~\ref {Gripper_2D} show the design domains for a 2D inverter and gripper mechanisms, respectively. An actuating load is applied at the midpoint of the left edge of each mechanism. The top-left and bottom-left corners are fixed. In the case of an inverter CM, it is desired that a point on the right edge corresponds to the input load, and should achieve maximum deformation in the opposite direction. Whereas the gripper mechanism should perform the gripping action. Two jaws are provided to facilitate the same (Fig.~\ref{Gripper_2D}). These jaws are indicated via non-design passive solid elements. As mentioned in Sec.~\ref{subsec5}, we minimize $\left(-\alpha \frac{\mathbf{L}^T\mathbf{U}}{\mathbf{F}^T\mathbf{U}}\right)$ to get the optimized CMs. ${F}$ is set to $10^{-3}$ unit in $x-$direction (Fig.~\ref{Compliant_mechanism_2D}). $\alpha$ is set to $100$. $\mathbf{L}$ is a force vector consisting of all zero entries except at the output DOF, where its value is 1.  $\mathbf{L}^T\mathbf{U}$ gives us the output displacement.  Stiffness of the output spring ($k_{\text{o}}$) is taken here as 0.05 for both inverter and gripper problems. Similar to SS cases, parameters mentioned in Table~\ref{3Mtable},  Table~\ref{7Mtable}, Table~\ref{15Mtable} and Table~\ref{24Mtable} are used, and upto 5D and 4D material elements are used to optimize mechanisms for inverter and gripper, respectively.

\subsubsection{Inverter mechanism}\label{subsec10}
The design domain of the inverter CM is parametrized along the \(x\) and \(y\) directions by \(N_\text{ex}\)$=160$ and \(N_\text{ey}\)$=160$ FEs, respectively, and a symmetric half model is used to obtain the optimized results. Filter radius $r_{\text{min}}$ is taken to 6.4. The value of $\eta_p =0.5$ is considered. $\beta_p$ starts from 1 and is doubled after every 50 iterations until it reaches 128. Figure~\ref{I2DSF1D} depicts the optimized mechanism obtained using 1D and 2D material elements. For the former, $v_f=0.3$ and $v_f=0.4$ are considered, and optimized mechanisms are reported in Fig.~\ref{I2DSF1D1Mvf3} and Fig.~\ref{I2DSF1D1Mvf4}, respectively. Optimized inverter mechanisms with 2 and 3 materials are reported in Fig.~\ref{I2DSF2D2Mvf2} and Fig.~\ref{I2DSF2D3Mvf2} using 2D SFs (2D material element). $v_f=0.2$ is considered for each material.

Using the 3D material element, results for 4, 5, 6, and 7 materials are depicted in Figs.~\ref{I2DSF3D4Mvf08}, \ref{I2DSF3D5Mvf08}, \ref{I2DSF3D6Mvf08}, and \ref{I2DSF3D7Mvf08}, respectively. Further, to achieve optimized mechanisms up to 15 material, 4D SFs are employed and results are reported in  Figs.~\ref{I2DSF4D8Mvf04}, \ref{I2DSF4D9Mvf04}, \ref{I2DSF4D10Mvf04}, \ref{I2DSF4D11Mvf04}, \ref{I2DSF4D12Mvf04}, \ref{I2DSF4D13Mvf04}, \ref{I2DSF4D14Mvf04}, and \ref{I2DSF4D15Mvf04}, respectively.
5D SFs are used to achieve optimized mechanisms with 16, 18, 20, and 24 materials as shown in Figs.~\ref{I2DSF5D16Mvf025}, \ref{I2DSF5D18Mvf025}, \ref{I2DSF5D20Mvf025}, \ref{I2DSF5D24Mvf025}, respectively. The volume fraction is set to 0.025 for each material while using 5D SFs.

We can see that the same material occupies the input and output portions. The material with the highest Young's modulus occupies those regions in the 1 and 3 material cases.
Fig.~\ref{I2DSF3D} represents the optimized results using 3D material elements. The volume fraction, $v_f=0.08$, is used for each candidate material. Here, too, one notices that the same material is distributed in each optimized result's input and output regions. 
Further, using 4D material elements, the optimized results are obtained for 8 to 15 materials (Fig.~\ref{I2DSF4D}). The volume fraction of each candidate material is set to 0.04. We observe that the same material is distributed at each result's input and output regions, by and large.

The objective convergence plots for all the optimized inverter CMs are shown in Fig.~\ref{I2D_all_graphs}. These plots converge with some jumps in the middle due to updation in $\beta_p$. Additionally, one notes that the objective converges in 300 iterations for all the cases presented. It is noted that volume constraints remain active at the end of the optimization.

\begin{figure}
	\centering
	\begin{subfigure}[H]{0.2\textwidth}
		\centering
		\includegraphics[width=\textwidth]{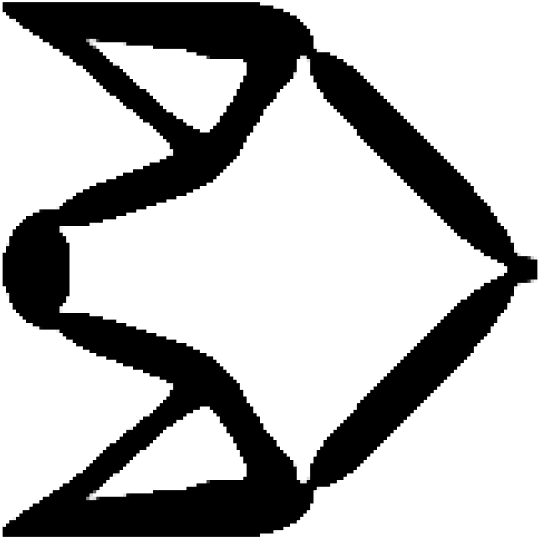}
		\captionsetup{justification=centering, labelformat=empty}
		\caption{\parbox{\linewidth}{\centering (a) 1 material,\\ $v_f= 0.30$,\\ $f_0$$= -28.2$}}
		\label{I2DSF1D1Mvf3}
	\end{subfigure}
	\hfill
	\begin{subfigure}[H]{0.2\textwidth}
		\centering
		\includegraphics[width=\textwidth]{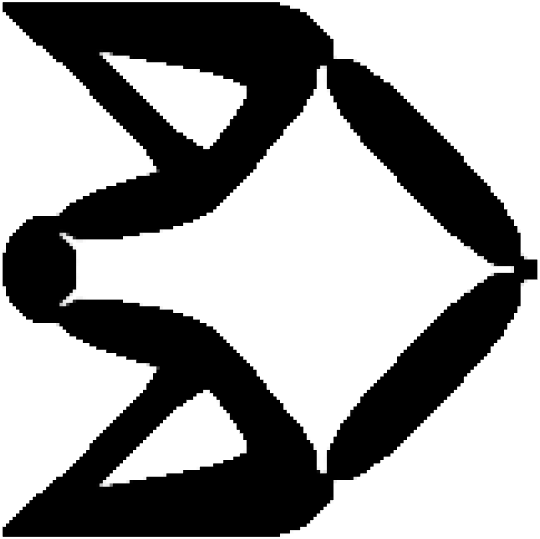}
		\captionsetup{justification=centering, labelformat=empty}
		\caption{\parbox{\linewidth}{\centering (b) 1 material,\\ $v_f= 0.40$,\\ $f_0$$= -32.2$}}
		\label{I2DSF1D1Mvf4}
	\end{subfigure}
	\hfill
	\begin{subfigure}[H]{0.2\textwidth}
		\centering
		\includegraphics[width=\textwidth]{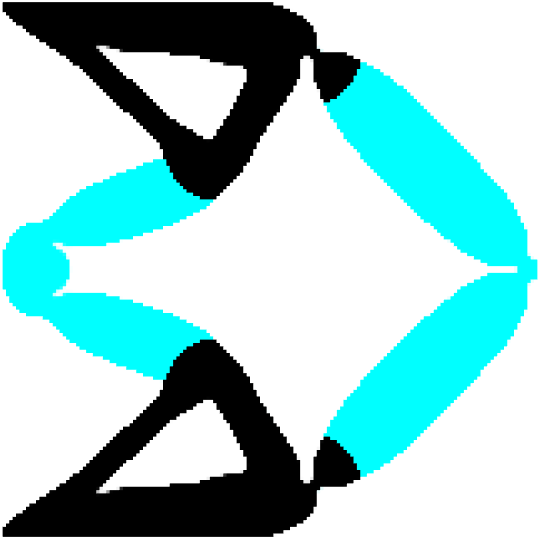}
		\captionsetup{justification=centering, labelformat=empty}
		\caption{\parbox{\linewidth}{\centering (c) 2 materials,\\ $v_f= 0.40$,\\ $f_0$$= -112.8$}}
		\label{I2DSF2D2Mvf2}
	\end{subfigure}
	\hfill
	\begin{subfigure}[H]{0.2\textwidth}
		\centering
		\includegraphics[width=\textwidth]{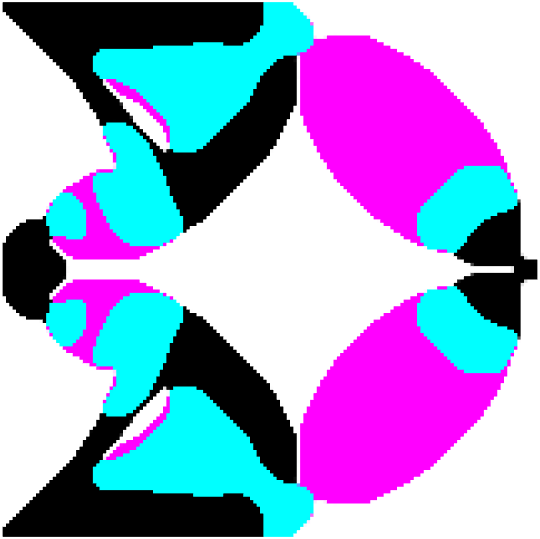}
		\captionsetup{justification=centering, labelformat=empty}
		\caption{\parbox{\linewidth}{\centering (d) 3 materials,\\ $v_f= 0.60$,\\ $f_0$$= -99.8$}}
		\label{I2DSF2D3Mvf2}
	\end{subfigure}
	\caption{Optimized results using 1D and 2D SFs}
	\label{I2DSF1D}
\end{figure}
\begin{figure}
	\centering
	\begin{subfigure}[t]{0.2\textwidth}
		\centering
		\includegraphics[width=\textwidth]{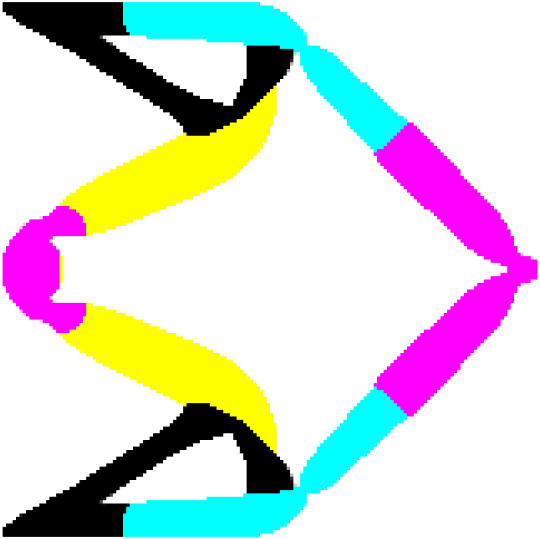}
		\captionsetup{justification=centering, labelformat=empty}
		\caption{\parbox{\linewidth}{\centering (a) 4 materials,\\ $v_f= 0.32$,\\ $f_0$$= -406.4$}}
		\label{I2DSF3D4Mvf08}
	\end{subfigure}
	\hfill
	\begin{subfigure}[t]{0.2\textwidth}
		\centering
		\includegraphics[width=\textwidth]{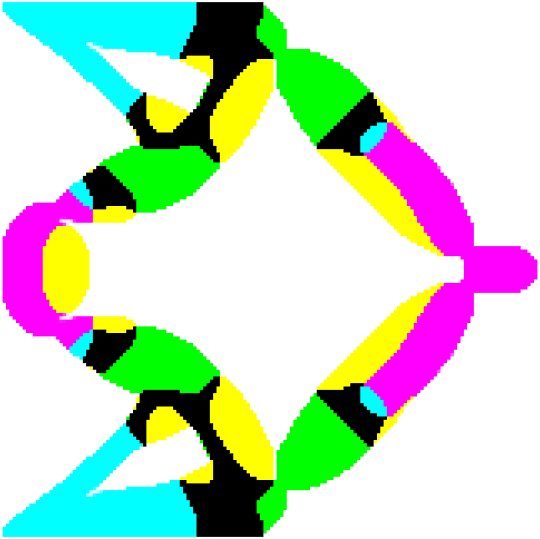}
		\captionsetup{justification=centering, labelformat=empty}
		\caption{\parbox{\linewidth}{\centering (b) 5 materials,\\ $v_f= 0.40$,\\ $f_0$$= -350.4$}}
		\label{I2DSF3D5Mvf08}
	\end{subfigure}
	\hfill
	\begin{subfigure}[t]{0.2\textwidth}
		\centering
		\includegraphics[width=\textwidth]{I2D6Mbeta50k05}
		\captionsetup{justification=centering, labelformat=empty}
		\caption{\parbox{\linewidth}{\centering (c) 6 materials,\\ $v_f= 0.48$,\\ $f_0$$= -350.0$}}
		\label{I2DSF3D6Mvf08}
	\end{subfigure}
	\hfill
	\begin{subfigure}[t]{0.2\textwidth}
		\centering
		\includegraphics[width=\textwidth]{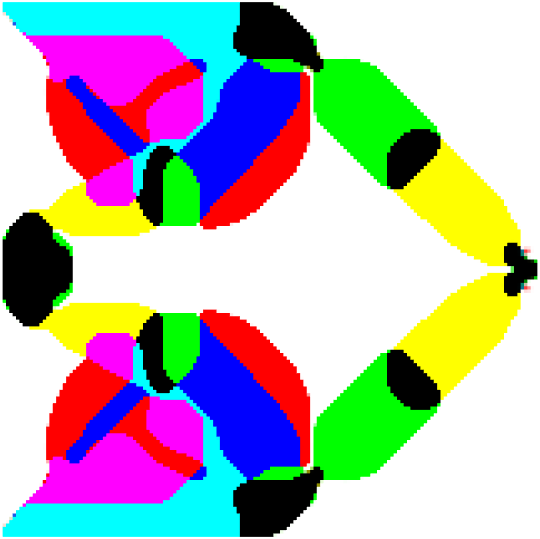}
		\captionsetup{justification=centering, labelformat=empty}
		\caption{\parbox{\linewidth}{\centering (d) 7 materials,\\ $v_f= 0.56$,\\ $f_0$$= -349.6$}}
		\label{I2DSF3D7Mvf08}
	\end{subfigure}
	\caption{Optimized results using 3D SFs}
	\label{I2DSF3D}
\end{figure}
\begin{figure}
	\centering
	\begin{subfigure}[t]{0.2\textwidth}
		\centering
		\includegraphics[width=\textwidth]{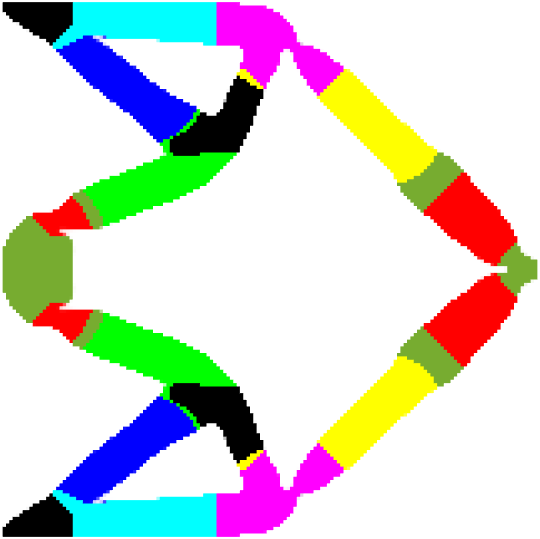}
		\captionsetup{justification=centering, labelformat=empty}
		\caption{\parbox{\linewidth}{\centering (a) 8 materials,\\ $v_f= 0.32$,\\ $f_0$$= -1551.1$}}
		\label{I2DSF4D8Mvf04}
	\end{subfigure}
	\hfill
	\begin{subfigure}[t]{0.2\textwidth}
		\centering
		\includegraphics[width=\textwidth]{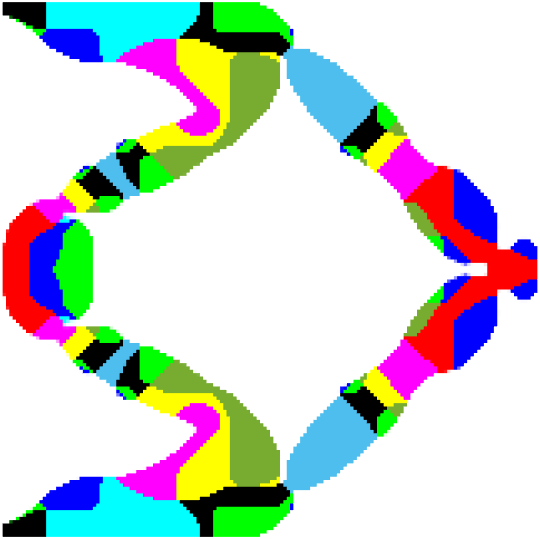}
		\captionsetup{justification=centering, labelformat=empty}
		\caption{\parbox{\linewidth}{\centering (b) 9 materials,\\ $v_f= 0.36$,\\ $f_0$$= -1424.4$}}
		\label{I2DSF4D9Mvf04}
	\end{subfigure}
	\hfill
	\begin{subfigure}[t]{0.2\textwidth}
		\centering
		\includegraphics[width=\textwidth]{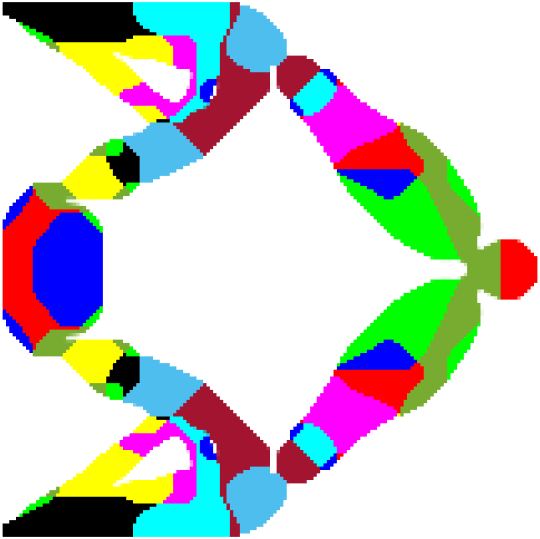}
		\captionsetup{justification=centering, labelformat=empty}
		\caption{\parbox{\linewidth}{\centering (c) 10 materials,\\ $v_f= 0.40$,\\ $f_0$$= -1347.2$}}
		\label{I2DSF4D10Mvf04}
	\end{subfigure}
	\hfill
	\begin{subfigure}[t]{0.2\textwidth}
		\centering
		\includegraphics[width=\textwidth]{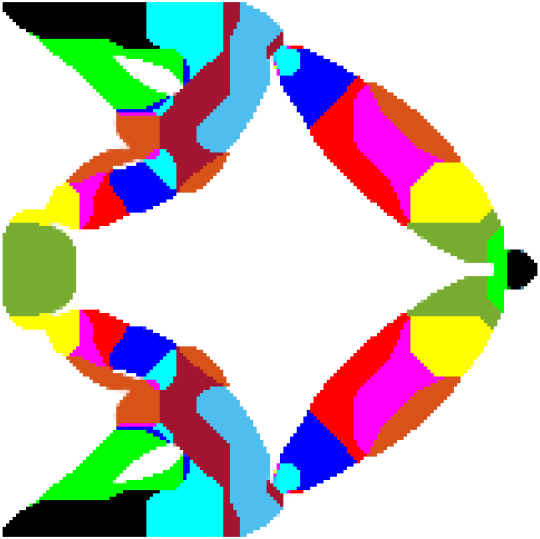}
		\captionsetup{justification=centering, labelformat=empty}
		\caption{\parbox{\linewidth}{\centering (d) 11 materials,\\ $v_f= 0.44$,\\ $f_0$$= -1377.4$}}
		\label{I2DSF4D11Mvf04}
	\end{subfigure}
	\begin{subfigure}[t]{0.2\textwidth}
		\centering
		\includegraphics[width=\textwidth]{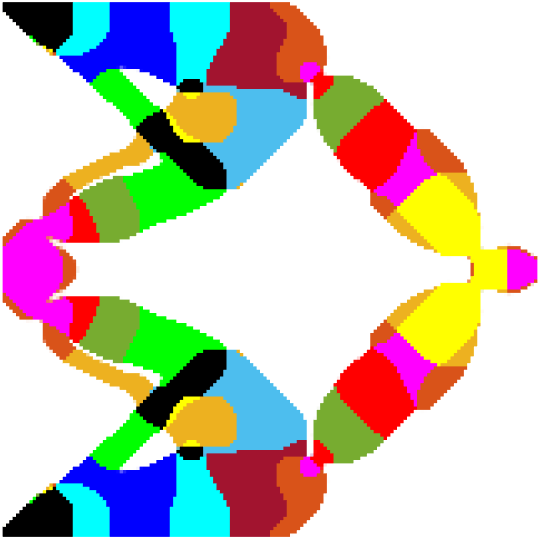}
		\captionsetup{justification=centering, labelformat=empty}
		\caption{\parbox{\linewidth}{\centering (e) 12 materials,\\ $v_f= 0.48$,\\ $f_0$$= -1428.4$}}
		\label{I2DSF4D12Mvf04}
	\end{subfigure}
	\hfill
	\begin{subfigure}[t]{0.2\textwidth}
		\centering
		\includegraphics[width=\textwidth]{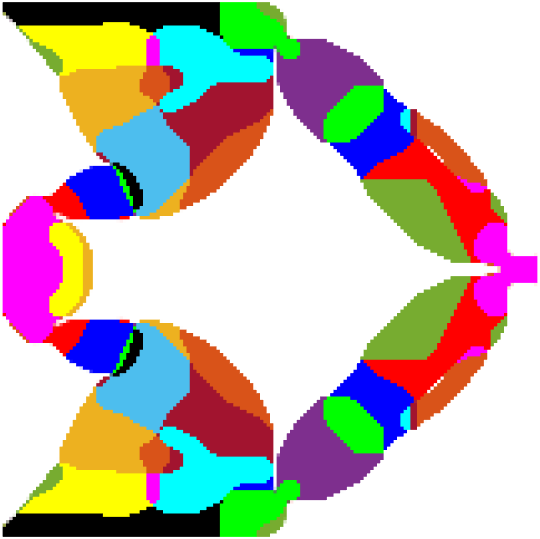}
		\captionsetup{justification=centering, labelformat=empty}
		\caption{\parbox{\linewidth}{\centering (f) 13 materials,\\ $v_f= 0.52$,\\ $f_0$$= -1341.1$}}
		\label{I2DSF4D13Mvf04}
	\end{subfigure}
	\hfill
	\begin{subfigure}[t]{0.2\textwidth}
		\centering
		\includegraphics[width=\textwidth]{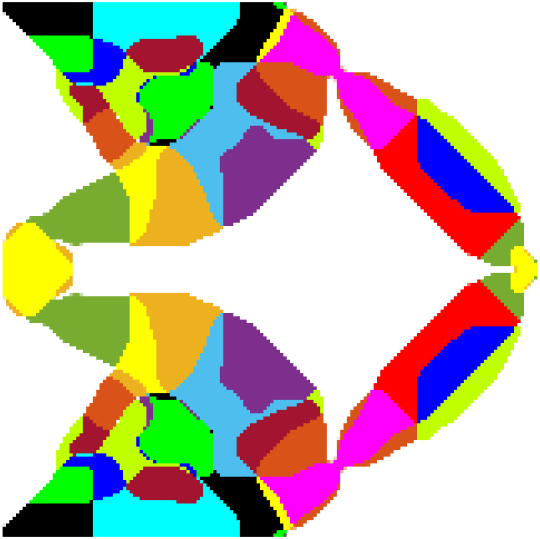}
		\captionsetup{justification=centering, labelformat=empty}
		\caption{\parbox{\linewidth}{\centering (g) 14 materials,\\ $v_f= 0.56$,\\ $f_0$$= -1434.8$}}
		\label{I2DSF4D14Mvf04}
	\end{subfigure}
	\hfill
	\begin{subfigure}[t]{0.2\textwidth}
		\centering
		\includegraphics[width=\textwidth]{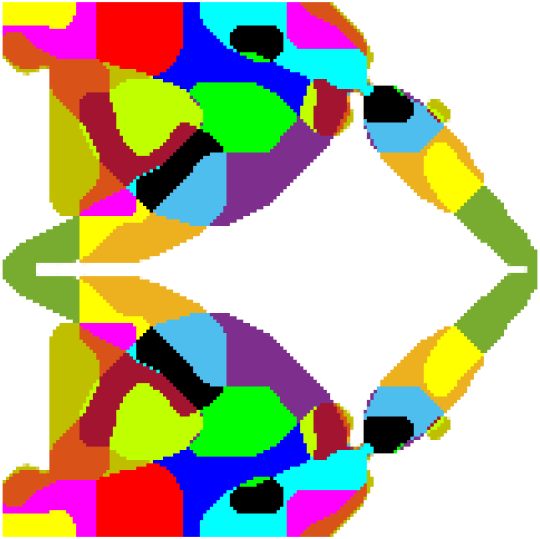}
		\captionsetup{justification=centering, labelformat=empty}
		\caption{\parbox{\linewidth}{\centering (h) 15 materials,\\ $v_f= 0.60$,\\ $f_0$$= -1241.0$}}
		\label{I2DSF4D15Mvf04}
	\end{subfigure}
	\caption{Optimized results using 4D SFs}
	\label{I2DSF4D}
\end{figure}

\begin{figure}
	\centering
    \begin{subfigure}[t]{0.2\textwidth}
		\centering
		\includegraphics[width=\textwidth]{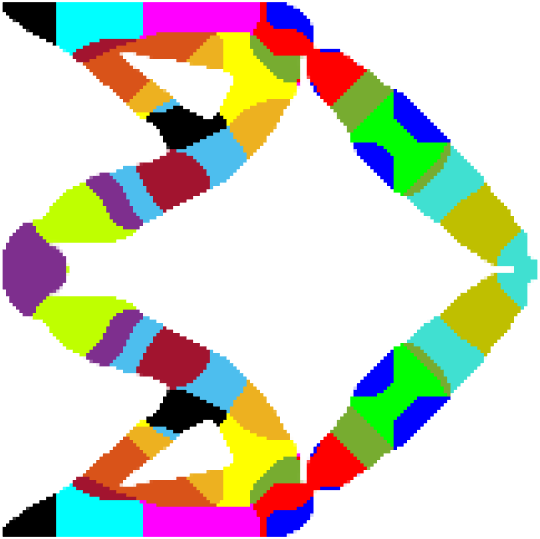}
		\captionsetup{justification=centering, labelformat=empty}
		\caption{\parbox{\linewidth}{\centering (a) 16 materials,\\ $v_f= 0.40$,\\ $f_0$$= -6893.6$}}		
        \label{I2DSF5D16Mvf025}
	\end{subfigure}
	\hfill
	\begin{subfigure}[t]{0.2\textwidth}
		\centering
		\includegraphics[width=\textwidth]{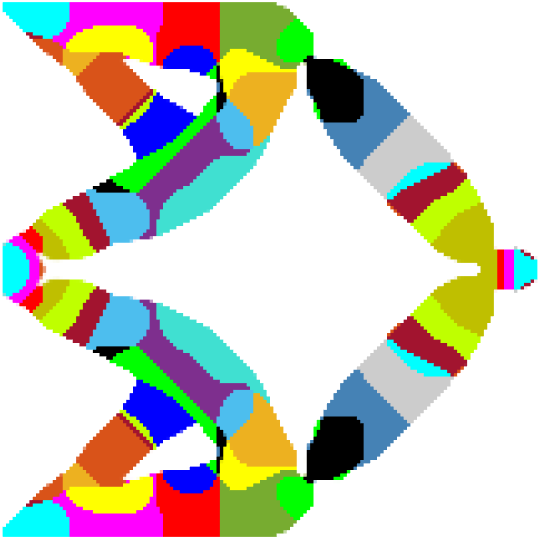}
		\captionsetup{justification=centering, labelformat=empty}
		\caption{\parbox{\linewidth}{\centering (b) 18 materials,\\ $v_f= 0.45$,\\ $f_0$$= -7041.5$}}
		\label{I2DSF5D18Mvf025}
	\end{subfigure}
	\hfill
	\begin{subfigure}[t]{0.2\textwidth}
		\centering
		\includegraphics[width=\textwidth]{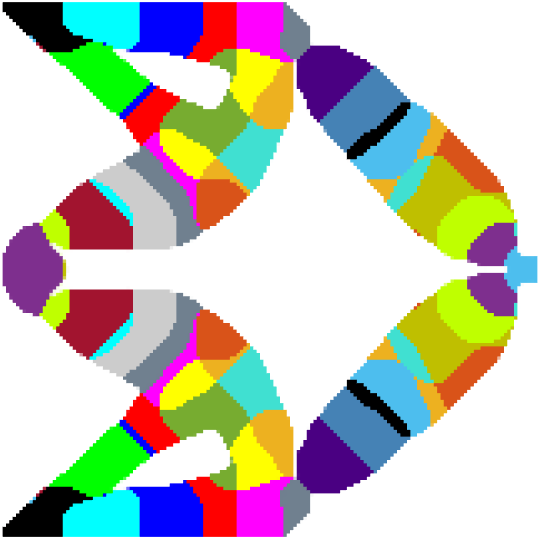}
		\captionsetup{justification=centering, labelformat=empty}
		\caption{\parbox{\linewidth}{\centering (c) 20 materials,\\ $v_f= 0.50$,\\ $f_0$$= -6201.7$}}
		\label{I2DSF5D20Mvf025}
	\end{subfigure}
	\hfill
	\begin{subfigure}[t]{0.2\textwidth}
		\centering
		\includegraphics[width=\textwidth]{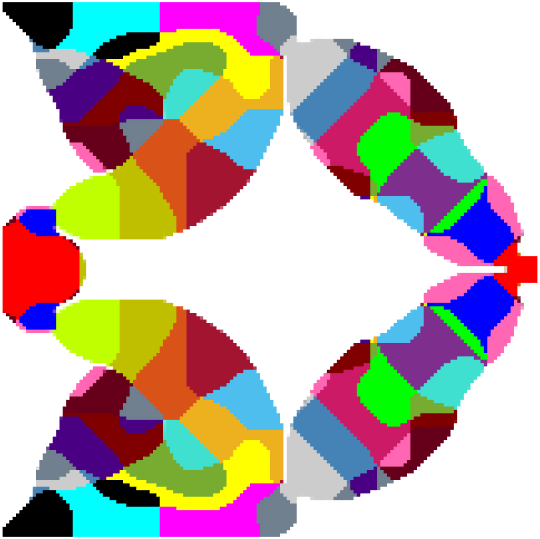}
		\captionsetup{justification=centering, labelformat=empty}
		\caption{\parbox{\linewidth}{\centering (d) 24 materials,\\ $v_f= 0.60$,\\ $f_0$$= -6051.3$}}
		\label{I2DSF5D24Mvf025}
	\end{subfigure}
	\caption{Optimized results using 5D SFs}
	\label{I2DSF5D}
\end{figure}

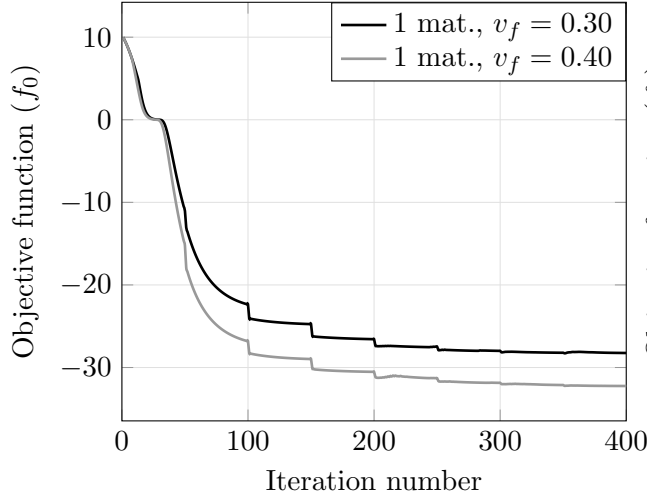
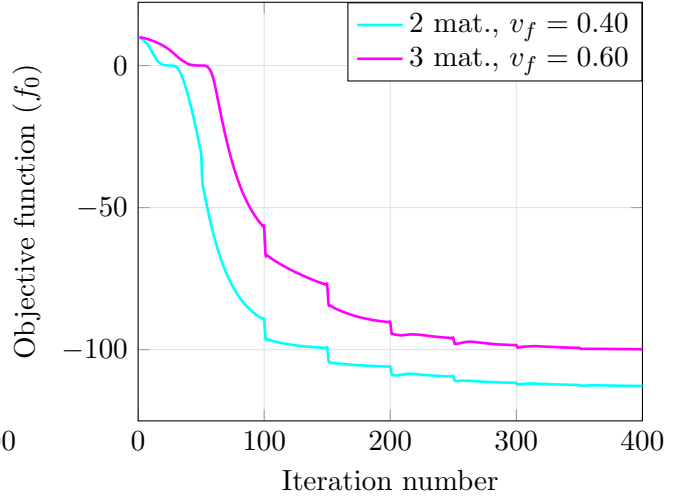
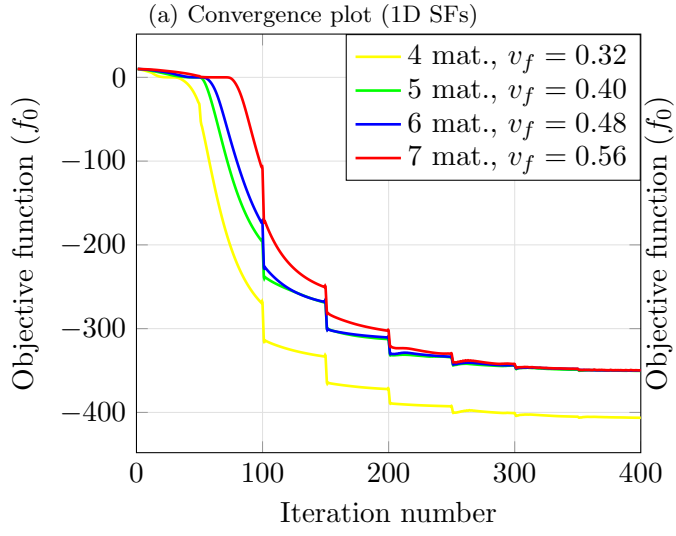
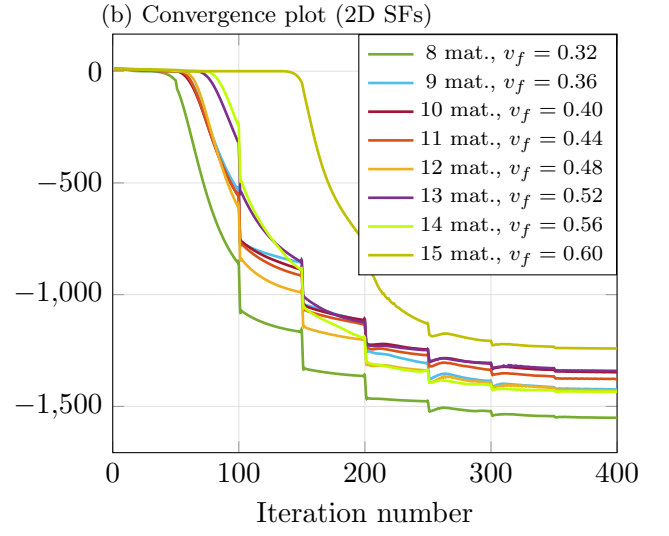
\begin{figure}[]
	\centering
	\begin{subfigure}[h!]{0.48\textwidth} 
		\centering
		\begin{tikzpicture}[scale=1]
			\pgfplotsset{compat=1.9}
			\begin{axis}[
				width = 1\textwidth,
				xlabel=  Iteration number,
				ylabel=  Objective function $(f_0)$,
				xmin=0,xmax=400,
                scaled y ticks=false, 
                y tick label style={/pgf/number format/fixed, /pgf/number format/precision=3},
				grid=both,
				major grid style={line width=0.2pt, draw=gray!30},
				legend style={at={(1.0,1.0)}, anchor=north east}]	
				\pgfplotstableread{I2D1Mvf30beta50k05graph.txt}\mydata;
				\addplot[smooth,{black}, line width=1pt, mark = none]
				table {\mydata};
				\addlegendentry{1 mat., $v_f=0.30$}				
				\pgfplotstableread{I2D1Mvf40beta50k05graph.txt}\mydata;
				\addplot[smooth,{gray}, line width=1pt, mark = none]
				table {\mydata};
				\addlegendentry{1 mat., $v_f=0.40$}			
			\end{axis}
		\end{tikzpicture}
		\caption{Convergence plot (1D SFs)}
		\label{I2D_SF1D_graphs}
	\end{subfigure}
	\begin{subfigure}[h!]{0.48\textwidth} 
		\centering
		\begin{tikzpicture}[scale=1]
			\pgfplotsset{compat=1.9}
			\begin{axis}[
				width = 1\textwidth,
				xlabel=  Iteration number,
				ylabel=  Objective function $(f_0)$,
				xmin=0,xmax=400,
                scaled y ticks=false, 
                y tick label style={/pgf/number format/fixed, /pgf/number format/precision=3},
				grid=both,
				major grid style={line width=0.2pt, draw=gray!30},
				legend style={at={(1.0,1.0)}, anchor=north east}]
				\pgfplotstableread{I2D2Mbeta50k05graph.txt}\mydata;
				\addplot[smooth,{cyan}, line width=1pt, mark = none]
				table {\mydata};
				\addlegendentry{2 mat., $v_f=0.40$}			
				\pgfplotstableread{I2D3Mbeta50k05graph.txt}\mydata;
				\addplot[smooth,{magenta}, line width=1pt, mark = none]
				table {\mydata};
				\addlegendentry{3 mat., $v_f=0.60$}			
			\end{axis}
		\end{tikzpicture}
		\caption{Convergence plot (2D SFs)}
		\label{I2D_SF2D_graphs}
	\end{subfigure}
	\begin{subfigure}[h!]{0.48\textwidth} 
		\centering
		\begin{tikzpicture}[scale=1]
			\pgfplotsset{compat=1.9}
			\begin{axis}[
				width = 1\textwidth,
				xlabel=  Iteration number,
				ylabel=  Objective function $(f_0)$,
				xmin=0,xmax=400,
				grid=both,
				major grid style={line width=0.2pt, draw=gray!30},
				legend style={at={(1.0,1.0)}, anchor=north east}]	
				\pgfplotstableread{I2D4Mbeta50k05graph.txt}\mydata;
				\addplot[smooth,{yellow}, line width=1pt, mark = none]
				table {\mydata};
				\addlegendentry{4 mat., $v_f=0.32$}				
				\pgfplotstableread{I2D5Mbeta50k05graph.txt}\mydata;
				\addplot[smooth,{green}, line width=1pt, mark = none]
				table {\mydata};
				\addlegendentry{5 mat., $v_f=0.40$}				
				\pgfplotstableread{I2D6Mbeta50k05graph.txt}\mydata;
				\addplot[smooth,{blue}, line width=1pt, mark = none]
				table {\mydata};
				\addlegendentry{6 mat., $v_f=0.48$}				
				\pgfplotstableread{I2D7Mbeta50k05graph.txt}\mydata;
				\addplot[smooth,{red}, line width=1pt, mark = none]
				table {\mydata};
				\addlegendentry{7 mat., $v_f=0.56$}		
			\end{axis}
		\end{tikzpicture}
		\caption{Convergence plot (3D SFs)}
		\label{I2D_SF3D_graphs}
	\end{subfigure}
	\begin{subfigure}[h!]{0.48\textwidth} 
		\centering
		\begin{tikzpicture}[scale=1]
			\pgfplotsset{compat=1.9}
			\begin{axis}[
				width = 1\textwidth,
				xlabel=  Iteration number,
				ylabel=  Objective function $(f_0)$,
				xmin=0,xmax=400,
				grid=both,
				major grid style={line width=0.2pt, draw=gray!30},
				legend style={at={(1.0,1.0)}, anchor=north east, font=\scriptsize}]	
				\pgfplotstableread{I2D8Mbeta50k05graph.txt}\mydata;
				\addplot[smooth,{shafgreen}, line width=1pt, mark = none]
				table {\mydata};
				\addlegendentry{8 mat., $v_f=0.32$}				
				\pgfplotstableread{I2D9Mbeta50k05graph.txt}\mydata;
				\addplot[smooth,{skyblue}, line width=1pt, mark = none]
				table {\mydata};
				\addlegendentry{9 mat., $v_f=0.36$}				
				\pgfplotstableread{I2D10Mbeta50k05graph.txt}\mydata;
				\addplot[smooth,{brownred}, line width=1pt, mark = none]
				table {\mydata};
				\addlegendentry{10 mat., $v_f=0.40$}				
				\pgfplotstableread{I2D11Mbeta50k05graph.txt}\mydata;
				\addplot[smooth,{redorange}, line width=1pt, mark = none]
				table {\mydata};
				\addlegendentry{11 mat., $v_f=0.44$}				
				\pgfplotstableread{I2D12Mbeta50k05graph.txt}\mydata;
				\addplot[smooth,{yellowochre}, line width=1pt, mark = none]
				table {\mydata};
				\addlegendentry{12 mat., $v_f=0.48$}				
				\pgfplotstableread{I2D13Mbeta50k05graph.txt}\mydata;
				\addplot[smooth,{violet}, line width=1pt, mark = none]
				table {\mydata};
				\addlegendentry{13 mat., $v_f=0.52$}				
				\pgfplotstableread{I2D14Mbeta50k05graph.txt}\mydata;
				\addplot[smooth,{greenyellow}, line width=1pt, mark = none]
				table {\mydata};
				\addlegendentry{14 mat., $v_f=0.56$}				
				\pgfplotstableread{I2D15Mbeta50k05graph.txt}\mydata;
				\addplot[smooth,{olivegreen}, line width=1pt, mark = none]
				table {\mydata};
				\addlegendentry{15 mat., $v_f=0.60$}		
			\end{axis}
		\end{tikzpicture}
		\caption{Convergence plot (4D SFs)}
		\label{I2D_SF4D_graphs}
	\end{subfigure}
	\caption{Convergence plots (2D Inverter)}
	\label{I2D_all_graphs}
\end{figure}

\begin{figure}[]
	\centering
	\begin{subfigure}[H]{0.2\textwidth}
		\centering
		\includegraphics[width=\textwidth]{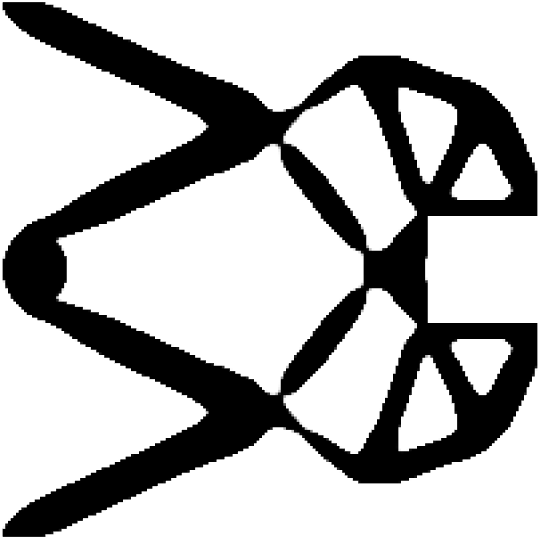}
		\captionsetup{justification=centering, labelformat=empty}
		\caption{\parbox{\linewidth}{\centering (a) 1 material,\\ $v_f= 0.30$,\\ $f_0$$= -271.2$}}
		\label{Gr2DSF1D1Mvf3}
	\end{subfigure}
	\hfill
	\begin{subfigure}[H]{0.2\textwidth}
		\centering
		\includegraphics[width=\textwidth]{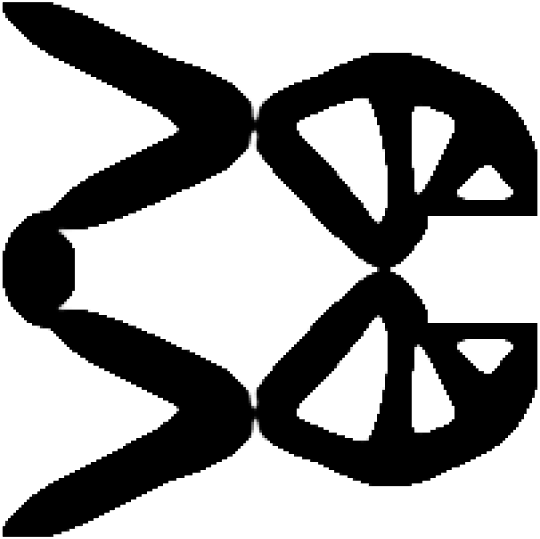}
		\captionsetup{justification=centering, labelformat=empty}
		\caption{\parbox{\linewidth}{\centering (b) 1 material,\\ $v_f= 0.40$,\\ $f_0$$= -299.2$}}
		\label{Gr2DSF1D1Mvf4}
	\end{subfigure}
	\hfill
	\begin{subfigure}[H]{0.2\textwidth}
		\centering
		\includegraphics[width=\textwidth]{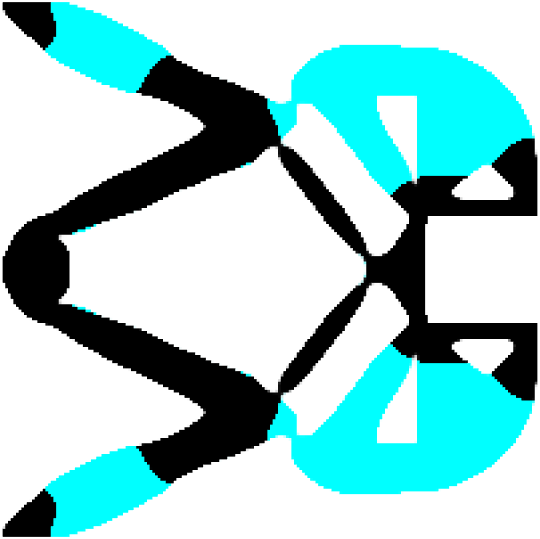}
		\captionsetup{justification=centering, labelformat=empty}
		\caption{\parbox{\linewidth}{\centering (c) 2 materials,\\ $v_f= 0.40$,\\ $f_0$$= -1053.5$}}
		\label{Gr2DSF2D2Mvf2}
	\end{subfigure}
	\hfill
	\begin{subfigure}[H]{0.2\textwidth}
		\centering
		\includegraphics[width=\textwidth]{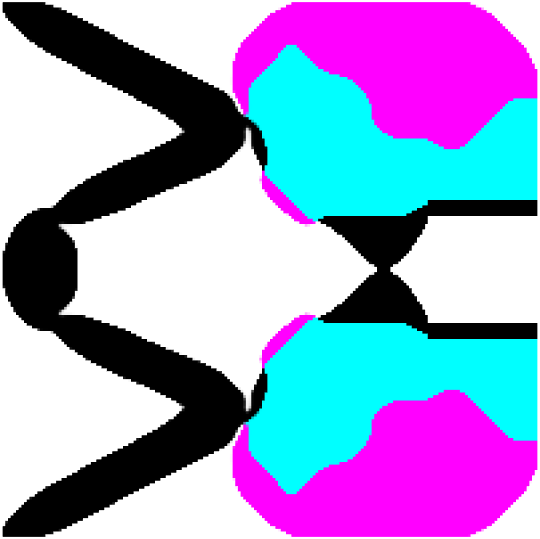}
		\captionsetup{justification=centering, labelformat=empty}
		\caption{\parbox{\linewidth}{\centering (d) 3 materials,\\ $v_f= 0.60$,\\ $f_0$$= -904.8$}}
		\label{Gr2DSF2D3Mvf2}
	\end{subfigure}
	\caption{Optimized results using 1D and 2D SFs}
	\label{Gr2DSF1D}
\end{figure}


\begin{figure}
	\centering
	\begin{subfigure}[t]{0.2\textwidth}
		\centering
		\includegraphics[width=\textwidth]{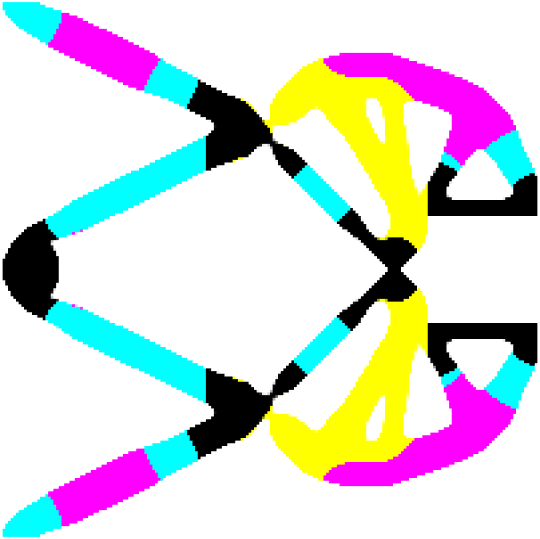}
		\captionsetup{justification=centering, labelformat=empty}
		\caption{\parbox{\linewidth}{\centering (a) 4 materials,\\ $v_f= 0.32$,\\ $f_0$$= -3320.4$}}
		\label{Gr2DSF3D4Mvf08}
	\end{subfigure}
	\hfill
	\begin{subfigure}[t]{0.2\textwidth}
		\centering
		\includegraphics[width=\textwidth]{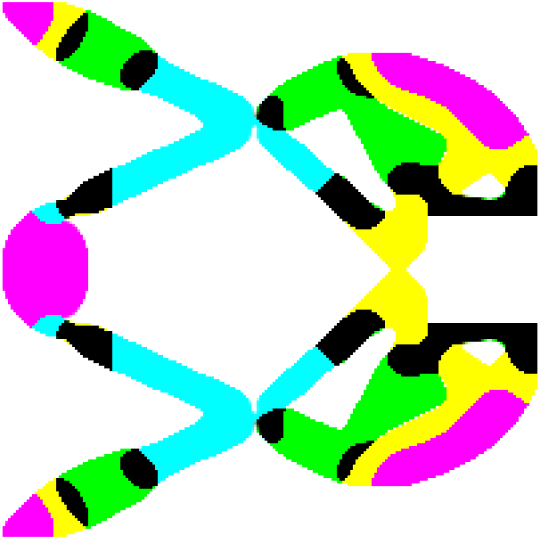}
		\captionsetup{justification=centering, labelformat=empty}
		\caption{\parbox{\linewidth}{\centering (b) 5 materials,\\ $v_f= 0.40$,\\ $f_0$$= -2939.0$}}
		\label{Gr2DSF3D5Mvf08}
	\end{subfigure}
	\hfill
	\begin{subfigure}[t]{0.2\textwidth}
		\centering
		\includegraphics[width=\textwidth]{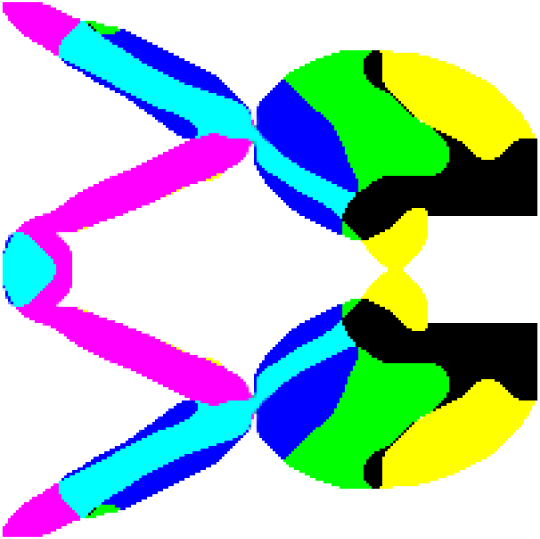}
		\captionsetup{justification=centering, labelformat=empty}
		\caption{\parbox{\linewidth}{\centering (c) 6 materials,\\ $v_f= 0.48$,\\ $f_0$$= -2852.7$}}
		\label{Gr2DSF3D6Mvf08}
	\end{subfigure}
	\hfill
	\begin{subfigure}[t]{0.2\textwidth}
		\centering
		\includegraphics[width=\textwidth]{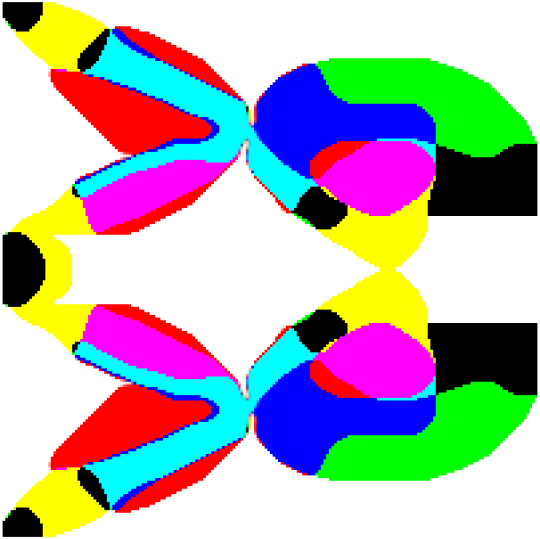}
		\captionsetup{justification=centering, labelformat=empty}
		\caption{\parbox{\linewidth}{\centering (d) 7 materials,\\ $v_f= 0.56$,\\ $f_0$$= -2852.8$}}
		\label{Gr2DSF3D7Mvf08}
	\end{subfigure}
	\caption{Optimized results using 3D SFs}
	\label{Gr2DSF3D}
\end{figure}
\begin{figure}
	\centering
	\begin{subfigure}[t]{0.2\textwidth}
		\centering
		\includegraphics[width=\textwidth]{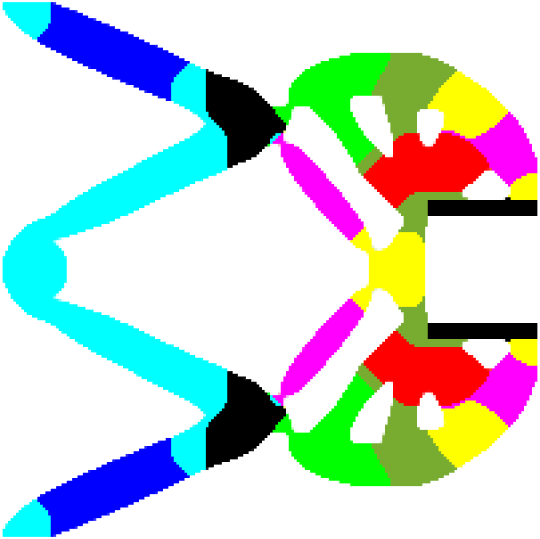}
		\captionsetup{justification=centering, labelformat=empty}
		\caption{\parbox{\linewidth}{\centering (a) 8 materials,\\ $v_f= 0.32$,\\ $f_0$$= -14911.0$}}
		\label{Gr2DSF4D8Mvf04}
	\end{subfigure}
	\hfill
	\begin{subfigure}[t]{0.2\textwidth}
		\centering
		\includegraphics[width=\textwidth]{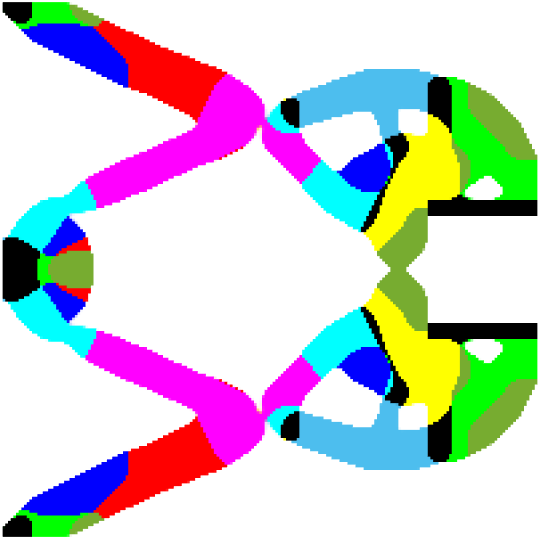}
		\captionsetup{justification=centering, labelformat=empty}
		\caption{\parbox{\linewidth}{\centering (b) 9 materials,\\ $v_f= 0.36$,\\ $f_0$$= -12794.8$}}
		\label{Gr2DSF4D9Mvf04}
	\end{subfigure}
	\hfill
	\begin{subfigure}[t]{0.2\textwidth}
		\centering
		\includegraphics[width=\textwidth]{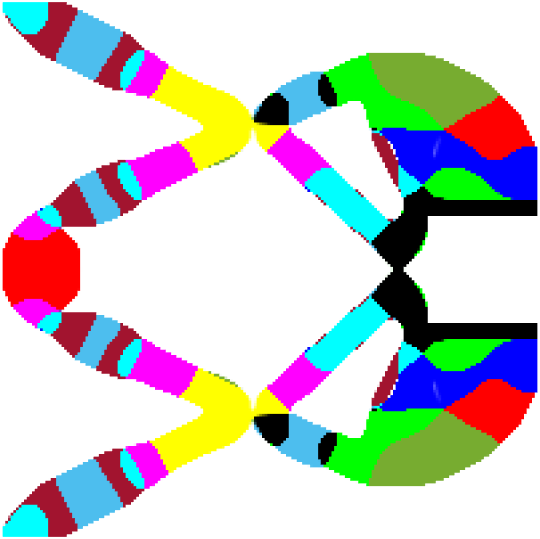}
		\captionsetup{justification=centering, labelformat=empty}
		\caption{\parbox{\linewidth}{\centering (c) 10 materials,\\ $v_f= 0.40$,\\ $f_0$$= -11486.5$}}
		\label{Gr2DSF4D10Mvf04}
	\end{subfigure}
	\hfill
	\begin{subfigure}[t]{0.2\textwidth}
		\centering
		\includegraphics[width=\textwidth]{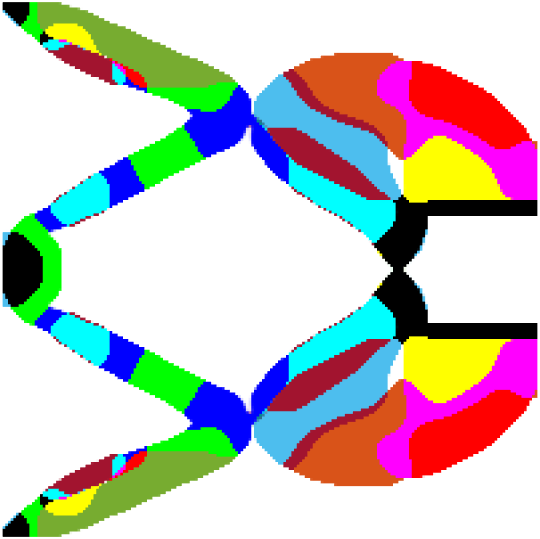}
		\captionsetup{justification=centering, labelformat=empty}
		\caption{\parbox{\linewidth}{\centering (d) 11 materials,\\ $v_f= 0.44$,\\ $f_0$$= -11771.7$}}
		\label{Gr2DSF4D11Mvf04}
	\end{subfigure}
	\begin{subfigure}[t]{0.2\textwidth}
		\centering
		\includegraphics[width=\textwidth]{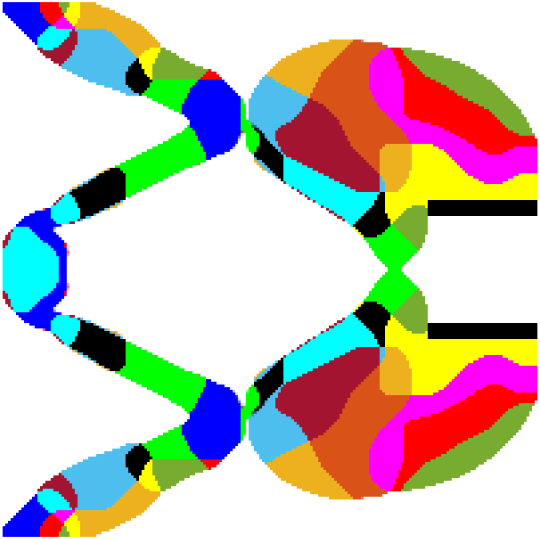}
		\captionsetup{justification=centering, labelformat=empty}
		\caption{\parbox{\linewidth}{\centering (e) 12 materials,\\ $v_f= 0.48$,\\ $f_0$$= -11077.8$}}
		\label{Gr2DSF4D12Mvf04}
	\end{subfigure}
	\hfill
	\begin{subfigure}[t]{0.2\textwidth}
		\centering
		\includegraphics[width=\textwidth]{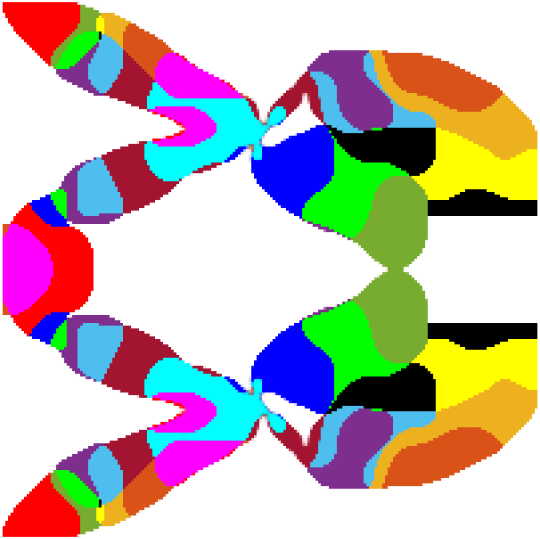}
		\captionsetup{justification=centering, labelformat=empty}
		\caption{\parbox{\linewidth}{\centering (f) 13 materials,\\ $v_f= 0.52$,\\ $f_0$$= -10541.4$}}
		\label{Gr2DSF4D13Mvf04}
	\end{subfigure}
	\hfill
	\begin{subfigure}[t]{0.2\textwidth}
		\centering
		\includegraphics[width=\textwidth]{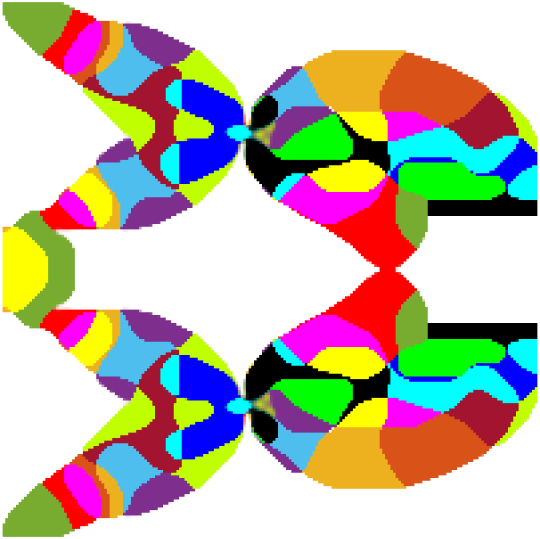}
		\captionsetup{justification=centering, labelformat=empty}
		\caption{\parbox{\linewidth}{\centering (g) 14 materials,\\ $v_f= 0.56$,\\ $f_0$$= -10364.0$}}
		\label{Gr2DSF4D14Mvf04}
	\end{subfigure}
	\hfill
	\begin{subfigure}[t]{0.2\textwidth}
		\centering
		\includegraphics[width=\textwidth]{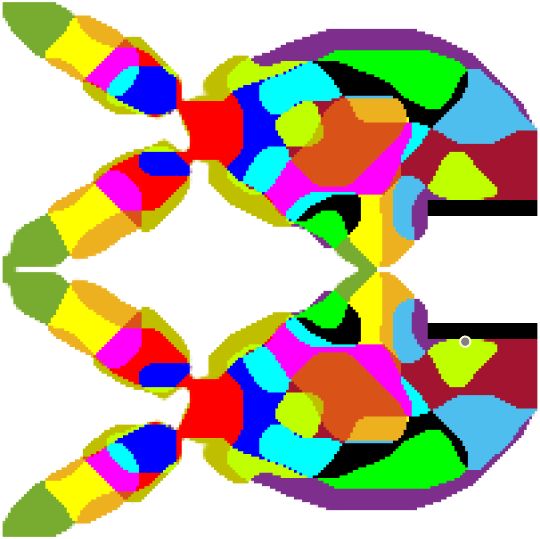}
		\captionsetup{justification=centering, labelformat=empty}
		\caption{\parbox{\linewidth}{\centering (h) 15 materials,\\ $v_f= 0.60$,\\ $f_0$$= -8968.2$}}
		\label{Gr2DSF4D15Mvf04}
	\end{subfigure}
	\caption{Optimized results using 4D SFs}
	\label{Gr2DSF4D}
\end{figure}

\clearpage

\subsection{Gripper mechanism}\label{subsec11}
The gripper's domain along the \(x\) and \(y\) directions is parameterized by \(N_\text{ex}\)$=200$ and \(N_\text{ey}\)$=200$ FEs, respectively, and the symmetric half model is used to get the optimized gripper CMs. Filter radius is set as $r_{\text{min}}=8$, and $\eta_p=0.5$. $\beta_p$ is taken as 1 at the start of the iteration and doubled after every 50 iterations.

Results with 1D and 2D material elements are shown in Fig.~\ref{Gr2DSF1D}. For single material with volume fraction $v_f=0.3$ and $v_f=0.4$, results are depicted in Fig.~\ref{Gr2DSF1D1Mvf3} and Fig.~\ref {Gr2DSF1D1Mvf4}, respectively. Optimized grippers resemble those obtained using the conventional density-based method with SIMP formulation~\cite{sigmund2013topology}.
Fig.~\ref{Gr2DSF2D2Mvf2} and Fig.~\ref{Gr2DSF2D3Mvf2} depict the optimized grippers with $v_f=0.2$ for each material in 2 and 3 material cases (2D material elements), respectively.
Fig.~\ref{Gr2DSF3D} shows the optimized results using 3D material elements. Figs.~\ref{Gr2DSF3D4Mvf08}, \ref{Gr2DSF3D5Mvf08}, \ref{Gr2DSF3D6Mvf08}, and \ref{Gr2DSF3D7Mvf08} illustrate the optimized gripper CMs with 4, 5, 6, and 7 materials, respectively. Each of the results consists of $v_f=0.08$ for each material. Using 4D material elements, the gripper mechanisms obtained are depicted in Fig.~\ref{Gr2DSF4D}. Figs.~\ref{Gr2DSF4D8Mvf04}, \ref{Gr2DSF4D9Mvf04}, \ref{Gr2DSF4D10Mvf04}, \ref{Gr2DSF4D11Mvf04}, \ref{Gr2DSF4D12Mvf04}, \ref{Gr2DSF4D13Mvf04}, \ref{Gr2DSF4D14Mvf04}, and \ref{Gr2DSF4D15Mvf04} show the optimized design with 8, 9, 10, 11, 12, 13, 14, and 15 materials, respectively. The volume fraction for each material is 0.04.


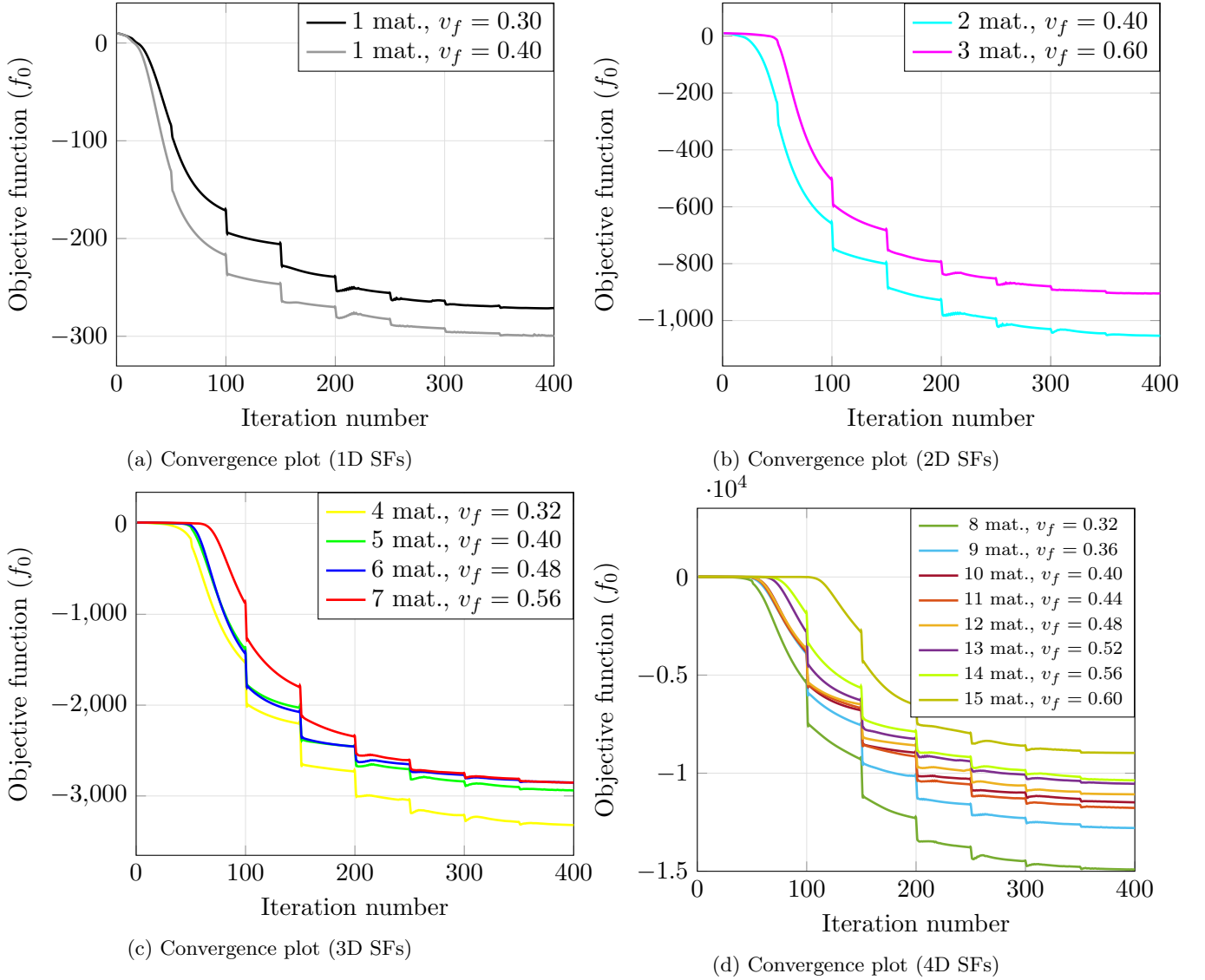
\begin{figure}[h!]
	\centering
	\begin{subfigure}[h!]{0.48\textwidth} 
		\centering
		\begin{tikzpicture}[scale=1]
			\pgfplotsset{compat=1.9}
			\begin{axis}[
				width = 1\textwidth,
				xlabel=  Iteration number,
				ylabel=  Objective function $(f_0)$,
				xmin=0,xmax=400,
				grid=both,
				major grid style={line width=0.2pt, draw=gray!30},
				legend style={at={(1.0,1.0)}, anchor=north east}]	
				\pgfplotstableread{Gr2D1Mvf30beta50k05graph.txt}\mydata;
				\addplot[smooth,{black}, line width=1pt, mark = none]
				table {\mydata};
				\addlegendentry{1 mat., $v_f=0.30$}				
				\pgfplotstableread{Gr2D1Mvf40beta50k05graph.txt}\mydata;
				\addplot[smooth,{gray}, line width=1pt, mark = none]
				table {\mydata};
				\addlegendentry{1 mat., $v_f=0.40$}			
			\end{axis}
		\end{tikzpicture}
		\caption{Convergence plot (1D SFs)}
		\label{Gr2D_SF1D_graphs}
	\end{subfigure}
	\hfill
	\begin{subfigure}[h!]{0.48\textwidth} 
		\centering
		\begin{tikzpicture}[scale=1]
			\pgfplotsset{compat=1.9}
			\begin{axis}[
				width = 1\textwidth,
				xlabel=  Iteration number,
				ylabel=  Objective function $(f_0)$,
				xmin=0,xmax=400,
				grid=both,
				major grid style={line width=0.2pt, draw=gray!30},
				legend style={at={(1.0,1.0)}, anchor=north east}]
				\pgfplotstableread{Gr2D2Mbeta50k05graph.txt}\mydata;
				\addplot[smooth,{cyan}, line width=1pt, mark = none]
				table {\mydata};
				\addlegendentry{2 mat., $v_f=0.40$}			
				\pgfplotstableread{Gr2D3Mbeta50k05graph.txt}\mydata;
				\addplot[smooth,{magenta}, line width=1pt, mark = none]
				table {\mydata};
				\addlegendentry{3 mat., $v_f=0.60$}			
			\end{axis}
		\end{tikzpicture}
		\caption{Convergence plot (2D SFs)}
		\label{Gr2D_SF2D_graphs}
	\end{subfigure}
	\begin{subfigure}[h!]{0.48\textwidth} 
		\centering
		\begin{tikzpicture}[scale=1]
			\pgfplotsset{compat=1.9}
			\begin{axis}[
				width = 1\textwidth,
				xlabel=  Iteration number,
				ylabel=  Objective function $(f_0)$,
				xmin=0,xmax=400,
				grid=both,
				major grid style={line width=0.2pt, draw=gray!30},
				legend style={at={(1.0,1.0)}, anchor=north east}]	
				\pgfplotstableread{Gr2D4Mbeta50k05graph.txt}\mydata;
				\addplot[smooth,{yellow}, line width=1pt, mark = none]
				table {\mydata};
				\addlegendentry{4 mat., $v_f=0.32$}				
				\pgfplotstableread{Gr2D5Mbeta50k05graph.txt}\mydata;
				\addplot[smooth,{green}, line width=1pt, mark = none]
				table {\mydata};
				\addlegendentry{5 mat., $v_f=0.40$}				
				\pgfplotstableread{Gr2D6Mbeta50k05graph.txt}\mydata;
				\addplot[smooth,{blue}, line width=1pt, mark = none]
				table {\mydata};
				\addlegendentry{6 mat., $v_f=0.48$}				
				\pgfplotstableread{Gr2D7Mbeta50k05graph.txt}\mydata;
				\addplot[smooth,{red}, line width=1pt, mark = none]
				table {\mydata};
				\addlegendentry{7 mat., $v_f=0.56$}		
			\end{axis}
		\end{tikzpicture}
		\caption{Convergence plot (3D SFs)}
		\label{Gr2D_SF3D_graphs}
	\end{subfigure}
	\hfill
	\begin{subfigure}[h!]{0.48\textwidth} 
		\centering
		\begin{tikzpicture}[scale=1]
			\pgfplotsset{compat=1.9}
			\begin{axis}[
				width = 1\textwidth,
				xlabel=  Iteration number,
				ylabel=  Objective function $(f_0)$,
				xmin=0,xmax=400,
                ymin=-15000,ymax=3500,
				grid=both,
				major grid style={line width=0.2pt, draw=gray!30},
				legend style={at={(1.0,1.0)}, anchor=north east, font=\scriptsize}]	
				\pgfplotstableread{Gr2D8Mbeta50k05graph.txt}\mydata;
				\addplot[smooth,{shafgreen}, line width=1pt, mark = none]
				table {\mydata};
				\addlegendentry{8 mat., $v_f=0.32$}				
				\pgfplotstableread{Gr2D9Mbeta50k05graph.txt}\mydata;
				\addplot[smooth,{skyblue}, line width=1pt, mark = none]
				table {\mydata};
				\addlegendentry{9 mat., $v_f=0.36$}				
				\pgfplotstableread{Gr2D10Mbeta50k05graph.txt}\mydata;
				\addplot[smooth,{brownred}, line width=1pt, mark = none]
				table {\mydata};
				\addlegendentry{10 mat., $v_f=0.40$}				
				\pgfplotstableread{Gr2D11Mbeta50k05graph.txt}\mydata;
				\addplot[smooth,{redorange}, line width=1pt, mark = none]
				table {\mydata};
				\addlegendentry{11 mat., $v_f=0.44$}				
				\pgfplotstableread{Gr2D12Mbeta50k05graph.txt}\mydata;
				\addplot[smooth,{yellowochre}, line width=1pt, mark = none]
				table {\mydata};
				\addlegendentry{12 mat., $v_f=0.48$}				
				\pgfplotstableread{Gr2D13Mbeta50k05graph.txt}\mydata;
				\addplot[smooth,{violet}, line width=1pt, mark = none]
				table {\mydata};
				\addlegendentry{13 mat., $v_f=0.52$}				
				\pgfplotstableread{Gr2D14Mbeta50k05graph.txt}\mydata;
				\addplot[smooth,{greenyellow}, line width=1pt, mark = none]
				table {\mydata};
				\addlegendentry{14 mat., $v_f=0.56$}				
				\pgfplotstableread{Gr2D15Mbeta50k05graph.txt}\mydata;
				\addplot[smooth,{olivegreen}, line width=1pt, mark = none]
				table {\mydata};
				\addlegendentry{15 mat., $v_f=0.60$}		
			\end{axis}
		\end{tikzpicture}
		\caption{Convergence plot (4D SFs)}
		\label{Gr2D_SF4D_graphs}
	\end{subfigure}
	\caption{Convergence plots (2D Gripper)}
	\label{Gr2D_all_graphs}
\end{figure}

\noindent Fig.~\ref{Gr2D_all_graphs} shows the convergence plots for all the 2D optimized gripper CMs. One notes similar observations as those of the inverter CMs. The volume constraints are predominantly active at the end of optimization for all the cases.

For compliant mechanisms, material placement is driven by finding a trade-off between the flexibility measure (e.g., output deformation) and structural stiffness (e.g., strain energy). In other words, to achieve the maximum deformation while maintaining adequate stiffness, the optimizer seeks to balance the region of high-stiffness materials that transmit force efficiently with the region of low-stiffness materials that allow elastic deformation. We observe that stiffer materials often appear near supports and output ports to prevent parasitic deformation, which would otherwise waste energy and reduce the desired output movement. This layout is a physical requirement to ensure that input energy is efficiently converted into output motion~\cite{sigmund1997design}.

\subsection{Parameter study}	
This section presents the effects of different parameters on the optimized multi-material designs in several of the above-mentioned problems. 
\subsubsection{Evolution of the objective function with increasing number of materials}
Here, we present the effect of increasing the number of candidate materials on the objective value using the same material interpolation elements. The MBB beam (Fig.~\ref{MBB_Beam_2D}) problem is selected. 

\begin{table}[h!]
	\caption{Performance trends of the objective value against the number of materials. \textbf{SF}, $m$, $v_f$, and $f_0$ represent the shape functions used, the number of materials, volume fraction of each material, and the objective value.}\label{Tab:ObjvsNumberofmaterial}
	\centering
	\begin{tabular}{|C{1cm}|C{0.8cm}|C{0.8cm}|C{7cm}|C{1.5cm}|}
		\hline
		\textbf{SF} & $m$ & $v_f$ & Optimized Design & $f_0$ \\ \hline
		\noalign{\hrule height 1.5pt}
		\multirow{2}{*}{2D} & 2 & 0.30 & \vspace{0.2cm} \includegraphics[scale=0.60]{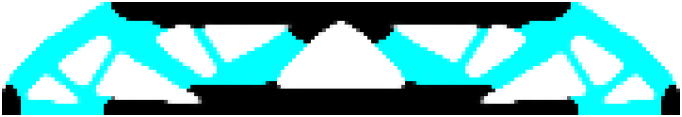} & $0.0093$
		\\ \cline{2-5}
		& 3 & 0.20 & \vspace{0.2cm} \includegraphics[scale=0.60]{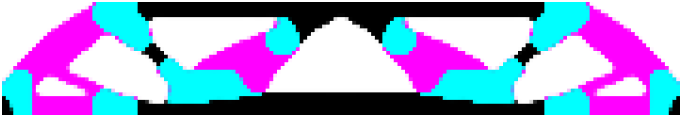} & $0.0112$ \\ \hline
		\multirow{3}{*}{3D} & 4 & 0.15 & \vspace{0.2cm} \includegraphics[scale=0.60]{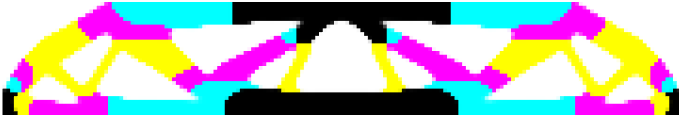} & $0.0098$
		\\ \cline{2-5}
		& 5 & 0.12 & \vspace{0.2cm} \includegraphics[scale=0.60]{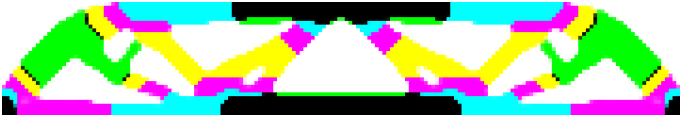} & $0.0106$ \\ \cline{2-5}
		& 6 & 0.10 & \vspace{0.2cm} \includegraphics[scale=0.60]{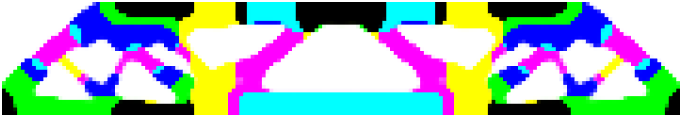} & $0.0121$ \\ \hline
		\multirow{3}{*}{4D} & 10 & 0.06 & \vspace{0.2cm} \includegraphics[scale=0.60]{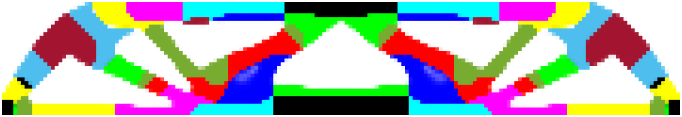} & $0.0108$
		\\ \cline{2-5}
		& 12 & 0.05 & \vspace{0.2cm} \includegraphics[scale=0.60]{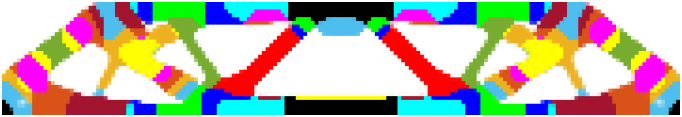} & $0.0124$ \\ \cline{2-5}
		& 15 & 0.04 & \vspace{0.2cm} \includegraphics[scale=0.60]{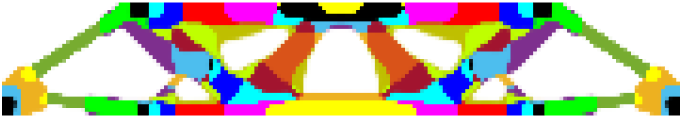} & $0.0151$ \\ \hline
	\end{tabular}
\end{table}

All employed parameters remain the same as those mentioned in Sec.~\ref{subsec8} unless specified.  The total volume fraction of the design is fixed at $0.6$, equally shared among the candidate materials. 2D, 3D and 4D SFs are employed for obtaining the optimized designs with \{2,\,3\}, \{4,\,5,\,6\}, and \{10,\,12,\,15\}-candidate materials, respectively.

The results for the 2D, 3D, and 4D SF cases are reported in Table~\ref{Tab:ObjvsNumberofmaterial}. In the 2D SF case, compliance increases as the number of candidate materials increases from 2 to 3. This behavior occurs because the introduction of additional materials includes lower-stiffness phases, with the total volume fraction shared equally. For example, in the 2-material case, Material-1 (Black) and Material-2 (cyan) are each assigned a volume fraction of 0.30, with normalized Young's modulus values of $E_1 = 1$ and $E_2 = 0.6667$. In contrast, the 3-material case employs Material-1 (Black), Material-2 (cyan), and Material-3 (magenta) with volume fractions of 0.20 each, and normalized Young’s modulus values of $E_1 = 1, E_2 = 0.6667,$ and $E_3 = 0.3333$. Because the material properties are normalized relative to the highest Young's modulus and the total volume fraction is shared equally, adding more materials introduces relatively softer phases. This naturally reduces the overall structural stiffness, leading to the higher compliance values observed in Table~\ref{Tab:ObjvsNumberofmaterial}. A similar trend is evident in the 3D SF cases (\{4,\,5,\,6\} materials) and the 4D SF cases (\{10,\,12,\,15\} materials). Therefore, the proposed gSF approach performs as expected as the number of materials increases, with the total volume fraction distributed equally. 

\subsubsection{Different initial guesses}
Here, we present the effect of different initial guesses on the optimized designs, i.e., using different natural coordinates associated with the shape functions (material density) for the optimization process. We select the cantilever beam problem (Fig.~\ref{Cantilever_2D}). All employed parameters remain the same as those mentioned in Sec.~\ref{subsec9} unless specified. 2D SFs for $\{m=2,\,3\}$ material, and 3D SFs for $\{m=4,\,5\}$ material are considered.

Results are depicted in Table~\ref{Tab:Diffiniguess2D} and Table~\ref{Tab:Diffiniguess3D} for 2D SFs ($\{m=2,\,3\}$) and 3D SFs ($\{m=4,\,5\}$), respectively. For different initial guesses, the approach yields different topologies, as expected, since the multi-material topology optimization introduces significant non-convexity to the optimization problem. The final objective values and $M\textsubscript{nd}$ values alter with different initial guesses (Table~\ref{Tab:Diffiniguess2D} and Table~\ref{Tab:Diffiniguess3D}). 
\begin{table}[h!]
\caption{Performance trends of the objective value for the different initial guesses for 2D SF. $\eta_1$ and $\eta_2$ are the design variable vectors.}\label{Tab:Diffiniguess2D}
\centering
\begin{tabular}{|C{1.6cm}|C{2.85cm}C{2.85cm}C{2.85cm}C{2.85cm}|}
\hline
\makecell{\textbf{Design}\\ \textbf{Variables}} & \makecell{${\eta_1=0}$ \\ $\eta_2=0$} & \makecell{${\eta_1=-0.75}$ \\ ${\eta_2=0.75}$} & \makecell{${\eta_1=-0.5}$ \\ ${\eta_2=0.5}$} & \makecell{${\eta_1=0.2}$ \\ ${\eta_2=-0.3}$} \\ \hline
\noalign{\hrule height 1.5pt}
\makecell{Opt.\\Design} & \vspace{0.2cm} \includegraphics[scale=0.25]{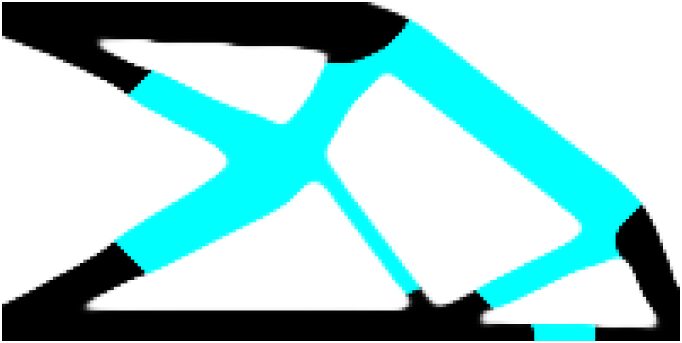} & \vspace{0.2cm} \includegraphics[scale=0.25]{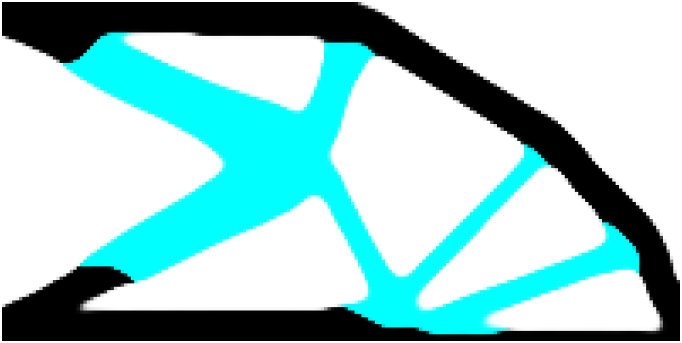} & \vspace{0.2cm} \includegraphics[scale=0.25]{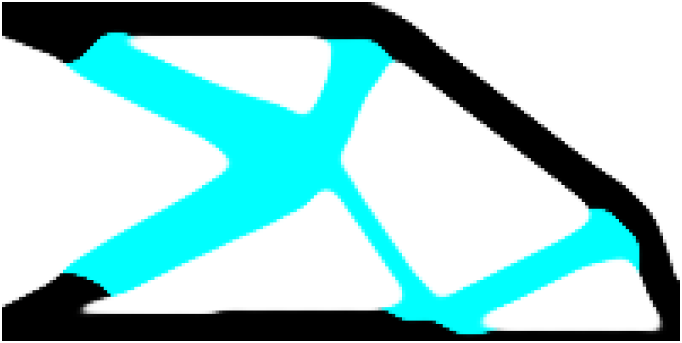} & \vspace{0.2cm} \includegraphics[scale=0.25]{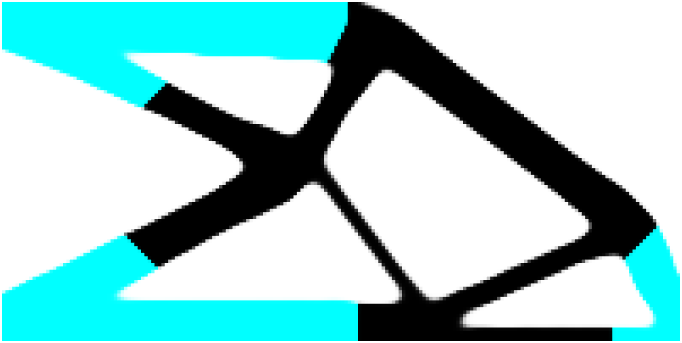} \\ \hline
$f_0$ & $0.0049$ & $0.0049$ & $0.0049$ & $0.0055$ \\ \hline
$M\textsubscript{nd}$ & $2.64\%$ & $2.84\%$ & $2.57\%$ & $2.61\%$\\ \hline
\noalign{\hrule height 1.5pt}
\makecell{Opt.\\Design} & \vspace{0.2cm} \includegraphics[scale=0.25]{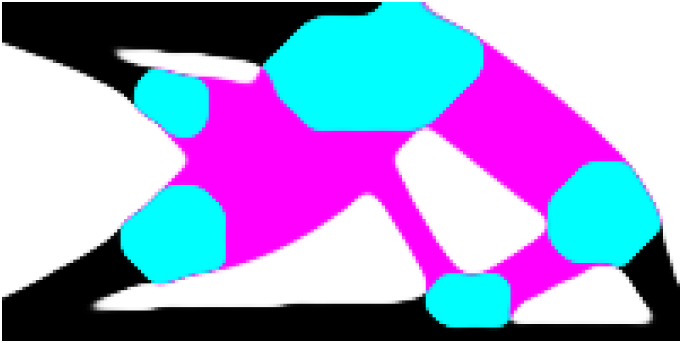} & \vspace{0.2cm} \includegraphics[scale=0.25]{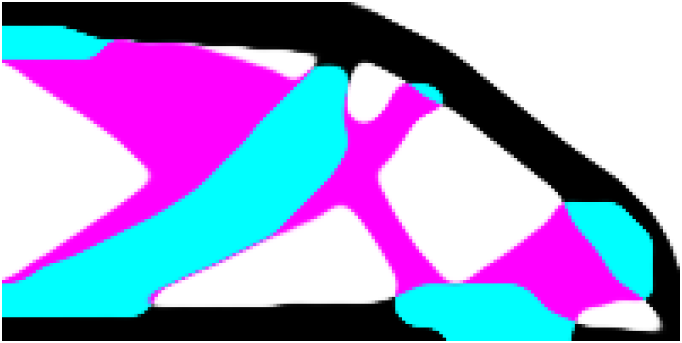} & \vspace{0.2cm} \includegraphics[scale=0.25]{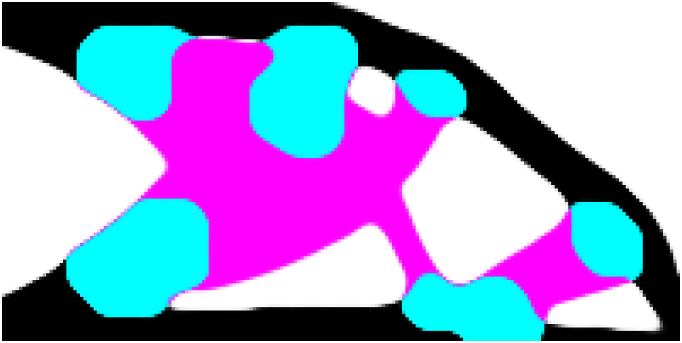} & \vspace{0.2cm} \includegraphics[scale=0.25]{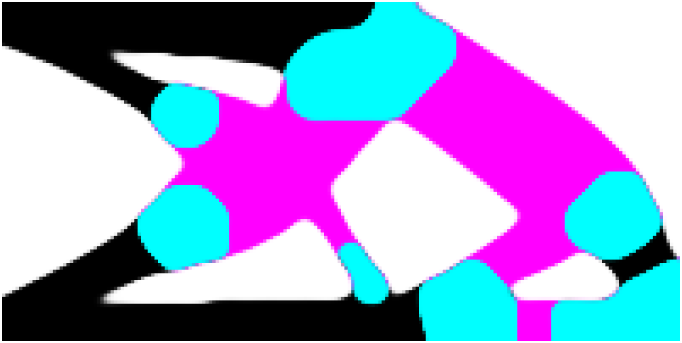} \\ \hline
$f_0$ & $0.0042$ & $0.0042$ & $0.0041$ & $0.0044$ \\ \hline
$M\textsubscript{nd}$ & $2.43\%$ & $2.59\%$ & $2.25\%$ & $2.32\%$ \\ \hline
\end{tabular}
\end{table}

\begin{table}[]
\caption{Performance trends of the objective value for the different initial guesses for 3D SF. $\eta_1$, $\eta_2$ and $\eta_3$ are the design variable vectors.}\label{Tab:Diffiniguess3D}
\centering
\begin{tabular}{|C{1.6cm}|C{2.85cm}C{2.85cm}C{2.85cm}C{2.85cm}|}
\hline
\makecell{\textbf{Design}\\ \textbf{Variables}} & \makecell{${\eta_1=0}$ \\ ${\eta_2=0}$\\ ${\eta_3=0}$} & \makecell{${\eta_1=-0.75}$ \\ ${\eta_2=0}$\\ ${\eta_3=0.75}$} & \makecell{${\eta_1=-0.5}$ \\ ${\eta_2=0.25}$\\ ${\eta_3=0.5}$} & \makecell{${\eta_1=0.2}$ \\ ${\eta_2=-0.3}$\\ ${\eta_3=-0.6}$} \\ \hline
\noalign{\hrule height 1.5pt}
\makecell{Opt.\\Design} & \vspace{0.2cm} \includegraphics[scale=0.25]{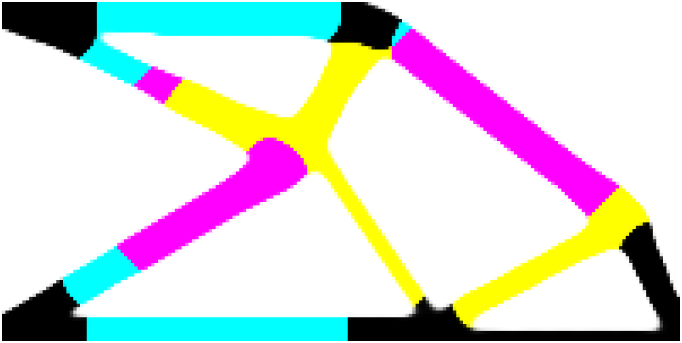} & \vspace{0.2cm} \includegraphics[scale=0.25]{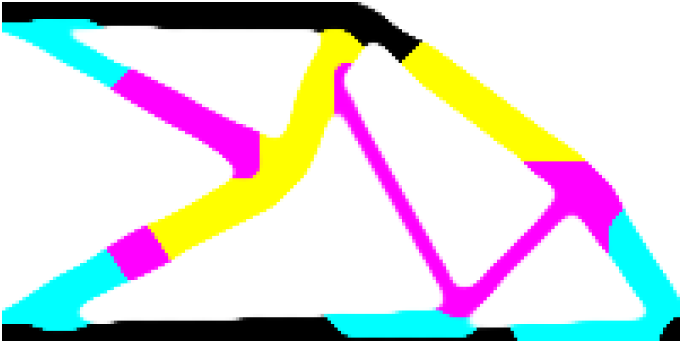} & \vspace{0.2cm} \includegraphics[scale=0.25]{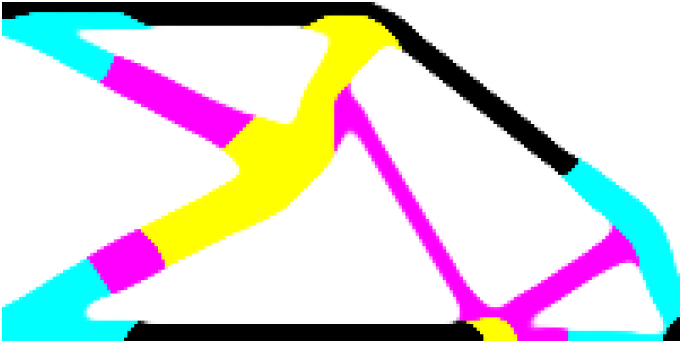} & \vspace{0.2cm} \includegraphics[scale=0.25]{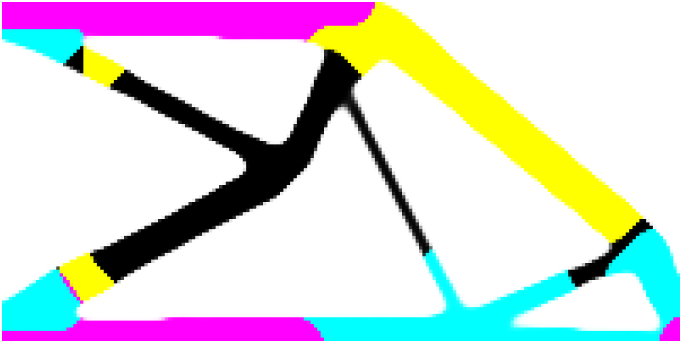} \\ \hline
$f_0$ & $0.0063$ & $0.0066$ & $0.0067$ & $0.0070$ \\ \hline
$M\textsubscript{nd}$ & $2.27\%$ & $2.40\%$ & $2.41\%$ & $2.53\%$\\ \hline
\noalign{\hrule height 1.5pt}
\makecell{Opt.\\Design} & \vspace{0.2cm} \includegraphics[scale=0.25]{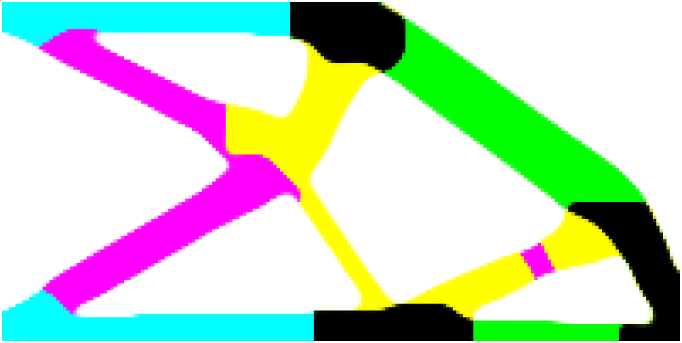} & \vspace{0.2cm} \includegraphics[scale=0.25]{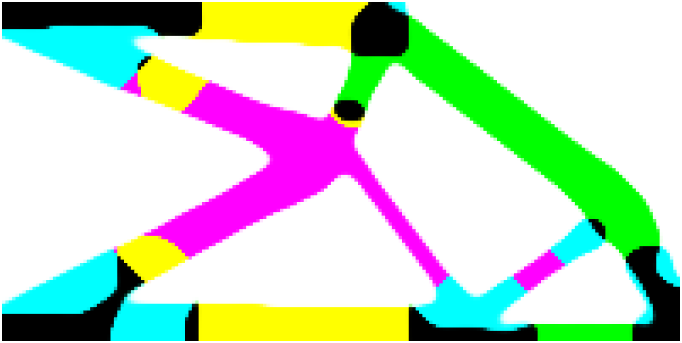} & \vspace{0.2cm} \includegraphics[scale=0.25]{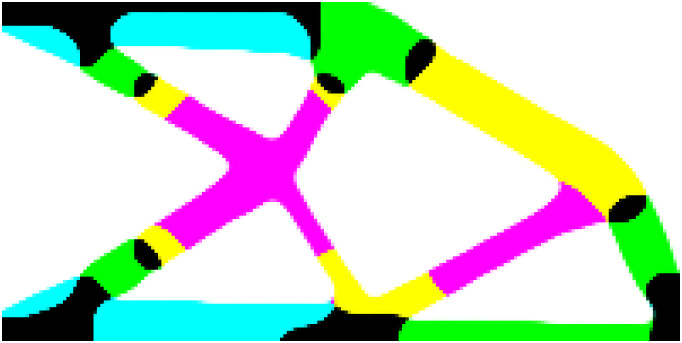} & \vspace{0.2cm} \includegraphics[scale=0.25]{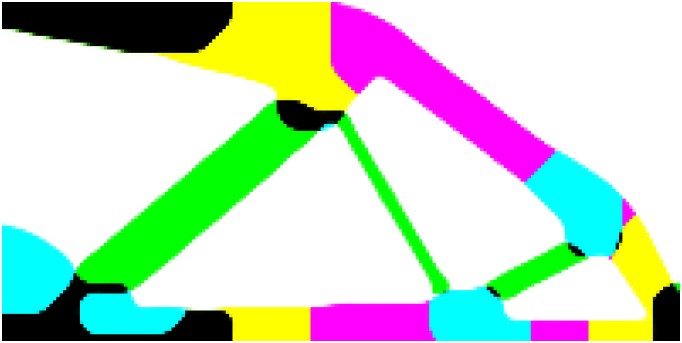} \\ \hline
$f_0$ & $0.0060$ & $0.0061$ & $0.0058$ & $0.0061$ \\ \hline
$M\textsubscript{nd}$ & $2.41\%$ & $2.38\%$ & $2.21\%$ & $2.03\%$ \\ \hline
\end{tabular}
\end{table}

\subsubsection{Material-specific volume fraction constraints}
In this section, we present the effect of material-specific volume fraction constraints. We select the MBB beam for the study. All the used parameters remain the same as those in Sec.~\ref{subsec8}, unless specified. 4-material case (3D SFs) is taken. The initial guess $(\eta_1, \eta_2, \eta_3) = (0, 0, 0)$ is set. Young's modulii for Material-1, Material-2, Material-3 and Material-4 are set to $E_1 = 1$, $E_2 = 0.8571$, $E_3 = 0.7143$ $E_4 = 0.5714$, respectively.

\begin{table}[]
\caption{Comparison with different volume fractions for 4 material TO.}\label{Tab:DiffVolfrac4M}
\centering
\begin{tabular}{|C{0.7cm}|C{1.6cm}|C{2.85cm}C{2.85cm}C{2.85cm}C{2.85cm}|}
\hline
$m$ & \makecell{\textbf{Volume}\\ \textbf{fractions}} & \makecell{\textbf{yellow} $=0.06$\\ \textbf{magenta} $=0.08$\\ \textbf{cyan} $=0.10$\\ \textbf{black} $=0.12$} & \makecell{\textbf{yellow} $=0.04$\\ \textbf{magenta} $=0.15$\\ \textbf{cyan} $=0.10$\\ \textbf{black} $=0.07$} & \makecell{\textbf{yellow} $=0.12$\\ \textbf{magenta} $=0.10$\\ \textbf{cyan} $=0.08$\\ \textbf{black} $=0.06$} & \makecell{\textbf{yellow} $=0.09$\\ \textbf{magenta} $=0.09$\\ \textbf{cyan} $=0.09$\\ \textbf{black} $=0.09$} \\ \hline
\noalign{\hrule height 1.5pt}
\multirow{3}{*}{4} & \makecell{Opt.\\Design} & \vspace{0.2cm} \includegraphics[scale=0.25]{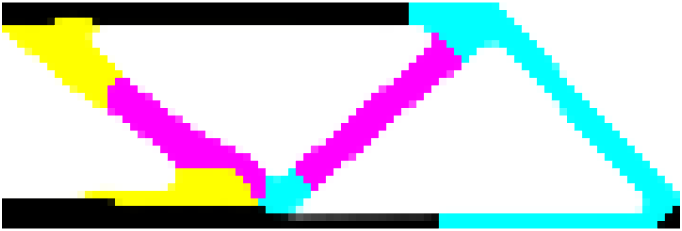} & \vspace{0.2cm} \includegraphics[scale=0.25]{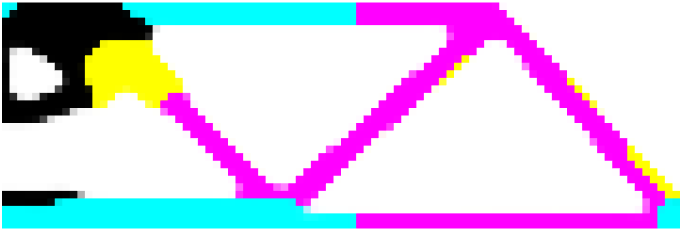} & \vspace{0.2cm} \includegraphics[scale=0.25]{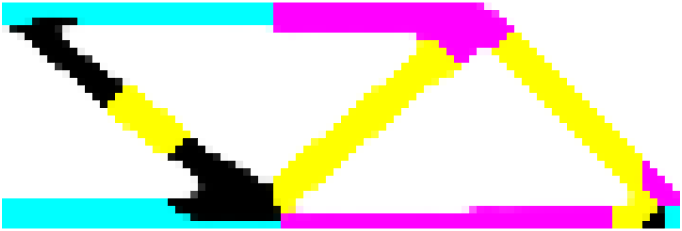} & \vspace{0.2cm} \includegraphics[scale=0.25]{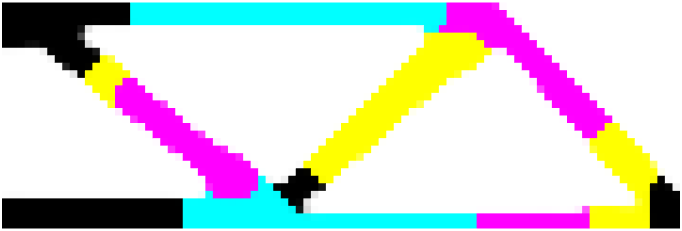} \\ \cline{2-6}
 & $f_0$ & $0.0149$ & $0.0167$ & $0.0168$ & $0.0158$ \\ \cline{2-6}
 & $M\textsubscript{nd}$ & $1.89\%$ & $1.82\%$ & $1.37\%$ & $1.39\%$\\ \hline
\end{tabular}
\end{table}

The results are depicted in Table~\ref{Tab:DiffVolfrac4M}. Regions with Material-1, Material-2, Material-3, and Material-4 are indicated by black, cyan, magenta, and yellow, respectively, in the optimized designs. Different volume fractions of different materials give different topologies. The cardinal reason is that different volume constraints yield distinct search directions, leading to distinct optimized designs. Additionally, the final locations of material vary with different volume fractions.

\subsubsection{Effect of output spring stiffness}
In compliant mechanism optimization problems, the output spring drives the TO algorithm to ensure a material layout between the output location and the input region. Here, we present the effect of output spring stiffness on the optimized inverter designs at three output stiffness. We select the inverter with six materials. All employed parameters are taken to be the same as those mentioned in Sec~\ref{subsec10} unless specified.

\begin{table}[]
\caption{Comparison with different spring constants for a 6-material optimized inverter.}\label{Tab:Diffsprinconst6M}
\centering
\begin{tabular}{|C{1.5cm}|C{2.25cm}C{2.25cm}C{2.25cm}C{2.25cm}C{2.25cm}|}
\hline
\makecell{\textbf{Spring}\\ \textbf{Stiffness}} & $k_{\text{o}}=0.05$ & $k_{\text{o}}=0.10$ & $k_{\text{o}}=0.20$ & $k_{\text{o}}=0.50$ & $k_{\text{o}}=1.00$ \\ \hline
\noalign{\hrule height 1.5pt}
\makecell{Opt.\\Design} & \vspace{0.2cm} \includegraphics[scale=0.25]{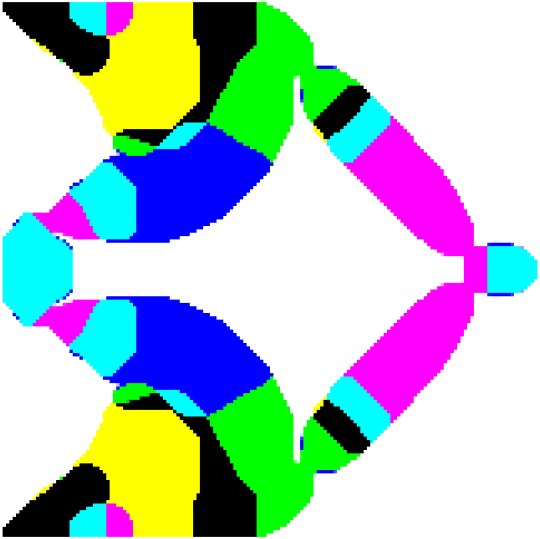} & \vspace{0.2cm} \includegraphics[scale=0.25]{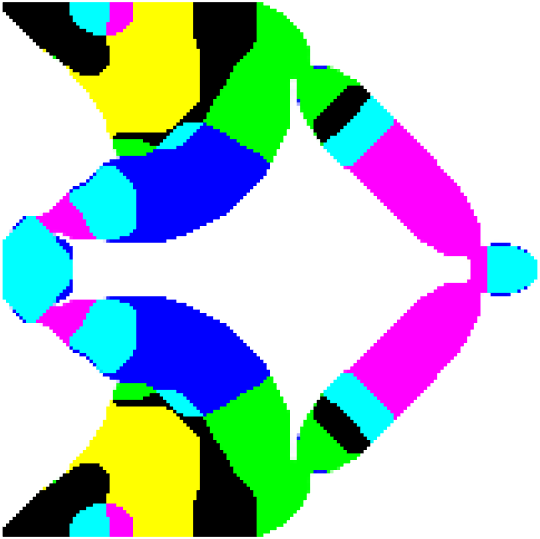} & \vspace{0.2cm} \includegraphics[scale=0.25]{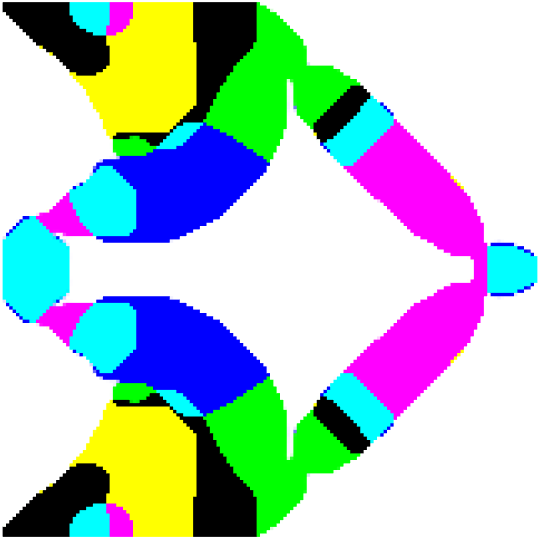} & \vspace{0.2cm} \includegraphics[scale=0.25]{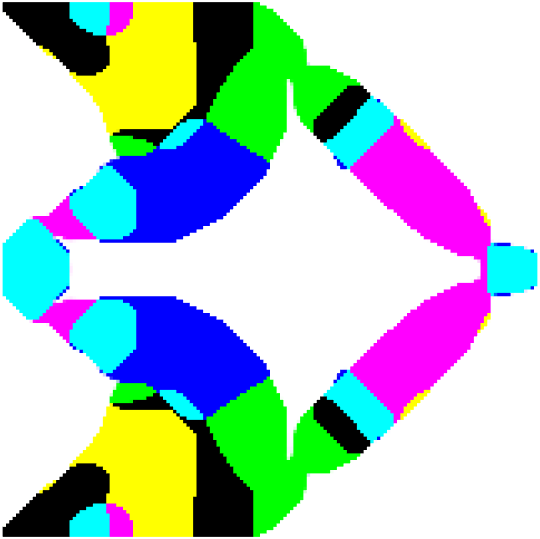} & \vspace{0.2cm} \includegraphics[scale=0.25]{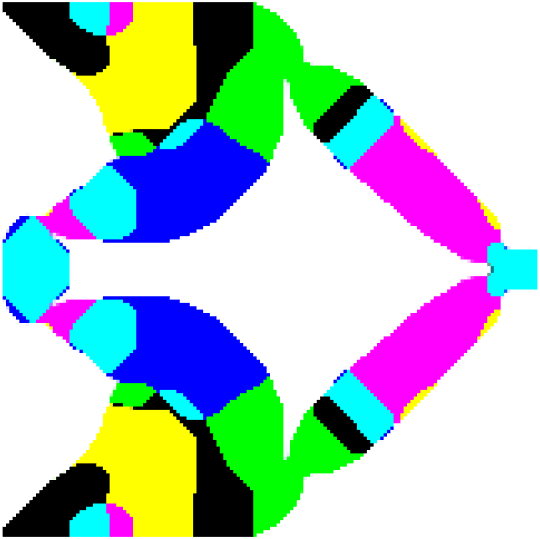} \\ \hline
$U_{out}$ & $-0.0147$ & $-0.0081$ & $-0.0043$ & $-0.0019$ & $-0.0010$ \\ \hline
$M\textsubscript{nd}$ & $0.16\%$ & $0.16\%$ & $0.19\%$ & $0.18\%$ & $0.18\%$ \\ \hline
\end{tabular}
\end{table}

The results for the five different output springs with stiffness 0.05, 0.10, 0.20, 0.50, and 1.00 are reported in Table~\ref{Tab:Diffsprinconst6M}. The solutions obtained using the symmetric half-domain design are suitably converted to the corresponding full mechanisms. The output deformation decreases with increasing output spring value, as expected. Additionally, the $M\textsubscript{nd}$ changes with different spring stiffness. The trend noted here is similar to when using only a single material~\cite{deepak2009comparative}. Having established the proposed method for 2D stiff-structure and compliant mechanism problems with up to 24 materials, we next demonstrate the versatility of the gSF on 3D problems.
\subsection{3D problems}\label{subsec12}
We have already demonstrated the success and versatility of the present approach for 2D designs involving SS and CMs with many materials in the above section. This section presents 3D problems for SS and CMs, showing that the method can readily be extended to 3D problems.

We solve 3D MBB beam and 3D inverter problems. Their respective design domains with boundary and (input) force conditions are shown in Fig.~\ref{MBB_Beam_3D} and Fig.~\ref{Inverter_3D}, respectively. Parameters mentioned in Table~\ref{3Mtable}, Table~\ref{7Mtable}, and Table~\ref{15Mtable} are also used here for the elastic modulus, volume fractions, and color scheme. Other parameters are kept same as those used for 2D problems.
\begin{figure}
\centering	
\begin{subfigure} {0.45\textwidth}
	\begin{tikzpicture}[scale=0.55]
		\tikzset{
			LineSpace/.store in=\LineSpace,
			LineSpace=2pt,
			LineThickness/.store in=\LineThickness,
			LineThickness=0.4pt
		}
		
		\draw[color=black, very thick](9.0,-0.2) circle (0.2);
        \draw[color=black, very thick](12.0,1) arc [start angle=90, end angle=-90, radius=0.2];
		\draw[black, very thick] (9.0,-0.4) -- (12.0,0.6);
		\fill [gray!50] (0,0) rectangle (9,3);
		\begin{scope}
			\clip (0,3) -- (9,3) -- (12,4) -- (3,4) -- cycle;
			\fill[gray!50] (0,3) rectangle (12,4);
		\end{scope}
		\begin{scope}
			\clip (9,0) -- (9,3) -- (12,4) -- (12,1) -- cycle;
			\fill[gray!50] (9,0) rectangle (12,4);
		\end{scope}
		
		\draw[black, very thick] (0,0) -- (9,0);
		\draw[black, very thick] (9,0) -- (9,3);
		\draw[black, very thick] (9,3) -- (0,3);
		\draw[black, very thick] (0,3) -- (0,0);
		
		\draw[dashed, thick] (3,1) -- (12,1);
		\draw[black, very thick] (12,1) -- (12,4);
		\draw[black, very thick] (12,4) -- (3,4);
		\draw[dashed, thick] (3,4) -- (3,1);
		
		\draw[black, very thick] (0,3) -- (3,4);
		\draw[dashed, thick] (0,0) -- (3,1);
		\draw[black, very thick] (9,3) -- (12,4);
		\draw[black, very thick] (9,0) -- (12,1);
		
		\filldraw[black] (0,0) circle (0.05);
		\draw[thick] (-0.3,-0.4) -- (0,0) -- (0.3,-0.4) -- cycle;		
		
		\draw[dash pattern=on 6pt off 2pt on 1pt off 2pt] (4.5,3) -- (7.5,4);
		\draw[red, very thick, -Stealth] (4.5,4.4) -- (4.5,3);
		\draw[red, very thick, -Stealth] (5.25,5.25/3+2.9) -- (5.25,5.25/3+1.5);
		\draw[red, very thick, -Stealth] (6,6/3+2.9) -- (6,6/3+1.5);
		\draw[red, very thick, -Stealth] (6.75,6.75/3+2.9) -- (6.75,6.75/3+1.5);
		\draw[red, very thick, -Stealth] (7.5,5.4) -- (7.5,4);
		
		\dimline[color=blue,extension start length=5pt, extension end length=5pt,label style={fill=none, yshift=6pt}]{(7.5,4.8)}{(12,4.8)}{$L_x/2$};
		\dimline[color=blue,extension start length=2pt, extension end length=2pt,label style={fill=none, yshift=-6pt}]{(0,-1)}{(9,-1)}{$L_x$};	
		\dimline[color=blue,extension start length=2pt, extension end length=2pt,label style={fill=none, yshift=6pt}]{(12.4,4)}{(12.4,1)}{$L_y$};
		\dimline[color=blue,extension start length=2pt, extension end length=2pt,label style={fill=none, yshift=-6pt}]{(9,-1)}{(12.3,0.1)}{$L_z$};
		\node at (4.1,4.4) {$F$};
		\node at (4.85,5.25/3+2.9) {$F$};
		\node at (5.6,6/3+2.9) {$F$};
		\node at (6.35,6.75/3+2.9) {$F$};
		\node at (7.1,5.4) {$F$};
		
		\fill[pattern=custom hatch] (-0.3,-0.6) rectangle (0.3,-0.4);
		\draw[black, very thick] (0.3,-0.4) -- (1.5,0);
		\begin{scope}
			\clip (9,-0.6) -- (12,0.4) -- (12,0.6) -- (9,-0.4) -- cycle;
			\fill[pattern=north east lines] (9,-0.6) rectangle (12,1);
		\end{scope}
		\clip (0.3,-0.6) -- (1.5,-0.2) -- (1.5,0) -- (0.3,-0.4) -- cycle;
		\fill[pattern=north east lines] (0.3,-0.6) rectangle (3.3,1);
	\end{tikzpicture}
	\caption{Design domain of a 3D MBB beam}
	\label{MBB_Beam_3D}
\end{subfigure}
\begin{subfigure}{0.45\textwidth}
	\begin{tikzpicture}[scale=0.55]
		\tikzset{
			LineSpace/.store in=\LineSpace,
			LineSpace=2pt,
			LineThickness/.store in=\LineThickness,
			LineThickness=0.4pt
		}
		
		\fill [gray!50] (0,0) rectangle (5,5);
		\begin{scope}
			\clip (0,5) -- (5,5) -- (8,6) -- (3,6) -- cycle;
			\fill[gray!50] (0,5) rectangle (8,6);
		\end{scope}
		\begin{scope}
			\clip (5,0) -- (5,5) -- (8,6) -- (8,1) -- cycle;
			\fill[gray!50] (5,0) rectangle (8,6);
		\end{scope}
		
		\draw[red, very thick, -Stealth] (-0.8,3) -- (1.5,3);
		\node at (-0.5,3.5) {$F$};
		
		\draw[black, very thick] (0,0) -- (5,0);
		\draw[black, very thick] (5,0) -- (5,5);
		\draw[black, very thick] (5,5) -- (0,5);
		\draw[black, very thick] (0,5) -- (0,0);
		
		\draw[dashed, thick] (3,1) -- (8,1);
		\draw[black, very thick] (8,1) -- (8,6);
		\draw[black, very thick] (8,6) -- (3,6);
		\draw[dashed, thick] (3,6) -- (3,1);
		
		\draw[black, very thick] (0,5) -- (3,6);
		\draw[dashed, thick] (0,0) -- (3,1);
		\draw[black, very thick] (5,5) -- (8,6);
		\draw[black, very thick] (5,0) -- (8,1);
		
		\draw[thick] (-0.4,0.4) -- (0,0) -- (-0.4,-0.4) -- cycle;	
		\draw[thick] (2.6,1.4) -- (3,1) -- (2.6,0.6) -- cycle;
		\draw[thick] (-0.4,5.4) -- (0,5) -- (-0.4,4.6) -- cycle;
		\draw[thick] (2.6,6.4) -- (3,6) -- (2.6,5.6) -- cycle;	
		
		\draw[dash pattern=on 6pt off 2pt on 1pt off 2pt] (0,2.5) -- (3,3.5);
		\draw[dash pattern=on 6pt off 2pt on 1pt off 2pt] (1.5,0.5) -- (1.5,5.5);
		\draw[dash pattern=on 6pt off 2pt on 1pt off 2pt] (5,2.5) -- (8,3.5);
		\draw[dash pattern=on 6pt off 2pt on 1pt off 2pt] (6.5,0.5) -- (6.5,5.5);
		
		\dimline[color=blue,extension start length=2pt, extension end length=2pt,label style={fill=none, yshift=-6pt}]{(0,-1)}{(5,-1)}{$L_z$};	
		\dimline[color=blue,extension start length=15pt, extension end length=15pt,label style={fill=none, yshift=6pt}]{(9.4,6)}{(9.4,1)}{$L_y$};
		\dimline[color=blue,extension start length=2pt, extension end length=2pt,label style={fill=none, yshift=-6pt}]{(5,-1)}{(8.3,0.1)}{$L_x$};
		\draw[green, very thick, -Stealth] (8.8,3) -- (6.5,3);
        \draw[decorate, decoration={coil, aspect=0.3, segment length=1.5mm, amplitude=2mm}] (6.5,3) -- (8.8,3);
        \fill [black] (8.8,2.8) rectangle (9,3.2);
        \node at (8.5,3.7) {$k_{\text{o}}$};
		\node at (6.8,2.3) {$d$};
		
		\fill[pattern=north east lines] (-0.6,-0.4) rectangle (-0.4,0.4);
		\fill[pattern=north east lines] (-0.6,4.6) rectangle (-0.4,5.4);
		\fill[pattern=north east lines] (2.4,0.6) rectangle (2.6,1.4);
		\fill[pattern=north east lines] (2.4,5.6) rectangle (2.6,6.4);
	\end{tikzpicture}
	\caption{Design domain of a 3D inverter}
	\label{Inverter_3D}
\end{subfigure}
	\caption{3D design domains}
	\label{3D_problems}
\end{figure}
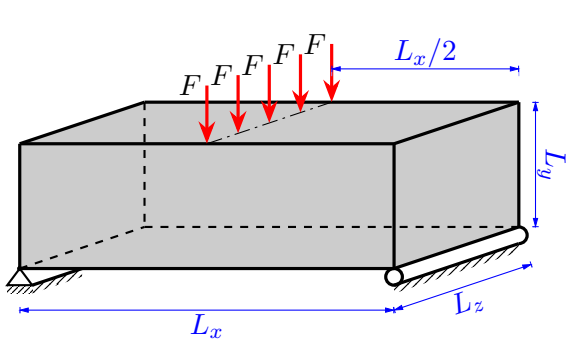
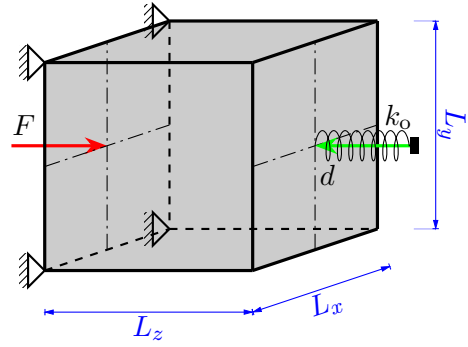

\subsubsection{3D MBB beam}
The design domain for the 3D MBB problem is parameterized by \(N_\text{ex}\)$=120$, \(N_\text{ey}\)$=20$ and \(N_\text{ez}\)$=10$ FEs along the \(x\), \(y\) and \(z\) directions, respectively. We use both symmetry conditions to achieve the optimized 3D-MBB beam. Filter radius ($r_{\text{min}}$) is set to 2.  $\eta_p$ and $\beta_p$ are same as that in 2D-MBB beam problem.

\begin{figure}[]
	\centering
	\begin{subfigure}[t]{0.45\textwidth}
		\centering
		\includegraphics[width=\textwidth]{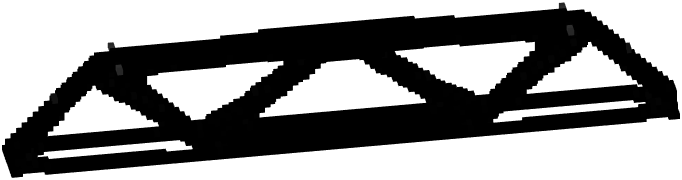}
		\caption{1 material, $v_f=0.2$, $f_0$$= 0.1504$}
		\label{MBB3DSF1D1Mvf2}
	\end{subfigure}
	\hspace{0.25cm}
	\begin{subfigure}[t]{0.45\textwidth}
		\centering
		\includegraphics[width=\textwidth]{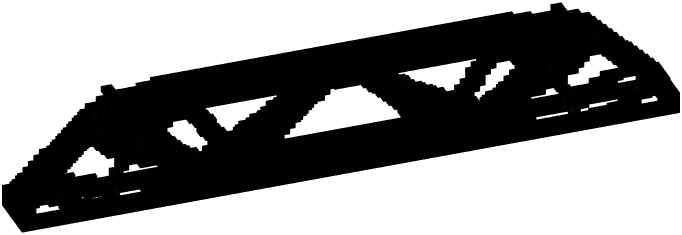}
		\caption{1 material, $v_f=0.3$, $f_0$$= 0.0989$}
		\label{MBB3DSF1D1Mvf3}
	\end{subfigure}
	\caption{Optimized results using 1D SF}
	\label{MBB3DSF1D}
\end{figure}

\begin{figure}[]
	\centering
	\begin{subfigure}[t]{0.45\textwidth}
		\centering
		\includegraphics[width=\textwidth]{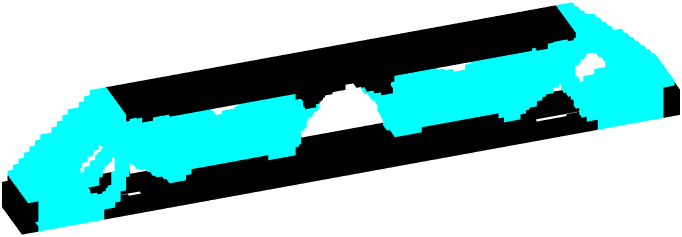}
		\caption{2 materials, $v_f= 0.40$, $f_0$$= 0.0875$}
		\label{MBB3DSF2D2Mvf2}
	\end{subfigure}
	\hspace{0.25cm}
	\begin{subfigure}[t]{0.45\textwidth}
		\centering
		\includegraphics[width=\textwidth]{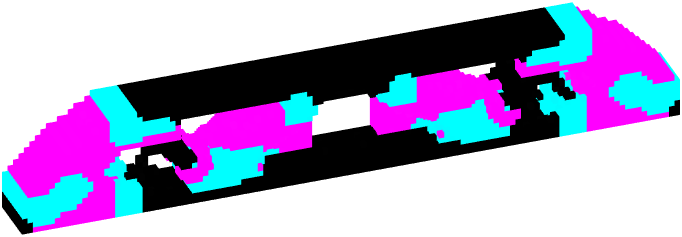}
		\caption{3 materials, $v_f= 0.60$, $f_0$$= 0.0804$}
		\label{MBB3DSF2D3Mvf2}
	\end{subfigure}
	\caption{Optimized results using 2D SFs}
	\label{MBB3DSF2D}
\end{figure}
Figure~\ref{MBB3DSF1D} depicts the optimized beam obtained using a 1D material element, wherein results in Fig.~\ref{MBB3DSF1D1Mvf2} and Fig.~\ref{MBB3DSF1D1Mvf3} show the design with $v_f=0.2$, and  $v_f=0.3$, respectively. The optimized beam obtained by using 2D material elements is shown in Fig.~\ref{MBB3DSF2D}.

\begin{figure}[]
	\centering
	\begin{subfigure}[t]{0.45\textwidth}
		\centering
		\includegraphics[width=\textwidth]{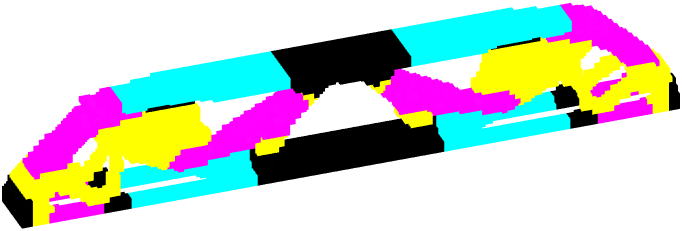}
		\caption{4 materials, $v_f= 0.32$, $f_0$$= 0.1165$}
		\label{MBB3DSF3D4Mvf08}
	\end{subfigure}
	\hspace{0.25cm}
	\begin{subfigure}[t]{0.45\textwidth}
		\centering
		\includegraphics[width=\textwidth]{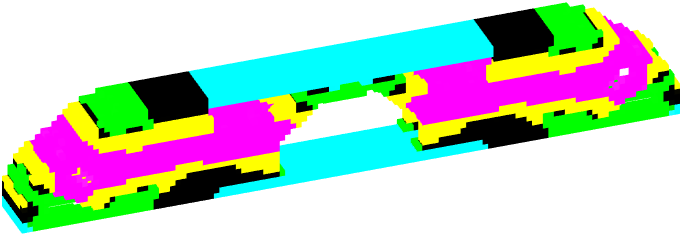}
		\caption{5 materials, $v_f= 0.40$, $f_0$$= 0.1139$}
		\label{MBB3DSF3D5Mvf08}
	\end{subfigure}
	\hspace{0.25cm}
	\begin{subfigure}[t]{0.45\textwidth}
		\centering
		\includegraphics[width=\textwidth]{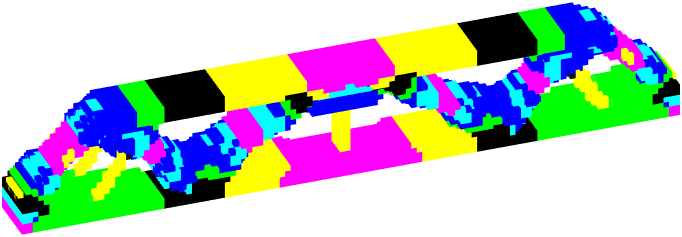}
		\caption{6 materials, $v_f= 0.48$, $f_0$$= 0.1153$}
		\label{MBB3DSF3D6Mvf08}
	\end{subfigure}
	\hspace{0.25cm}
	\begin{subfigure}[t]{0.45\textwidth}
		\centering
		\includegraphics[width=\textwidth]{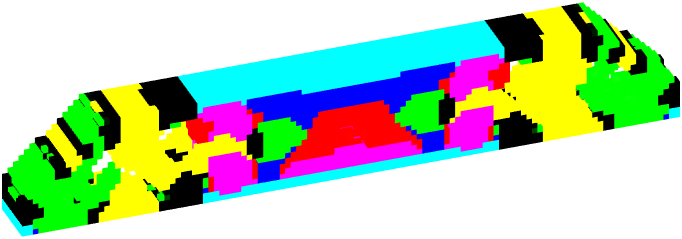}
		\caption{7 materials, $v_f= 0.56$, $f_0$$= 0.0983$}
		\label{MBB3DSF3D7Mvf08}
	\end{subfigure}	
	\caption{Optimized results using 3D SFs}
	\label{MBB3DSF3D}
\end{figure}

\begin{figure}
	\centering
	\begin{subfigure}[t]{0.45\textwidth}
		\centering
		\includegraphics[width=\textwidth]{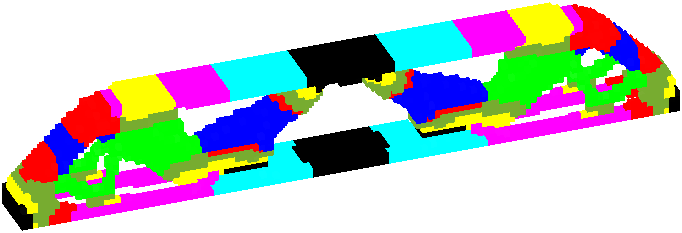} 
		\caption{8 Materials, $v_f= 0.32$, $f_0$$= 0.0422$}
		\label{MBB3DSF4D8Mvf04}
	\end{subfigure}
	\hspace{0.25cm}
	\begin{subfigure}[t]{0.45\textwidth}
		\centering
		\includegraphics[width=\textwidth]{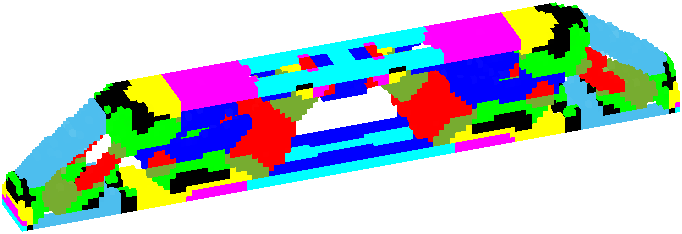} 
		\caption{9 Materials, $v_f= 0.36$, $f_0$$= 0.0446$}
		\label{MBB3DSF4D9Mvf04}
	\end{subfigure}
	\hspace{0.25cm}
	\begin{subfigure}[t]{0.45\textwidth}
		\centering
		\includegraphics[width=\textwidth]{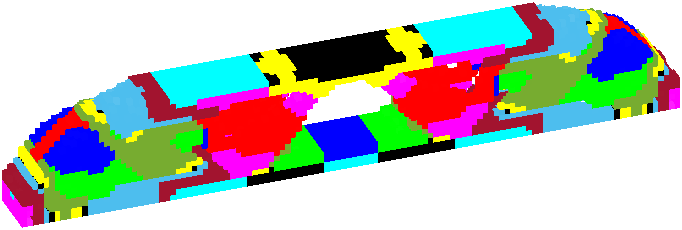}
		\caption{10 Materials, $v_f= 0.40$, $f_0$$= 0.0398$}
		\label{MBB3DSF4D10Mvf04}
	\end{subfigure}
	\hspace{0.25cm}
	\begin{subfigure}[t]{0.45\textwidth}
		\centering
		\includegraphics[width=\textwidth]{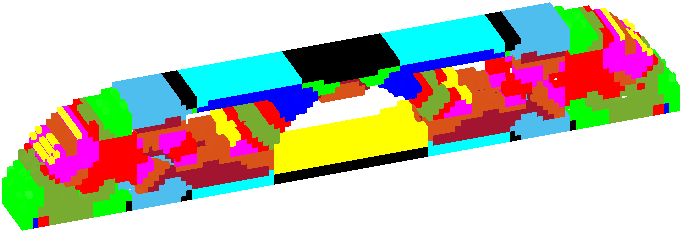}
		\caption{11 Materials, $v_f= 0.44$, $f_0$$= 0.0423$}
		\label{MBB3DSF4D11Mvf04}
	\end{subfigure}
	\begin{subfigure}[t]{0.45\textwidth}
		\centering
		\includegraphics[width=\textwidth]{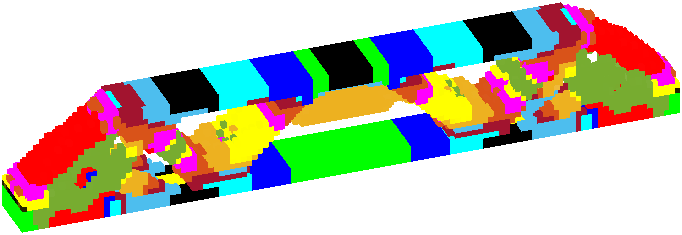}
		\caption{12 Materials, $v_f= 0.48$, $f_0$$= 0.0455$}
		\label{MBB3DSF4D12Mvf04}
	\end{subfigure}
	\hspace{0.25cm}
	\begin{subfigure}[t]{0.45\textwidth}
		\centering
		\includegraphics[width=\textwidth]{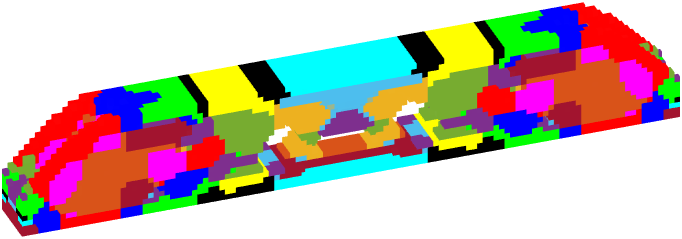}
		\caption{13 Materials, $v_f= 0.52$, $f_0$$= 0.0419$}
		\label{MBB3DSF4D13Mvf04}
	\end{subfigure}
	\hspace{0.25cm}
	\begin{subfigure}[t]{0.45\textwidth}
		\centering
		\includegraphics[width=\textwidth]{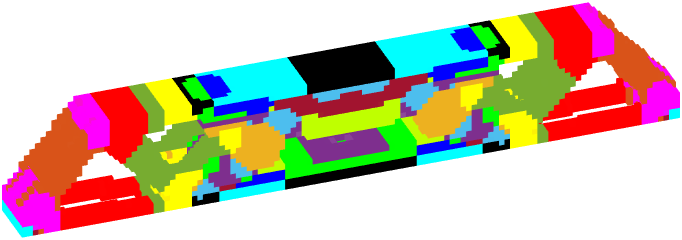}
		\caption{14 Materials, $v_f= 0.56$, $f_0$$= 0.0435$}
		\label{MBB3DSF4D14Mvf04}
	\end{subfigure}
	\hspace{0.25cm}
	\begin{subfigure}[t]{0.45\textwidth}
		\centering
		\includegraphics[width=\textwidth]{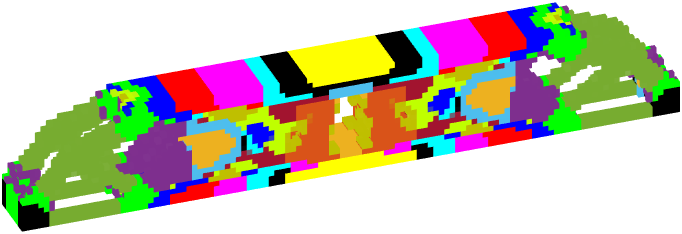}
		\caption{15 Materials, $v_f= 0.60$, $f_0$$= 0.0475$}
		\label{MBB3DSF4D15Mvf04}
	\end{subfigure}
	\caption{Optimized results using 4D SFs}
	\label{MBB3DSF4D}
\end{figure}
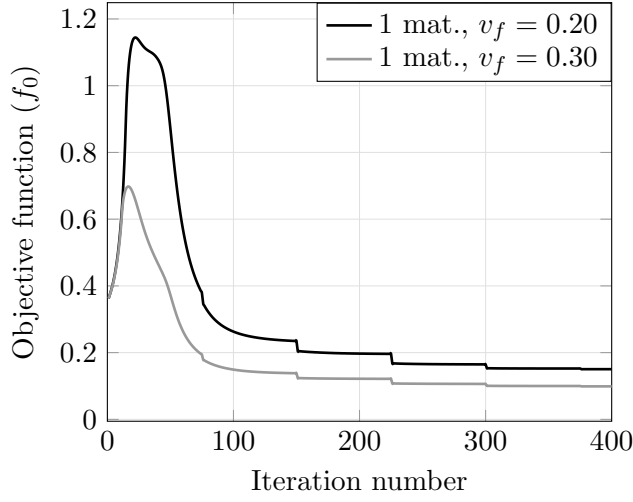
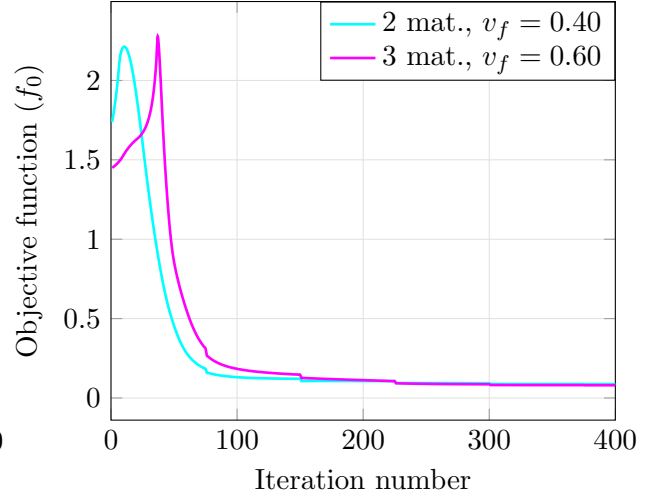
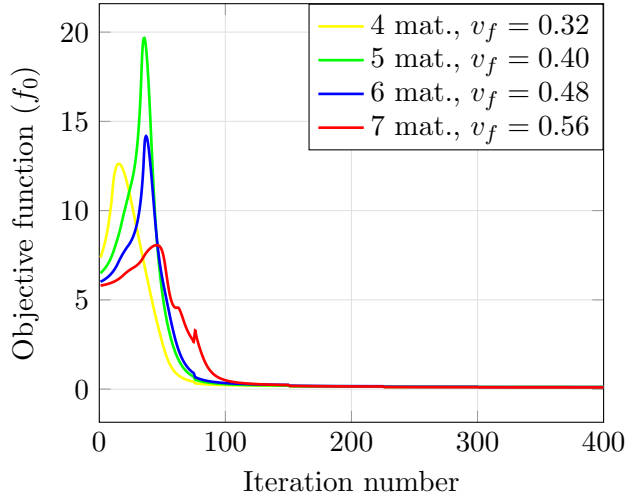
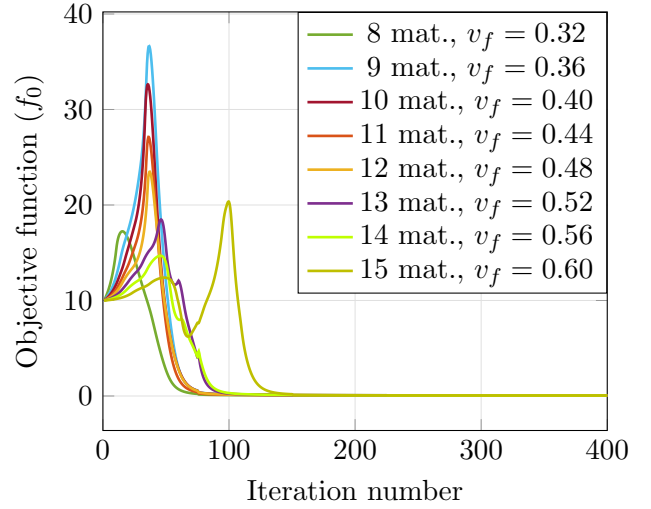
\begin{figure}
	\centering
	\begin{subfigure}[h!]{0.48\textwidth} 
		\centering
		\begin{tikzpicture}[scale=1]
			\pgfplotsset{compat=1.9}
			\begin{axis}[
				width = 1\textwidth,
				xlabel=  Iteration number,
				ylabel=  Objective function $(f_0)$,
				xmin=0,xmax=400,
				grid=both,
				major grid style={line width=0.2pt, draw=gray!30},
				legend style={at={(1.0,1.0)}, anchor=north east}]	
				\pgfplotstableread{MBB3D1Mvf20x1graph.txt}\mydata;
				\addplot[smooth,{black}, line width=1pt, mark = none]
				table {\mydata};
				\addlegendentry{1 mat., $v_f=0.20$}				
				\pgfplotstableread{MBB3D1Mvf30x1graph.txt}\mydata;
				\addplot[smooth,{gray}, line width=1pt, mark = none]
				table {\mydata};
				\addlegendentry{1 mat., $v_f=0.30$}			
			\end{axis}
		\end{tikzpicture}
		\caption{Convergence plot (1D SFs)}
		\label{MBB3D_SF1D_graphs}
	\end{subfigure}
	\begin{subfigure}[h!]{0.48\textwidth} 
		\centering
		\begin{tikzpicture}[scale=1]
			\pgfplotsset{compat=1.9}
			\begin{axis}[
				width = 1\textwidth,
				xlabel=  Iteration number,
				ylabel=  Objective function $(f_0)$,
				xmin=0,xmax=400,
				grid=both,
				major grid style={line width=0.2pt, draw=gray!30},
				legend style={at={(1.0,1.0)}, anchor=north east}]
				\pgfplotstableread{MBB3D2Mgraph.txt}\mydata;
				\addplot[smooth,{cyan}, line width=1pt, mark = none]
				table {\mydata};
				\addlegendentry{2 mat., $v_f=0.40$}			
				\pgfplotstableread{MBB3D3Mgraph.txt}\mydata;
				\addplot[smooth,{magenta}, line width=1pt, mark = none]
				table {\mydata};
				\addlegendentry{3 mat., $v_f=0.60$}			
			\end{axis}
		\end{tikzpicture}
		\caption{Convergence plot (2D SFs)}
		\label{MBB3D_SF2D_graphs}
	\end{subfigure}
	\begin{subfigure}[h!]{0.48\textwidth} 
		\centering
		\begin{tikzpicture}[scale=1]
			\pgfplotsset{compat=1.9}
			\begin{axis}[
				width = 1\textwidth,
				xlabel=  Iteration number,
				ylabel=  Objective function $(f_0)$,
				xmin=0,xmax=400,
				grid=both,
				major grid style={line width=0.2pt, draw=gray!30},
				legend style={at={(1.0,1.0)}, anchor=north east}]	
				\pgfplotstableread{MBB3D4Mgraph.txt}\mydata;
				\addplot[smooth,{yellow}, line width=1pt, mark = none]
				table {\mydata};
				\addlegendentry{4 mat., $v_f=0.32$}				
				\pgfplotstableread{MBB3D5Mgraph.txt}\mydata;
				\addplot[smooth,{green}, line width=1pt, mark = none]
				table {\mydata};
				\addlegendentry{5 mat., $v_f=0.40$}				
				\pgfplotstableread{MBB3D6Mgraph.txt}\mydata;
				\addplot[smooth,{blue}, line width=1pt, mark = none]
				table {\mydata};
				\addlegendentry{6 mat., $v_f=0.48$}				
				\pgfplotstableread{MBB3D7Mgraph.txt}\mydata;
				\addplot[smooth,{red}, line width=1pt, mark = none]
				table {\mydata};
				\addlegendentry{7 mat., $v_f=0.56$}		
			\end{axis}
		\end{tikzpicture}
		\caption{Convergence plot (3D SFs)}
		\label{MBB3D_SF3D_graphs}
	\end{subfigure}
	\begin{subfigure}[h!]{0.48\textwidth} 
		\centering
		\begin{tikzpicture}[scale=1]
			\pgfplotsset{compat=1.9}
			\begin{axis}[
				width = 1\textwidth,
				xlabel=  Iteration number,
				ylabel=  Objective function $(f_0)$,
				xmin=0,xmax=400,
				grid=both,
				major grid style={line width=0.2pt, draw=gray!30},
				legend style={at={(1.0,1.0)}, anchor=north east}]	
				\pgfplotstableread{MBB3D8Mgraph.txt}\mydata;
				\addplot[smooth,{shafgreen}, line width=1pt, mark = none]
				table {\mydata};
				\addlegendentry{8 mat., $v_f=0.32$}				
				\pgfplotstableread{MBB3D9Mgraph.txt}\mydata;
				\addplot[smooth,{skyblue}, line width=1pt, mark = none]
				table {\mydata};
				\addlegendentry{9 mat., $v_f=0.36$}				
				\pgfplotstableread{MBB3D10Mgraph.txt}\mydata;
				\addplot[smooth,{brownred}, line width=1pt, mark = none]
				table {\mydata};
				\addlegendentry{10 mat., $v_f=0.40$}				
				\pgfplotstableread{MBB3D11Mgraph.txt}\mydata;
				\addplot[smooth,{redorange}, line width=1pt, mark = none]
				table {\mydata};
				\addlegendentry{11 mat., $v_f=0.44$}				
				\pgfplotstableread{MBB3D12Mgraph.txt}\mydata;
				\addplot[smooth,{yellowochre}, line width=1pt, mark = none]
				table {\mydata};
				\addlegendentry{12 mat., $v_f=0.48$}				
				\pgfplotstableread{MBB3D13Mgraph.txt}\mydata;
				\addplot[smooth,{violet}, line width=1pt, mark = none]
				table {\mydata};
				\addlegendentry{13 mat., $v_f=0.52$}				
				\pgfplotstableread{MBB3D14Mgraph.txt}\mydata;
				\addplot[smooth,{greenyellow}, line width=1pt, mark = none]
				table {\mydata};
				\addlegendentry{14 mat., $v_f=0.56$}				
				\pgfplotstableread{MBB3D15Mgraph.txt}\mydata;
				\addplot[smooth,{olivegreen}, line width=1pt, mark = none]
				table {\mydata};
				\addlegendentry{15 mat., $v_f=0.60$}		
			\end{axis}
		\end{tikzpicture}
		\caption{Convergence plot (4D SFs)}
		\label{MBB3D_SF4D_graphs}
	\end{subfigure}
	\caption{Convergence plots (3D MBB)}
	\label{MBB3D_all_graphs}
\end{figure}
Fig.~\ref{MBB3DSF2D2Mvf2} and Fig.~\ref{MBB3DSF2D3Mvf2} show the results with $v_f=0.2$ for each material, for the 2-material and 3-material cases, respectively. Similar to the 2D results, one notes that the regions with roller and hinge supports and load application are occupied by material with the highest Young's modulus.
Results using 3D material elements are illustrated in Fig.~\ref{MBB3DSF3D}. Figs.~\ref{MBB3DSF3D4Mvf08}, \ref{MBB3DSF3D5Mvf08}, \ref{MBB3DSF3D6Mvf08}, and \ref{MBB3DSF3D7Mvf08} show the optimized design with 4, 5, 6, and 7 materials, respectively, having a volume fraction of 0.08 for each material.  Fig.~\ref{MBB3DSF4D} shows the optimized beams with  4D material elements. Results with 8, 9, 10, 11, 12, 13, 14, and 15 materials are shown in Figs.~ \ref{MBB3DSF4D8Mvf04}, \ref{MBB3DSF4D9Mvf04}, \ref{MBB3DSF4D10Mvf04}, \ref{MBB3DSF4D11Mvf04}, \ref{MBB3DSF4D12Mvf04}, \ref{MBB3DSF4D13Mvf04}, \ref{MBB3DSF4D14Mvf04}, and \ref{MBB3DSF4D15Mvf04}, respectively. One notes a similar observation to the 2D MBB beam regarding the material placement. Next, the objective convergence plots are shown in Fig.~\ref{MBB3D_all_graphs}.  These plots show the converging behaviors with some jumps in the middle, which are attributed to $\beta_p$ updation during optimization. The volume constraints are found to be active at the end of the optimization.

\subsection{3D Inverter CM}\label{subsec13}
The 3D inverter design domain is parameterized using \(N_\text{ex}\)$=10$, \(N_\text{ey}\)$=40$ and \(N_\text{ez}\)$=40$ FEs along the \(x\), \(y\) and \(z\) directions, respectively. Only a symmetric quarter domain is utilized to optimize the inverter mechanism by exploiting all symmetric conditions. Filter radius $r_{\text{min}}= 1.5$ is taken, whereas $\eta_p$ and $\beta_p$ are kept similar to 2D inverter mechanism. The output spring stiffness ($k_{\text{o}}$) is set as 0.05.

The optimized results obtained using 1D and 2D material elements are shown in Fig.~\ref{I3DSF1D2D}. Fig.~\ref{I3DSF1D1Mvf3} and Fig.~\ref{I3DSF1D1Mvf4} show the optimized mechanism with 1 material having $v_f =0.30$  and $v_f=0.4$, respectively. Optimized grippers with 2 and 3 materials are depicted in Fig.~\ref{I3DSF2D2Mvf2} and Fig.~\ref{I3DSF2D3Mvf2}. For both cases, $v_f$ for each material is  0.2. Fig.~\ref{I3DSF3D} illustrates the optimized results using 3D material elements. The gripper mechanisms with 4, 5, 6, and 7 materials are demonstrated in Fig.~\ref{I3DSF3D4Mvf08}, \ref{I3DSF3D5Mvf08}, \ref{I3DSF3D6Mvf08}, and \ref{I3DSF3D7Mvf08}, respectively, having a volume fraction of 0.08 for each material.


\begin{figure}
	\centering
	\begin{subfigure}{0.2\textwidth}
		\centering
		\includegraphics[width=\textwidth]{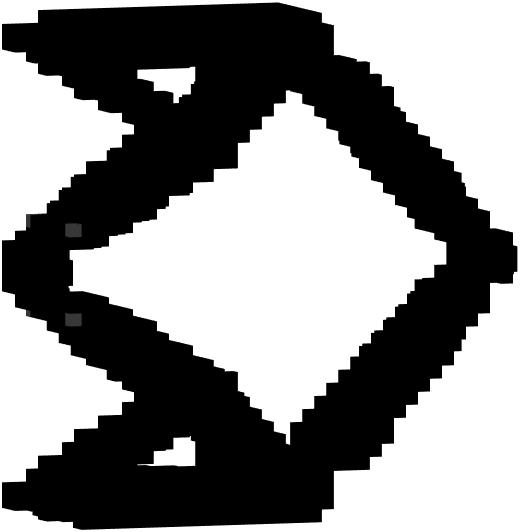}
		\captionsetup{justification=centering, labelformat=empty}
		\caption{\parbox{\linewidth}{\centering (a) 1 material,\\ $v_f= 0.30$,\\ $f_0$$= -186.5$}}
		\label{I3DSF1D1Mvf3}
	\end{subfigure}
	\hfill
	\begin{subfigure}{0.2\textwidth}
		\centering
		\includegraphics[width=\textwidth]{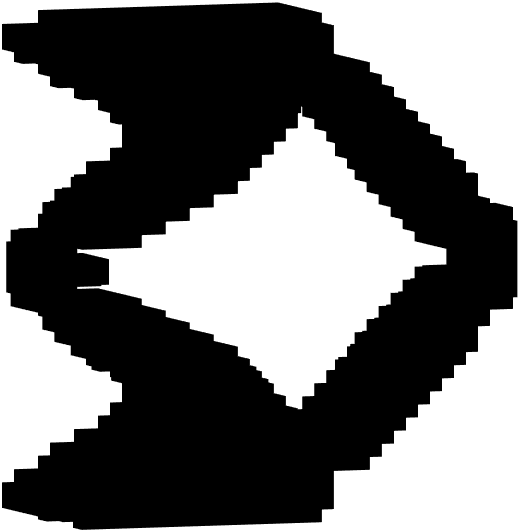}
		\captionsetup{justification=centering, labelformat=empty}
		\caption{\parbox{\linewidth}{\centering (b) 1 material,\\ $v_f= 0.40$,\\ $f_0$$= -199.1$}}
		\label{I3DSF1D1Mvf4}
	\end{subfigure}
	\hfill
	\begin{subfigure}{0.2\textwidth}
		\centering
		\includegraphics[width=\textwidth]{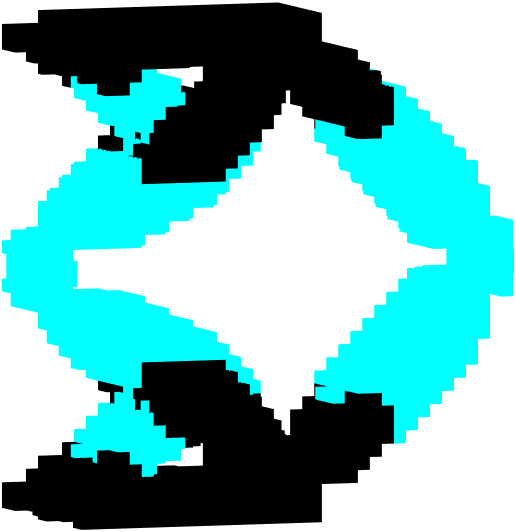}
		\captionsetup{justification=centering, labelformat=empty}
		\caption{\parbox{\linewidth}{\centering (c) 2 materials,\\ $v_f= 0.40$,\\ $f_0$$= -607.3$}}
		\label{I3DSF2D2Mvf2}
	\end{subfigure}
	\hfill
	\begin{subfigure}{0.2\textwidth}
		\centering
		\includegraphics[width=\textwidth]{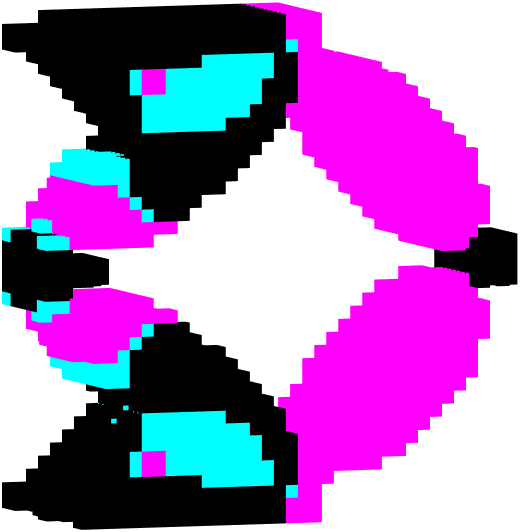}
		\captionsetup{justification=centering, labelformat=empty}
		\caption{\parbox{\linewidth}{\centering (d) 3 materials,\\ $v_f= 0.60$,\\ $f_0$$= -534.2$}}
		\label{I3DSF2D3Mvf2}
	\end{subfigure}	
	\caption{Optimized results using 1D and 2D SFs}
	\label{I3DSF1D2D}
\end{figure}

\begin{figure}
	\centering
	\begin{subfigure}[t]{0.2\textwidth}
		\centering
		\includegraphics[width=\textwidth]{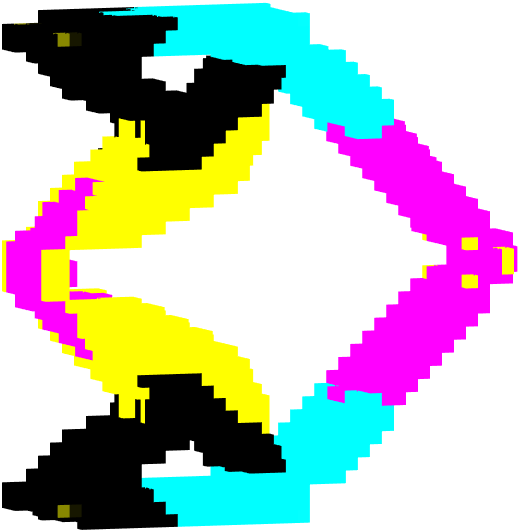}
		\captionsetup{justification=centering, labelformat=empty}
		\caption{\parbox{\linewidth}{\centering (a) 4 materials,\\ $v_f= 0.32$,\\ $f_0$$= -2347.2$}}
		\label{I3DSF3D4Mvf08}
	\end{subfigure}
	\hfill
	\begin{subfigure}[t]{0.2\textwidth}
		\centering
		\includegraphics[width=\textwidth]{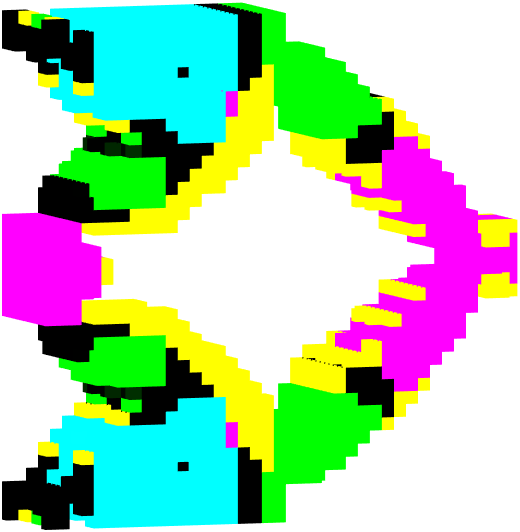}
		\captionsetup{justification=centering, labelformat=empty}
		\caption{\parbox{\linewidth}{\centering (b) 5 materials,\\ $v_f= 0.40$,\\ $f_0$$= -1990.2$}}
		\label{I3DSF3D5Mvf08}
	\end{subfigure}
	\hfill
	\begin{subfigure}[t]{0.2\textwidth}
		\centering
		\includegraphics[width=\textwidth]{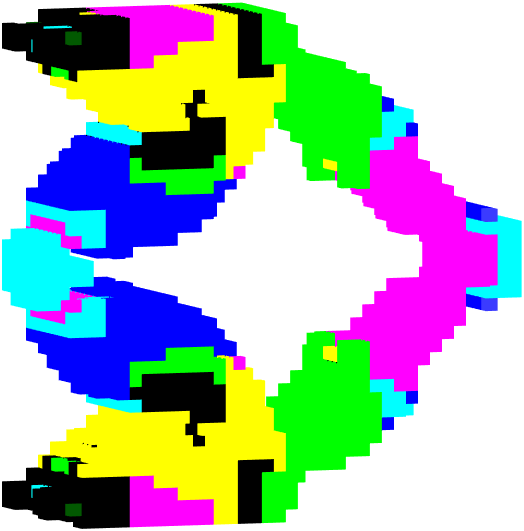}
		\captionsetup{justification=centering, labelformat=empty}
		\caption{\parbox{\linewidth}{\centering (c) 6 materials,\\ $v_f= 0.48$,\\ $f_0$$= -1923.7$}}
		\label{I3DSF3D6Mvf08}
	\end{subfigure}
	\hfill
	\begin{subfigure}[t]{0.2\textwidth}
		\centering
		\includegraphics[width=\textwidth]{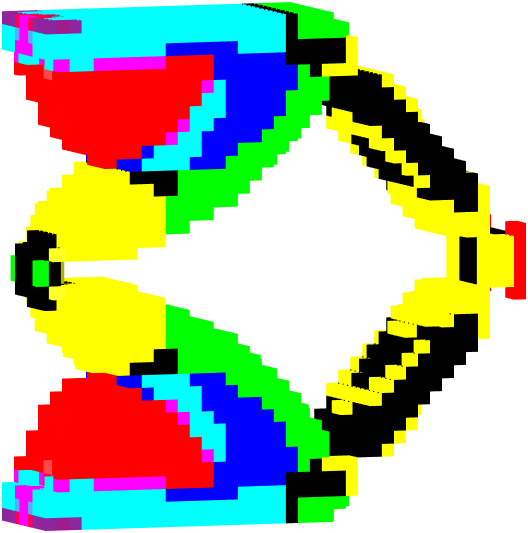}
		\captionsetup{justification=centering, labelformat=empty}
		\caption{\parbox{\linewidth}{\centering (d) 7 materials,\\ $v_f= 0.56$,\\ $f_0$$= -2181.2$}}
		\label{I3DSF3D7Mvf08}
	\end{subfigure}
	\caption{Optimized results using 3D SF}
	\label{I3DSF3D}
\end{figure}

\begin{figure}
	\centering
	\begin{subfigure}[t]{0.2\textwidth}
		\centering
		\includegraphics[width=\textwidth]{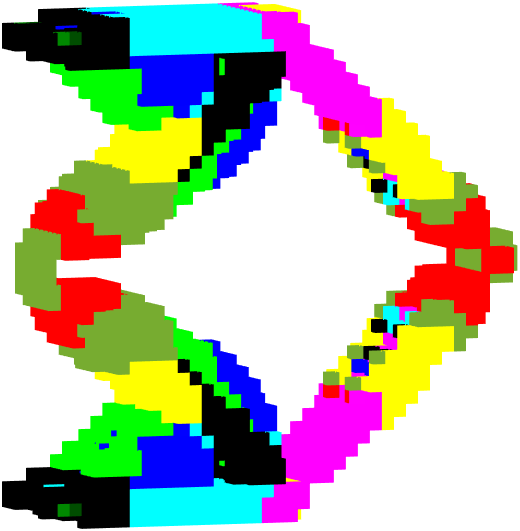}
		\captionsetup{justification=centering, labelformat=empty}
		\caption{\parbox{\linewidth}{\centering (a) 8 materials,\\ $v_f= 0.32$,\\ $f_0$$= -8322.2$}}
		\label{I3DSF4D8Mvf04}
	\end{subfigure}
	\hfill
	\begin{subfigure}[t]{0.2\textwidth}
		\centering
		\includegraphics[width=\textwidth]{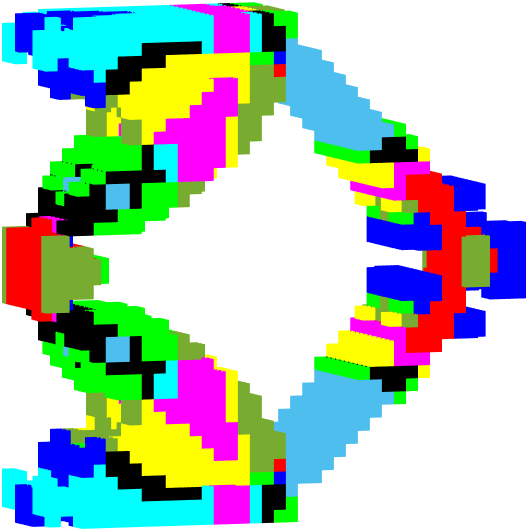}
		\captionsetup{justification=centering, labelformat=empty}
		\caption{\parbox{\linewidth}{\centering (b) 9 materials,\\ $v_f= 0.36$,\\ $f_0$$= -8019.4$}}
		\label{I3DSF4D9Mvf04}
	\end{subfigure}
	\hfill
	\begin{subfigure}[t]{0.2\textwidth}
		\centering
		\includegraphics[width=\textwidth]{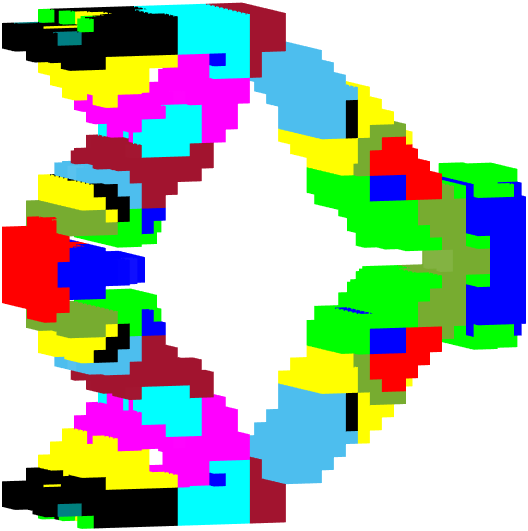}
		\captionsetup{justification=centering, labelformat=empty}
		\caption{\parbox{\linewidth}{\centering (c) 10 materials,\\ $v_f= 0.40$,\\ $f_0$$= -7176.4$}}
		\label{I3DSF4D10Mvf04}
	\end{subfigure}
	\hfill
	\begin{subfigure}[t]{0.2\textwidth}
		\centering
		\includegraphics[width=\textwidth]{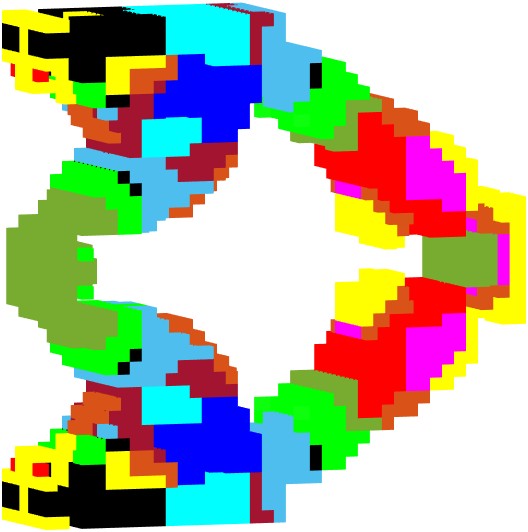}
		\captionsetup{justification=centering, labelformat=empty}
		\caption{\parbox{\linewidth}{\centering (d) 11 materials,\\ $v_f= 0.44$,\\ $f_0$$= -7158.9$}}
		\label{I3DSF4D11Mvf04}
	\end{subfigure}
	\begin{subfigure}[t]{0.2\textwidth}
		\centering
		\includegraphics[width=\textwidth]{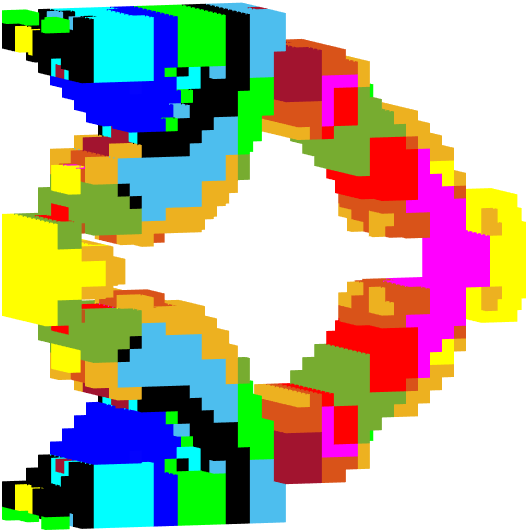}
		\captionsetup{justification=centering, labelformat=empty}
		\caption{\parbox{\linewidth}{\centering (e) 12 materials,\\ $v_f= 0.48$,\\ $f_0$$= -7755.1$}}
		\label{I3DSF4D12Mvf04}
	\end{subfigure}
	\hfill
	\begin{subfigure}[t]{0.2\textwidth}
		\centering
		\includegraphics[width=\textwidth]{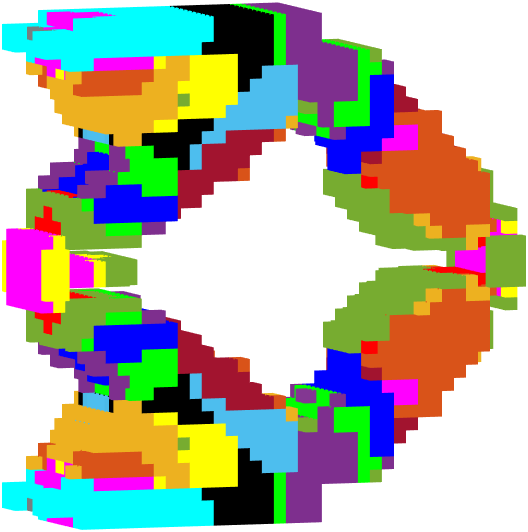}
		\captionsetup{justification=centering, labelformat=empty}
		\caption{\parbox{\linewidth}{\centering (f) 13 materials,\\ $v_f= 0.52$,\\ $f_0$$= -7387.3$}}
		\label{I3DSF4D13Mvf04}
	\end{subfigure}
	\hfill
	\begin{subfigure}[t]{0.2\textwidth}
		\centering
		\includegraphics[width=\textwidth]{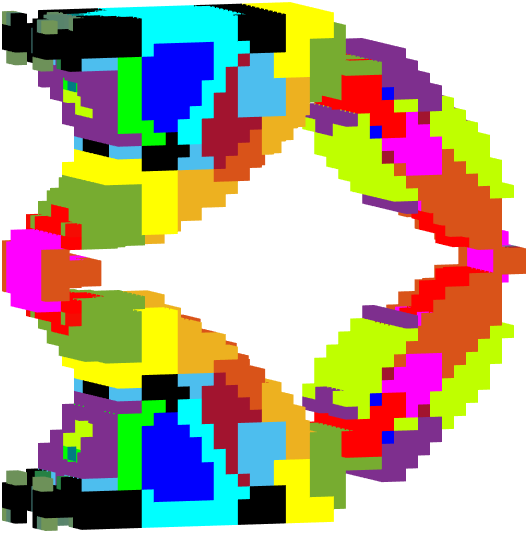}
		\captionsetup{justification=centering, labelformat=empty}
		\caption{\parbox{\linewidth}{\centering (g) 14 materials,\\ $v_f= 0.56$,\\ $f_0$$= -8077.6$}}
		\label{I3DSF4D14Mvf04}
	\end{subfigure}
	\hfill
	\begin{subfigure}[t]{0.2\textwidth}
		\centering
		\includegraphics[width=\textwidth]{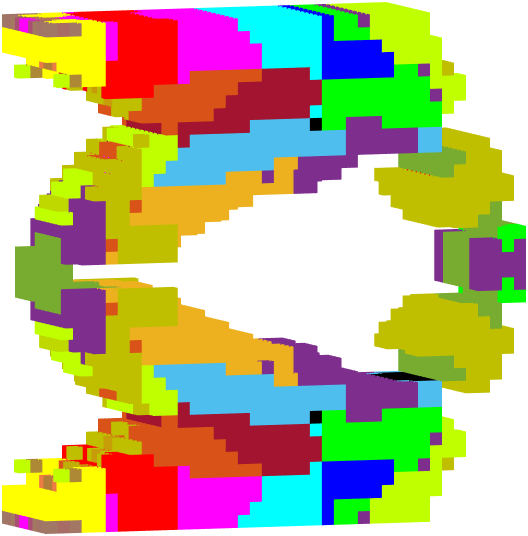}
		\captionsetup{justification=centering, labelformat=empty}
		\caption{\parbox{\linewidth}{\centering (h) 15 materials,\\ $v_f= 0.60$,\\ $f_0$$= -6585.6$}}
		\label{I3DSF4D15Mvf04}
	\end{subfigure}
	\caption{Optimized results using 4D SF}
	\label{I3DSF4D}
\end{figure}

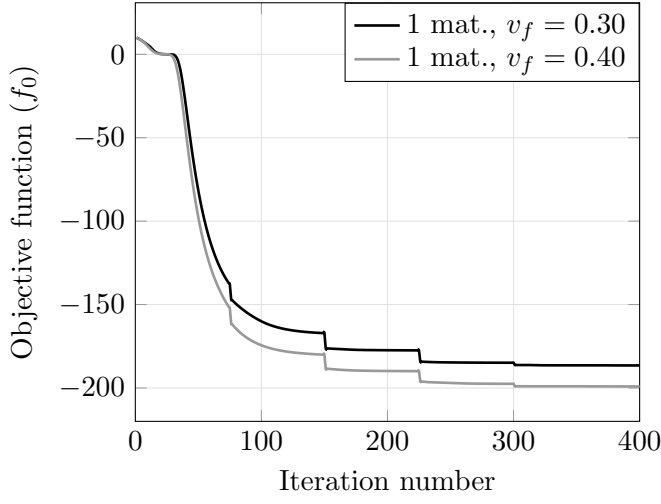
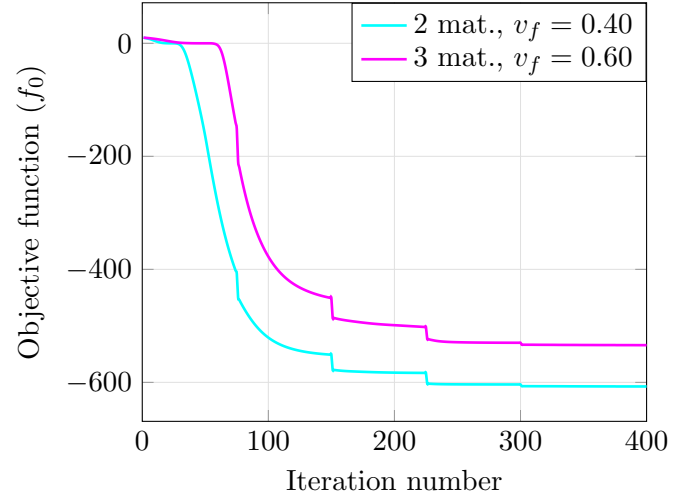
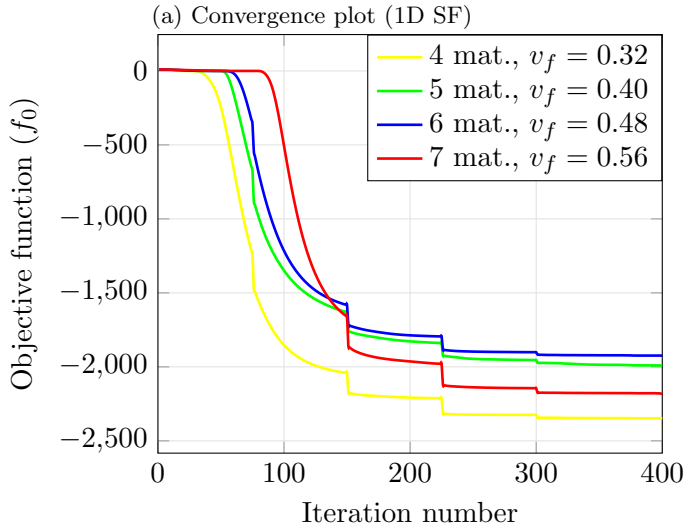
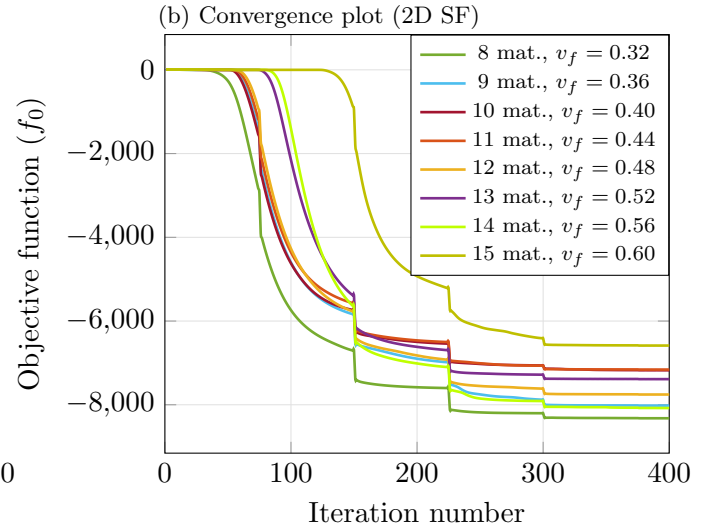
\begin{figure}
	\centering
	\begin{subfigure}[h!]{0.48\textwidth} 
		\centering
		\begin{tikzpicture}[scale=1]
			\pgfplotsset{compat=1.9}
			\begin{axis}[
				width = 1\textwidth,
				xlabel=  Iteration number,
				ylabel=  Objective function $(f_0)$,
				xmin=0,xmax=400,
				grid=both,
				major grid style={line width=0.2pt, draw=gray!30},
				legend style={at={(1.0,1.0)}, anchor=north east}]	
				\pgfplotstableread{I3D1Mvf30x1graph.txt}\mydata;
				\addplot[smooth,{black}, line width=1pt, mark = none]
				table {\mydata};
				\addlegendentry{1 mat., $v_f=0.30$}				
				\pgfplotstableread{I3D1Mvf40x1graph.txt}\mydata;
				\addplot[smooth,{gray}, line width=1pt, mark = none]
				table {\mydata};
				\addlegendentry{1 mat., $v_f=0.40$}			
			\end{axis}
		\end{tikzpicture}
		\caption{Convergence plot (1D SF)}
		\label{I3D_SF1D_graphs}
	\end{subfigure}
	\hfill
	\begin{subfigure}[h!]{0.48\textwidth} 
		\centering
		\begin{tikzpicture}[scale=1]
			\pgfplotsset{compat=1.9}
			\begin{axis}[
				width = 1\textwidth,
				xlabel=  Iteration number,
				ylabel=  Objective function $(f_0)$,
				xmin=0,xmax=400,
				grid=both,
				major grid style={line width=0.2pt, draw=gray!30},
				legend style={at={(1.0,1.0)}, anchor=north east}]
				\pgfplotstableread{I3D2Mgraph.txt}\mydata;
				\addplot[smooth,{cyan}, line width=1pt, mark = none]
				table {\mydata};
				\addlegendentry{2 mat., $v_f=0.40$}			
				\pgfplotstableread{I3D3Mgraph.txt}\mydata;
				\addplot[smooth,{magenta}, line width=1pt, mark = none]
				table {\mydata};
				\addlegendentry{3 mat., $v_f=0.60$}			
			\end{axis}
		\end{tikzpicture}
		\caption{Convergence plot (2D SF)}
		\label{I3D_SF2D_graphs}
	\end{subfigure}
	\begin{subfigure}[h!]{0.48\textwidth} 
		\centering
		\begin{tikzpicture}[scale=1]
			\pgfplotsset{compat=1.9}
			\begin{axis}[
				width = 1\textwidth,
				xlabel=  Iteration number,
				ylabel=  Objective function $(f_0)$,
				xmin=0,xmax=400,
				grid=both,
				major grid style={line width=0.2pt, draw=gray!30},
				legend style={at={(1.0,1.0)}, anchor=north east}]	
				\pgfplotstableread{I3D4Mgraph.txt}\mydata;
				\addplot[smooth,{yellow}, line width=1pt, mark = none]
				table {\mydata};
				\addlegendentry{4 mat., $v_f=0.32$}				
				\pgfplotstableread{I3D5Mgraph.txt}\mydata;
				\addplot[smooth,{green}, line width=1pt, mark = none]
				table {\mydata};
				\addlegendentry{5 mat., $v_f=0.40$}				
				\pgfplotstableread{I3D6Mgraph.txt}\mydata;
				\addplot[smooth,{blue}, line width=1pt, mark = none]
				table {\mydata};
				\addlegendentry{6 mat., $v_f=0.48$}				
				\pgfplotstableread{I3D7Mgraph.txt}\mydata;
				\addplot[smooth,{red}, line width=1pt, mark = none]
				table {\mydata};
				\addlegendentry{7 mat., $v_f=0.56$}		
			\end{axis}
		\end{tikzpicture}
		\caption{Convergence plot (3D SF)}
		\label{I3D_SF3D_graphs}
	\end{subfigure}
	\hfill
	\begin{subfigure}[h!]{0.48\textwidth} 
		\centering
		\begin{tikzpicture}[scale=1]
			\pgfplotsset{compat=1.9}
			\begin{axis}[
				width = 1\textwidth,
				xlabel=  Iteration number,
				ylabel=  Objective function $(f_0)$,
				xmin=0,xmax=400,
				grid=both,
				major grid style={line width=0.2pt, draw=gray!30},
				legend style={at={(1.0,1.0)}, anchor=north east, font=\scriptsize}]	
				\pgfplotstableread{I3D8Mgraph.txt}\mydata;
				\addplot[smooth,{shafgreen}, line width=1pt, mark = none]
				table {\mydata};
				\addlegendentry{8 mat., $v_f=0.32$}				
				\pgfplotstableread{I3D9Mgraph.txt}\mydata;
				\addplot[smooth,{skyblue}, line width=1pt, mark = none]
				table {\mydata};
				\addlegendentry{9 mat., $v_f=0.36$}				
				\pgfplotstableread{I3D10Mgraph.txt}\mydata;
				\addplot[smooth,{brownred}, line width=1pt, mark = none]
				table {\mydata};
				\addlegendentry{10 mat., $v_f=0.40$}				
				\pgfplotstableread{I3D11Mgraph.txt}\mydata;
				\addplot[smooth,{redorange}, line width=1pt, mark = none]
				table {\mydata};
				\addlegendentry{11 mat., $v_f=0.44$}				
				\pgfplotstableread{I3D12Mgraph.txt}\mydata;
				\addplot[smooth,{yellowochre}, line width=1pt, mark = none]
				table {\mydata};
				\addlegendentry{12 mat., $v_f=0.48$}				
				\pgfplotstableread{I3D13Mgraph.txt}\mydata;
				\addplot[smooth,{violet}, line width=1pt, mark = none]
				table {\mydata};
				\addlegendentry{13 mat., $v_f=0.52$}				
				\pgfplotstableread{I3D14Mgraph.txt}\mydata;
				\addplot[smooth,{greenyellow}, line width=1pt, mark = none]
				table {\mydata};
				\addlegendentry{14 mat., $v_f=0.56$}				
				\pgfplotstableread{I3D15Mgraph.txt}\mydata;
				\addplot[smooth,{olivegreen}, line width=1pt, mark = none]
				table {\mydata};
				\addlegendentry{15 mat., $v_f=0.60$}		
			\end{axis}
		\end{tikzpicture}
		\caption{Convergence plot (4D SF)}
		\label{I3D_SF4D_graphs}
	\end{subfigure}
	\caption{Convergence plots (3D Inverter)}
	\label{I3D_all_graphs}
\end{figure}
\noindent Fig.~\ref{I3DSF4D} represents the optimized mechanism obtained using 4D material elements. Fig.~\ref{I3DSF4D8Mvf04}, \ref{I3DSF4D9Mvf04}, \ref{I3DSF4D10Mvf04}, \ref{I3DSF4D11Mvf04}, \ref{I3DSF4D12Mvf04}, \ref{I3DSF4D13Mvf04}, \ref{I3DSF4D14Mvf04}, and \ref{I3DSF4D15Mvf04} show the optimized CMs with 8, 9, 10, 11, 12, 13, 14, and 15 materials, respectively. All of the results obtained from 4D material elements have $v_f=0.04$ for each material. The objective convergence plots are shown in  Fig.~\ref{I3D_all_graphs}. These plots have a converging nature. Volume constraints are predominantly found to be active at the end of the optimization.



 To the best of the author's knowledge, the SSs and CMs in 2D and 3D  up to 24 and 15 distinct materials, respectively, presented in this paper are the first of their kind. This achievement highlights the robustness and versatility of the proposed method in effectively handling complex multi-material optimization problems. The results demonstrate that the approach can readily manage and design structures comprising numerous materials, providing a way to develop advanced multi-material systems for other applications. The TO problems formulated herein are highly non-convex, particularly due to the use of SIMP parameter $p=3$, density filtering, and projection schemes. Additionally, the non-convex nature of the problems becomes even more pronounced in multi-material cases as the number of candidate materials increases. Consequently, different parameter choices and initial guesses will lead to distinct local minima.

  \section{Conclusions}\label{sec5}
This paper proposes a generalized shape function (gSF) approach for multi-material topology optimization, wherein building upon 1D (linear), 2D (bilinear), and 3D (trilinear)  shape functions, generalized nD (n-linear) shape functions are proposed. The natural coordinates of these functions serve as the design variables used to determine material densities. These densities are mathematically proven to satisfy essential barycentric properties, guaranteeing a physically valid material interpolation space. Furthermore, we demonstrate that applying density filtering directly to the natural coordinates is mathematically equivalent to filtering the densities themselves, and that the tailored projection scheme preserves these vital barycentric properties in the projected states. The method's efficacy, robustness, and success are demonstrated by solving 2D and 3D stiff-structure (SS) and compliant mechanism (CM) problems with up to 24 materials (25 phases) and 15 materials (16 phases), respectively, using 1D to 5D material elements. In SS optimization, the boundary and force regions are predominantly surrounded by material with the highest elastic modulus. In inverter CM optimized designs, the input and output regions are surrounded by the same material.

Strain energy is minimized to obtain optimized SSs, whereas the ratio of output deformation to compliance is minimized to obtain optimized CMs. Sensitivities are determined using the adjoint-variable method. The objective history plots converge smoothly at the end of the optimization across all numerical experiments. The noted jumps in the objective convergence plots are due to updates to the projection parameter $\beta_p$. Additionally, the volume constraint associated with each candidate material remains active at the end of optimization, by and large. The approach helps reduce the number of design variables, making it possible to design with any number of materials using a few design variables. The approach uses $n$ variables to provide a design with up to $(2^n-1)$ materials or $2^n$ phase systems. The associated manifold acts as a well-conditioned, direct mathematical highway connecting these vertices. Additionally, it provides continuous, non-vanishing gradients throughout the interior of the design domain, ensuring the optimizer can freely navigate the search space and reach superior solutions.

This study is the first to report the SSs and CMs in 2D and 3D up to 24 and 15 distinct materials, respectively. This shows the gSF approach's ability to effectively handle TO with many materials, furnishing a novel way to develop advanced multi-material designs for different applications in the near future. Though the material space of the gSF is lower-dimensional than that of the corresponding SIMP/DMO methods, the proposed method efficiently provides optimized results across many materials that converge
to objective values that are either identical to or marginally better than those of full-space formulations. In addition, considering advanced constraints, such as stress/buckling, with the presented gSF approach is an exciting next avenue for research across various applications.
Furthermore, multi-material numerical investigations using the gSF approach for multiphysics problems for different applications will be solved in near-future work.


  \newpage
  
  \appendix
  \section{Theoretical framework for the \lowercase{g}SF mapping}\label{AppA}
  The material space of gSF defines an $n$-dimensional hypercube $\mathcal{H} = [-1, 1]^n$ where the state of any point is governed by the natural coordinate vector $\boldsymbol{\chi} = [\chi_1, \chi_2, \dots, \chi_n]^\top$. The space corresponding to the classical material simplex $\mathcal{S}$ represents the set of all density vectors $\boldsymbol{\rho} \in \mathbb{R}^{2^n}$ with $\rho_m \geq 0$ and $\sum_{m=1}^{2^n} \rho_m = 1$. In the simplex, the set of pure material states $\mathcal{V}$ corresponds to $2^n$ vertices wherein a density component related to a particular material is unity, and all others are zero. Consider $\Phi: \mathcal{H} \to \mathcal{S}$ uses the generalized $n$-linear shape functions such that the density of the $m$-th material phase is given by $\rho_m(\boldsymbol{\chi}) = \frac{1}{2^n} \prod_{i=1}^{n} (1 + x_{im} \chi_i)$ with $x_{im} \in \{-1, 1\}$ denotes the fixed coordinate of the $m$-th vertex of the $n$-dimensional hypercube. A schematic diagram for the $n$-linear mapping $\Phi: \mathcal{H} \to \mathcal{S}$ for $n=2$ design variables and $m=4$ material phases is depicted in Fig.~\ref{Fig:AppScheSimplex}.
  \begin{figure}[h]
  	\centering
  	\includegraphics[scale =1]{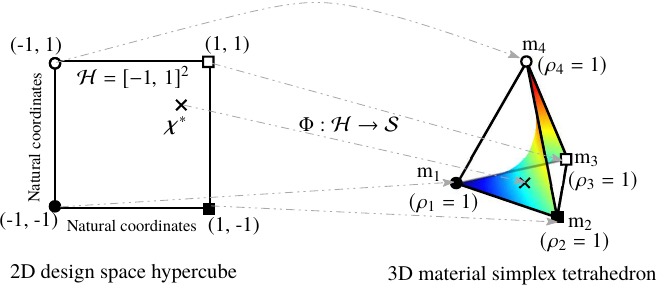}
  	\caption{Schematic of the $n$-linear mapping $\Phi: \mathcal{H} \to \mathcal{S}$ for $n=2$ design variables and $m=4$ material phases. The mapping transforms the 2D design hypercube (left) into a 3D material simplex (right). Each vertex of the design space is uniquely mapped to a pure material state in the simplex, as indicated by the dashed-dot-dotted projection lines, ensuring a vertex-complete manifold for topology optimization. The coloured surface within the tetrahedron represents the admissible manifold $\mathcal{M}$.}\label{Fig:AppScheSimplex}
  \end{figure}
  
  To establish the theoretical completeness of the gSF approach, we demonstrate that the mapping $\Phi$ is surjective onto the set of pure material states $\mathcal{V}$. For a pure material $k$, we select a particular point $\boldsymbol{\chi}^*$ in the design space such that each design variable $\chi_j^*$ is set equal to the corresponding coordinate of the $k$-th hypercube vertex, $x_{kj}$. Now, evaluating the density function for the $k$-th material at this point provides $\rho_k(\boldsymbol{\chi}^*) = \frac{1}{2^n} \prod_{j=1}^{n} (1 + x_{kj}^2)=\frac{1}{2^n} (2^n) = 1$ as $x_{kj} \in \{-1, 1\}$, every term in the product becomes 2. For any other material $i \neq k$, the vertex coordinate vector must differ from the $k$-th vector in at least one dimension $m$, that is $x_{im} = -x_{km}$, due to which the $m$-th term of the product $(1 + x_{im} \chi_m^*) =0$. Consequently, all other densities $\rho_{i \neq k} =0$, indicating that the mapping can uniquely reach every pure material vertex in the simplex.
  
  As depicted in Fig.~\ref{Fig:AppScheSimplex}, the embedded manifold $\mathcal{M}$ forms a continuous, smooth, and monotonic hypersurface that spans directly across the vertices of the simplex, establishing path completeness. In conventional formulations, such as extended SIMP or DMO, the optimizer searches over a higher-dimensional volume of the full simplex; however, superior solutions should always lead the design variables to the vertices, otherwise either material mixing or artificial material states will be achieved, which are not desirable. The gSF $n$-linear manifold acts as a well-conditioned, direct mathematical highway connecting these vertices (Fig.~\ref{Fig:AppScheSimplex}). Additionally, it provides continuous, non-vanishing gradients throughout the interior of the design domain $\mathcal{H}$, ensuring the optimizer can freely navigate the search space and reach superior solutions.
  
  \section{A comparative study between the proposed \lowercase{g}SF approach and several existing methods}\label{AppB}
  Herein, we present a comparative study between the proposed gSF approach and several existing methods, such as extended SIMP~\cite{sigmund1997design}, DMO~\cite{stegmann2005discrete}, ordered SIMP~\cite{zuo2017multi}, and AAP~\cite{tavakoli2014alternating}  to evaluate their relative performances with respect to the final objective values, discreteness measure, number of design variables per element, and convergence plots. We develop in-house MATLAB implementations for the extended SIMP and DMO methods for MMTO. Although a MATLAB code for DMO using polygonal elements is publicly available in~\cite{sanders2018polymat}, we write our own version for consistency. In addition, publicly available MATLAB codes from Refs.~\cite{da2022some} and \cite{tavakoli2014alternating} are adopted for the ordered SIMP and AAP methods, respectively. The AAP code is modified to work with the conventional projection filter~\cite{sigmund2013topology} and MMA optimizer~\cite{svanberg1987method}. These appropriate modifications are performed to ensure, as far as possible, a consistent set of parameters across all approaches and optimizers employed. The MBB beam problem (Fig.~\ref{MBB_Beam_2D}) is selected for this study. $90\times 30$ quad finite elements are used to discretize the symmetric half design domain. Filter radius $r =3.6$ is set. The projection parameter $\beta$ is updated every 75 iterations up to 32. Equal volume fractions of each candidate material are used. The maximum number of MMA iterations is fixed to 400. We normalize the objective, and thus sensitivity using a multiplication factor in every iteration, $n_f =\left.\min \left(\frac{10}{\mathrm{obj}},1 \times 10^2\right)\right|_{\mathrm{loop}=1}$, wherein $\mathtt{loop}$ indicates MMA iteration number. The colour scheme with the corresponding material elastic stiffness values is adopted from Sec.~\ref{sec4}. As the existing methods operate on different principles, the following $M\textsubscript{nd}$ expressions are considered to determine the corresponding measures of discreteness for the optimized designs. $M_\mathrm{{nd}^{SIMP/DMO}} =\frac{\displaystyle \sum_{j=1}^{N_e} 4\rho_{j}^1\left(1 - \rho_{j}^1\right)}{N_e}\times 100\%,\, \rho_{j}^1\,$ defines solid/void state of element $j$; $M_\mathrm{{nd}^{OSIMP}} =\frac{\displaystyle \sum_{e=1}^{N^{\mathrm{int}}}4\,\bar{w}_e \left(1 - \bar{w}_e \right)}{N^{\mathrm{int}}}\times 100\%,\,\bar{w}_e$ and $N^\mathrm{int}$ are the relative position of the physical density and the number of finite elements within the considered interval~\cite{da2022some}, respectively; and $M_\mathrm{{nd}^{AAP}} = \frac{\displaystyle \sum_{j=1}^{N_e}\left(\frac{m+1}{m}\right)^{m+1}\prod_{i=1}^{m+1} \left(1 - \rho_{ij}\right)}{N_e}\times 100\%, \rho_{ij}$ is the density of $i$\textsuperscript{th} phase in $j$\textsuperscript{th} element \cite{yan2024multi}, $m$ is the number of materials; represent $M\textsubscript{nd}$ for SIMP/DMO, O-SIMP and AAP approaches, respectively.

  The comparative study is demonstrated in Table~\ref{Tab:ComStudDiffMethods}. Results are presented for up to six distinct materials using the existing approaches and the gSF. The extended SIMP and DMO methods require a number of design variables per element that scales linearly with the number of candidate materials. In contrast, the ordered SIMP uses a single variable per element, irrespective of the number of candidate materials (Table~\ref{Tab:ComStudDiffMethods}). Since the AAP approach considers a void region (element) as a material, it requires $m+1$ variables per element for $m$ materials (Table~\ref{Tab:ComStudDiffMethods}). For the proposed gSF approach, the design variables correspond to the natural coordinates associated with the $n$-dimensional material element, requiring only $n$ design variables per element while providing $2^n-1$ distinct materials. Consequently, it requires fewer design variables per element compared to extended SIMP, DMO, and AAP approaches (Table~\ref{Tab:ComStudDiffMethods}).

  \begin{table}[]
  	\caption{A comparative study between the proposed, generalized shape function (gSF) approach and existing methods, such as the extended SIMP (Ex-SIMP), discrete material optimization (DMO), ordered-SIMP (O-SIMP), and alternative active phase (AAP)  approaches, is presented. ndv, $v_f$, OD, Obj, and $M_\text{nd}$ indicate the number of design variables per element, volume fraction of each material, optimized design, objective value, and the discrete measure value. The MBB beam is solved with  $90\times 30$  quad finite elements. A filter radius of $r = 3.6$ is used. The projection parameter $\beta$ is updated every 75 iterations up to 32. The maximum number of MMA iterations, $\text{IterMax} = 400$. We normalize the objective, and thus sensitivity using $n_f = \min\left({\frac{10}{\text{obj}},\,\num{1e2}}\right)|_{\mathtt{loop} =1}$, wherein $\mathtt{loop}$ indicates MMA iteration loop. Color scheme with respective material elastic stiffness values is adopted from Sec.~\ref{sec4}}\label{Tab:ComStudDiffMethods}
  	\centering
  	\begin{tabular}{|C{1.6cm}|C{2.75cm}|C{2.75cm}|C{2.75cm}|C{2.75cm}|C{2.75cm}|}
  		\hline
  		\makecell{\textbf{Method/}\\ \textbf{Variables}} & \textbf{gSF} & \textbf{SIMP / Ex-SIMP} & \textbf{DMO} & \textbf{O-SIMP} & \textbf{AAP} \\ \hline
  		\multicolumn{6}{|c|}{Number of material = 1, $v_f=0.30$} \\ \hline
  		ndv & 1 & 1 & 1 & 1 & 2 \\ \hline
  		\makecell{OD} & \vspace{0.2cm} \includegraphics[scale=0.25]{MBB2D1Mbeta75.png} & \vspace{0.2cm} \includegraphics[scale=0.25]{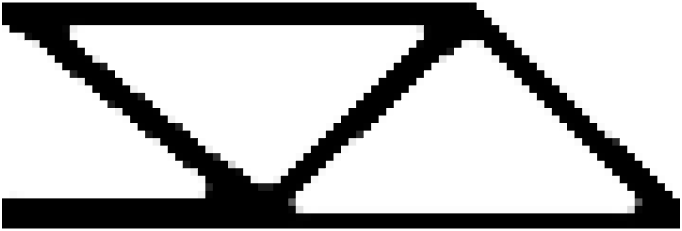} & \vspace{0.2cm} \includegraphics[scale=0.25]{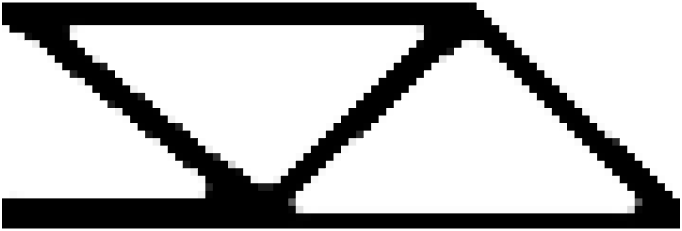} & \vspace{0.2cm} \includegraphics[scale=0.25]{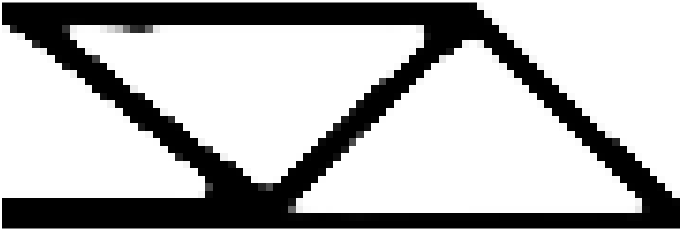} & \vspace{0.2cm} \includegraphics[scale=0.25]{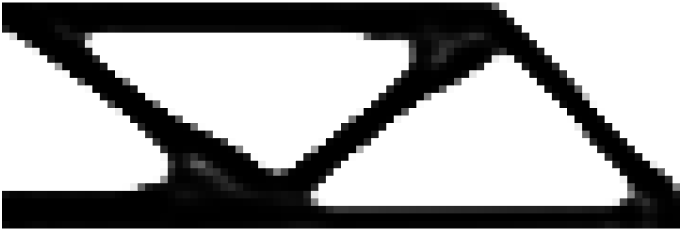}  \\ \hline
  		$f_0$ & $0.0149$ & $0.0148$ & $0.0148$ & $0.0296$ & $0.0235$ \\ \hline
  		$M_\text{nd}$ & $1.57\%$ & $0.91\%$ & $0.91\%$ & $1.32\%$ & $5.22\%$ \\ \hline \hline
  		\multicolumn{6}{|c|}{Number of material = 2, \,$v_f=0.20$} \\ \hline
  		ndv & 2 & 2 & 2 & 1 & 3 \\ \hline
  		\makecell{OD} & \vspace{0.2cm} \includegraphics[scale=0.25]{MBB2D2Mbeta75.png} & \vspace{0.2cm} \includegraphics[scale=0.25]{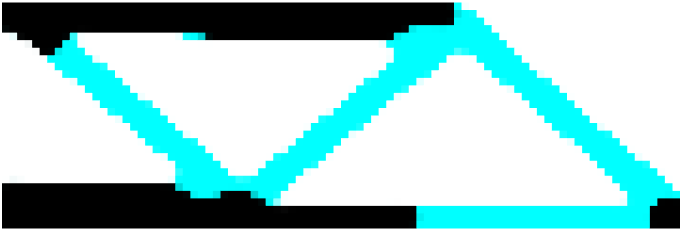} & \vspace{0.2cm} \includegraphics[scale=0.25]{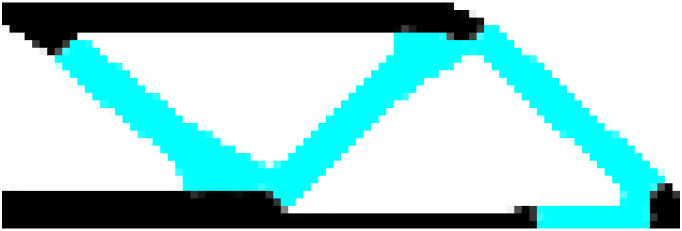} & \vspace{0.2cm} \includegraphics[scale=0.25]{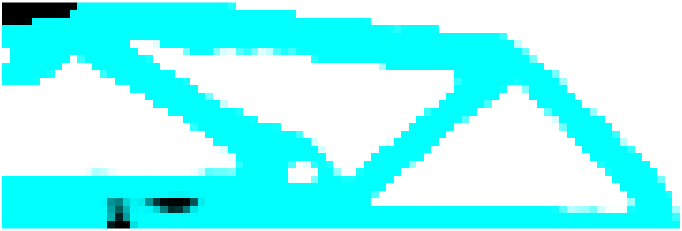} & \vspace{0.2cm} \includegraphics[scale=0.25]{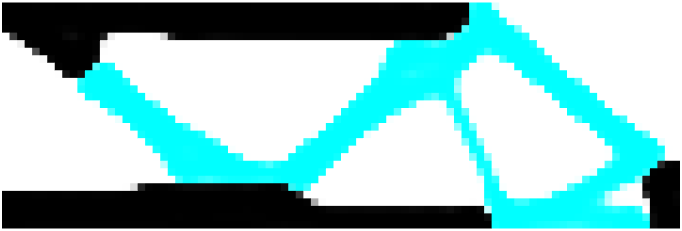} \\ \hline
  		$f_0$ & $0.0130$ & $0.0131$ & $0.0130$ & $0.0316$ & $0.0237$ \\ \hline
  		$M\textsubscript{nd}$ & $1.39\%$ & $0.90\%$ & $1.70\%$ & $3.08\%$ & $3.41\%$ \\ \hline \hline
  		\multicolumn{6}{|c|}{Number of material = 3, \,$v_f=0.20$} \\ \hline
  		ndv & 2 & 3 & 3 & 1 & 4 \\ \hline
  		\makecell{OD} & \vspace{0.2cm} \includegraphics[scale=0.25]{MBB2D3Mbeta75.png} & \vspace{0.2cm} \includegraphics[scale=0.25]{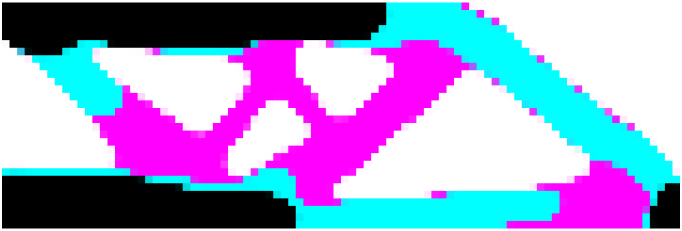} & \vspace{0.2cm} \includegraphics[scale=0.25]{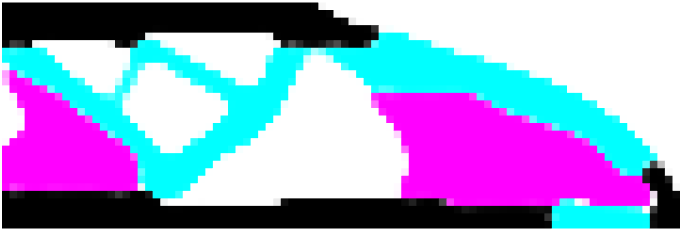} & \vspace{0.2cm} \includegraphics[scale=0.25]{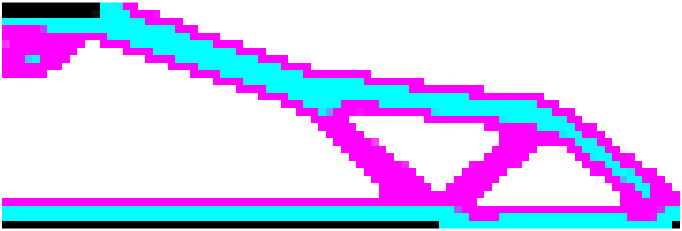} & \vspace{0.2cm} \includegraphics[scale=0.25]{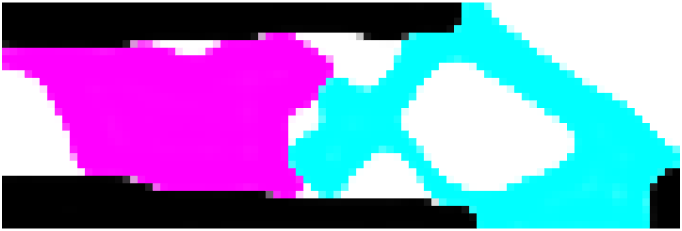} \\ \hline
  		$f_0$ & $0.0112$ & $0.0109$ & $0.0113$ & $0.0501$ & $0.0192$ \\ \hline
  		$M\textsubscript{nd}$ & $2.03\%$ & $1.28\%$ & $4.28\%$ & $0.68\%$ & $2.95\%$ \\ \hline \hline
  		\multicolumn{6}{|c|}{Number of material = 4, \,$v_f=0.08$} \\ \hline
  		ndv & 3 & 4 & 4 & 1 & 5 \\ \hline
  		\makecell{OD} & \vspace{0.2cm} \includegraphics[scale=0.25]{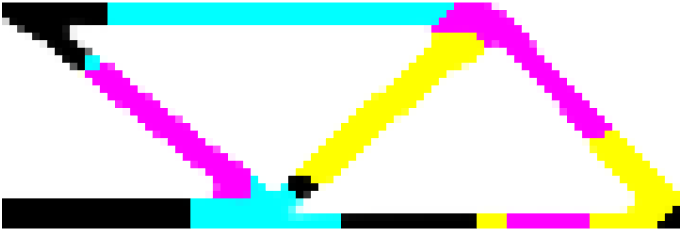} & \vspace{0.2cm} \includegraphics[scale=0.25]{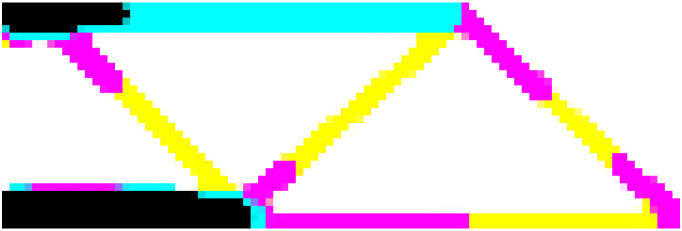} & \vspace{0.2cm} \includegraphics[scale=0.25]{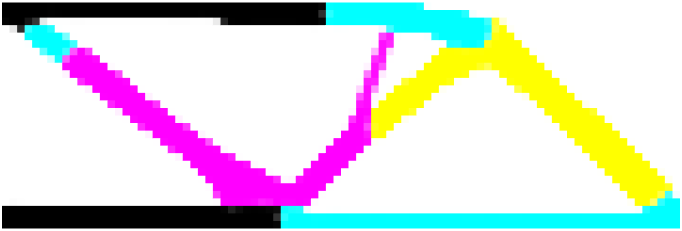} & \vspace{0.2cm} \includegraphics[scale=0.25]{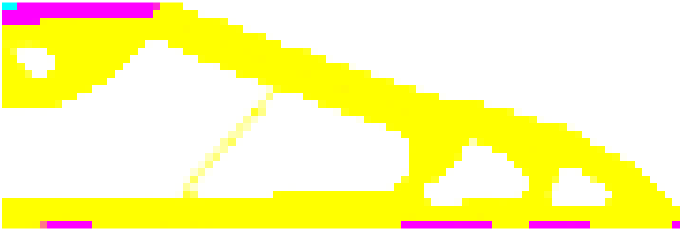} & \vspace{0.2cm} \includegraphics[scale=0.25]{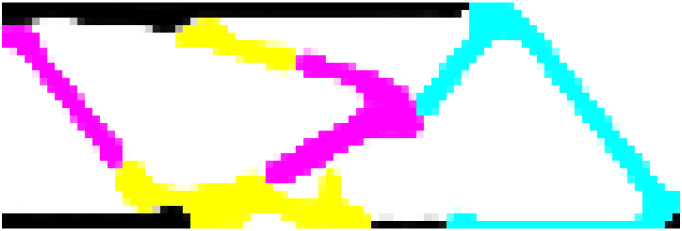} \\ \hline
  		$f_0$ & $0.0172$ & $0.0182$ & $0.0177$ & $0.0518$ & $0.0440$ \\ \hline
  		$M\textsubscript{nd}$ & $1.60\%$ & $1.23\%$ & $2.49\%$ & $1.75\%$ & $2.96\%$ \\ \hline \hline
  		\multicolumn{6}{|c|}{Number of material = 5, \,$v_f=0.08$} \\ \hline
  		ndv & 3 & 5 & 5 & 1 & 6 \\ \hline
  		\makecell{OD} & \vspace{0.2cm} \includegraphics[scale=0.25]{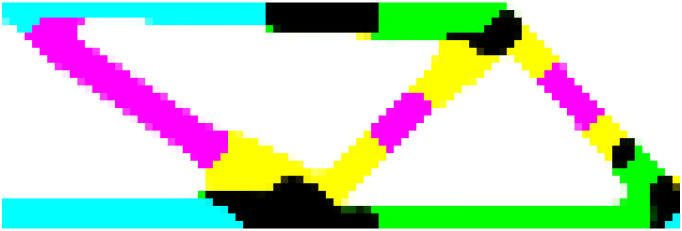} & \vspace{0.2cm} \includegraphics[scale=0.25]{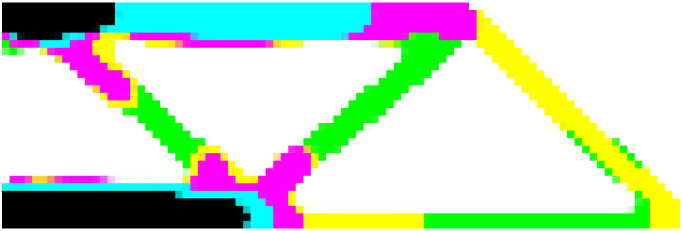} & \vspace{0.2cm} \includegraphics[scale=0.25]{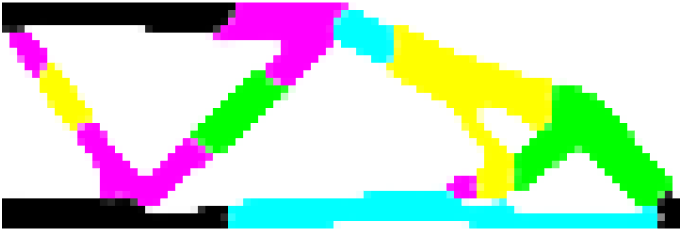} & \vspace{0.2cm} \includegraphics[scale=0.25]{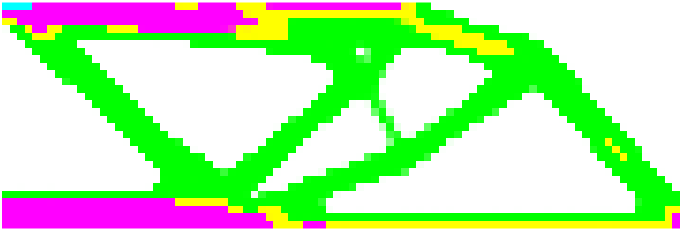} & \vspace{0.2cm} \includegraphics[scale=0.25]{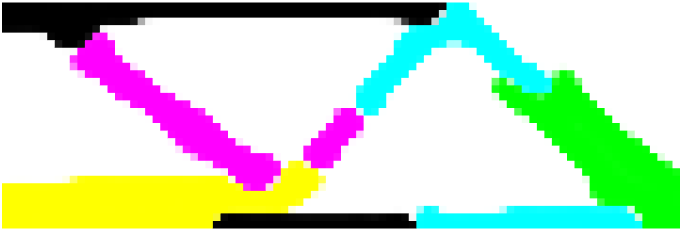} \\ \hline
  		$f_0$ & $0.0167$ & $0.0145$ & $0.0155$ & $0.0370$ & $0.0378$ \\ \hline
  		$M\textsubscript{nd}$ & $1.60\%$ & $1.12\%$ & $3.33\%$ & $1.29\%$ & $1.96\%$ \\ \hline \hline
  		\multicolumn{6}{|c|}{Number of material = 6, \,$v_f=0.08$} \\ \hline
  		ndv & 3 & 6 & 6 & 1 & 7 \\ \hline
  		\makecell{OD} & \vspace{0.2cm} \includegraphics[scale=0.25]{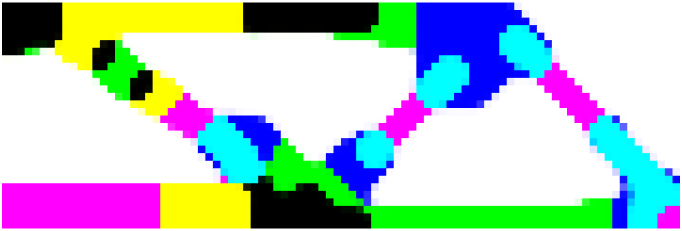} & \vspace{0.2cm} \includegraphics[scale=0.25]{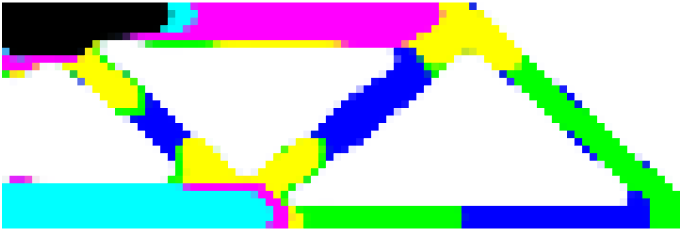} & \vspace{0.2cm} \includegraphics[scale=0.25]{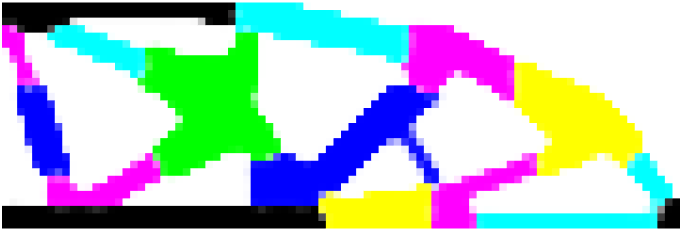} & \vspace{0.2cm} \includegraphics[scale=0.25]{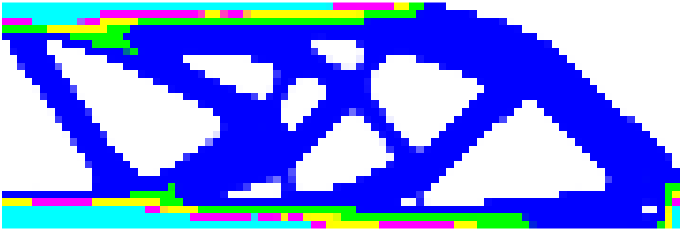} & \vspace{0.2cm} \includegraphics[scale=0.25]{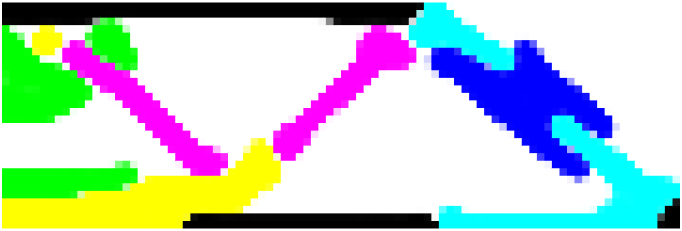} \\ \hline
  		$f_0$ & $0.0162$ & $0.0101$ & $0.0144$ & $0.0344$ & $0.0380$ \\ \hline
  		$M\textsubscript{nd}$ & $1.55\%$ & $0.98\%$ & $5.21\%$ & $1.73\%$ & $2.53\%$ \\ \hline
  	\end{tabular}
  \end{table}
  
  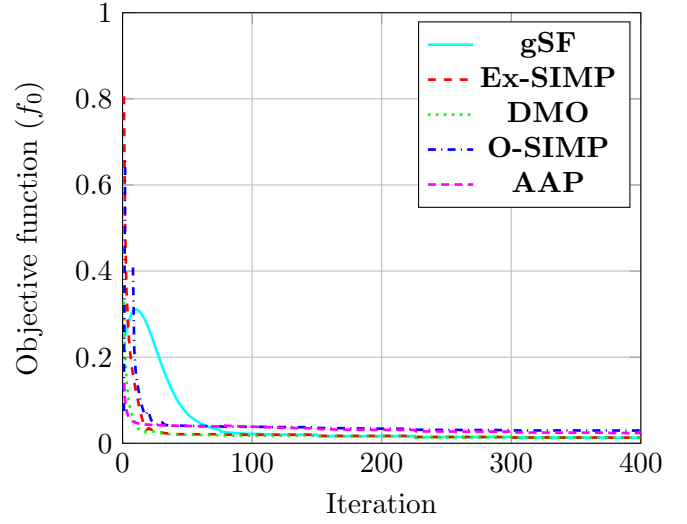
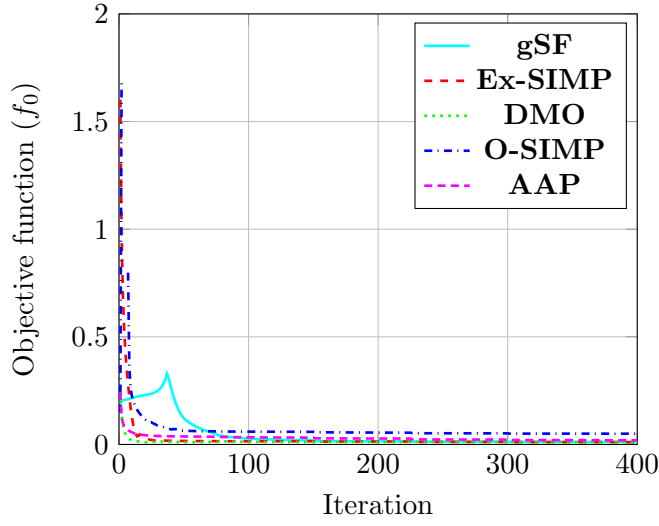
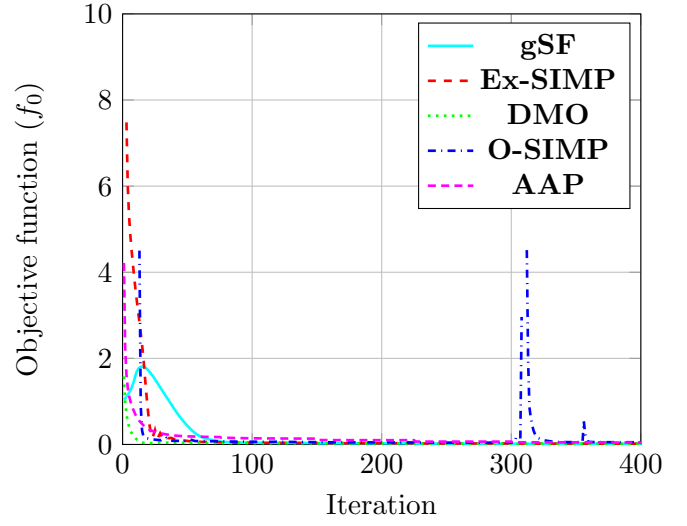
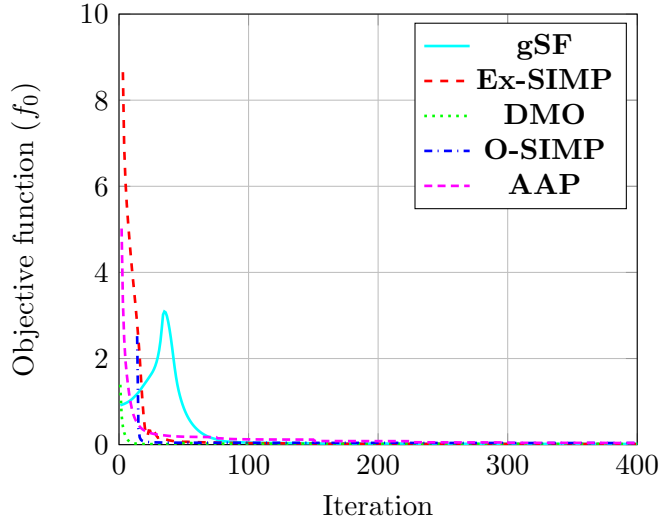
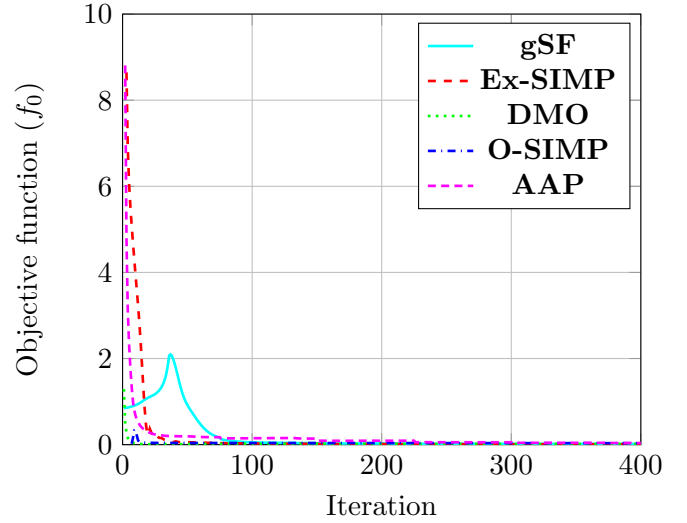
\begin{figure}[]
  	\centering
  	\begin{subfigure}[t]{0.40\textwidth}
  		\centering
  		\begin{tikzpicture}[scale=1]
  			\begin{axis}[
  				xlabel={MMA iteration},
  				ylabel={Objective function $(f_0)$},
  				xmin=0,xmax=400,
  				grid=major,
  				]    
  				\addplot[solid, {cyan}, line width=1pt] table[x index=0, y index=1] {MBB2D1Mvf30beta75graph.txt}; \addlegendentry{\textbf{gSF}}
  				\addplot[dashed, {red}, line width=1pt] table[x index=0, y index=1] {SIMP1Mvf30.txt};
  				\addlegendentry{\textbf{Ex-SIMP}}
  				\addplot[dotted, {green}, line width=1pt] table[x index=0, y index=1] {DMO1Mvf30.txt};
  				\addlegendentry{\textbf{DMO}}
  				\addplot[dashdotted, {blue}, line width=1pt] table[x index=0, y index=1] {OrderedSIMP1Mvf30.txt};
  				\addlegendentry{\textbf{O-SIMP}}
  				\addplot[densely dashed, {magenta}, line width=1pt] table[x index=0, y index=1] {AAP3MMA1Mvf30.txt};
  				\addlegendentry{\textbf{AAP}}
  			\end{axis}
  		\end{tikzpicture}
  		\caption{Single material, $m =1$}
  	\end{subfigure}
  	\hspace{2 cm}
  	\begin{subfigure}[t]{0.40\textwidth}
  		\centering
  		\begin{tikzpicture}[scale=1]
  			\begin{axis}[
  				xlabel={Iteration},
  				ylabel={Objective function $(f_0)$},
  				xmin=0,xmax=400,
  				grid=major,
  				ymin=0, ymax=1,
  				restrict y to domain=0:1
  				]    
  				\addplot[solid, {cyan}, line width=1pt] table[x index=0, y index=1] {MBB2D2Mbeta75graph.txt}; \addlegendentry{\textbf{gSF}}
  				\addplot[dashed, {red}, line width=1pt] table[x index=0, y index=1] {ExSIMP2M.txt};
  				\addlegendentry{\textbf{Ex-SIMP}}
  				\addplot[dotted, {green}, line width=1pt] table[x index=0, y index=1] {DMO2M.txt};
  				\addlegendentry{\textbf{DMO}}
  				\addplot[dashdotted, {blue}, line width=1pt] table[x index=0, y index=1] {OrderedSIMP2M.txt};
  				\addlegendentry{\textbf{O-SIMP}}
  				\addplot[densely dashed, {magenta}, line width=1pt] table[x index=0, y index=1] {AAP3MMA2M.txt};
  				\addlegendentry{\textbf{AAP}}
  			\end{axis}
  		\end{tikzpicture}
  		\caption{Two materials, $m=2$}
  	\end{subfigure}
  	
  	\begin{subfigure}[t]{0.40\textwidth}
  		\centering
  		\begin{tikzpicture}[scale=1]
  			\begin{axis}[
  				xlabel={Iteration},
  				ylabel={Objective function $(f_0)$},
  				xmin=0,xmax=400,
  				grid=major,
  				ymin=0, ymax=2,
  				restrict y to domain=0:2
  				]    
  				\addplot[solid,{cyan}, line width=1pt] table[x index=0, y index=1] {MBB2D3Mbeta75graph.txt}; \addlegendentry{\textbf{gSF}}
  				\addplot[dashed,{red}, line width=1pt, mark = none] table[x index=0, y index=1] {ExSIMP3M.txt};
  				\addlegendentry{\textbf{Ex-SIMP}}
  				\addplot[dotted,{green}, line width=1pt] table[x index=0, y index=1] {DMO3M.txt};
  				\addlegendentry{\textbf{DMO}}
  				\addplot[dashdotted,{blue}, line width=1pt] table[x index=0, y index=1] {OrderedSIMP3M.txt};
  				\addlegendentry{\textbf{O-SIMP}}
  				\addplot[densely dashed, {magenta}, line width=1pt] table[x index=0, y index=1] {AAP3MMA3M.txt};
  				\addlegendentry{\textbf{AAP}}
  			\end{axis}
  		\end{tikzpicture}
  		\caption{Three materials, $m=3$}
  	\end{subfigure}
  	\hspace{2cm}
  	\begin{subfigure}[t]{0.40\textwidth}
  		\centering
  		\begin{tikzpicture}[scale=1]
  			\begin{axis}[
  				xlabel={Iteration},
  				ylabel={Objective function $(f_0)$},
  				xmin=0,xmax=400,
  				grid=major,
  				ymin=0, ymax=10,
  				restrict y to domain=0:10
  				]    
  				\addplot[solid,{cyan}, line width=1pt] table[x index=0, y index=1] {MBB2D4Mgraph.txt}; \addlegendentry{\textbf{gSF}}
  				\addplot[dashed,{red}, line width=1pt, mark = none] table[x index=0, y index=1] {ExSIMP4M.txt};
  				\addlegendentry{\textbf{Ex-SIMP}}
  				\addplot[dotted,{green}, line width=1pt] table[x index=0, y index=1] {DMO4M.txt};
  				\addlegendentry{\textbf{DMO}}
  				\addplot[dashdotted,{blue}, line width=1pt] table[x index=0, y index=1] {OrderedSIMP4M.txt};
  				\addlegendentry{\textbf{O-SIMP}}
  				\addplot[densely dashed, {magenta}, line width=1pt] table[x index=0, y index=1] {AAP3MMA4M.txt};
  				\addlegendentry{\textbf{AAP}}
  			\end{axis}
  		\end{tikzpicture}
  		\caption{Four Materials, $m=4$}
  	\end{subfigure}
  	
  	\begin{subfigure}[t]{0.40\textwidth}
  		\centering
  		\begin{tikzpicture}[scale=1]
  			\begin{axis}[
  				xlabel={Iteration},
  				ylabel={Objective function $(f_0)$},
  				xmin=0,xmax=400,
  				grid=major,
  				ymin=0, ymax=10,
  				restrict y to domain=0:10
  				]    
  				\addplot[solid,{cyan}, line width=1pt] table[x index=0, y index=1] {MBB2D5Mbeta75graph.txt}; \addlegendentry{\textbf{gSF}}
  				\addplot[dashed,{red}, line width=1pt, mark = none] table[x index=0, y index=1] {ExSIMP5M.txt};
  				\addlegendentry{\textbf{Ex-SIMP}}
  				\addplot[dotted,{green}, line width=1pt] table[x index=0, y index=1] {DMO5M.txt};
  				\addlegendentry{\textbf{DMO}}
  				\addplot[dashdotted,{blue}, line width=1pt] table[x index=0, y index=1] {OrderedSIMP5M.txt};
  				\addlegendentry{\textbf{O-SIMP}}
  				\addplot[densely dashed, {magenta}, line width=1pt] table[x index=0, y index=1] {AAP3MMA5M.txt};
  				\addlegendentry{\textbf{AAP}}
  			\end{axis}
  		\end{tikzpicture}
  		\caption{Five materials, $m=5$}
  	\end{subfigure}
  	\hspace{2cm}
  	\begin{subfigure}[t]{0.40\textwidth}
  		\centering
  		\begin{tikzpicture}[scale=1]
  			\begin{axis}[
  				xlabel={Iteration},
  				ylabel={Objective function $(f_0)$},
  				xmin=0,xmax=400,
  				grid=major,
  				ymin=0, ymax=10,
  				restrict y to domain=0:10
  				]    
  				\addplot[solid,{cyan}, line width=1pt] table[x index=0, y index=1] {MBB2D6Mbeta75graph.txt}; \addlegendentry{\textbf{gSF}}
  				\addplot[dashed,{red}, line width=1pt, mark = none] table[x index=0, y index=1] {ExSIMP6M.txt};
  				\addlegendentry{\textbf{Ex-SIMP}}
  				\addplot[dotted,{green}, line width=1pt] table[x index=0, y index=1] {DMO6M.txt};
  				\addlegendentry{\textbf{DMO}}
  				\addplot[dashdotted,{blue}, line width=1pt] table[x index=0, y index=1] {OrderedSIMP6M.txt};
  				\addlegendentry{\textbf{O-SIMP}}
  				\addplot[densely dashed, {magenta}, line width=1pt] table[x index=0, y index=1] {AAP3MMA6M.txt};
  				\addlegendentry{\textbf{AAP}}
  			\end{axis}
  		\end{tikzpicture}
  		\caption{Six materials, $m=6$}
  	\end{subfigure}
  	\caption{Convergence plots for the optimized results with $m =1\cdots6$ depicted in Table~\ref{Tab:ComStudDiffMethods} for gSF, Ex-SIMP, DMO, O-SIMP, and AAP approaches.} \label{Fig:ComConverPlot}
  \end{figure}
  
  By and large, the final topologies provided by these methods differ. In some cases, similarities are noted between a subset of approaches, for example, for $m=1$, final topologies provided by all methods are similar, for $m=2$, gSF, Ex-SIMP, and DMO provide similar topologies, and for $m=4,\,5,\,6$ the gSF and Ex-SIMP have the same final topologies (Table~\ref{Tab:ComStudDiffMethods}). A similar trend is observed in the location allocation for different materials in the optimized designs. The ordered SIMP does not always utilize all candidate materials in the final design, as it relies on a single volume constraint (Table~\ref{Tab:ComStudDiffMethods}). In contrast, the other methods enforce the specified volume fractions for each distinct material. This indicates that applying material-specific volume constraints encourages the optimizer to utilize the available material set more effectively. Using different volume constraints for different materials may give the user more control. In terms of objective value, the gSF demonstrates competitive or improved performance, indicating the effective exploration of the design space even with a comparatively lower number of design variables. Likewise, the gSF approach promotes a higher, or more competitive, discreteness level for the optimized designs, indicating that the obtained solutions are close to 0-1 (Table~\ref{Tab:ComStudDiffMethods}). The objective convergence plots are depicted in Fig.~\ref{Fig:ComConverPlot}, the gSF convergence is competitive and well. In the case of the ordered SIMP, we observe that the objective values rise very high; however, after a few iterations, they decrease and converge well. Overall, the study indicates the efficiency, scalability, success, and robustness of the proposed approach, particularly for problems involving a large number of material phases with only a few design variables.
  The gSF method consistently attains objective values that match or slightly improve upon those of full-space formulations and substantially outperform the ordered SIMP and AAP approaches, resulting in structures with lower compliance. This consistently confirms that the ``missing" interior volume of the full simplex contains only redundant (Fig.~\ref{Fig:AppScheSimplex}), non-optimal material mixtures rather than superior physical solutions.
  
  \subsection{Computational cost comparison}
  To demonstrate the computational efficiency of the proposed gSF approach, we compare its performance with the conventional extended SIMP method for the MBB beam problem (Fig.~\ref{MBB_Beam_2D}) for candidate material $m$ from 2 to 7. A 64-bit operating system machine with 32.0 GB RAM and 13th Gen Intel(R) Core(TM) i7-13700K (3.40 GHz) is used. The MBB design domain is discretized by $180\times 60$ FEs.
  \begin{table}[h]
  	\caption{Cost comparison table between the extended SIMP and the proposed gSF approach is presented. The MBB design domain is discretized by $180\times 60$ FEs. t\textsuperscript{Ex-SIMP} and t\textsuperscript{gSF} represent the time taken by the system for the first 10 iterations using the extended SIMP and proposed gSF approach, respectively. $m$ denotes the number of materials considered for optimization. A 64-bit operating system machine with 32.0 GB RAM and 13th Gen Intel(R) Core(TM) i7-13700K (3.40 GHz) is used.}
  	\label{CC_comparison}
  	\centering
  	\begin{tabular}{|C{3.0cm}|C{2.75cm}|C{2.5cm}|C{4.0cm}|}
  		\hline
  		$m$ & t\textsuperscript{Ex-SIMP} (sec) & t\textsuperscript{gSF} (sec) & t\textsuperscript{Ex-SIMP}/t\textsuperscript{gSF} \\ \hline
  		2 & 1.22 & 1.16 & 1.151 \\ \hline
  		3 & 1.56 & 1.06 & 1.472 \\ \hline
  		4 & 2.08 & 1.37 & 1.518 \\ \hline
  		5 & 3.46 & 1.66 & 2.084 \\ \hline
  		6 & 4.12 & 1.56 & 2.641 \\ \hline
  		7 & 5.17 & 1.63 & 3.172 \\ \hline
  	\end{tabular}
  \end{table}
  
  Table~\ref{CC_comparison} shows comparison in computational time requirements. For the extended SIMP, as $m$ increases, computational time quickly increases. This occurs because the extended SIMP requires a separate design variable for each material phase per element, thereby increasing the size of the total design vector linearly. Thus, a high $m$ means more computation time for the extended SIMP. In contrast, the computational time required by the gSF approach remains flat and stable as the number of materials increases. The gSF approach uses a clever hypercube mapping that requires far fewer design variables to represent the same number of materials. At $m=7$, the gSF method runs more than 3 times faster than the extended SIMP method.
  
  \subsection{Differences between the gSF approach and SFP or Wachspress-type interpolation method}
  For 2D and 3D material interpolation elements (domains), the SFP and gSF use the same bilinear and trilinear interpolation shape functions. However, the latter is also equipped with density filtering and a modified projection filter, while retaining the barycentric characteristics of the design variables, i.e., the filtered and projected variables preserve barycentricity. Additionally, we generalize the classical shape functions to $n$-dimensional shape functions (material elements), thereby accommodating a wider range of materials ($2^n-1$), as demonstrated in the results and discussion section. Note that both the SFP and gSF have orthogonal material interpolation domains and, consequently, are bounded by the limit constraints on the design variables, which are relatively easy to handle in a TO setting. We also note that the material manifold $\mathcal{M}$ of the proposed approach is lower-dimensional than the classical material simplex; however, the gSF approach facilitates physical design variables to reach close to the vertices of the simplex for the pure material states, i.e., to achieve solutions close to 0-1.
  
  A convex polytope-based material interpolation scheme, such as Wachspress shape functions-based, is not limited to the orthogonal interpolation domain, though it can permit any number of candidate materials as the proposed gSF can; however, due to a non-orthogonal interpolation domain, many linear constraints are required to keep the design variables within the polytope apart from the resource or other physical constraints. Therefore, as the dimension of the polytope increases, the number of constraints also increases, which may not be a good sign for the optimization problem's convergence. Readers can refer to the paper \cite{cherriere2022multi} for a detailed comparative study of the SFP and Wachspress-type interpolation formulations.
  
  \subsection{Differences between the gSF approach and the colour level set method}
  The colour level-set approach is proposed within the level-set framework by Wang and Wang~\cite{wang2004color} and is primarily applied to compliance minimization problems involving up to three materials. In contrast, the proposed generalized shape function (gSF) approach adopts a density-based formulation. It solves both compliance minimization and compliant mechanism problems in 2D and 3D, accommodating up to 24 distinct materials. Thus, the underlying framework and working principles of the two approaches are fundamentally different.
  
  The colour level-set uses $n$ level-set functions to implicitly represent up to $2^n$ material phases (Sec. 2.3~\cite{wang2004color}). In contrast, the proposed approach utilizes the natural coordinates of an $n|_{1,\,2,\,3,\cdots,\,n}-$dimensional material element to optimize designs with up to $2^n$ material combinations. Furthermore, the former requires a careful coordination between the rectilinear grid spacing (to ensure level-set accuracy) and the finite element mesh resolution (Sec. 7.1~\cite{wang2004color}). In contrast, the proposed approach decouples the material interpolation from the finite element discretization, permitting the mesh resolution to remain independent of the material elements. The current paper demonstrates this flexibility through 2D examples employing 1D to 5D material interpolating elements successfully in Sec.~\ref{sec4}.
  
  The level-set method requires the careful approximation of Dirac delta and Heaviside functions in numerical implementation and accuracy, introducing additional numerical parameters, e.g., the bandwidth $\xi$ (Sec. 7.3~\cite{wang2004color}). The proposed approach, on the other hand, uses the standard paradigm in which the optimized designs depend on parameters related to the modified SIMP method, the filter radius, and the projection parameter $\beta$ to achieve solutions near 0-1 with many materials.
  
  The level-set approach requires the time step to satisfy the Courant-Friedrichs-Lewy condition for stability (Sec. 7.3~\cite{wang2004color}). No such restrictions are required for the stability of the proposed approach.
  
\end{document}